\documentclass[acmsmall,nonacm]{acmart}
\usepackage{subcaption}
\usepackage{multirow}
\usepackage{amsmath}
\usepackage{xcolor}
\usepackage{colortbl}
\usepackage{adjustbox}
\usepackage{fancyhdr}
\usepackage{setspace}
\usepackage{array,multirow}
\usepackage{makecell, tabularx}
\usepackage{color,soul}
\usepackage[most]{tcolorbox}
\usepackage{rotating, graphicx}
\usepackage{algpseudocode}
\usepackage{titlesec}
\usepackage{enumitem}

\setlist[itemize]{noitemsep, topsep=0pt}
\setlist[enumerate]{noitemsep, topsep=0pt}

\tcbset{
    left=8pt,
    right=8pt,
    enhanced,
    halign=center,
    colback=gray!16,
    colframe=black,
    boxrule=0.8pt,
    width=\dimexpr\textwidth\relax,
}

\usepackage{color}
\definecolor{darkred}{rgb}{0.55, 0.0, 0.0}
\definecolor{codegray}{rgb}{0.5,0.5,0.5}
\definecolor{mygreen}{RGB}{68,85,37}

\AtBeginDocument{%
  \providecommand\BibTeX{{%
    \normalfont B\kern-0.5em{\scshape i\kern-0.25em b}\kern-0.8em\TeX}}}

\setcopyright{acmcopyright}
\copyrightyear{2022}
\acmYear{2022}
\acmDOI{}

\acmJournal{POMACS}
\acmVolume{6}
\acmNumber{1}
\acmArticle{}
\acmMonth{2}

\newcommand{\SparseP}{\emph{SparseP}}
\newcommand{\spmv}{SpMV}
\newcommand{\equallySized}{\emph{equally-sized}}
\newcommand{\equallyWidth}{\emph{equally-wide}}
\newcommand{\variableSized}{\emph{variable-sized}}
\newcommand{\dpuActive}{2528 DPUs}

\newcommand\arxiv[1]{\noindent{\color{purple}{#1}}} 
\newcommand{\jgl}[1]{\textcolor{teal}{}}
\newcommand{\christina}[1]{\textcolor{blue}{}}

\newcommand\camone[1]{\noindent{\color{black}{#1}}} 
\newcommand\camtwo[1]{\noindent{\color{black}{#1}}} 
\begin{document}

\title{\SparseP: Towards Efficient Sparse Matrix Vector Multiplication on Real Processing-In-Memory Systems}

\author{Christina Giannoula}
\email{christina.giann@gmail.com}
\affiliation{%
  \institution{ETH Z{\"u}rich}
  \country{Switzerland}
}  
\affiliation{%
  \institution{National Technical University of Athens}
  \country{Greece}
}

\author{Ivan Fernandez}
\affiliation{%
  \institution{ETH Z{\"u}rich}
  \country{Switzerland}
}
\affiliation{%
  \institution{University of Malaga}
  \country{Spain}
}

\author{Juan Gómez-Luna}
\affiliation{%
  \institution{ETH Z{\"u}rich}
  \country{Switzerland}
}

\author{Nectarios Koziris}
\affiliation{%
  \institution{National Technical University of Athens}
  \country{Greece}
}

\author{Georgios Goumas}
\affiliation{%
  \institution{National Technical University of Athens}
  \country{Greece}
}

\author{Onur Mutlu}
\affiliation{%
  \institution{ETH Z{\"u}rich}
  \country{Switzerland}
}

\renewcommand{\shortauthors}{Christina Giannoula, et al.}


\begin{abstract}

Several manufacturers have already started to commercialize near-bank Processing-In-Memory (PIM) architectures, after decades of research efforts. Near-bank PIM architectures place simple cores close to DRAM banks. Recent research demonstrates that they can yield significant performance and energy improvements in parallel applications by alleviating data access costs. Real PIM systems can provide high levels of parallelism, large aggregate memory bandwidth and low memory access latency, thereby being a good fit to accelerate the Sparse Matrix Vector Multiplication (\spmv{}) kernel. \spmv{} has been characterized as one of the most significant and thoroughly
studied scientific \camone{computation} kernels. \camone{It} is primarily a memory-bound kernel \camone{with intensive} memory accesses due its algorithmic nature, the compressed matrix format used, and the sparsity patterns of the input matrices given.

This paper provides the first comprehensive analysis of \spmv{} on a real-world PIM architecture, and presents \SparseP{}, the first \spmv{} library for real PIM architectures. We make three key contributions. First, we implement a wide variety of software strategies on \spmv{} for a multithreaded PIM core, including (1) various compressed matrix formats, (2) load balancing schemes \camone{across} parallel threads and (3) synchronization approaches, and characterize the computational limits of a single multithreaded PIM core. Second, we design various load balancing schemes \camone{across} multiple PIM cores, and two types of data partitioning techniques to execute \spmv{} on \camone{thousands of} PIM cores: (1) \camone{1D-partitioned} kernels to perform the complete \spmv{} computation only using PIM cores, and (2) \camone{2D-partitioned} kernels to strive a balance between computation and data transfer costs to PIM-enabled memory. \camone{Third}, we compare \spmv{} execution on a real-world PIM system with 2528 PIM cores to an Intel Xeon CPU and an NVIDIA Tesla V100 GPU to study the performance and energy efficiency of various devices, i.e., both memory-centric PIM systems and conventional processor-centric CPU/GPU systems, for the \spmv{} kernel. \SparseP{} software package provides 25 \spmv{} kernels for real PIM systems supporting the four most widely used compressed matrix formats, i.e., CSR, COO, BCSR and BCOO, and a wide range of data types. \SparseP{} is publicly and freely available at \url{https://github.com/CMU-SAFARI/SparseP}. Our extensive evaluation using 26 matrices with various sparsity patterns provides new insights and recommendations for software designers and hardware architects to efficiently accelerate the \spmv{} kernel on real PIM systems.

\end{abstract}


\maketitle

\section{Introduction} 

Sparse Matrix Vector Multiplication (\spmv) is a fundamental linear algebra kernel for important applications from the scientific computing, machine learning, and graph analytics domains. In commodity systems, it has been repeatedly reported to achieve only a small fraction of the peak performance~\cite{Elafrou2018SparseX,Shengen2014YaSpMV,Liu2018Towards,Elafrou2017PerformanceAA,Goumas2009Performance,Karakasis2009Performance,Vuduc2005Fast,im2004sparsity,Vuduc2003PhD,Vuduc2002Performance,Goumas2008Understanding,Elafrou2017PerformanceXeon} due to its algorithmic nature, the employed compressed matrix storage format, and the sparsity pattern of the input matrix. \spmv{} performs indirect memory references as a result of storing the matrix in a compressed format, and irregular memory accesses to the input vector due to sparsity. The matrices involved are very sparse, i.e., the vast majority of elements are zeros~\cite{Kanellopoulos2019SMASH,Elafrou2018SparseX,Elafrou2017PerformanceAA,YouTubeGraph,FacebookGraph,Goumas2008Understanding,White97Improving,Helal2021ALTO,Pelt2014Medium}. For example, the matrices that represent Facebook’s and YouTube’s network connectivity contain 0.0003\%~\cite{YouTubeGraph,Kanellopoulos2019SMASH} and 2.31\%~\cite{FacebookGraph,Kanellopoulos2019SMASH} non-zero elements, respectively. Therefore, in processor-centric systems, \spmv{} is a \camone{memory-}bandwidth-bound kernel for the majority of \camone{real} sparse matrices, and is bottlenecked by data movement between memory and processors~\cite{Gomez2021Benchmarking,Elafrou2018SparseX,Elafrou2017PerformanceAA,Elafrou2019Conflict,Xie2021SpaceA,Goumas2009Performance,Pal2018OuterSpace,Gomez2021Analysis,Vuduc2002Performance,Vuduc2003PhD,Vuduc2005Fast,Karakasis2009Performance,dongarra1996sparse,im2004sparsity,Liu2018Towards,Kourtis2011CSX,Goumas2008Understanding,Kourtis2008Optimizing,Elafrou2017PerformanceXeon}.

One promising way to alleviate the data movement bottleneck is the Processing-In-Memory (PIM) paradigm~\cite{Gomez2021Benchmarking,ahn2015scalable,tensordimm,gao2017tetris,Lee2021HardwareAA,fernandez2020natsa,devaux2019,Gao2015Practical,ke2019recnmp,Kwon2021Function,Hadi2016Chameleon,Gomez2021Analysis,Giannoula2021SynCron,mutlu2020modern,Mutlu2019Processing,Ghose2019Workload,Mutlu2019Enabling,Gagandeep2019Near,Oliveira2021Damov,Boroumand2018Google,Zhang2018GraphP,Lockerman2020Livia,bostanci2022drstrange,Kim2019DRange,Alser2020Accelerating,Cali2020GenASM,Kim2017GrimFilter,Gao2016HRL,Farmahini2015NDA,Dai2018GraphH,Nair2015Active,choe2019concurrent,liu2017concurrent,Boroumand2021Google,Nag2021OrderLight,Gu2021DLUX,Aga2019coml,Shin2018MCDRAM,Cho2020MCDRAM,Yazdanbakhsh2018InDRAM,Farmahini2015DRAMA,Alian2019NetDIMM,Kautz1969Cellular,Stone1970Logic,Ahn2015PIMenabled,Hsieh2016TOM,hashemi2016accelerating,Singh2019Napel,Singh2020NEROAN,seshadri2020indram,Olgun2021QuacTrng,Alves2015Opportunities,hashemi2016continuous,huangfu2019medal,Seshadri2017Ambit,Aga2017Compute,Eckert2018Neural,Fujiki2019Duality,Kang2014Energy,Li2016Pinatubo,Seshadri2013RowClone,Angizi2019GraphiDe,Chang2016LISA,Gao2019ComputeDRAM,Xin2020ELP2IM,li2017drisa,Deng2018DrAcc,Hajinazar2021SIMDRAM,Rezaei2020NoM,Wang2020Figaro,Ali2020InMemory,levi2014Loci,Kvatinsky2014Magic,Shafiee2016ISAAC,Kvatinsky2011Memristor,Gaillardon2016Programmable,Bhattacharjee2017ReVAMP,Hamdioui2015Memristor,Xie2015FastBL,Song2018GraphR,Ankit2020Panther,Ankit2019PUMA,Chi2016PRIME,Xi2021Memory,Zheng2016RRAM,Hamdioui2017Memristor,Yu2018Memristive,Kim2018PUF,orosa2021codic,ferreira2021pluto,Sun2021ABCDIMMAT,olgun2021pidram,Wu2021Sieve,Yuan2021FORMS,Khan2020Survey}. PIM moves computation close to application data by equipping memory chips with processing capabilities~\cite{mutlu2020modern,Mutlu2019Enabling}. Prior works~\cite{ahn2015scalable,fernandez2020natsa,Gao2015Practical,gao2017tetris,Giannoula2021SynCron,Boroumand2019Conda,Dai2018GraphH,Nai2017GraphPIM,Youwei2019GraphQ,Lee2008Adaptive,Lockerman2020Livia,choe2019concurrent,liu2017concurrent,boroumand2017lazypim,Drumond2017mondrian,Hsieh2016accelerating,Kim2021Functionality,Huang2019active,Boroumand2018Google,Zhang2018GraphP,Lockerman2020Livia,Gao2016HRL,pugsley2014ndc,Zhang2014TOPPIM,Nair2015Active,Farmahini2015DRAMA} propose PIM architectures wherein a processor logic layer is tightly integrated with DRAM memory layers using 2.5D/3D-stacking technologies~\cite{HMC,HBM,Lee2016Simultaneous}. Nonetheless, the 2.5D/3D integration itself \camone{might not always be able to} provide significantly higher memory bandwidth for processors than standard DRAM~\cite{Lee2021HardwareAA,Hadi2016Chameleon}. To provide \camone{even} higher bandwidth for the \camone{in-memory} processors, \textit{near-bank} PIM designs have been explored~\cite{Lee2021HardwareAA,upmem,Hadi2016Chameleon,devaux2019,Kwon2021Function,Gomez2021Analysis,Gomez2021Benchmarking,Gu2020iPIM,Cho2020Near,Cho2021Accelerating,Kumar2020Parallel,Nag2021OrderLight,Park2021Trim,Sadredini2021Sunder,Gu2021DLUX,Aga2019coml,Shin2018MCDRAM,Cho2020MCDRAM,Yazdanbakhsh2018InDRAM,Alves2015Opportunities,li2017drisa}. \textit{Near-bank} PIM designs tightly couple a PIM core with each DRAM bank, exploiting bank-level parallelism to expose high on-chip memory bandwidth of standard DRAM to processors. Moreover, manufacturers of near-bank PIM architectures avoid disturbing the key components (i.e., subarray and bank) of commodity DRAM to provide a cost-efficient and practical way for silicon materialization. Two \textit{real} near-bank PIM architectures are Samsung's FIMDRAM~\cite{Lee2021HardwareAA,Kwon2021Function} and \camtwo{the UPMEM PIM system~\cite{upmem2018,devaux2019,Gomez2021Analysis,Gomez2021Benchmarking}.}

Most near-bank PIM architectures~\cite{Lee2021HardwareAA,upmem,Hadi2016Chameleon,devaux2019,Kwon2021Function,Gomez2021Analysis,Gomez2021Benchmarking,Gu2020iPIM,Cho2020Near,Cho2021Accelerating,Kumar2020Parallel,Nag2021OrderLight,Park2021Trim} support several PIM-enabled memory chips connected to a host CPU via memory channels. Each memory chip \camone{comprises} multiple PIM cores, which are low-area and low-power cores with relatively low \camone{computation} capability~\cite{Gomez2021Benchmarking,Gomez2021Analysis}, and each of them is located close to a DRAM bank~\cite{Lee2021HardwareAA,upmem,Hadi2016Chameleon,devaux2019,Kwon2021Function,Gomez2021Analysis,Gomez2021Benchmarking,Gu2020iPIM,Cho2020Near,Cho2021Accelerating,Kumar2020Parallel,Nag2021OrderLight,Park2021Trim}. \camone{Each PIM core} can access data located on their local DRAM banks, and typically there is no direct communication channel among PIM cores. Overall, near-bank PIM architectures provide high levels of parallelism and \camone{very large} memory bandwidth, thereby being a very promising computing platform to accelerate memory-bound kernels. Recent works leverage near-bank PIM architectures to provide high performance and energy benefits on bioinformatics~\cite{lavenier2020Variant,Gomez2021Benchmarking,Gomez2021Analysis,Lavenier2016DNA}, skyline
computation~\cite{Zois2018Massively}, compression~\cite{Nider2020Processing} and neural network~\cite{Lee2021HardwareAA,Gomez2021Benchmarking,Gomez2021Analysis,Gu2020iPIM,Cho2021Accelerating} kernels. \camtwo{A recent study~\cite{Gomez2021Analysis,Gomez2021Benchmarking} provides PrIM benchmarks~\cite{PrIMLibrary}, which are a collection of 16 kernels for evaluating near-bank PIM architectures, like the UPMEM PIM system.} However, there is \emph{no} prior work to thoroughly study the widely used, memory-bound \spmv{} kernel on a real PIM system.

Our work is the first to efficiently map the \spmv{} execution kernel on near-bank PIM systems, and understand its performance implications on a real PIM system. Specifically, our \textbf{goal} in this work is twofold: (i) design efficient \spmv{} algorithms to accelerate this kernel in current and future PIM systems, while covering a wide variety of sparse matrices with diverse sparsity patterns, and (ii) provide an extensive characterization analysis of the widely used \spmv{} kernel on a real PIM architecture. To this end, we provide a wide variety of \spmv{} implementations for real PIM architectures, and conduct a rigorous experimental analysis of \spmv{} kernels in \camone{the} UPMEM PIM system, the first publicly-available real-world PIM architecture.

We present \camone{the} \SparseP{} library~\cite{SparsePLibrary} that includes 25 \spmv{} kernels for real PIM systems, supporting various (1) data types, (2) data partitioning techniques of the sparse matrix to PIM-enabled memory, (3) compressed matrix formats, (4) load balancing schemes across PIM cores, (5) load balancing schemes across threads of a multithreaded PIM core, and (6) synchronization approaches among threads within PIM core. We support a wide range of data types, i.e., 8-bit integer, 16-bit integer, 32-bit integer, 64-bit integer, 32-bit float and 64-bit float data types to cover a \camone{wide} variety of real-world applications that employ \spmv{} as their underlying kernel. We design two types of well-crafted data partitioning techniques: (i) the 1D partitioning technique to perform the complete \spmv{} computation only using PIM cores, and (ii) the 2D partitioning technique to strive a balance between computation and data transfer costs to PIM-enabled memory. In the 1D partitioning technique, the matrix is horizontally partitioned across PIM cores, and the \textit{whole} input vector is copied \camone{into} the DRAM bank of \textit{each} PIM core, while PIM cores directly compute the elements of the final output vector. In the 2D partitioning technique, the matrix is split in 2D tiles, the number of which is equal to the number of PIM cores, and a \textit{subset} of the elements of the input vector is copied \camtwo{into} the DRAM bank of each PIM core. However, \camone{in the 2D partitioning technique,} PIM cores create a large number of partial results for the elements of the output vector which are gathered and merged by the host CPU cores to assemble the final output vector. We support the most popular compressed matrix formats, i.e., CSR~\cite{bjorck1996numerical,Pooch1973Survey},  COO~\cite{Pooch1973Survey,Shubhabrata2007Scan}, BCSR~\cite{Im1999Optimizing}, BCOO~\cite{Pooch1973Survey}, and for each compressed format we implement various load balancing schemes \camone{across} PIM cores to provide efficient \spmv{} execution for a wide variety of sparse matrices with diverse sparsity patterns. Finally, we design several load balancing schemes and synchronization approaches among parallel threads within a PIM core to cover \camone{a variety of} real PIM systems that provide multithreaded PIM cores.

We conduct an extensive characterization analysis of \SparseP{} kernels on \camone{the} UPMEM PIM system~\cite{upmem,Gomez2021Analysis,Gomez2021Benchmarking,devaux2019} analyzing the \spmv{} execution using (1) one single multithreaded PIM core, (2) thousands of PIM cores, and (3) comparing it with that achieved on conventional processor-centric CPU and GPU systems. First, we characterize the limits of a single multithreaded PIM core, and show that (i) high operation imbalance \camone{across} threads of a PIM core can impose high overhead in the core pipeline, and (ii) fine-grained synchronization approaches to increase parallelism cannot outperform a coarse-grained approach, if PIM hardware serializes accesses to the local DRAM bank. Second, we analyze the end-to-end \spmv{} execution of 1D and 2D partitioning techniques using \camone{thousands of} PIM cores. Our study indicates that the performance (i) of the 1D partitioning technique is limited by data transfer costs to \textit{broadcast} the whole input vector \camone{into} \textit{each} DRAM bank of PIM cores, and (ii) of the 2D partitioning technique \camone{is limited} by data transfer costs to \textit{gather} partial results for the elements of the output vector from PIM-enabled memory to the host CPU. Such data transfers incur high overheads, because they take place via the narrow memory bus. In addition, our detailed study across a wide variety of compressed matrix formats and sparse matrices with diverse sparsity patterns demonstrates that (i) the compressed matrix format determines the data partitioning \camone{strategy} across DRAM banks of PIM-enabled memory, thereby affecting the computation balance \camone{across} PIM cores with corresponding performance implications, and (ii) there is \textit{no \camone{one-size-fits-all}} solution. The load balancing scheme \camone{across} PIM cores (and \camone{across} threads within a PIM core) and data partitioning technique that provides the best-performing \spmv{} execution depends on the characteristics of the input matrix and the underlying PIM hardware. Finally, we compare the \spmv{} execution on \camone{a state-of-the-art} UPMEM PIM system with 2528 PIM cores to state-of-the-art CPU and GPU systems, and observe that \spmv{} on the UPMEM PIM system achieves \camone{a much higher fraction of \camtwo{the} machine's peak performance compared to that \camtwo{on the state-of-the-art} CPU and GPU systems}. Our extensive evaluation provides programming recommendations for software designers, and suggestions and hints for hardware and system designers of future PIM systems.

Our most significant recommendations for \camone{PIM} software designers \camone{are}:
\begin{enumerate}
    \item Design algorithms that provide high \camone{load} balance \camone{across} threads of PIM core in terms of computations, loop control iterations, synchronization points and memory accesses.
    \item Design compressed data structures that can be effectively partitioned across DRAM banks, \camone{with the goal of providing} high computation balance \camone{across} PIM cores.
    \item Design \textit{adaptive} algorithms that \camone{trade off computation balance across PIM cores for lower data transfer costs to PIM-enabled memory}, and adapt their configuration to the particular patterns of each input given, as well as the characteristics of the PIM hardware.
\end{enumerate}

Our most significant suggestions for \camone{PIM} hardware and system designers \camone{are}:
\begin{enumerate}
    \item Provide low-cost synchronization support and \camone{hardware support to enable concurrent memory accesses by multiple threads to the local DRAM bank} to increase parallelism in a multithreaded PIM core.
    \item Optimize the broadcast collective operation in data transfers from main memory to PIM-enabled memory to minimize overheads of copying the input data \camone{into} all DRAM banks \camtwo{in the PIM} system.
    \item Optimize the gather collective operation \textit{at DRAM bank granularity} for data transfers from PIM-enabled memory to \camtwo{the} host CPU to minimize overheads of retrieving the output results.
    \item Design high-speed communication channels and optimized libraries for data transfers to/from thousands of DRAM banks of PIM-enabled memory.
\end{enumerate}

Our \SparseP{} software package is freely and publicly available~\cite{SparsePLibrary} to enable further research on \spmv{} in current and future PIM systems. The main contributions of this work are as follows:
\begin{itemize}
    \item We present \SparseP{}, the first open-source \spmv{} software package for real PIM architectures. \SparseP{} includes 25 \spmv{} kernels, supporting the four most widely used compressed matrix formats and a wide range of data types. \SparseP{} is publicly available at~\cite{SparsePLibrary}, and can be useful for researchers to improve multiple aspects of future PIM hardware and software. 
    \item We perform the first comprehensive study of the widely used \spmv{} \camtwo{kernel} on \camone{the} UPMEM PIM architecture, the first real \camone{commercial} PIM architecture. We analyze performance implications of \spmv{} PIM execution using a wide variety of (1) compressed matrix formats, (2) data types, (3) data partitioning and load balancing techniques, and (4) 26 sparse matrices with diverse sparsity patterns.
    \item We compare the performance and energy of \spmv{} on \camone{the state-of-the-art} UPMEM PIM system with 2528 PIM cores to state-of-the-art CPU and GPU systems. \spmv{} execution achieves less than 1\% of the peak performance on processor-centric CPU and GPU systems, while it achieves on average 51.7\% of the peak performance on the UPMEM PIM system, thus \camone{better} leveraging the computation capabilities of underlying hardware. The UPMEM PIM system also provides high energy efficiency on the \spmv{} kernel.
\end{itemize}

\section{Background and Motivation}

\subsection{Sparse Matrix Vector Multiplication (\spmv{})}

\begin{figure}[H]
    \centering
    \includegraphics[width=0.9\linewidth]{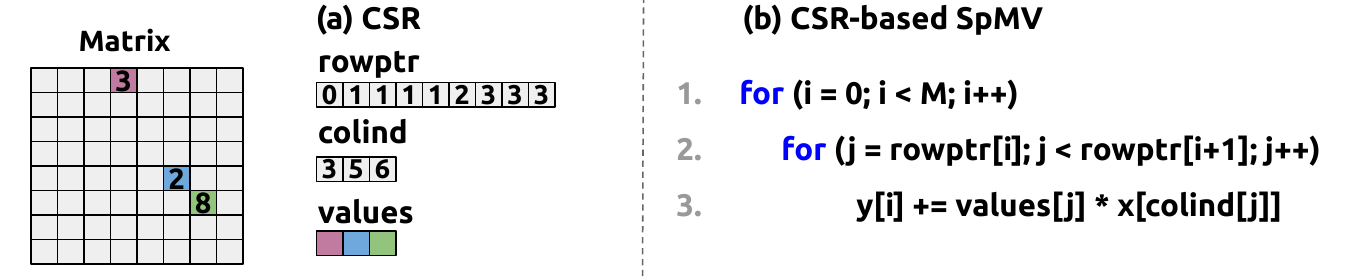}
    \vspace{-5pt}
    \caption{(a) CSR representation of a sparse matrix. (b) CSR-based \spmv{} implementation. }
    \label{fig:csr-spmv}
    \vspace{-4pt}
\end{figure}

The \spmv{} kernel multiples a sparse matrix of size $M\times N$ with a dense input vector of size $1\times N$ to compute \camone{an} output vector of size $M\times 1$. The \spmv{} kernel is widely used in a variety of applications including graph processing~\cite{Brin1998the,besta2017slimsell,Kanellopoulos2019SMASH,Giannoula2018Combining}, neural networks~\cite{liu2015sparse,Zhou2018CambriconS,Han2016EIE,Han2015Learning}, machine learning~\cite{dnn2018,linden2003amazon,recommenderFB2019,recommendfb2,Zhang2016CambriconX,Gupta2020DeepRecSys}, and high performance computing~\cite{solversGPU, Falgout2006an,dongarra1996sparse, falgout2002hypre, henon2002pastix,Cho2020Near,Elafrou2017PerformanceXeon}. These applications involve matrices with
very high sparsity~\cite{Kanellopoulos2019SMASH,Elafrou2018SparseX,Elafrou2017PerformanceAA,YouTubeGraph,FacebookGraph,Goumas2008Understanding,White97Improving,Helal2021ALTO,Pelt2014Medium}, i.e., a large fraction of zero elements. Thus, using a compression scheme is a straightforward approach to avoid unnecessarily storing zero elements and performing computations on them. For general sparse matrices, the most widely used storage format is the Compressed Sparse Row (CSR) format~\cite{bjorck1996numerical,Pooch1973Survey}. Figure~\ref{fig:csr-spmv} presents an example of a compressed matrix using the CSR format (left), and the CSR-based \spmv{} execution (right), assuming an input vector $x$ and an output vector $y$.

\subsubsection{\textbf{Compressed Matrix Storage Formats}} 
\begin{figure}[H]
    \centering
    \includegraphics[width=0.99\linewidth]{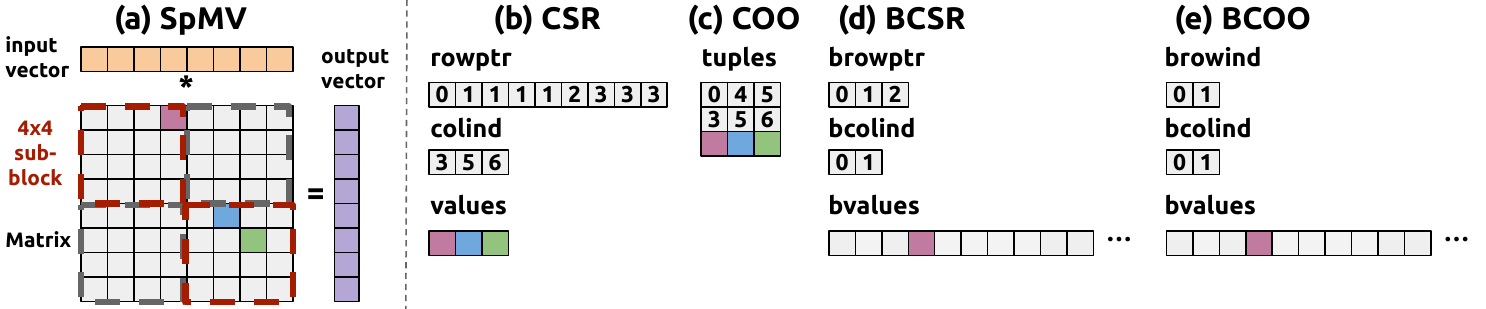}
    \vspace{-2pt}
    \caption{(a) \spmv{} with a dense matrix representation, and (b) CSR, (c) COO, (d) BCSR, (e) BCOO formats.}
    \label{fig:sparse_formats}
\end{figure}

\hfill  \\
Several prior works~\cite{Im1999Optimizing,Langr2016Evaluation,Liu2015CSR5,Pinar1999Improving,Vuduc2005Fast,Yang2014Optimization,Shengen2014YaSpMV,Kourtis2011CSX,Kourtis2008Optimizing,Belgin2009Pattern,LIL,ELL,bjorck1996numerical,Pooch1973Survey,Shubhabrata2007Scan,Changwan2018Efficient,Liu2013Efficient,Monakov2010Automatically,Saad1989Krylov,Buluc2009Parallel,Martone2014251,Martone2010Blas,Kreutzer2012Sparse,Benatia2018BestSF} propose compressed storage formats for sparse matrices, which are typically of two types~\cite{Kanellopoulos2019SMASH}. The first approach is to design general purpose compressed formats, such as CSR~\cite{Pooch1973Survey,bjorck1996numerical}, CSR5~\cite{Liu2015CSR5}, COO~\cite{Pooch1973Survey,Shubhabrata2007Scan}, BCSR~\cite{Im1999Optimizing}, and BCOO~\cite{Pooch1973Survey}. Such encodings are general in applicability and are highly-efficient in storage. The second approach is to leverage a certain known structure in a given type of sparse matrix. For example, the DIA format~\cite{Belgin2009Pattern} is effective in matrices where the non-zero elements are concentrated along the diagonals of the matrix. Such encodings aim to improve performance of sparse matrix computations by specializing to particular matrix patterns, \camone{but they sacrifice} generality. In this work, we explore with the four most widely used \textit{general} compressed formats (Figure~\ref{fig:sparse_formats}), which we describe in more detail next.

\noindent\textbf{Compressed Sparse Row (CSR)~\cite{bjorck1996numerical,Pooch1973Survey}.} The CSR format (Figure~\ref{fig:sparse_formats}b) sequentially stores values in \camtwo{a row-wise} order. \camone{A column index array (\texttt{colind[]}) and a value array (\texttt{values[]}) store the column index and value of each non-zero element, respectively. An array, named \texttt{rowptr[]}, stores the location of the first non-zero element of each row within the \texttt{values[]} array. 
\camtwo{The values of an adjacent pair of the \texttt{rowptr[]} array, i.e., \texttt{rowptr[i, i+1]}, represent a slice of the \texttt{colind[]} and \texttt{values[]} arrays.} 
The corresponding slice of the \texttt{colind[]} and \texttt{values[]} arrays stores the column indices and the values of the non-zero elements, respectively, for the i-th row \camtwo{of the original matrix}.}

\noindent\textbf{Coordinate Format (COO)~\cite{Pooch1973Survey,Shubhabrata2007Scan}.} The COO format (Figure~\ref{fig:sparse_formats}c) stores the non-zero elements as a series of tuples (\texttt{tuples[]} array). Each tuple includes the row index, column index, and value of the non-zero element.

\noindent\textbf{Block Compressed Sparse Row (BCSR)~\cite{Im1999Optimizing}.} The BCSR format (Figure~\ref{fig:sparse_formats}d) is a block representation of CSR. \camone{Instead of storing and indexing single non-zero elements, BCSR stores and indexes \camtwo{$r\times c$ sub-blocks} with at least one non-zero element. The original matrix is split \camtwo{into $r\times c$ sub-blocks.} Figure~\ref{fig:sparse_formats}d shows an example of BCSR assuming $4\times 4$ sub-blocks. The original matrix of Figure~\ref{fig:sparse_formats}a is split \camtwo{into} four sub-blocks, and two of them (highlighted with red color) contain at least one non-zero element. The \texttt{bvalues[]} array stores the values of all the \emph{non-zero sub-blocks} of the original matrix. Each non-zero sub-block is stored in the \texttt{bvalues[]} array with a dense representation, i.e., padding with zero values when needed. The \texttt{bcolind[]} array stores the block-column index of each non-zero sub-block. The \texttt{browptr[]} array stores \camtwo{the location of} the first non-zero sub-block of each block row within the \texttt{bcolind[]} array, assuming \camtwo{a block row represents $r$ consecutive rows of the original matrix, where $r$ is the vertical dimension of the sub-block.}}


\noindent\textbf{Block Coordinate Format (BCOO)~\cite{Pooch1973Survey}.} The BCOO format is the block counterpart of COO. The \texttt{browind[]}, \texttt{bcolind[]} and  \texttt{bvalues[]} arrays store the row indices, column indices and values of the non-zero sub-blocks, respectively. \camone{Figure~\ref{fig:sparse_formats}e shows an example of BCOO, assuming $4\times 4$ sub-blocks.} 

\subsubsection{\textbf{\spmv{} in Processor-Centric Systems}} \hfill  \\
Many prior works~\cite{Elafrou2018SparseX,Shengen2014YaSpMV,Elafrou2017PerformanceAA,Goumas2009Performance,Karakasis2009Performance,Vuduc2005Fast,im2004sparsity,Vuduc2003PhD,Vuduc2002Performance,White97Improving,Elafrou2017PerformanceXeon,Elafrou2019BASMAT} generally \camone{show} that \spmv{} performs poorly on commodity CPU and GPU systems, and achieves a small fraction of the peak performance (e.g., 10\% of the peak performance~\cite{Vuduc2003PhD}) due to its algorithmic nature,  the employed compressed matrix storage format and the sparsity pattern of the matrix. 

The \spmv{} kernel is highly bottlenecked by the memory subsystem in processor-centric CPU and GPU systems \camone{due to} three reasons. First, due to its algorithmic nature there is \textit{no} temporal locality in the input matrix. Unlike  traditional algebra kernels like Matrix Matrix Multiplication or LU decomposition, the elements of the matrix in \spmv{} are used only \textit{once}~\cite{Goumas2009Performance,Goumas2008Understanding}. Second, due to the sparsity of the matrix, the matrix is stored in a compressed \camone{format} (e.g., CSR) to avoid unnecessary computations and data accesses. Specifically, the non-zero elements of the matrix are stored contiguously in memory, while additional data structures assist in the proper traversal of the matrix, i.e., to discover the positions of the non-zero elements. For example, CSR uses the \texttt{rowptr[]} and \texttt{colind[]} arrays to discover the positions of the non-zero elements of the matrix. These additional data structures cause additional memory access operations, memory bandwidth pressure and \camone{contention with other requests in the} memory subsystem. Third, due to the sparsity of the \camone{input} matrix, \spmv{} causes irregular memory accesses to the elements of the input vector $x$. The memory accesses to the elements of the input vector are input driven, i.e., they follow the sparsity pattern of the input matrix. This irregularity results to poor data locality on the elements of the input vector and expensive data accesses, because it increases the \camone{average access} latency due to a high number of cache misses on commodity systems with deep cache hierarchies~\cite{Goumas2009Performance,Goumas2008Understanding}.
As a result, memory-centric near-bank PIM systems constitute a better fit for the widely used \spmv{} kernel, because they provide high levels of parallelism, large aggregate memory bandwidth and low memory access latency~\cite{Gomez2021Benchmarking,Gomez2021Analysis, upmem, Hadi2016Chameleon, Lee2021HardwareAA}.

\subsection{Near-Bank PIM Systems}
\begin{figure}[t]
    \centering
    \includegraphics[width=0.99\linewidth]{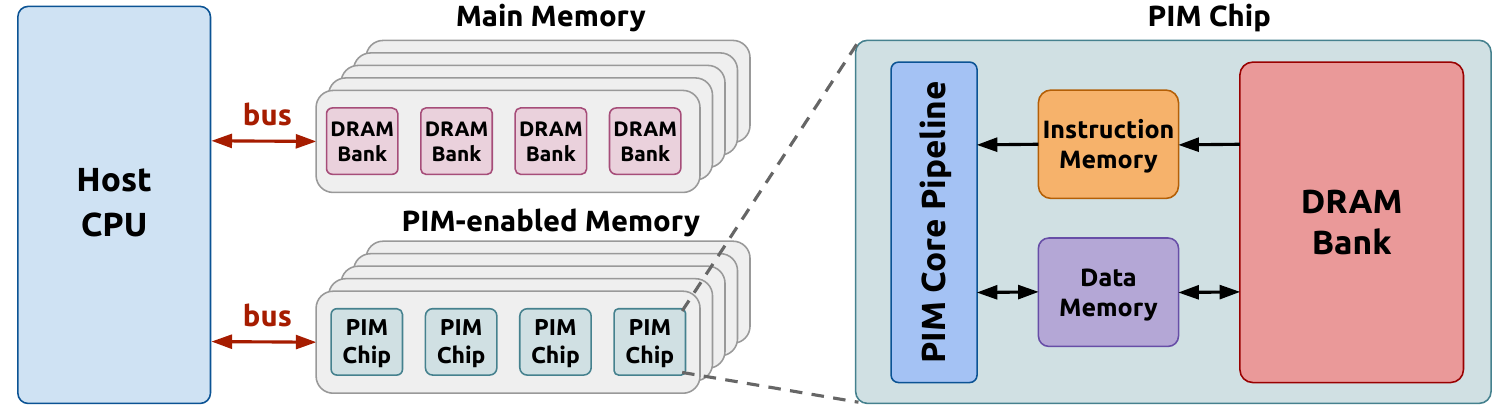}
    \vspace{-4pt}
    \caption{High-level organization of a near-bank PIM architecture.}
    \label{fig:near-bank-pim}
\end{figure}

Figure~\ref{fig:near-bank-pim} shows the baseline organization of a near-bank PIM system that we assume in this work. The PIM system consists of a host CPU, standard DRAM memory modules, and PIM-enabled memory modules. PIM-enabled modules are connected \camone{to the host CPU} using one or more memory channels, and include multiple PIM chips. A PIM chip (Figure~\ref{fig:near-bank-pim} right) tightly integrates a low-area PIM core with a DRAM bank. We assume that each PIM core can additionally include a small private instruction memory and a small data (scratchpad or cache) memory. PIM cores can access data located on their local DRAM bank, and typically there is no direct communication channel among PIM cores. The DRAM banks of PIM chips are accessible by the host CPU for copying input data and retrieving results via the memory bus.

\subsubsection{\textbf{The UPMEM PIM Architecture}} \hfill  \\
The UPMEM PIM system~\cite{Gomez2021Analysis,Gomez2021Benchmarking,devaux2019} includes the host CPU with standard main memory, and UPMEM PIM modules. An UPMEM PIM module is a standard DDR4-2400 DIMM~\cite{ddr4jedec} with 2 ranks. Each rank contains 64 PIM cores, which are called \underline{D}RAM \underline{P}rocessing \underline{U}nits (DPUs). In the current UPMEM PIM system, there are 20 double-rank PIM DIMMs with 2560 DPUs.\footnote{There are thirty two faulty DPUs in the system where we run our experiments. They cannot be used and do not affect the correctness of our results, but take away from the system’s full computational power of 2560 DPUs.}

\noindent\textbf{DPU Architecture and Interface.} Each DPU has exclusive access to a 24-KB instruction memory, \camone{called} \textbf{IRAM}, a 64-KB scratchpad memory, \camone{called} \textbf{WRAM}, and a 64-MB DRAM bank, \camone{called} \textbf{MRAM}. A DPU is a multithreaded in-order 32-bit RISC core that can potentially reach 500 MHz~\cite{upmem}. The DPU has 24 hardware threads, each of \camone{which} has 24 32-bit general purpose registers. The DPU pipeline has 14 stages, and supports a single cycle 8x8-bit multiplier. Multiplications on 64-bit integers, 32-bit floats and 64-bit floats are not supported in hardware, and require longer routines with a \camone{large} number of operations~\cite{Gomez2021Benchmarking,Gomez2021Analysis,upmem}. Threads share the IRAM and WRAM, and can access the MRAM by executing transactions at 64-bit granularity via a DMA engine, i.e., data can be accessed from/to MRAM as a multiple of 8 bytes, up to 2048 bytes. MRAM transactions are serialized in the DMA engine. The ISA provides DMA instructions to move instructions from MRAM to IRAM, or data between MRAM and WRAM. The DPU accesses the WRAM through 8-, 16-, 32- and 64-bit load/store instructions. DPUs use the \textit{Single Program Multiple Data} programming model, where software threads, called \textbf{tasklets}, execute the same code, but operate in different pieces of data, and can execute different control-flow paths during runtime. Tasklets can synchronize using mutexes, barriers, handshakes and semaphores provided by the UPMEM runtime library.

\noindent\textbf{CPU-DPU Data Transfers.} Standard main memory and PIM-enabled memory have different data layouts. The UPMEM SDK~\cite{upmem-guide} has a transposition library to execute necessary data shuffling when moving data between main memory and MRAM banks \camtwo{of PIM-enabled memory modules} via a programmer-transparent way. The CPU-DPU and DPU-CPU data transfers can be performed in parallel, i.e., concurrently across multiple MRAM banks, \camone{with the limitation that} \textit{the transfer sizes from/to all MRAM banks \camone{need to} be the same}. The UPMEM SDK provides two options: (i) perform parallel transfers to all MRAM banks of all ranks, or (ii) iterate over each rank to perform parallel transfers to MRAM banks of the same rank, and serialize data transfers across ranks.


\section{\camtwo{The} \SparseP{} Library}

This section describes the parallelization techniques that we explore for \spmv{} on real PIM architectures, and presents the \spmv{} implementations of our \SparseP{} package. Section~\ref{sec:pim_exec} describes \spmv{} execution on a real PIM system. Section~\ref{sec:lib_overview} presents an overview of the data partitioning techniques that we explore.
Section~\ref{sec:lib_1d-2d} and Section~\ref{sec:lib_1dpu} describe in detail the parallelization techniques \camone{across} PIM cores, and \camone{across} threads within a PIM core, respectively. Section~\ref{sec:kernel_impl} describes the kernel implementation for all compressed \camtwo{matrix storage} formats.

\subsection{\spmv{} Execution on a PIM System}\label{sec:pim_exec}
\begin{figure}[H]
    \centering
    \includegraphics[width=0.99\linewidth]{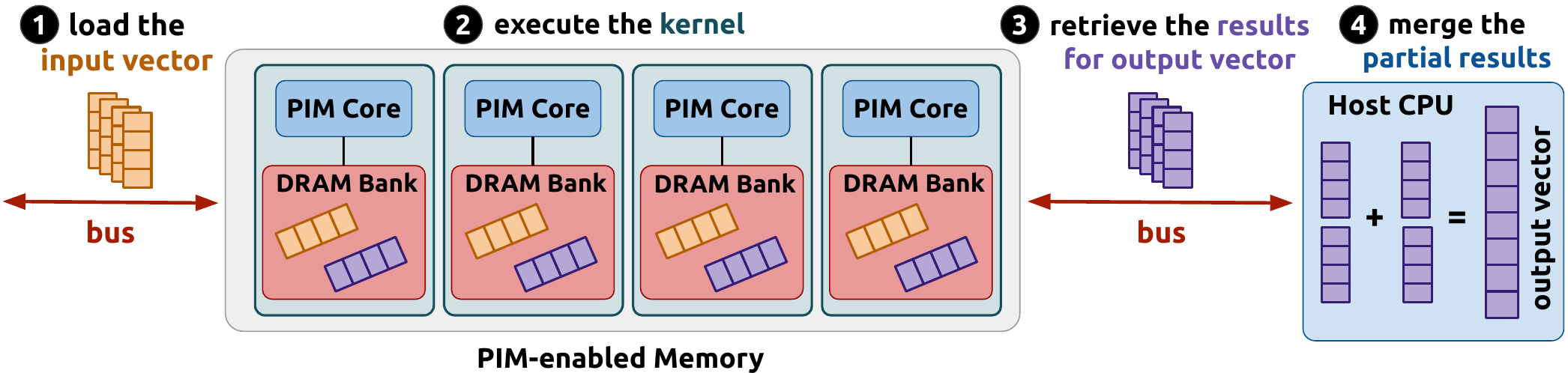}
    \vspace{-6pt}
    \caption{Execution of the \spmv{} kernel on a real PIM system.}
    \label{fig:spmv-pim-execution}
    \vspace{-6pt}
\end{figure}

Figure~\ref{fig:spmv-pim-execution} shows the \spmv{} execution on a \camone{real} PIM system, which is broken down in four steps: (1) the time to load the input vector \camone{into} DRAM banks of PIM-enabled memory (\texttt{\textbf{load}}), (2) the time to execute the \spmv{} \camone{kernel} on PIM cores (\texttt{\textbf{kernel}}), (3) the time to retrieve from DRAM banks to the host CPU results for the output vector (\texttt{\textbf{retrieve}}), and (4) the time to merge partial results and assemble the final output vector on the host CPU (\texttt{\textbf{merge}}). In our analysis, we omit the time to load the matrix \camone{into} PIM-enabled memory, since this step can typically be hidden in real-world applications (it can be overlapped with other computation performed by the application or amortized if the application performs multiple \spmv{} iterations on the same matrix).

\subsection{Overview of Data Partitioning Techniques}\label{sec:lib_overview}
\begin{figure}[H]
    \centering
    \includegraphics[width=0.78\linewidth]{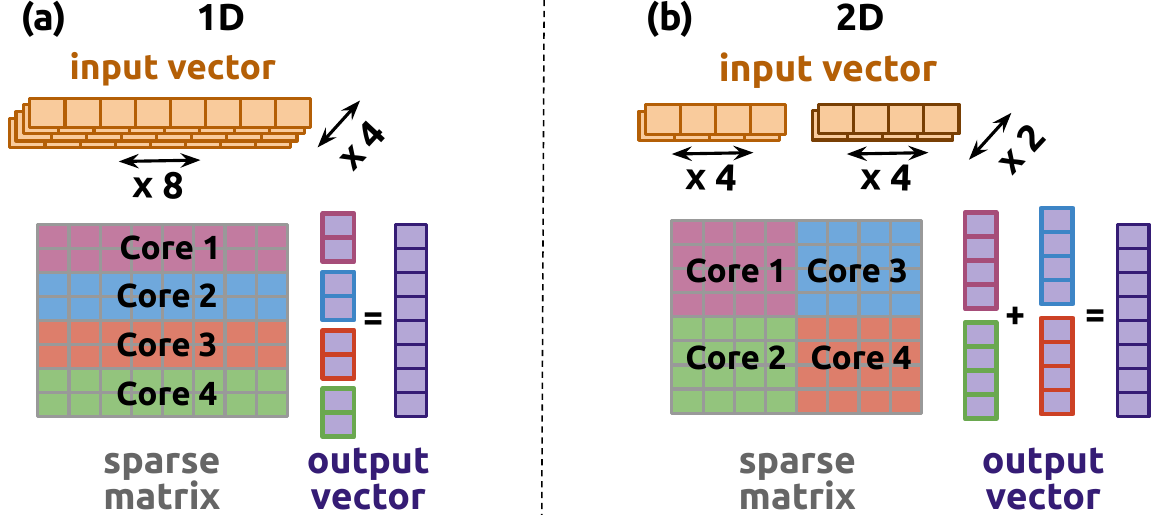}
    \vspace{-4pt}
    \caption{Data partitioning techniques of \camone{the} \SparseP{} package.}
    \label{fig:partitioning_overview}
    \vspace{-6pt}
\end{figure}

To parallelize the \spmv{} kernel, we implement well-crafted data partitioning schemes \camone{to} split the matrix across multiple DRAM banks of PIM cores. \SparseP{} supports two \camone{general} types of data partitioning techniques, shown in Figure~\ref{fig:partitioning_overview}.

First, we provide \camtwo{an} 1D partitioning technique (Figure~\ref{fig:partitioning_overview}a), \camone{where} the matrix is horizontally partitioned across PIM cores, and the whole input vector is copied \camone{into} the DRAM bank of each PIM core. With the 1D partitioning technique, \camone{almost the entire} \spmv{} computation is performed using only PIM cores, since the \texttt{merge} step in the host CPU is negligible: a very small number of partial results \camtwo{is} created, i.e., only for a few rows that are split across neighboring PIM cores. Thus, the number of partial elements of the output vector is at most equal to the number of PIM cores used. Second, we provide a 2D partitioning technique (Figure~\ref{fig:partitioning_overview}b), \camone{where} the matrix is partitioned \camone{into} 2D tiles, the number of which is equal to the number of PIM cores. With the 2D partitioning technique, we aim to strive a balance between computation and data transfer costs, since \camone{only} a subset of the elements of the input vector is copied \camone{into} the DRAM bank of each PIM core. However, PIM cores assigned to tiles \camone{that} horizontally overlap, i.e., tiles that share the same rows of the original matrix (rows that are split across multiple tiles), produce \emph{many} partial results for the elements of the output vector. These partial results are transferred to the host CPU, and merged by CPU cores, which assemble the final output vector. In \camone{the} \SparseP{} library, the \texttt{merge} step performed by the CPU cores is parallelized using the OpenMP API~\cite{Dagum98OpenMP}.

In both data partitioning schemes, matrices are stored in a row-sorted way, i.e., the non-zero elements are sorted in increasing order of their row indices. Therefore, each PIM core computes results for a \textit{continuous} subset of elements of the output vector. \camone{This way we minimize data transfer costs, since we only transfer necessary data to the host CPU, i.e., \textit{the values} of the elements of the output vector produced at PIM cores. If \camtwo{each} PIM core \camtwo{instead} computed results for \camtwo{a} \textit{non-continuous} \camtwo{subset of} elements of the output vector, an additional array \textit{per core}, \camtwo{which} would store \textit{the indices} of the \textit{non-continuous} elements within the output vector, would need to be transferred to the host CPU, causing additional data transfer overheads.}

\subsection{Parallelization Techniques Across PIM Cores}\label{sec:lib_1d-2d}
\camtwo{To parallelize \spmv{} across multiple PIM cores \SparseP{} supports various parallelization schemes for both 1D and 2D partitioning techniques.}

\subsubsection{\textbf{1D Partitioning Technique}} 
\begin{figure}[H]
    
    \centering
    \includegraphics[width=1\linewidth]{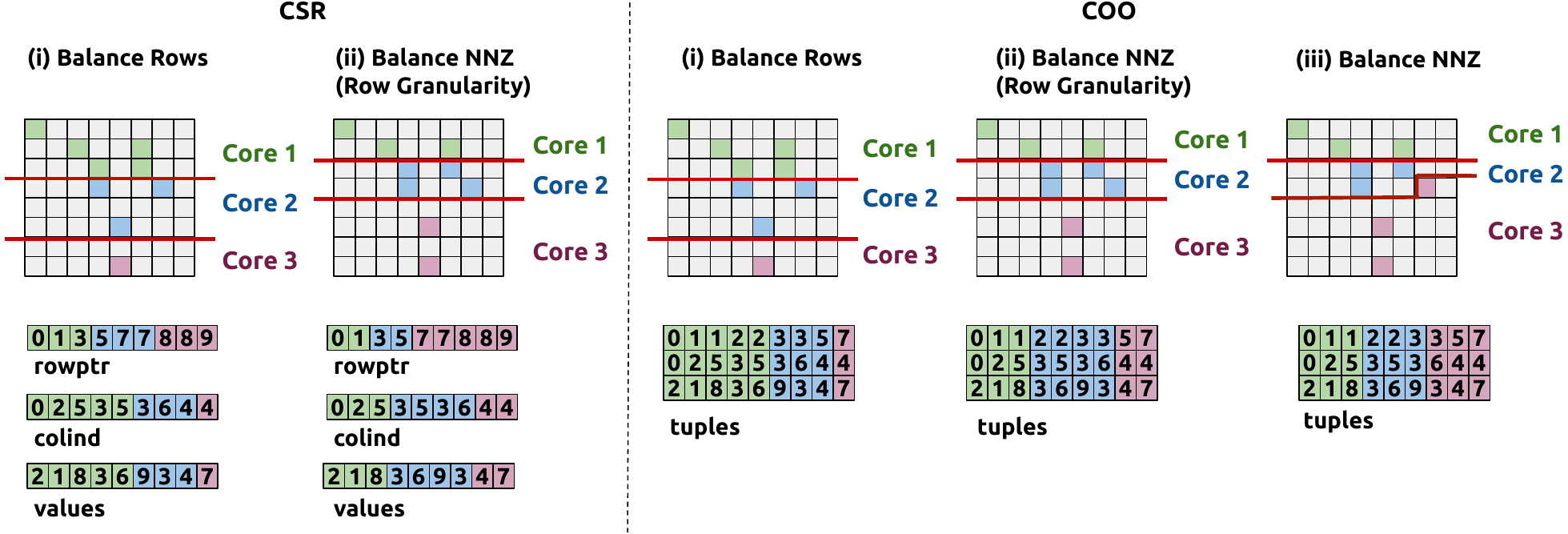}
    \vspace{-18pt}
    \caption{Load balancing schemes \camone{across} PIM cores for the CSR (left) and COO (right) formats with the 1D partitioning technique. The colored cells of the matrix represent non-zero elements.}
    \label{fig:1d_partitioning}
\end{figure}

\hfill  \\
To efficiently parallelize \spmv{} \camone{across} multiple PIM cores via the 1D partitioning technique, \SparseP{} provides various load balancing schemes for each supported compressed \camtwo{matrix} format. Figure~\ref{fig:1d_partitioning} presents an example of parallelizing \spmv{} \camone{across} multiple PIM cores using load balancing schemes \camone{for the} CSR and COO formats. For the CSR and COO formats, we balance either the rows, such that each PIM core processes almost the same number of rows, or the non-zero elements, such that each PIM core processes almost the same number of non-zero elements. In the CSR format, since the matrix is stored in row-order, i.e., the \texttt{rowptr[]} array stores the index pointers of the non-zero elements of \textit{each row}, \camtwo{and thus} balancing the non-zero elements \camone{across} PIM cores is performed at row granularity. In the COO format, the matrix is stored in non-zero order using the \texttt{tuples[]} array, \camtwo{and thus} balancing the non-zero elements can be performed either at row granularity, or by splitting a row across two neighboring PIM cores to provide a \camone{near-perfect} non-zero element balance \camone{across} cores. In the latter case, as mentioned, a small number of partial results for the output vector is merged by the host CPU: if the row is split between two neighboring PIM cores at most one element needs to be accumulated at the host CPU cores.

\begin{figure}[H]
    \vspace{-4pt}
    \centering
    \includegraphics[width=1\linewidth]{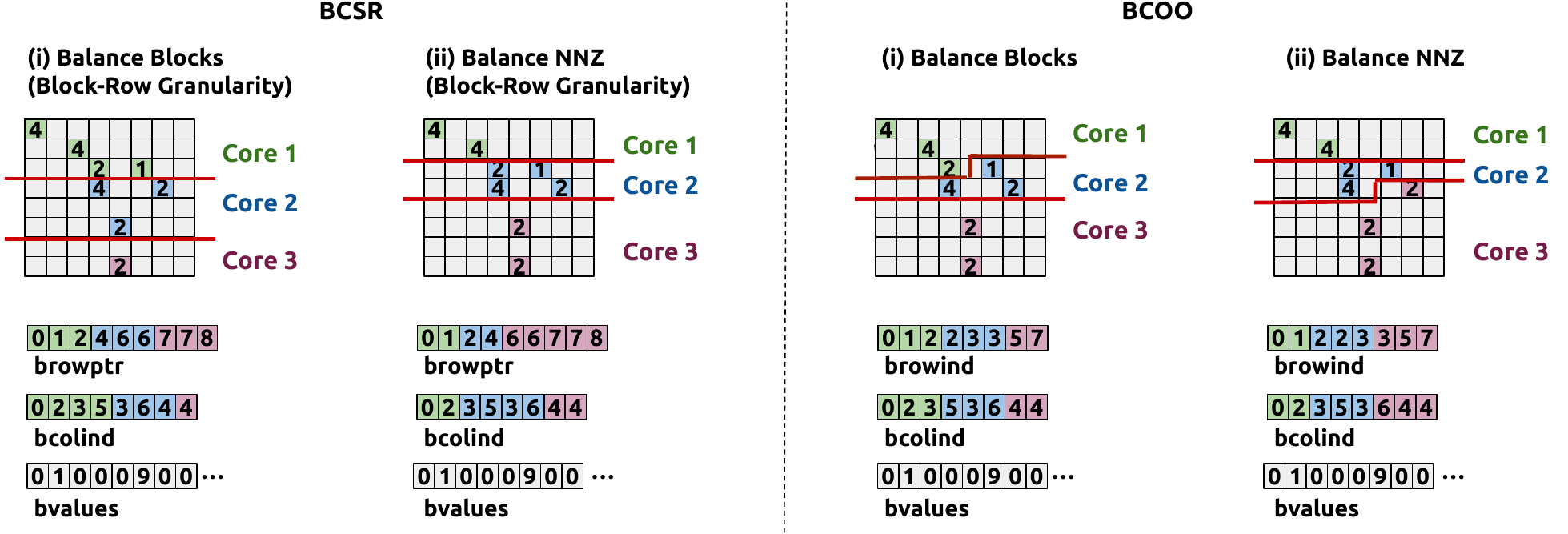}
    \vspace{-18pt}
    \caption{Load balancing schemes \camone{across} PIM cores for the BCSR (left) and BCOO (left) formats with the 1D partitioning technique. The cells of the matrix represent sub-blocks of size 4x4. The colored cells of the matrix represent non-zero sub-blocks, \camone{and the number inside a colored cell describes the number of non-zero elements of the corresponding sub-block.}}
    \label{fig:1d_partitioning2}
    \vspace{-8pt}
\end{figure}

Figure~\ref{fig:1d_partitioning2} presents an example of parallelizing \spmv{} \camone{across} multiple PIM cores using load balancing schemes of the BCSR and BCOO formats. \camone{In Figure~\ref{fig:1d_partitioning2}, the cells of the matrix represent sub-blocks of size 4x4: the \textit{grey} cells represent sub-blocks that do not have \textit{any} non-zero element, and the \textit{colored} cells represent sub-blocks that have \textit{$k$} non-zero elements, where $k$ is the number shown inside the colored cell.} In the BCSR and BCOO formats, since the matrix is stored in sub-blocks of non-zero elements, we balance either the blocks, such that each PIM core processes almost the same number of blocks, or the non-zero elements, such that each PIM core processes almost the same number of non-zero elements. Similarly to CSR, in the BCSR format, the matrix is stored in block-row-order,  i.e., the \texttt{browptr[]} array stores the index pointers of the non-zero blocks of \textit{each block row} (\camtwo{recall that a block row represents $r$ consecutive rows of the original matrix, where $r$ is the vertical dimension of the sub-block}), and thus balancing the blocks or the non-zero elements \camone{across} cores is limited to be performed at block-row granularity. In the BCOO format, given that a block-row might be split across two PIM cores, a small number of partial results for the output vector is merged by the host CPU: between two neighboring PIM cores at most block size \camtwo{$r$ elements ($r$ is the vertical dimension of the block size)} might need to be accumulated at the host CPU cores.

\subsubsection{\textbf{2D Partitioning Technique}} 
\begin{figure}[t]
    \centering
    \includegraphics[width=1\linewidth]{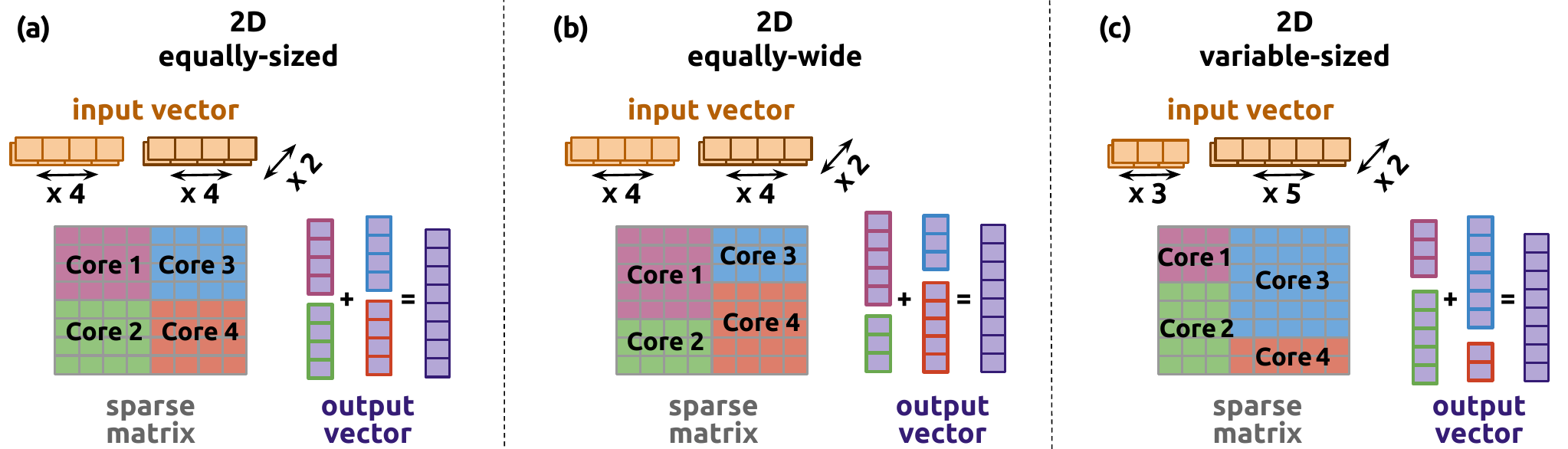}
    \vspace{-16pt}
    \caption{The 2D partitioning techniques of \SparseP{} package \camone{assuming}
    4 PIM cores and 2 vertical partitions.}
    \label{fig:2d_partitioning}
    \vspace{-12pt}
\end{figure}

\hfill  \\
\SparseP{} includes three 2D partitioning techniques, shown in Figure~\ref{fig:2d_partitioning}:
\begin{enumerate}
    \item \textbf{\equallySized} (Figure~\ref{fig:2d_partitioning}a): The 2D tiles are statically created to have the same height and width. This way the subsets of the elements for the input and output vectors have the same sizes \camone{across} all PIM cores.
    \item \textbf{\equallyWidth} (Figure~\ref{fig:2d_partitioning}b): The 2D tiles have the same width and variable height. This way the subset of the elements for the input vector has the same size \camone{across} PIM cores, while the subset of the elements for the output vector varies \camone{across} PIM cores. We balance the non-zero elements \camone{across} the tiles of the \textit{same} vertical partition, such that \camone{we can} provide high non-zero \camtwo{element} balance \camone{across} PIM cores assigned to the same vertical partition.
    \item \textbf{\variableSized} (Figure~\ref{fig:2d_partitioning}c): The 2D tiles have both variable width and height. We balance the non-zero elements both \camone{across} the vertical partitions and \camone{across} the tiles of the \textit{same} vertical partition. This way we can provide high non-zero \camtwo{element} balance \camone{across} all PIM cores.
\end{enumerate}

\SparseP{} provides various load balancing schemes \camone{across} PIM cores in the \equallyWidth{} and \variableSized{} techniques. In the \equallyWidth{} technique, for the CSR and COO formats, we balance the non-zero elements \camone{across} the tiles of the same vertical partition. Load balancing in the CSR format is performed at row-granularity, i.e., splitting the \texttt{rowptr[]} array \camone{across} PIM cores. For the BCSR and BCOO formats, we balance either the blocks or the non-zero elements \camone{across} the tiles of the same vertical partition. Load balancing in the BCSR format is performed at block-row granularity, i.e., splitting the \texttt{browptr[]} array \camone{across} PIM cores. In the \variableSized{} technique, we first balance the non-zero elements \camone{across} the vertical partitions, such that the vertical partitions include the same number of non-zero elements. Then, \camone{across} the tiles of the same vertical partition, we balance the non-zero elements for the CSR (at row-granularity) and COO formats, and either the blocks or the non-zero elements for the BCSR (at block-row granularity) and BCOO formats.

Table~\ref{table:library} summarizes the parallelization approaches \camone{across} PIM cores. Please also see Appendix~\ref{sec:appendix-sparseP} for all \spmv{} kernels provided by the \SparseP{} software package.  All kernels support a wide range of data types, i.e., 8-bit integer (\textbf{int8}), 16-bit integer (\textbf{int16}), 32-bit integer (\textbf{int32}), 64-bit integer (\textbf{int64}), 32-bit float (\textbf{fp32}), and 64-bit float (\textbf{fp64}) data types.

\begin{table}[t]
\centering
\begin{minipage}{0.62\linewidth}
\resizebox{1.0\linewidth}{!}{
\begin{tabular}{| l | l | l | l |} 
 \hline 
 \cellcolor{gray!15}\raisebox{-0.30\height}{\textbf{Partitioning}}& \cellcolor{gray!15}\raisebox{-0.30\height}{\textbf{Compressed}} & \cellcolor{gray!15}\raisebox{-0.30\height}{\textbf{Load Balancing}} \\ 
 \cellcolor{gray!15}\textbf{Technique} & \cellcolor{gray!15}\textbf{Format} & \cellcolor{gray!15}\textbf{Across PIM Cores}  \\ [0.5ex] \hline \hline
    \multirow{9}{*}{\hspace*{2.1em}\turnbox{0}{1D}} & \multirow{2}{*}{CSR} &  rows (\textbf{CSR.row})  \\
    & & nnz$^{\star}$ (\textbf{CSR.nnz}) \\ \cline{2-3}
    & \multirow{3}{*}{COO} &  rows (\textbf{COO.row}) \\
    & & nnz$^{\star}$ (\textbf{COO.nnz-rgrn})  \\
    & & nnz (\textbf{COO.nnz}) \\  \cline{2-3}
    & \multirow{2}{*}{BCSR} &  blocks$^{\dagger}$ (\textbf{BCSR.block})   \\
    & & nnz$^{\dagger}$ (\textbf{BCSR.nnz}) \\  \cline{2-3}
    & \multirow{2}{*}{BCOO} &  blocks (\textbf{BCOO.block}) \\
    & & nnz (\textbf{BCOO.nnz}) \\ \hline
    \multirow{4}{*}{\shortstack{2D \\ \equallySized}} & CSR (\textbf{DCSR}) & - \\  \cline{2-3}
    &  COO (\textbf{DCOO}) & -  \\  \cline{2-3}
    &  BCSR (\textbf{DBCSR}) & -  \\  \cline{2-3}
    &  BCOO (\textbf{DBCOO}) & - \\ \hline
  \multirow{6}{*}{\shortstack{2D \\ \equallyWidth}} &  CSR (\textbf{RBDCSR}) & nnz$^{\star}$ \\  \cline{2-3}
    &  COO (\textbf{RBDCOO}) & nnz  \\  \cline{2-3}
    & \multirow{2}{*}{BCSR} & blocks$^{\dagger}$ (\textbf{RBDBCSR}) \\
    &  & nnz$^{\dagger}$ \\ \cline{2-3}
    & \multirow{2}{*}{BCOO} & blocks (\textbf{RBDBCOO})  \\ 
    & & nnz  \\ \hline    
  \multirow{6}{*}{\shortstack{2D \\ \variableSized}} &  CSR (\textbf{BDCSR}) & nnz$^{\star}$  \\ \cline{2-3}
    &  COO (\textbf{BDCOO}) & nnz   \\  \cline{2-3}
    &  \multirow{2}{*}{BCSR} & blocks$^{\dagger}$ (\textbf{BDBCSR})  \\
    &  & nnz$^{\dagger}$ \\ \cline{2-3}
    &  \multirow{2}{*}{BCOO } & blocks (\textbf{BDBCOO}) \\ 
    &  &  nnz  \\ \hline     
\end{tabular}
}
\end{minipage} %
\vspace{4pt}
\caption{Parallelization techniques \camone{across} PIM cores of \camtwo{the} \SparseP{} library. $^{\star}$: row-granularity, $^{\dagger}$: block-row-granularity}
\label{table:library}
\vspace{-26pt}
\end{table}

\subsection{Parallelization Techniques Across Threads within a PIM Core}\label{sec:lib_1dpu}
PIM cores can support multiple hardware threads to exploit \camone{high memory bank} bandwidth~\cite{Gomez2021Analysis,Gomez2021Benchmarking}. To parallelize \spmv{} \camone{across} multiple threads within a multithreaded PIM core \SparseP{} supports various load balancing schemes for each compressed \camtwo{matrix} format, and three synchronization approaches to ensure correctness among threads of a PIM core.

\subsubsection{\textbf{Load Balancing Approaches}}  \hfill  \\
In a similar way as explained in Figure~\ref{fig:1d_partitioning}, for the CSR and COO formats, we balance either the rows, such that each thread processes almost the same number of rows, or the non-zero elements, such that each thread processes almost the same number of non-zero elements. In the CSR format, matrix is stored in row-order, \camtwo{and thus} load balancing \camone{across} threads is performed at row granularity. In the UPMEM PIM system, elements of the output vector are accessed at 64-bit granularity in DRAM memory. Thus, when balancing is performed at row granularity, we assign rows to threads in chunks of $8 / sizeof(data\_type)$ to ensure 8-byte alignment on the elements of the output vector. In the COO format, balancing the non-zero elements can be performed either at row granularity or by splitting the row \camone{between threads}, i.e., providing an almost perfect non-zero balance \camone{across} threads. In the latter case, synchronization among threads for write accesses on the elements of the output vector can be implemented with three synchronization approaches described in Section~\ref{sec:1dpu_sync}.

For the BCSR and BCOO formats, we balance either the blocks, such that each thread processes almost the same number of blocks, or the non-zero elements, such that each thread processes almost the same number of non-zero elements. In the BCSR format, the matrix is stored in block-row order, \camtwo{and thus} load balancing \camone{across} threads is performed at block row granularity. For both formats, the block sizes are \textit{configurable} in \SparseP{}. In our evaluation, we use block sizes of 4x4, since these are the most common dimensions to cover various sparse matrices~\cite{asgari2020copernicus,Elafrou2018SparseX,Karakasis2009Performance}. In the UPMEM PIM architecture, elements of the output vector are accessed at 64-bit granularity. Therefore, for the BCSR format, \camone{with an} 8-bit integer data type and small block sizes (4x4 or smaller), \camone{threads use synchronization primitives to ensure correctness when writing the elements of the output vector. This is because} different threads may write to the same 64-bit-aligned DRAM memory location. Synchronization among threads for writes to the elements of the output vector is necessary for all configurations of the BCOO format, and can be implemented with three approaches described next.

\subsubsection{\textbf{Synchronization Approaches}}\label{sec:1dpu_sync}  \hfill  \\
\SparseP{} provides three synchronization approaches.
\begin{enumerate}
\item \textbf{Coarse-Grained Locking (\textcolor{darkred}{lb-cg}).} One global mutex protects the elements of the \camone{entire} output vector.
\item\textbf{Fine-Grained Locking (\textcolor{darkred}{lb-fg}).} Multiple mutexes protect the elements of the output vector. \SparseP{} associates mutexes to the elements of the output vector in a round-robin manner. The UPMEM API supports up to 56 mutexes~\cite{upmem-guide}. In our evaluation, we use 32 mutexes such that \camone{we can} find the corresponding mutex for a particular element of the output vector only with a shift operation on the MRAM address, avoiding costly division operations.
\item\textbf{Lock-Free (\textcolor{darkred}{lf}).} Since the formats are row-sorted or block-row-sorted, 
race conditions in the elements of the output vector arise \textit{only in a few elements}, i.e., either when a row (or a block row for BCSR/BCOO) is split across threads, or when continuous elements of the output vector processed by different threads belong to the same 64-bit-aligned DRAM location in the UPMEM PIM system. In our proposed lock-free approach, threads temporarily store partial results for these few elements in the data (scratchpad) memory \camone{(i.e., WRAM in the UPMEM PIM system)}, and later one single thread merges the partial results, and writes the final result for the corresponding element of the output vector to the DRAM bank.
\end{enumerate}

Table~\ref{table:1DPU-impl} summarizes the parallelization techniques \camone{across} threads of a PIM core. All kernels support a wide range of data types, i.e., 8-bit integer (\textbf{int8}), 16-bit integer (\textbf{int16}), 32-bit integer (\textbf{int32}), 64-bit integer (\textbf{int64}), 32-bit float (\textbf{fp32}), and 64-bit float (\textbf{fp64}) data types.

\begin{table}[H]
\centering
\begin{minipage}{0.78\linewidth}
\resizebox{1.0\linewidth}{!}{
\begin{tabular}{| l | l | l |} 
 \hline
  \cellcolor{gray!15}\raisebox{-0.30\height}{\textbf{Compressed}} & \cellcolor{gray!15}\raisebox{-0.30\height}{\textbf{Load Balancing}} & \cellcolor{gray!15}\raisebox{-0.30\height}{\textbf{Synchronization}}  \\ 
  \cellcolor{gray!15}\textbf{Format} & \cellcolor{gray!15}\textbf{Across Threads} & \cellcolor{gray!15}\textbf{Approach}  \\
  [0.5ex] \hline \hline
   \multirow{2}{*}{CSR} & rows (\textbf{CSR.row}) & -  \\
   & nnz$^{\star}$ (\textbf{CSR.nnz}) & -   \\ \hline
   \multirow{3}{*}{COO} & rows (\textbf{COO.row}) & -   \\
   & nnz$^{\star}$ (\textbf{COO.nnz-rgrn}) & -   \\
   & nnz (\textbf{COO.nnz}) & lb-cg / lb-fg / lf   \\ \hline
   \multirow{2}{*}{BCSR} & blocks$^{\dagger}$ (\textbf{BCSR.block}) & lb-cg / lb-fg (only for int8 and small block sizes)   \\
   & nnz$^{\dagger}$ (\textbf{BCSR.nnz}) & lb-cg / lb-fg (only for int8 and small block sizes)  \\  \hline
   \multirow{2}{*}{BCOO} & blocks (\textbf{BCOO.block}) & lb-cg / lb-fg / lf \\
   & nnz (\textbf{BCOO.nnz}) & lb-cg / lb-fg / lf  \\
 \hline
\end{tabular}
}
\end{minipage}
\vspace{4pt}
\caption{Parallelization schemes \camone{across} threads of a PIM core. 
$^{\star}$: row-granularity, $^{\dagger}$: block-row-granularity}
\label{table:1DPU-impl}
\vspace{-16pt}
\end{table}

\subsection{Kernel Implementation}\label{sec:kernel_impl}
We briefly describe the \SparseP{} implementations for all compressed matrix formats, i.e., the way that threads access data involved in the kernel from/to the local DRAM bank. The \spmv{} kernels include three types of data structures: (i) the arrays that store the non-zero elements, i.e., the values (\texttt{values[]}) and the positions of the non-zero elements (\texttt{rowptr[]}, \texttt{colind[]} for CSR, \texttt{tuples[]} for COO, \texttt{browptr[]}, \texttt{bcolind[]} for BCSR, \texttt{browind[]}, \texttt{bcolind[]} for BCOO), (ii) the array that stores the elements of the input vector, and (iii) the array that stores the partial results created for the elements of the output vector.

First, \camone{\spmv{} performs streaming memory accesses to the arrays that store the non-zero elements and their positions. Therefore, to} exploit spatial locality and immense bandwidth in data (scratchpad or cache) memory, each thread reads the non-zero elements by fetching large chunks of bytes in a coarse-grained manner from DRAM to data memory \camone{(i.e., WRAM in the UPMEM PIM system)}. Then, it accesses elements through data memory in a fine-grained manner. In the UPMEM PIM system, we fetch chunks of 256-byte data to discover the non-zero elements, as suggested by the UPMEM API~\cite{upmem-guide}, since 256-byte transfer sizes highly exploit the available local bandwidth of DRAM bank~\cite{Gomez2021Benchmarking,Gomez2021Analysis}. 
For the BCSR and BCOO formats, only for \camone{the array that stores the} values of the non-zero elements (i.e., \texttt{bvalues[]}), we fetch from DRAM to data memory block size chunks, i.e., \camone{chunks of} \camtwo{$r\times c \times sizeof(data\_type)$ bytes, assuming that the matrix is stored in blocks of size $r\times c$.}

Second, \camtwo{\spmv{} causes irregular memory accesses to the elements of the input vector. Specifically, the accesses to the elements of the input vector are input-driven, i.e.,  they are determined by the column positions (column indexes) of the non-zero elements of each particular matrix. Given that matrices involved in \spmv{} are very sparse~\cite{Kanellopoulos2019SMASH,Elafrou2018SparseX,Elafrou2017PerformanceAA,YouTubeGraph,FacebookGraph,Goumas2008Understanding,White97Improving,Helal2021ALTO,Pelt2014Medium}, i.e., the column indexes of the non-zero elements significantly vary, memory accesses to the input vector incur poor data locality. Thus, in our \spmv{} implementations, }threads of a PIM core directly access elements of the input vector through DRAM \camone{bank at} fine-granularity~\cite{Gomez2021Benchmarking,upmem-guide,Gomez2021Analysis}, i.e., using the smallest possible granularity: for the CSR and COO formats at 64-bit granularity, and for the BCSR and BCOO formats at \camtwo{the granularity of $c\times sizeof(data\_type)$ bytes, where $c$ is the horizontal dimension of the block size.}

Third, \camone{regarding the output vector,} threads temporarily store partial results for the same elements of the output vector in data (scratchpad or cache) memory to exploit data locality, until all the non-zero elements of the \textit{same} row or the \textit{same} block row have been traversed \camone{(recall matrices are stored in a row-sorted way)}. Then, the produced results are written to DRAM \camone{bank at} fine-granularity~\cite{Gomez2021Benchmarking,upmem-guide,Gomez2021Analysis}: for the CSR and COO formats at 64-bit granularity, and for the BCSR and BCOO formats at \camtwo{the granularity of $r\times sizeof(data\_type)$ bytes, where $r$ is the vertical dimension of the block size.}


\section{Evaluation Methodology}\label{methodology}
We conduct our evaluation on \camtwo{an} UPMEM PIM system that includes a 2-socket Intel Xeon Silver 4110 CPU~\cite{intel4110} at 2.10 GHz (host CPU), standard main memory (DDR4-2400)~\cite{ddr4jedec} of 128 GB, and 20 UPMEM PIM DIMMs with 160 GB PIM-capable memory and 2560 DPUs.\footnote{There are thirty two faulty DPUs in the system where we run our experiments. They cannot be used and do not affect the correctness of our results, but take away from the system’s full computational power of 2560 DPUs.}

First, we evaluate \spmv{} execution using one single DPU and multiple tasklets (Section~\ref{1DPU}). Table~\ref{tab:small-matrices} shows our evaluated small matrices that fit in the 64 MB DRAM memory of a single DPU. The evaluated matrices vary in sparsity (i.e., NNZ / (rows x columns)), standard deviation of non-zero elements among rows (NNZ-r-std) and columns (NNZ-c-std). The highlighted matrices in Table~\ref{tab:small-matrices} with red color exhibit block pattern~\cite{Kourtis2011CSX,Elafrou2018SparseX}, \camone{i.e., they include \emph{a lot} of dense sub-blocks (almost all their non-zero elements fit in dense sub-blocks).}

\begin{table}[H]
\begin{center}
\centering
\resizebox{0.6\linewidth}{!}{
\begin{tabular}{|l||r|r|r|}
    \hline
    \cellcolor{gray!15}\raisebox{-0.10\height}{\textbf{Matrix Name}} & \cellcolor{gray!15}\raisebox{-0.10\height}{\textbf{Sparsity}} & \cellcolor{gray!15}\raisebox{-0.10\height}{\textbf{NNZ-r-std}} & \cellcolor{gray!15}\raisebox{-0.10\height}{\textbf{NNZ-c-std}} \\
    \hline \hline
    delaunay\_n13  & 	7.32e-04 &  1.343  & 1.343 \\ \hline 
    wing\_nodal &	1.26e-03 &  2.861 & 2.861 \\ 
    \hline
    
    \textcolor{darkred}{raefsky4} & 3.396e-03  & 15.956 & 15.956 \\ \hline
    \textcolor{darkred}{pkustk08} & 0.006542 &  61.537 &  61.537 \\     
    \hline
\end{tabular}
}
\end{center}
\caption{Small Matrix Dataset.}
\label{tab:small-matrices}
\vspace{-20pt}
\end{table}

Second, we evaluate \spmv{} execution using \emph{multiple} DPUs of the UPMEM PIM system (Section~\ref{MultipleDPUs}). We evaluate \spmv{} execution using both 1D (Section~\ref{1D}) and 2D (Section~\ref{2D}) partitioning techniques, and compare them (Section~\ref{1D-2D}) using a wide variety of sparse matrices with diverse sparsity patterns. We select 22 representative sparse matrices from the Sparse Suite Collection~\cite{davis2011florida}, the characteristics of which are shown in Table~\ref{tab:large-matrices}. 
As the values of the last two metrics increase (i.e., NNZ-r-std and NNZ-c-std), the matrix becomes very irregular~\cite{Namashivayam2021Variable,Tang2015Optimizing}, \camone{and is referred} to as \textit{scale-free} matrix. \camtwo{In our evaluation, we refer to all matrices between \texttt{hgc} to \texttt{bns} matrices of Table~\ref{tab:large-matrices} as \textit{regular} matrices. The  matrices in which NNZ-r-std is larger than 25, i.e., all matrices between \texttt{wbs} to \texttt{ask} in Table~\ref{tab:large-matrices}, we refer to as \textit{scale-free} matrices.} 
Please see Appendix~\ref{sec:appendix-matrix-dataset} for a complete description of our dataset of large sparse matrices.

\begin{table}[t]
\begin{center}
\centering
\resizebox{.74\linewidth}{!}{
\begin{tabular}{|l||r|r|r|}
    \hline
    \cellcolor{gray!15}\raisebox{-0.10\height}{\textbf{Matrix Name}} & \cellcolor{gray!15}\raisebox{-0.10\height}{\textbf{Sparsity}} & \cellcolor{gray!15}\raisebox{-0.10\height}{\textbf{NNZ-r-std}} & \cellcolor{gray!15}\raisebox{-0.10\height}{\textbf{NNZ-c-std}} \\
    \hline
    \hline
    hugetric-00020 (\textbf{hgc}) & 4.21e-07 & 0.031 & 0.031 \\ \hline
    
    mc2depi (\textbf{mc2}) & 7.59e-06 & 0.076 & 0.076 \\ \hline
    
    parabolic\_fem (\textbf{pfm}) & 	1.33e-05 & 0.153 &  0.153 \\ \hline 
    
    roadNet-TX (\textbf{rtn}) & 	1.98e-06 &  1.037 & 1.037 \\ \hline
    
    rajat31 (\textbf{rjt})  &  9.24e-07  & 1.106  & 1.106 \\ \hline
    
    \textcolor{darkred}{af\_shell1 (\textbf{ash})} & 	6.90e-05 & 1.275 & 1.275 \\ \hline  
    
    delaunay\_n19 (\textbf{del}) & 	1.14e-05 &  1.338 & 1.338 \\ \hline
    
    thermomech\_dK  (\textbf{tdk}) & 6.81e-05 & 1.431 &  1.431 \\ \hline
    
    memchip	(\textbf{mem}) & 	2.02e-06 &  2.062 & 1.173 \\ \hline
        
    amazon0601	(\textbf{amz}) & 	2.08e-05 &  2.79 & 15.29  \\ \hline
    
    FEM\_3D\_thermal2 (\textbf{fth}) & 	1.59e-04 & 4.481  & 4.481 \\ \hline

    web-Google (\textbf{wbg}) & 	6.08e-06 & 6.557 &  38.366 \\ \hline
    
    \textcolor{darkred}{ldoor (\textbf{ldr})} & 	5.13e-05  & 11.951 & 11.951 \\ \hline 
    
    poisson3Db (\textbf{psb}) & 	3.24e-04  & 14.712  & 14.712 \\ \hline
    
    \textcolor{darkred}{boneS10 (\textbf{bns})}  & 	6.63e-05 &  20.374 &  20.374 \\ \hline \hline  
    
    webbase-1M (\textbf{wbs}) & 	3.106e-06 & 25.345  & 36.890 \\ \hline  
    
    in-2004	(\textbf{in}) & 	8.846e-06  & 37.230 & 144.062 \\ \hline 
    
    \textcolor{darkred}{pkustk14 (\textbf{pks})} & 	6.428e-04 &  46.508  & 46.508 \\ \hline
    
    com-Youtube (\textbf{cmb}) & 4.639e-06  & 50.754  & 50.754 \\ \hline

    as-Skitter (\textbf{skt})  & 	7.71e-06  & 136.861 &  136.861 \\ \hline
        
    sx-stackoverflow (\textbf{sxw}) & 	5.352e-06 &  137.849 & 65.367 \\ \hline
    
    ASIC\_680k (\textbf{ask})  & 	8.303e-06  & 659.807 & 659.807 \\ \hline 
\end{tabular}
}

\end{center}
\vspace{4pt}
\caption{Large Matrix Dataset. Matrices are sorted by NNZ-r-std, i.e., based on their irregular pattern. \camtwo{The highlighted matrices with red color exhibit block pattern~\cite{Kourtis2011CSX,Elafrou2018SparseX}.}}
\label{tab:large-matrices}
\vspace{-20pt}
\end{table}

Third, we compare the performance and energy consumption
of \spmv{} execution on the UPMEM PIM system to those
\camone{on the} Intel Xeon Silver 4110 CPU~\cite{intel4110} and \camone{the} NVIDIA Tesla V100 GPU~\cite{nvidiaTeslaV100} (Section~\ref{cpu-gpu}).

In Section~\ref{recommendations}, we summarize our key takeaways and provide programming recommendations for software designers, and suggestions and hints for hardware and system designers of future PIM systems.

\section{Analysis of \spmv{} Execution on One DPU}\label{1DPU}
This \camone{section} characterizes \spmv{} performance with various load balancing schemes and compressed \camone{matrix} formats using multiple tasklets in a single DPU. Section~\ref{1DPU-MulTskl} compares load balancing schemes of each compressed matrix format, and Section~\ref{1DPU-Formats} compares the scalability of various compressed matrix formats.

\subsection{\textbf{Load Balancing Schemes Across Tasklets of One DPU}}\label{1DPU-MulTskl}

\begin{figure}[t]
\vspace{4pt}
\centering
\begin{minipage}{\textwidth}
\includegraphics[width=.245\textwidth]{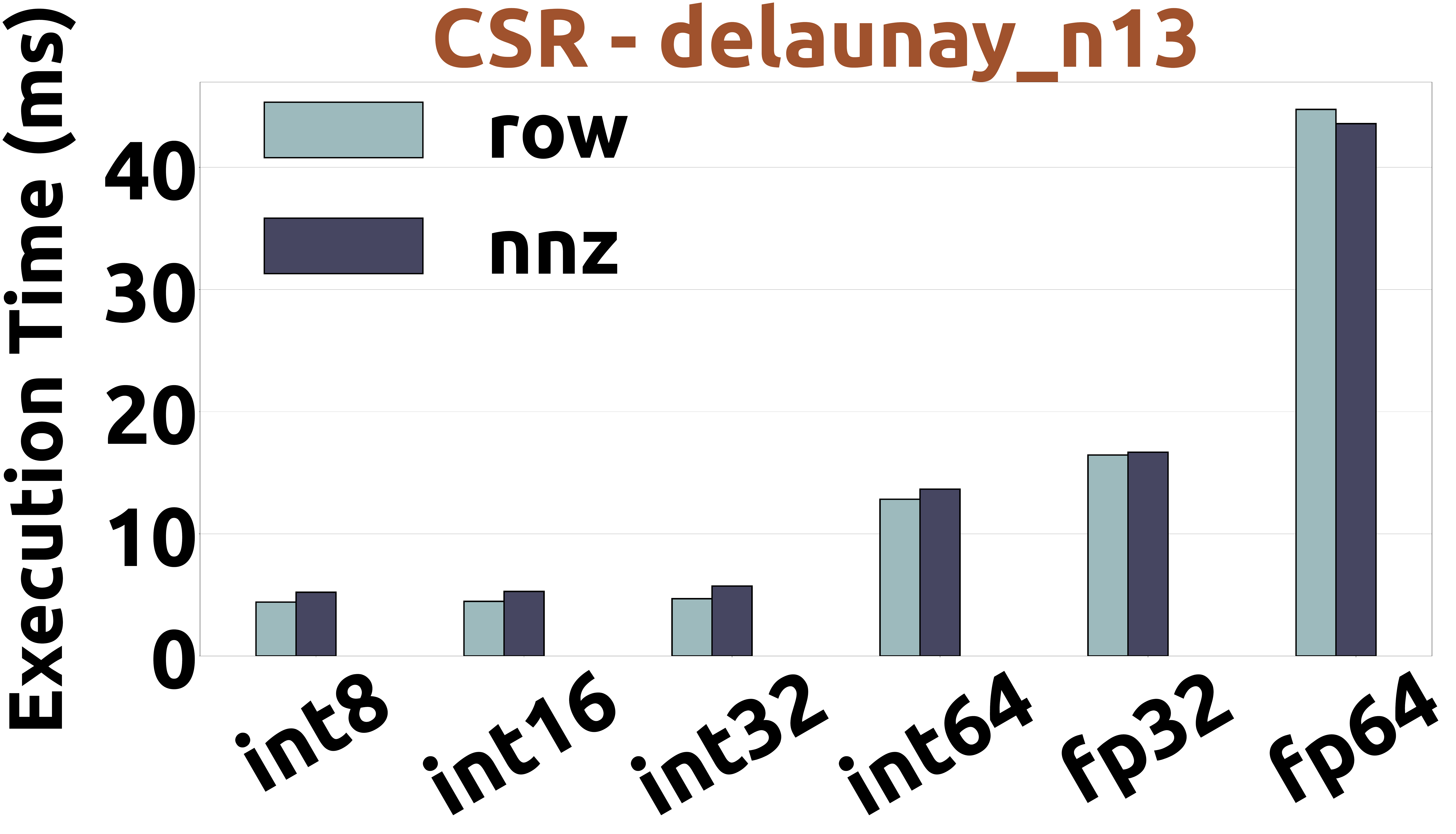}
\includegraphics[width=.245\textwidth]{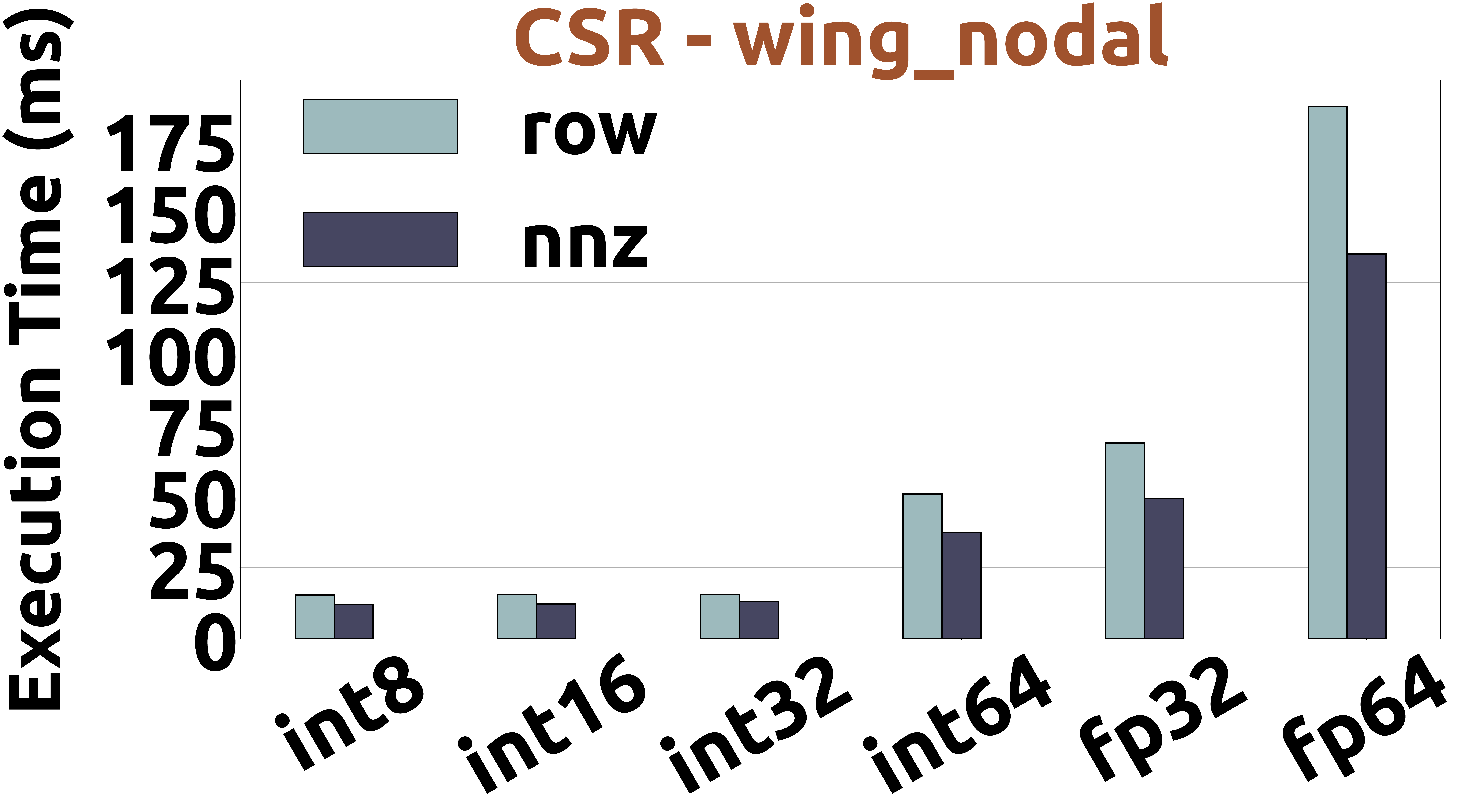}
\includegraphics[width=.245\textwidth]{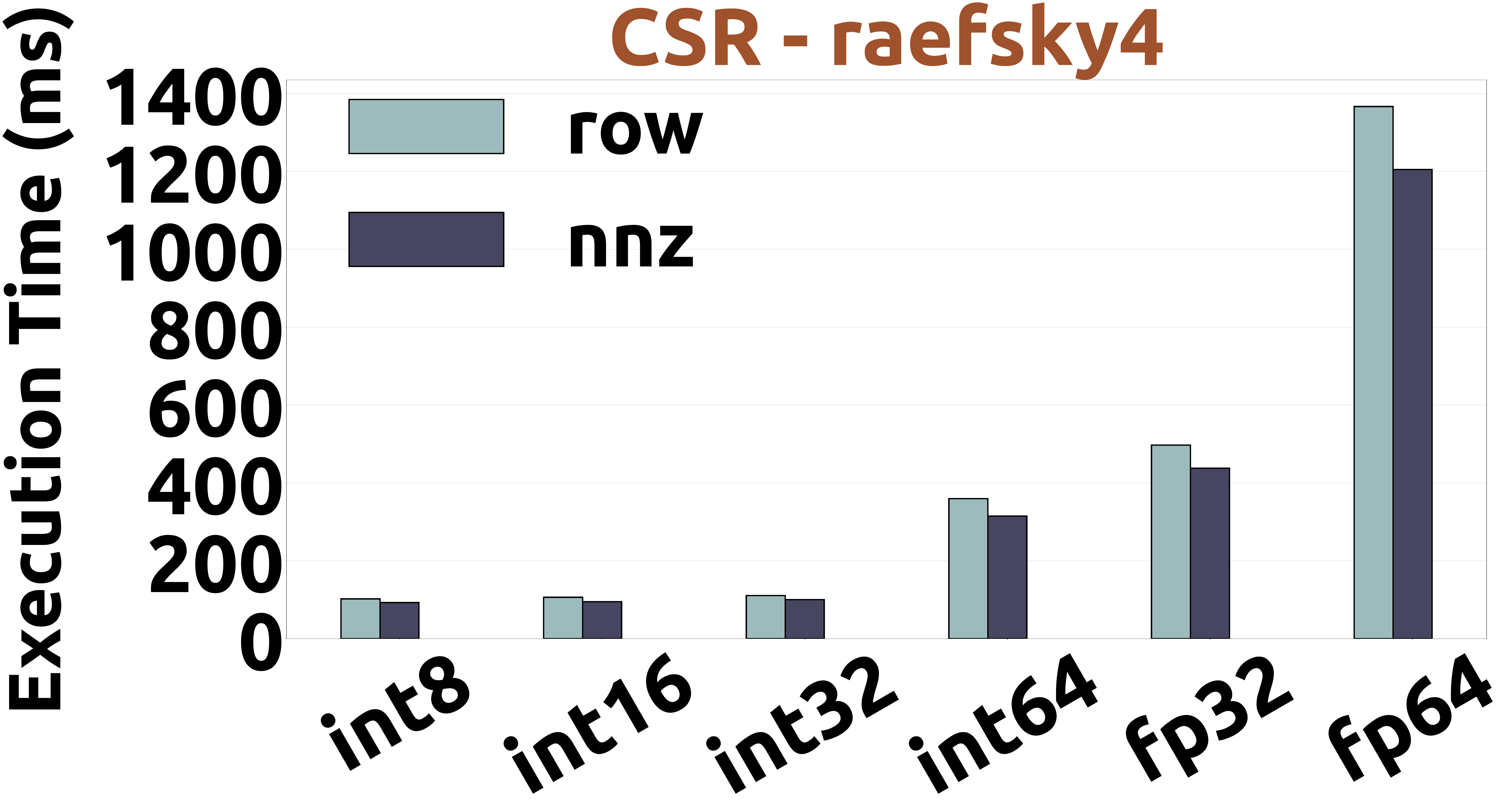}
\includegraphics[width=.245\textwidth]{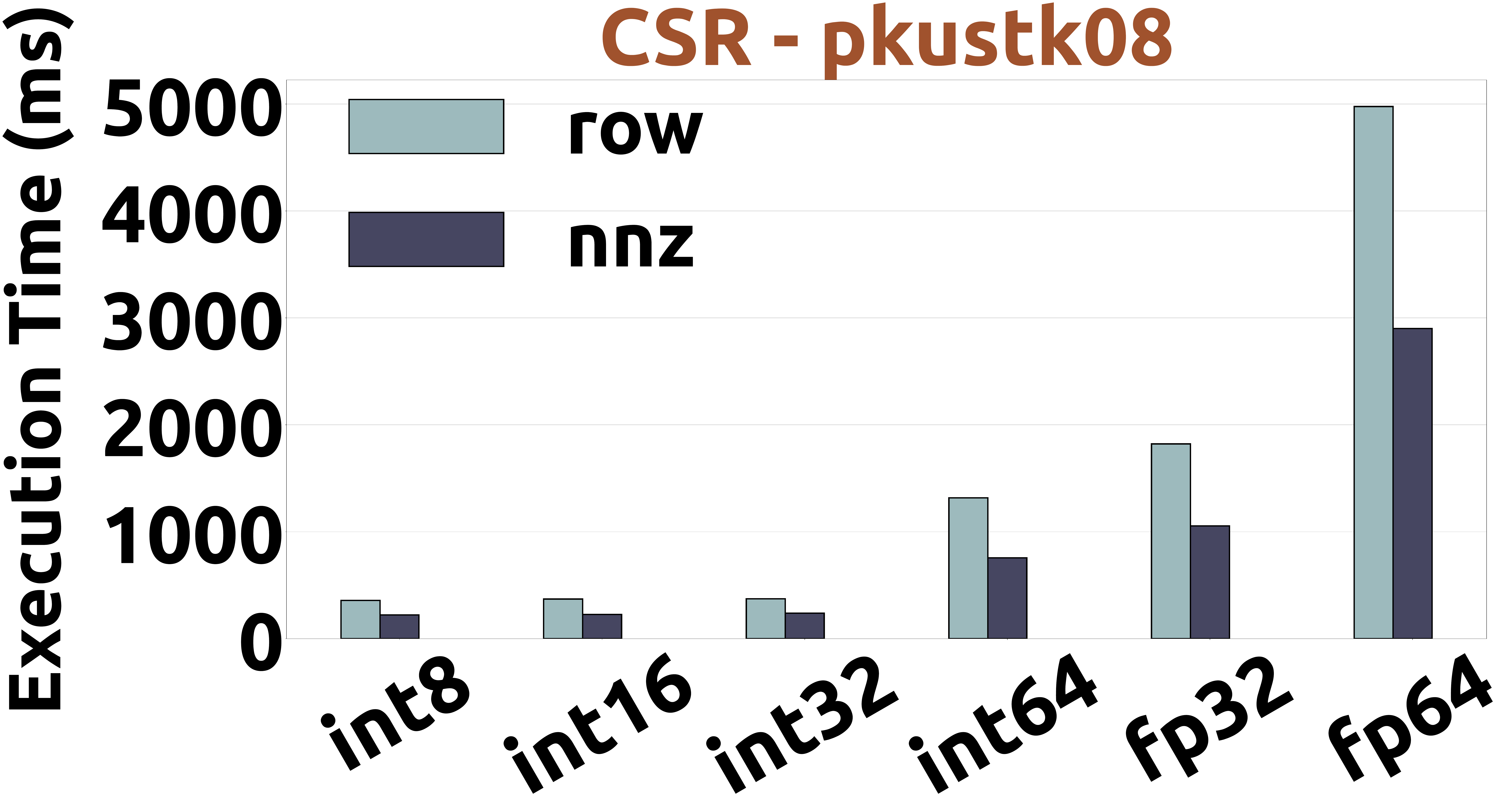}
\end{minipage}\hspace{2pt}%
\begin{minipage}{\textwidth}
\includegraphics[width=.245\textwidth]{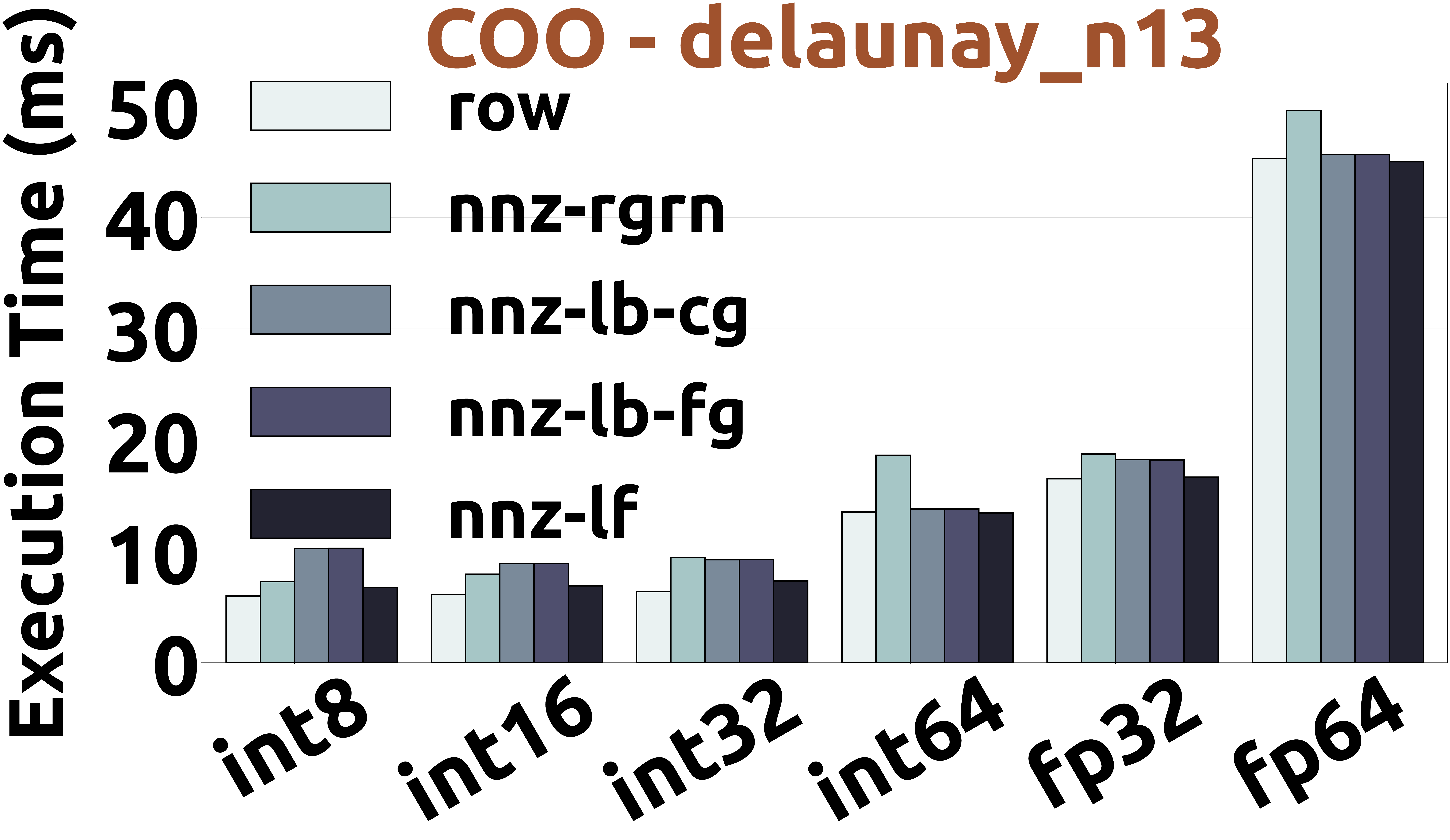}
\includegraphics[width=.245\textwidth]{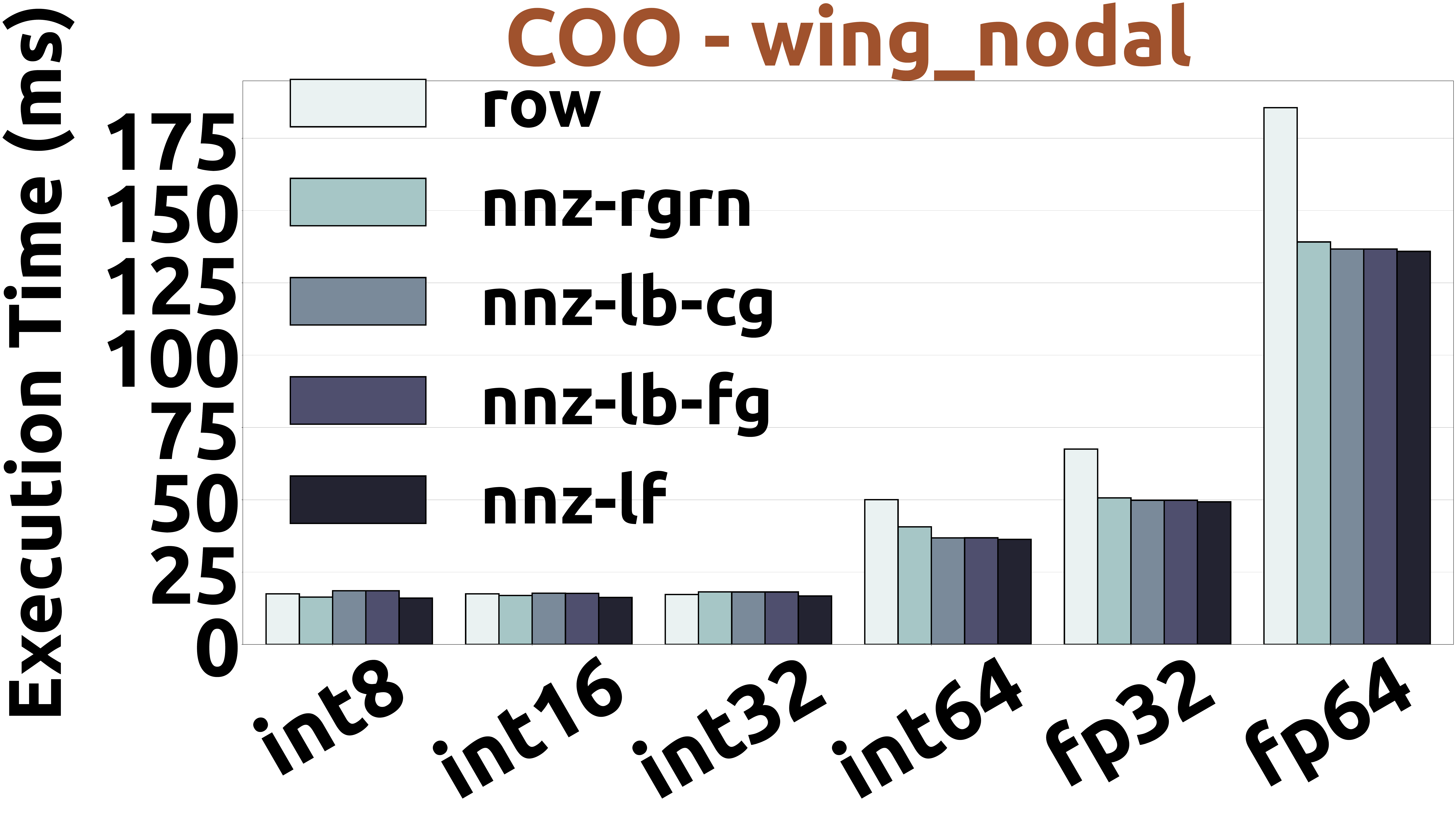}
\includegraphics[width=.245\textwidth]{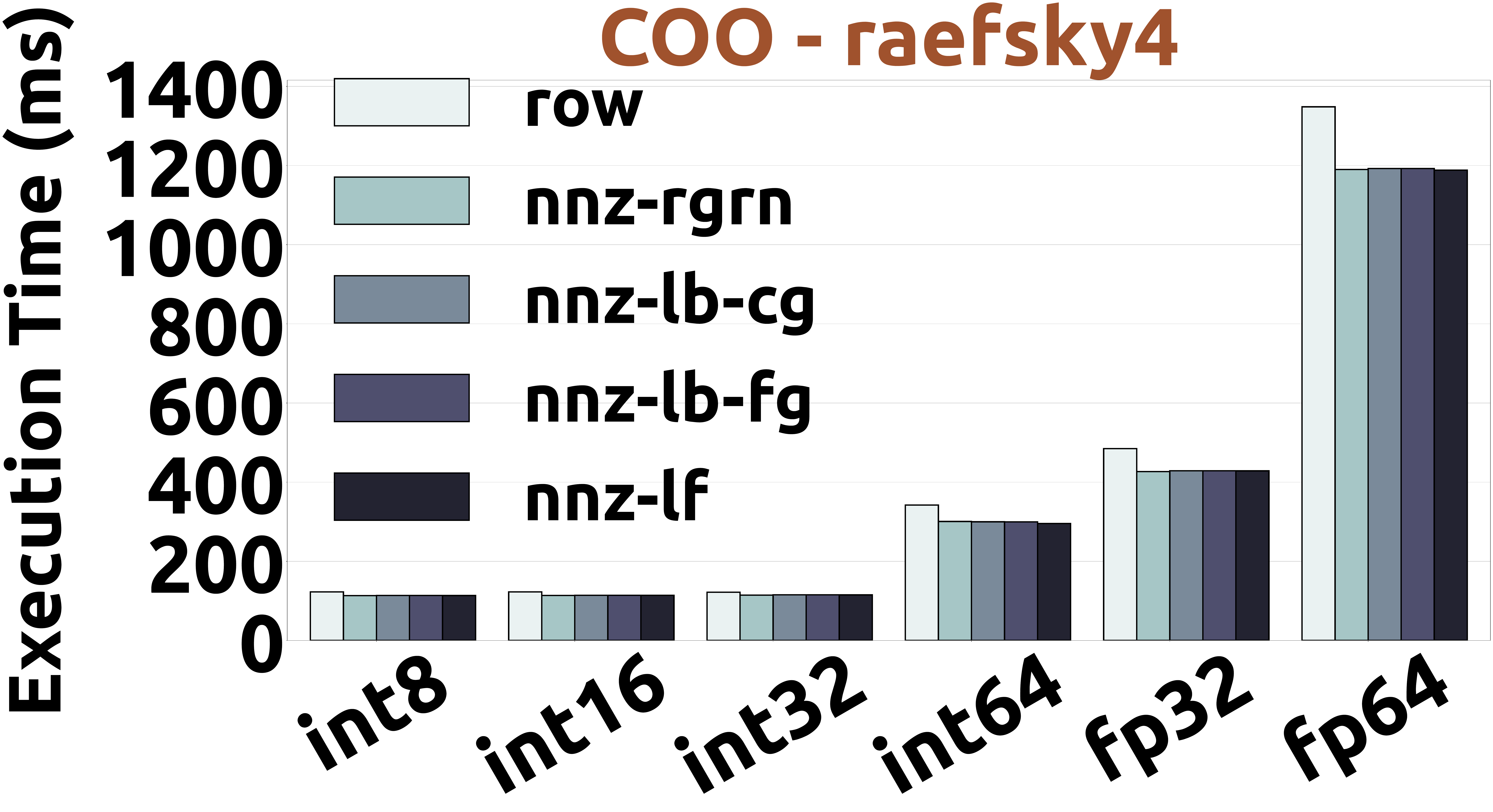}
\includegraphics[width=.245\textwidth]{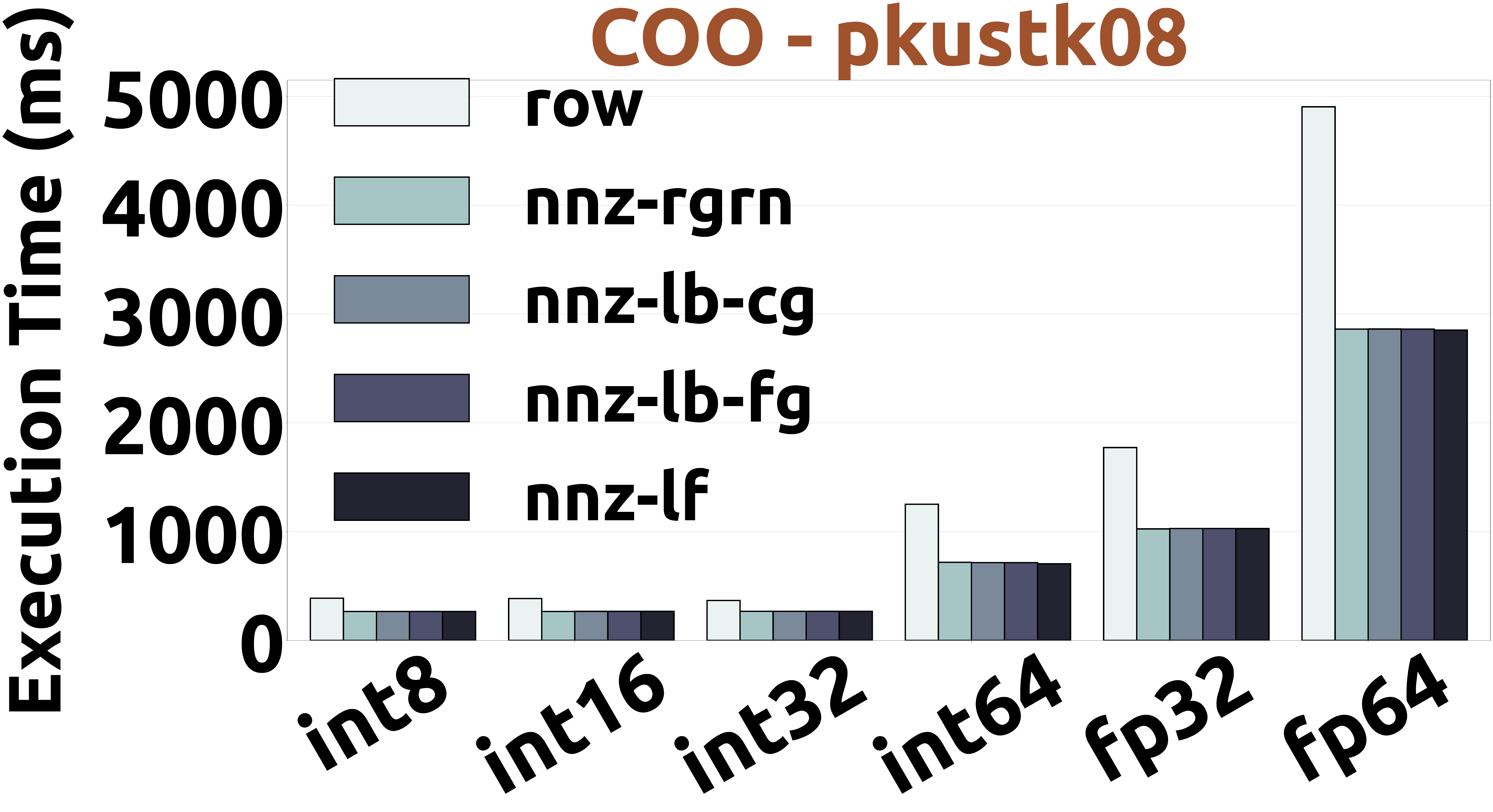}
\end{minipage}\hspace{2pt}%
\begin{minipage}{\textwidth}
\includegraphics[width=.245\textwidth]{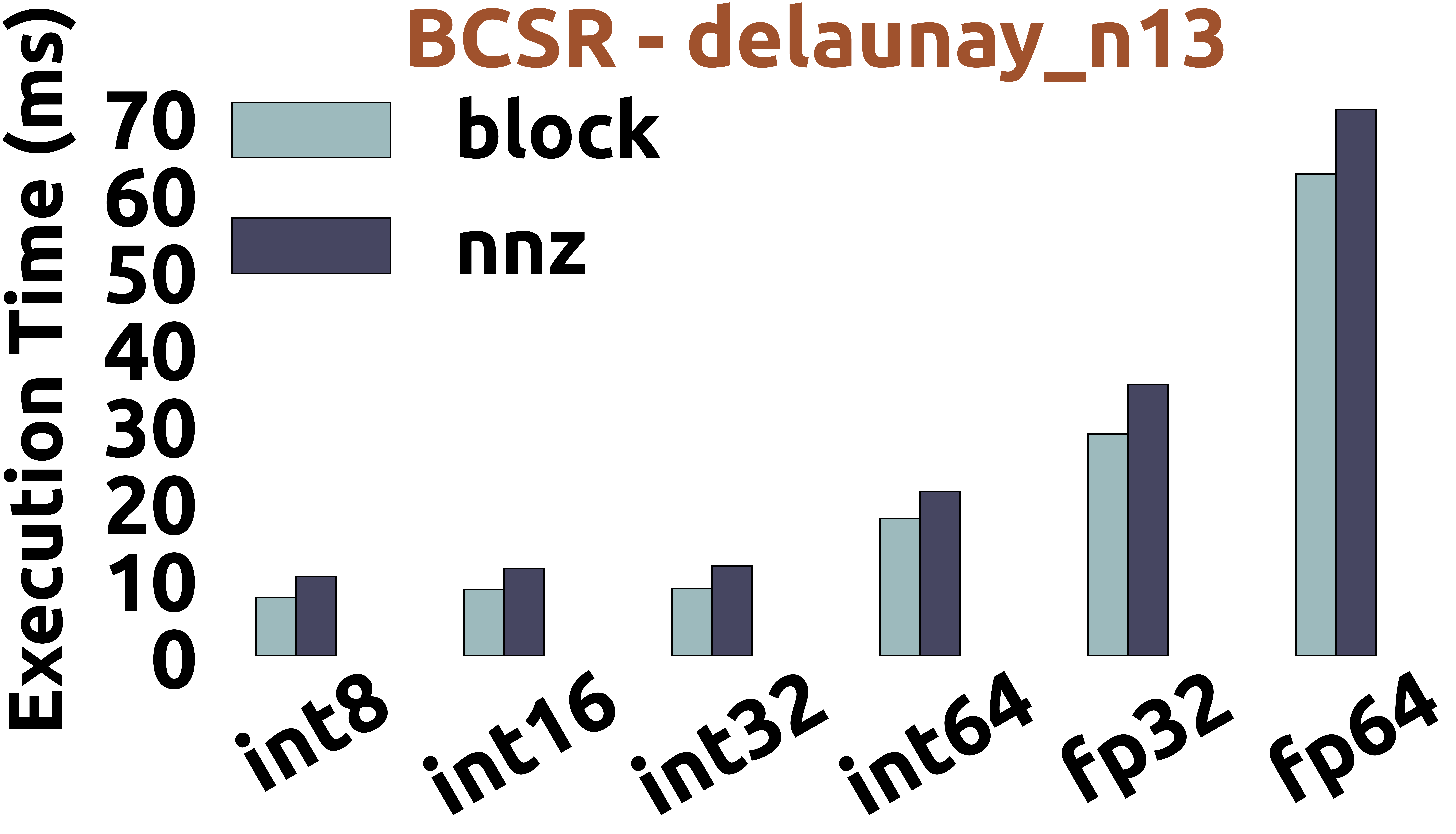}
\includegraphics[width=.245\textwidth]{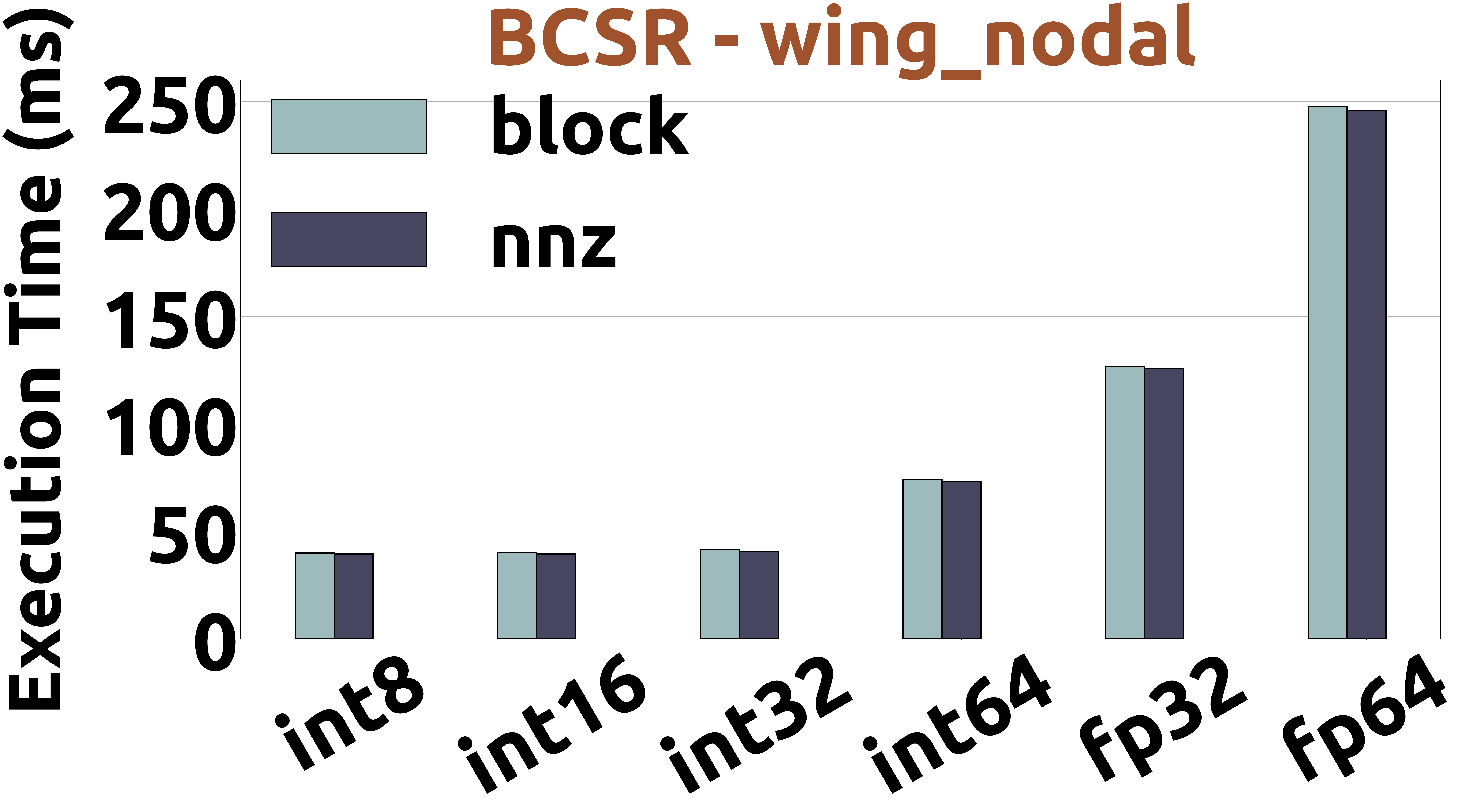}
\includegraphics[width=.245\textwidth]{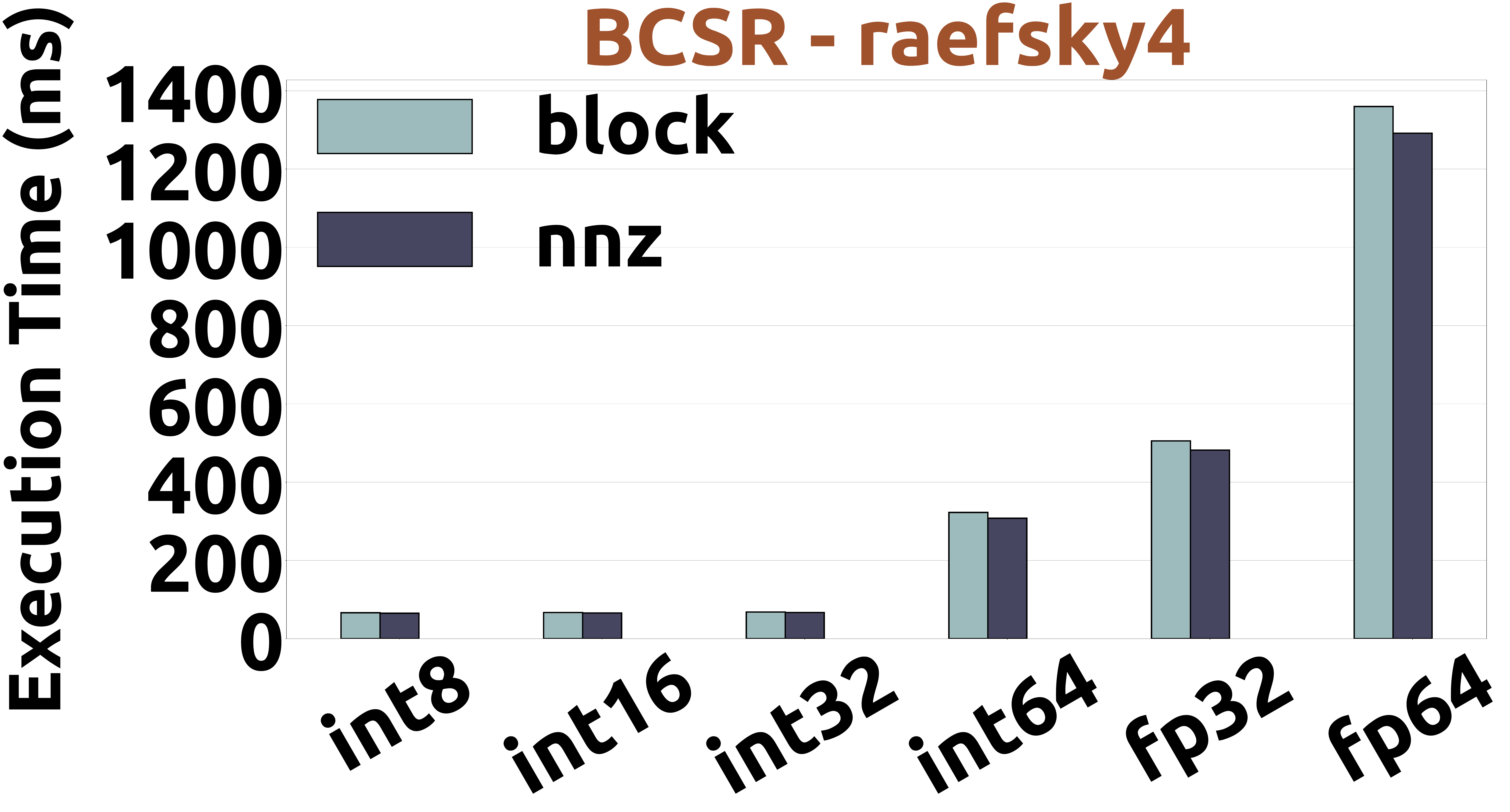}
\includegraphics[width=.245\textwidth]{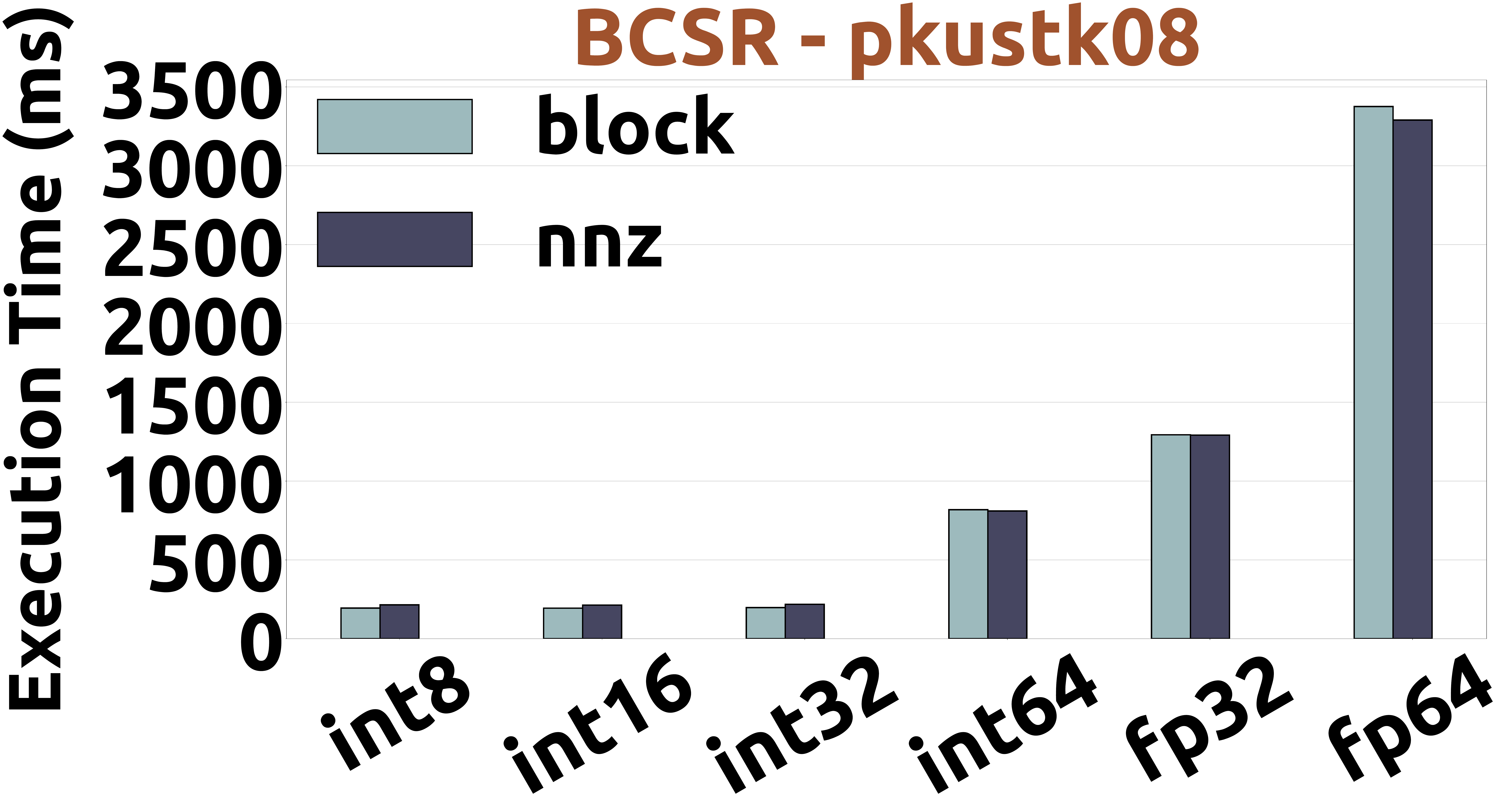}
\end{minipage}\hspace{2pt}%
\begin{minipage}{\textwidth}
\includegraphics[width=.245\textwidth]{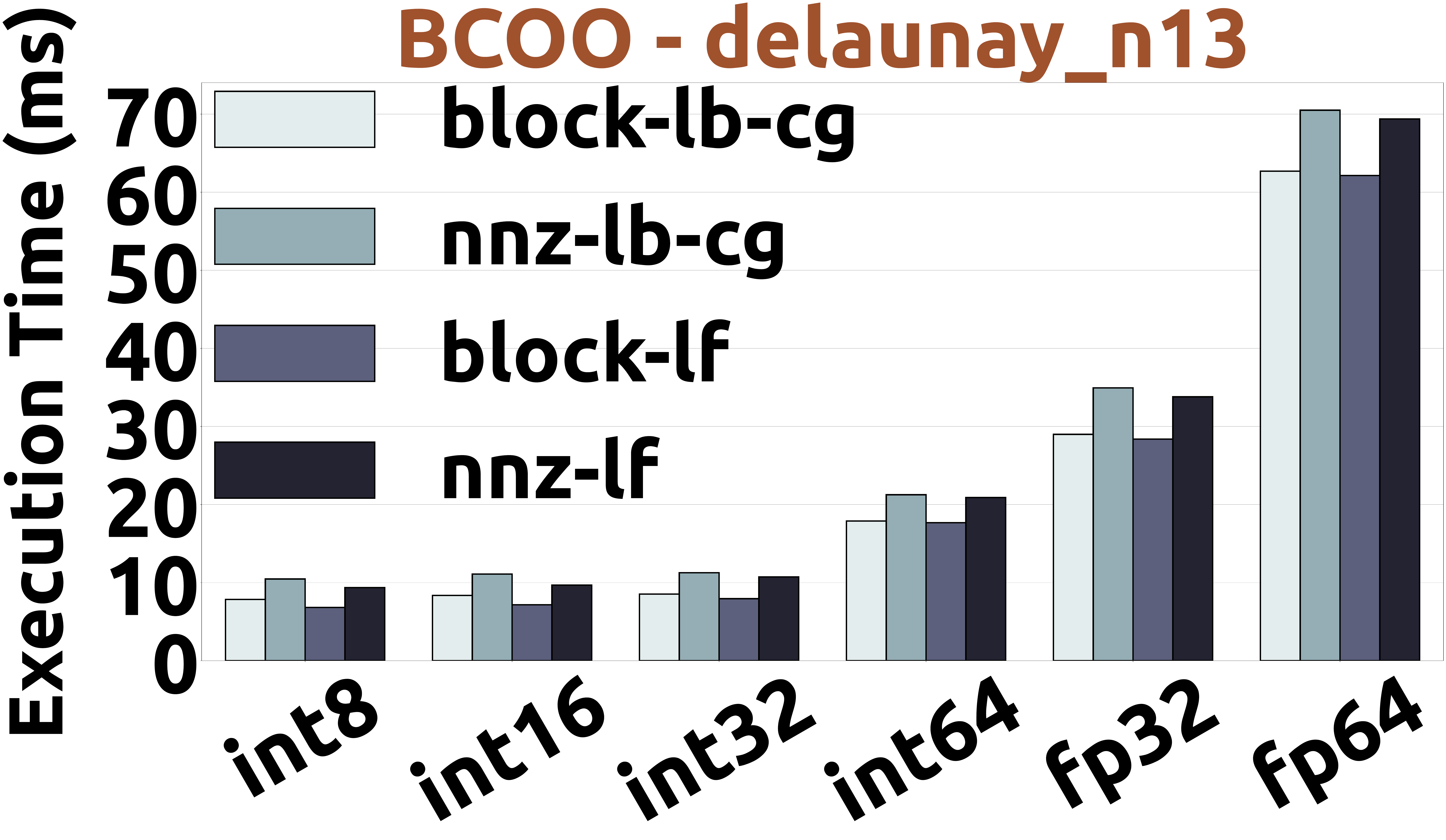}
\includegraphics[width=.245\textwidth]{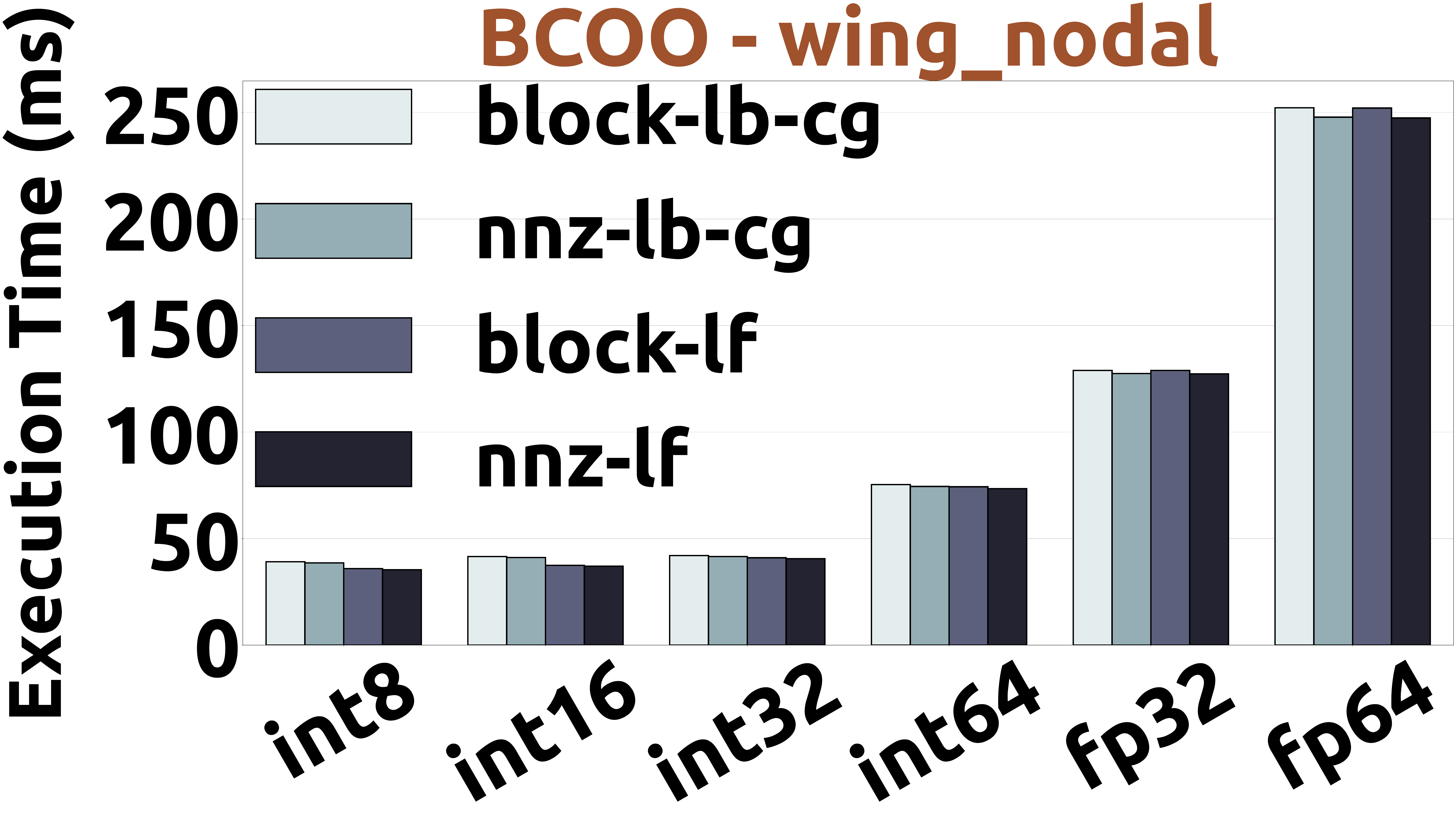}
\includegraphics[width=.245\textwidth]{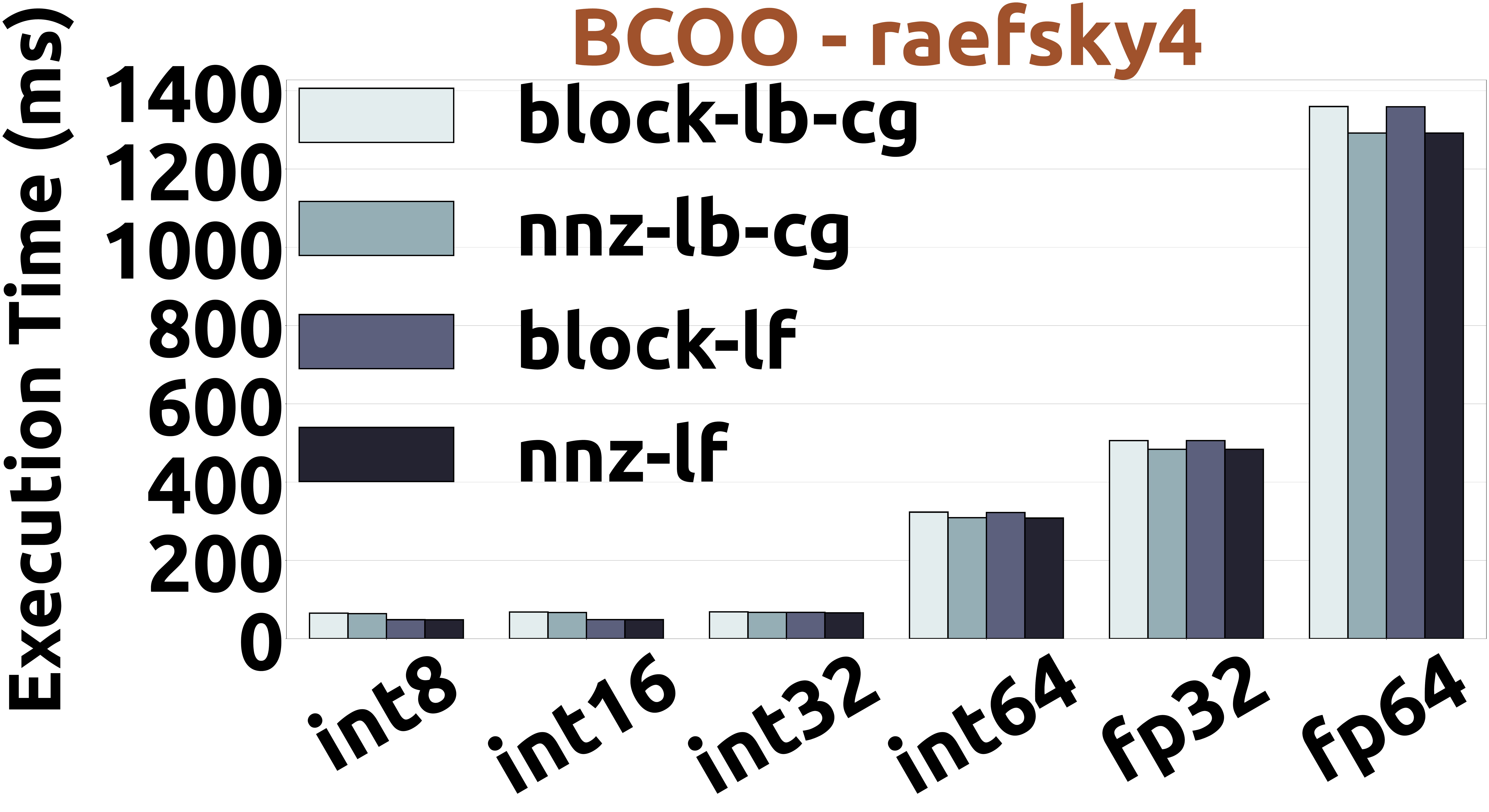}
\includegraphics[width=.245\textwidth]{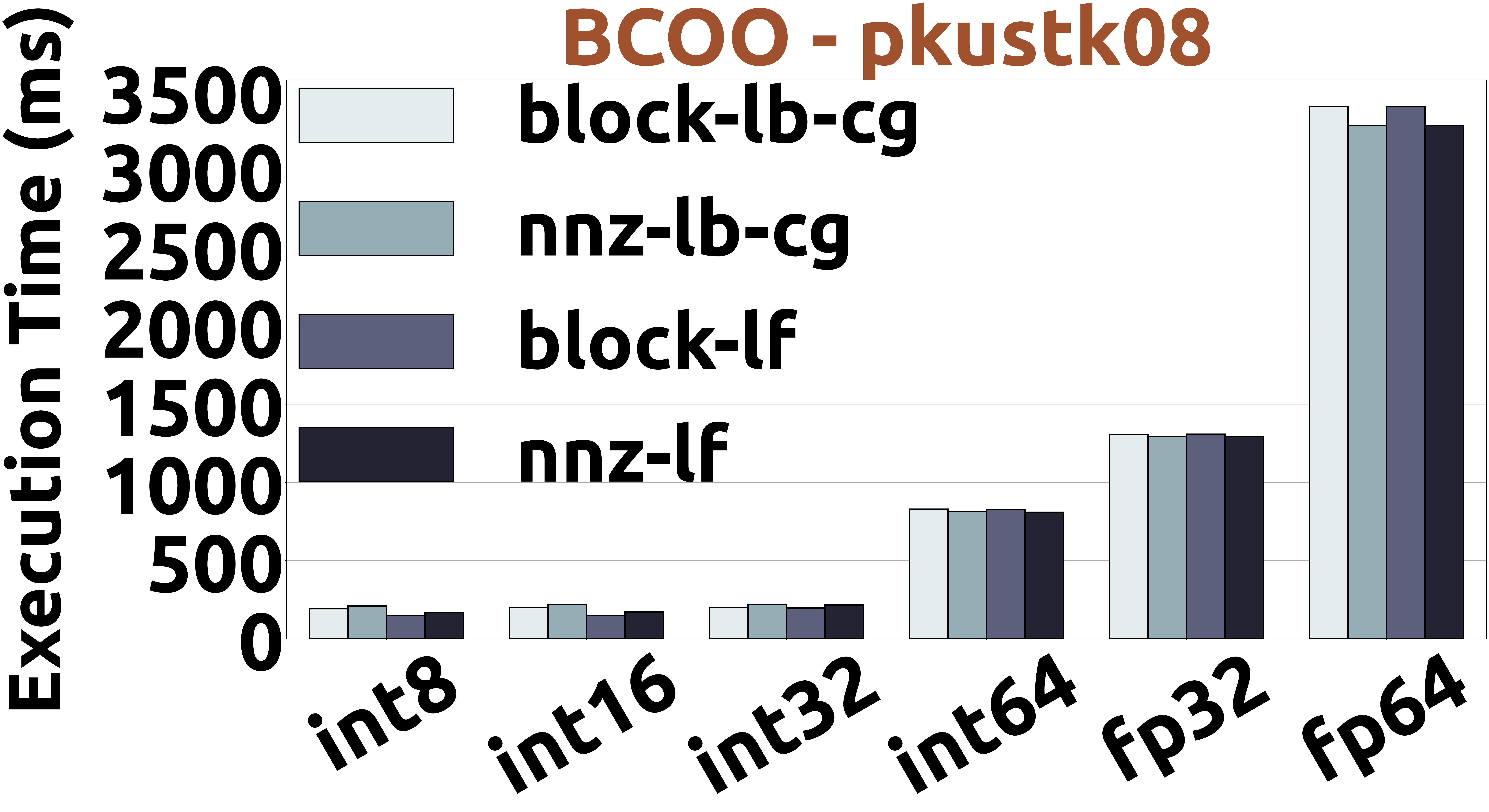}
\end{minipage}
\vspace{-6pt}
\caption{Execution time achieved by various load balancing schemes \camone{of each compressed matrix format using 16 tasklets of a single DPU}.}
\label{fig:1DPU-datatypes}
\vspace{-14pt}
\end{figure}

We compare the parallelization schemes of each compressed matrix format supported by \SparseP{} library (presented in Table~\ref{table:1DPU-impl}) \camone{across} multiple threads of a multithreaded PIM core.
Figure~\ref{fig:1DPU-datatypes} compares the load balancing schemes of each compressed matrix format using 16 tasklets in a single DPU. For the BCSR and BCOO formats, we omit results for the fine-grained locking approach, since it performs similarly with the coarse-grained locking approach: as we explain in Appendix ~\ref{sec:appendix-1DPU-BCOO}, fine-grained locking does not increase parallelism over coarse-grained, \camone{since in the UPMEM PIM hardware, DRAM memory accesses of the critical section are serialized in the DMA engine of the DPU~\cite{Gomez2021Analysis,upmem-guide,Gomez2021Benchmarking}.}

We draw four findings from Figure~\ref{fig:1DPU-datatypes}. First, we find that \spmv{} execution using int8, int16, and int32 data types achieves similar execution times \camone{across} them. This is because the multiplication operation of these data types is sufficiently supported by hardware~\cite{Gomez2021Benchmarking}. In contrast, execution time \camone{sharply} increases \camone{when using} more heavyweight data types, \camone{i.e., int64 and} floating point data types, in which multiplication is emulated in \camone{software using the 8x8-bit multiplier of the DPU~\cite{upmem-guide,Gomez2021Benchmarking,Gomez2021Analysis}.}

Second, we observe that balancing the non-zero elements \camone{across} tasklets typically outperforms balancing the rows for the CSR/COO formats or blocks for the BCSR/BCOO formats, since the non-zero element multiplications are computationally very expensive and can significantly affect load balance \camone{across} tasklets. However, in \texttt{delaunay\_n13 matrix}, balancing the non-zero elements causes high row/block imbalance \camone{across} tasklets, since one tasklet processes a significantly higher number of rows/blocks over the rest, thereby \camone{causing} high operation imbalance \camone{across} tasklets within the DPU core pipeline. As a result, balancing the rows/blocks outperforms balancing the non-zero elements due to the particular pattern of \texttt{delaunay\_n13 matrix}. \camone{In addition, performance benefits of balancing the blocks over balancing the non-zero elements are significant in the BCSR/BCOO formats,} because they operate at block granularity and incur high loop control costs.


Third, we observe that the lock-free approach (\texttt{COO.nnz-lf}) outperforms the lock-based approaches (\texttt{COO.nnz-lb-cg} \camone{and} \texttt{COO.nnz-lb-fg}) in \texttt{delaunay\_n13 matrix}, especially in data types where the multiplication operation is supported \camone{directly in} hardware. In \texttt{delaunay\_n13 matrix}, one tasklet processes a much \camone{larger} number of rows than the rest, i.e., it performs a much \camone{larger} number of critical sections than the rest. \camone{In other words, one tasklet performs a much \camone{larger} number of lock acquisitions/releases and memory instructions than the rest.} Thus, lock-based approaches cause high operation imbalance in the DPU core pipeline \camone{with significant performance costs. Instead, lock-free and lock-based approaches in the BCOO format perform similarly, since 
lock acquisition/release costs can be hidden due to BCOO's higher loop control costs and larger critical sections. Overall, based on the second and the third findings, we conclude that in matrices and formats, where the load balancing and/or the synchronization scheme used cause \emph{high} disparity in the number of non-zero elements/blocks/rows processed \camone{across} tasklets or the number of
lock acquisitions/lock releases/memory accesses performed \camone{across} tasklets, the DPU core pipeline can incur significant performance overheads.} 


\begin{tcolorbox}
\noindent\textbf{OBSERVATION 1:} \\
\textit{High operation imbalance} in computation, control, synchronization, or memory instructions executed by multiple threads of a PIM core can cause \textit{high performance overheads} in the compute-bound and area-limited PIM cores.
\end{tcolorbox}

Fourth, we find that the fine-grained locking approach (\texttt{COO.nnz-lb-fg}) performs similarly with the coarse-grained locking approach (\texttt{COO.nnz-lb-cg}). This is because the critical section includes memory accesses to the local DRAM bank, which, in the UPMEM PIM hardware, are serialized in the DMA engine of the DPU. Therefore, fine-grained locking does not increase execution parallelism over coarse-grained locking, since concurrent accesses to MRAM \camone{bank} are not supported in the UPMEM PIM hardware. \camone{Fine-grained locking does not improve performance over coarse-grained locking, also when using block-based formats (e.g., BCSR/BCOO formats), as we demonstrate in Appendix~\ref{sec:appendix-1DPU-BCOO}. Therefore, we recommend PIM hardware designers to provide lightweight synchronization mechanisms~\cite{Giannoula2021SynCron} for PIM cores, and/or enable concurrent accesses to local DRAM memory, e.g., supporting sub-array level parallelism~\cite{Kim2012Case,Seshadri2013RowClone,Seshadri2017Ambit,Hajinazar2021SIMDRAM,Seshadri2017Simple,Chang2016LISA,seshadri2020indram,Chang2014Improving} or multiple DRAM banks per PIM core.}

\begin{tcolorbox}
\noindent\textbf{OBSERVATION 2:} \\
\textit{Fine-grained} locking approaches to parallelizing critical sections that perform memory accesses to different DRAM memory locations cannot improve performance over \textit{coarse-grained} locking, when the PIM hardware does not support \textit{concurrent accesses to a DRAM bank}. 
\end{tcolorbox}

\subsection{\textbf{Analysis of Compressed Matrix Formats on One DPU}}\label{1DPU-Formats}
 
\begin{figure}[t]
\centering
\includegraphics[width=.86\textwidth]{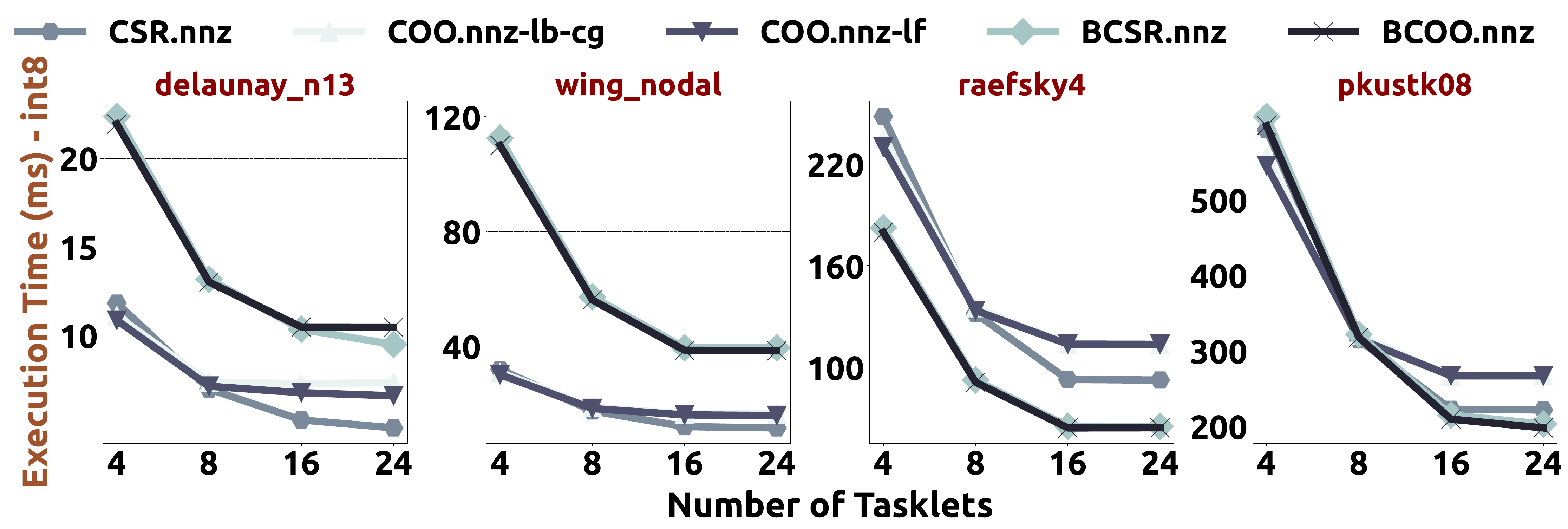}
\includegraphics[width=.86\textwidth]{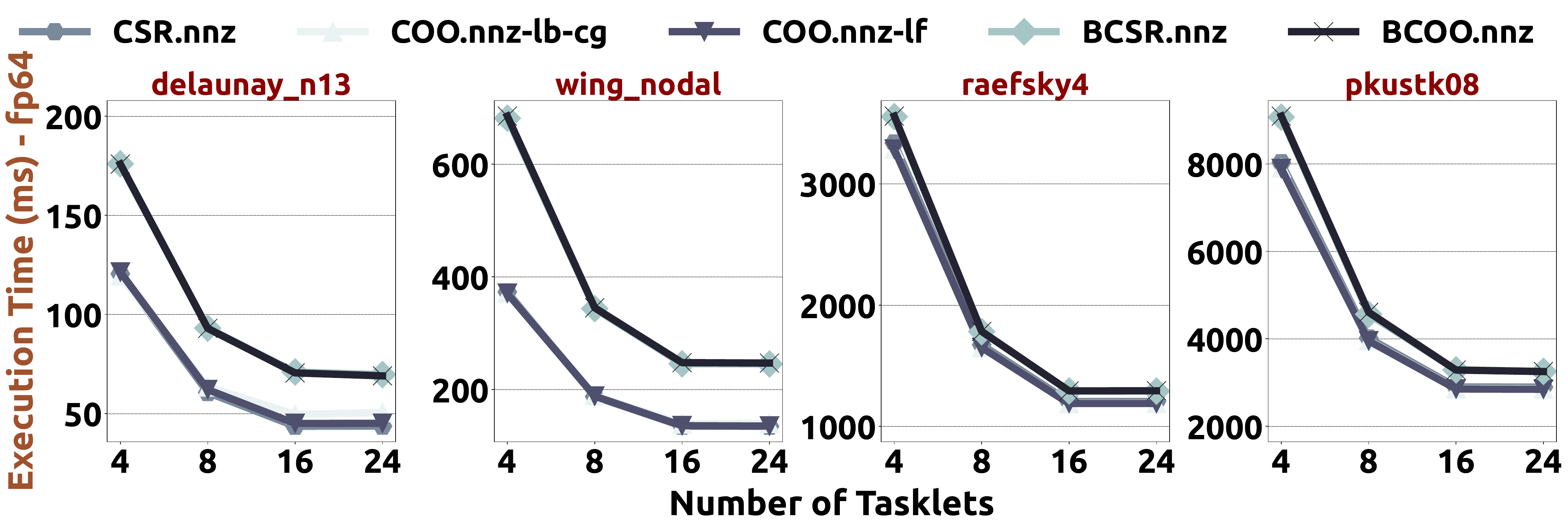}
\vspace{-5pt}
\caption{Scalability of all compressed formats for \camone{the} int8 (\camone{top graphs}) and fp64 (\camone{bottom graphs}) data types as the number of tasklets of a single DPU increases.}
\label{fig:1DPU-scalability}
\vspace{-8pt}
\end{figure}

We compare the scalability and the performance achieved by various compressed matrix formats. Figure~\ref{fig:1DPU-scalability} compares the supported compressed formats for \camone{the} int8 (\camone{top graphs}) and fp64 (\camone{bottom graphs}) data types when balancing the non-zero elements \camone{across} tasklets of a DPU.

We draw three findings. First, we find that even though a DPU supports 24 tasklets, \spmv{} execution typically scales up to 16 tasklets, since the DPU pipeline is fully utilized. In \texttt{delaunay\_n13} matrix, \texttt{CSR.nnz} scales up to 24 tasklets. \camone{In this matrix, when using 16 tasklets, performance of the \texttt{CSR.nnz} scheme is limited by memory accesses:  \textit{only} one tasklet processes 6 $\times$ more rows than the rest, i.e., it performs 6 $\times$ more memory accesses to fetch elements from the \texttt{rowptr[]} array. Thus, as we increase the number of tasklets from 16 to 24, the disparity in the number of rows \camone{across} tasklets decreases, and the performance of the \texttt{CSR.nnz} scheme improves.} 
Second, we observe that for \camone{the} data types with hardware-supported multiplication operation (e.g., int8 data type), CSR achieves the highest scalability, since it provides a better balance between memory access and computation. In contrast, in \camone{the} floating point data types (e.g., fp64 data type), the DPU is significantly bottlenecked by the expensive software-emulated multiplication operations, and thus all formats scale similarly. Third, we observe that the BCSR and BCOO formats outperform the CSR and COO formats in matrices that exhibit block pattern (i.e., \texttt{raefsky4} and \texttt{pkustk08} matrices), only when multiplication is supported by hardware (e.g., int8 data type). This is  because they exploit spatial and temporal locality in data memory (i.e., WRAM) in the accesses of the elements of the input vector. Instead, in \camone{the} fp64 data type, performance is severely bottlenecked by computation, thus the BCSR/BCOO formats perform worse than the CSR/COO formats, since they incur higher indexing costs to discover the positions of the non-zero elements~\cite{asgari2020copernicus,Kanellopoulos2019SMASH}.

\begin{tcolorbox}
\noindent\textbf{OBSERVATION 3:} \\
\camone{Block-based} formats (e.g., BCSR/BCOO) and can provide high performance gains over \camone{non-block-based} formats (e.g., CSR/COO) in matrices that exhibit block pattern, if the multiplication operation is supported by hardware. Otherwise, the state-of-the-art CSR and COO formats can provide high \camone{performance and scalability}.
\end{tcolorbox}

\section{Analysis of \spmv{} Execution on Multiple DPUs}\label{MultipleDPUs}

This section analyzes \spmv{} execution using multiple DPUs in the UPMEM PIM system using the large matrix data set of Table~\ref{tab:large-matrices}. 

Section~\ref{1D} evaluates the 1D partitioning \camone{schemes}. Section~\ref{1D-Kernel} evaluates the actual kernel time of \spmv{} by comparing (a) all load balancing schemes of each compressed matrix format, and (b) the performance of all compressed matrix formats. Section~\ref{1D-EndToEnd} characterizes end-to-end \spmv{} execution \camone{time} of the 1D partitioning technique including the data transfer costs for the input and output vectors. 

Section~\ref{2D} evaluates the 2D partitioning techniques. Section~\ref{2D-Studies} presents three characterization studies on (a) performing fine-grained data transfers to transfer the elements of the input and output vectors to/from PIM-enabled memory, (b) the scalability of 2D partitioning techniques to thousands of DPUs, and (c) the number of vertical partitions to perform on the matrix. Section~\ref{2D-Formats} compares the end-to-end performance of all compressed matrix formats for each of the three types of 2D partitioning techniques. Section~\ref{2D-Comparison} compares the best-performing \spmv{} implementations of all three types of 2D partitioning techniques. 

Section~\ref{1D-2D} compares the best-performing (\camone{on average across all matrices and data types}) \spmv{} implementations of the 1D and 2D partitioning techniques.

\subsection{Analysis of \spmv{} Execution Using 1D Partitioning Techniques}\label{1D}
We evaluate the 1D partitioning \camone{schemes} highlighted in bold in Table~\ref{table:library}. Specifically, for \texttt{COO.nnz}, we present the coarse-grained locking (\texttt{COO.nnz-lb}) and lock-free (\texttt{COO.nnz-lf}) approaches, since the fine-grained locking approach performs similarly with the coarse-grained locking approach, as shown in the previous section (Section~\ref{1DPU-MulTskl}). Similarly, for the BCSR (int8 data type) and BCOO formats, we present only the coarse-grained locking approach, since all synchronization approaches perform similarly (Section~\ref{1DPU-MulTskl}). Finally, in all experiments presented henceforth, we use 16 tasklets and \camone{load-balance} the non-zero elements \camone{across} tasklets within the DPU, since this load balancing scheme provides the highest performance benefits on average across all matrices and data types, according to our evaluation shown in Section~\ref{1DPU}.

\subsubsection{\textbf{Analysis of \texttt{Kernel} Time}}\label{1D-Kernel}  \hfill  \\
We compare the \texttt{kernel} time of \spmv{} achieved by various load balancing schemes for each particular compressed \camone{matrix} format, and then we compare the \texttt{kernel} time of the compressed \camone{matrix} formats.

\begin{figure}[t]
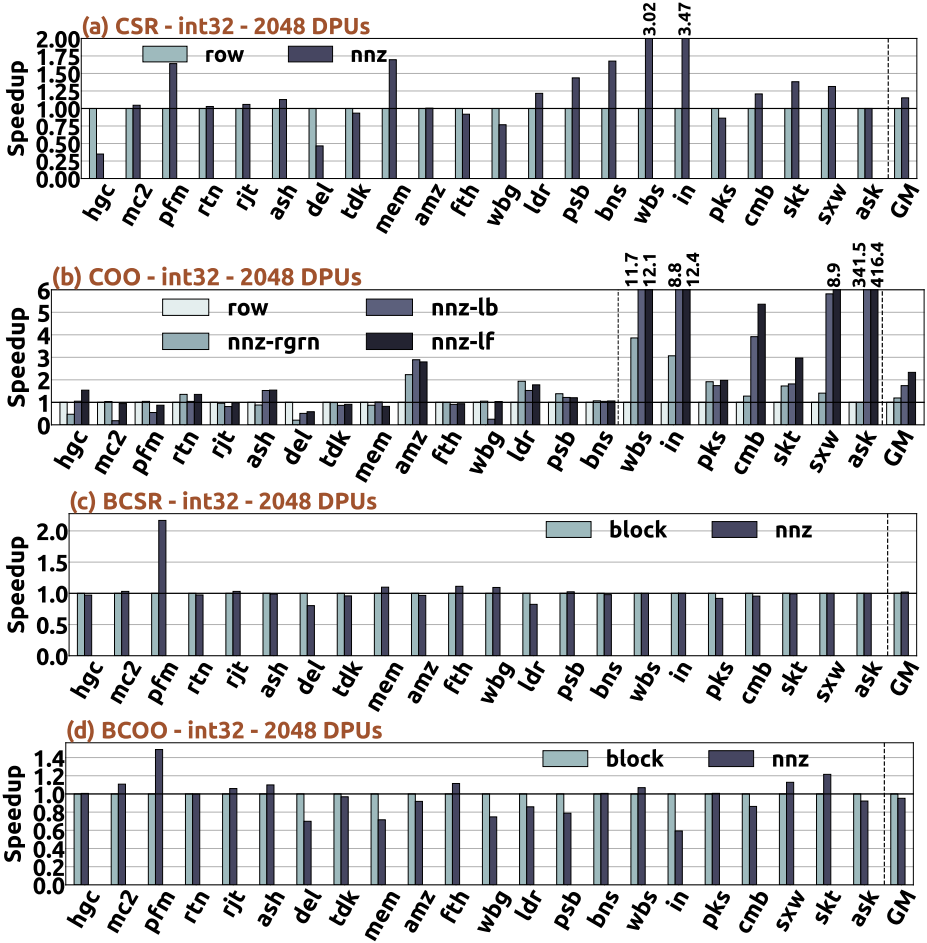

\centering
\includegraphics[width=.88\textwidth]{1D-partitioning/csr_balance_dpu_2048_int32.pdf}
\includegraphics[width=.88\textwidth]{1D-partitioning/coo_balance_dpu_2048_int32.pdf}
\includegraphics[width=.88\textwidth]{1D-partitioning/bcsr_balance_dpu_2048_int32.pdf}
\includegraphics[width=.88\textwidth]{1D-partitioning/bcoo_balance_dpu_2048_int32.pdf}
\vspace{-7pt}
\caption{Performance comparison of load balancing techniques for each particular compressed format using 2048 DPUs and the int32 data type.}
\label{fig:1D_balance}
\vspace{-12pt}
\end{figure}

\noindent\textbf{Analysis of Load Balancing Schemes Across DPUs.}
Figure~\ref{fig:1D_balance} compares load balancing techniques for each compressed \camone{matrix} format using 2048 DPUs and the int32 data type. 

We draw four findings. First, we observe that \texttt{CSR.nnz} and \texttt{COO.nnz-rgrn}, i.e., balancing the non-zero elements \camone{across} DPUs (at row granularity), either outperform or perform similarly to \texttt{CSR.row} and \texttt{COO.row}, respectively, i.e., balancing the rows \camone{across} DPUs, \camone{except for} \texttt{hgc} and \texttt{del} matrices. In these two matrices, \texttt{CSR.nnz} and \texttt{COO.nnz-rowgrn} incur a high \camone{disparity} in rows assigned to DPUs, i.e., only one DPU processes $4\times$ and $11\times$ more rows than the rest, for \texttt{hgc} and \texttt{del} matrices, respectively. This in turn creates a high \camone{disparity} in the elements of the output vector processed \camone{across} DPUs, \camone{causing} performance \camone{to be} limited by the DPU that processes the \camone{largest} number of rows. Thus, \camone{we find that} adaptive load balancing approaches and selection methods based on the characteristics of each input matrix need to be developed \camone{to achieve high performance across all matrices}.

\begin{tcolorbox}
\noindent\textbf{OBSERVATION 4:} \\
\textit{Adaptive} load balancing schemes and selection methods for the balancing scheme on rows/blocks/non-zero elements based on the characteristics of each input matrix need to be developed to provide best performance across all matrices.
\end{tcolorbox}

Second, we find that \texttt{COO.nnz-lb} and \texttt{COO.nnz-lf}, which provide an almost perfect non-zero element balance \camone{across} DPUs, significantly outperform \texttt{COO.row} and \texttt{COO.nnz-rgrn} in \textit{scale-free} matrices (i.e., from \texttt{wbs} to \texttt{ask} matrices) \camone{by on average 6.73$\times$}. Scale-free matrices have only a few rows, that include a much \camone{larger} number of non-zero elements \camone{compared to} the remaining rows of the matrix. Therefore, perfectly balancing the non-zero elements \camone{across} DPUs provides high performance gains. 

\begin{tcolorbox}
\noindent\textbf{OBSERVATION 5:} \\
\textit{Perfectly balancing the non-zero elements} \camone{across} PIM cores can provide significant performance benefits in \textit{highly irregular, scale-free matrices.}
\end{tcolorbox}

Third, we find that the lock-free \texttt{COO.nnz-lf} scheme outperforms the lock-based \texttt{COO.nnz-lb} scheme by 1.34$\times$ on average, and provides high performance benefits when there is a high row imbalance \camone{across} tasklets within the DPU. When one tasklet processes a much \camone{larger} number of rows \camone{versus} the rest, it executes a much \camone{larger} number of critical sections. As a result, the core pipeline incurs high imbalance in lock acquisitions/releases, \camone{causing} the lock-based approach \camone{to incur} high performance overheads in relatively compute-bound DPUs~\cite{Gomez2021Benchmarking,Gomez2021Analysis}.

\begin{tcolorbox}
\noindent\textbf{OBSERVATION 6:} \\
\textit{Lock-free} approaches can provide high performance benefits over \textit{lock-based} approaches in PIM architectures, \camone{because they minimize synchronization overheads in PIM cores.}
\end{tcolorbox}

Finally, in the BCSR and BCOO formats, balancing the blocks \camone{across DPUs} performs similarly (on average across all matrices) to balancing the non-zero elements \camone{across DPUs}.

\begin{figure}[H]
\vspace{-4pt}
\begin{minipage}{\textwidth}
\centering
\includegraphics[width=.41\textwidth]{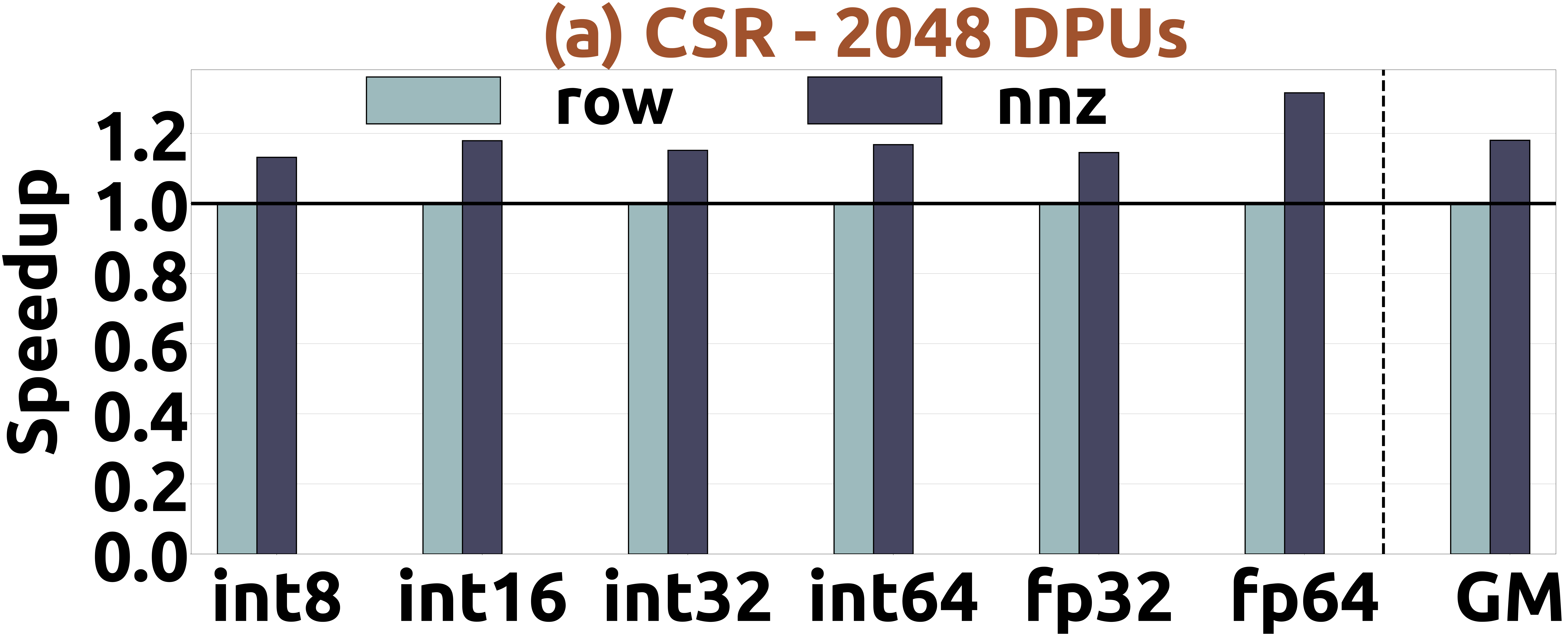} \hspace{10pt}
\includegraphics[width=.41\textwidth]{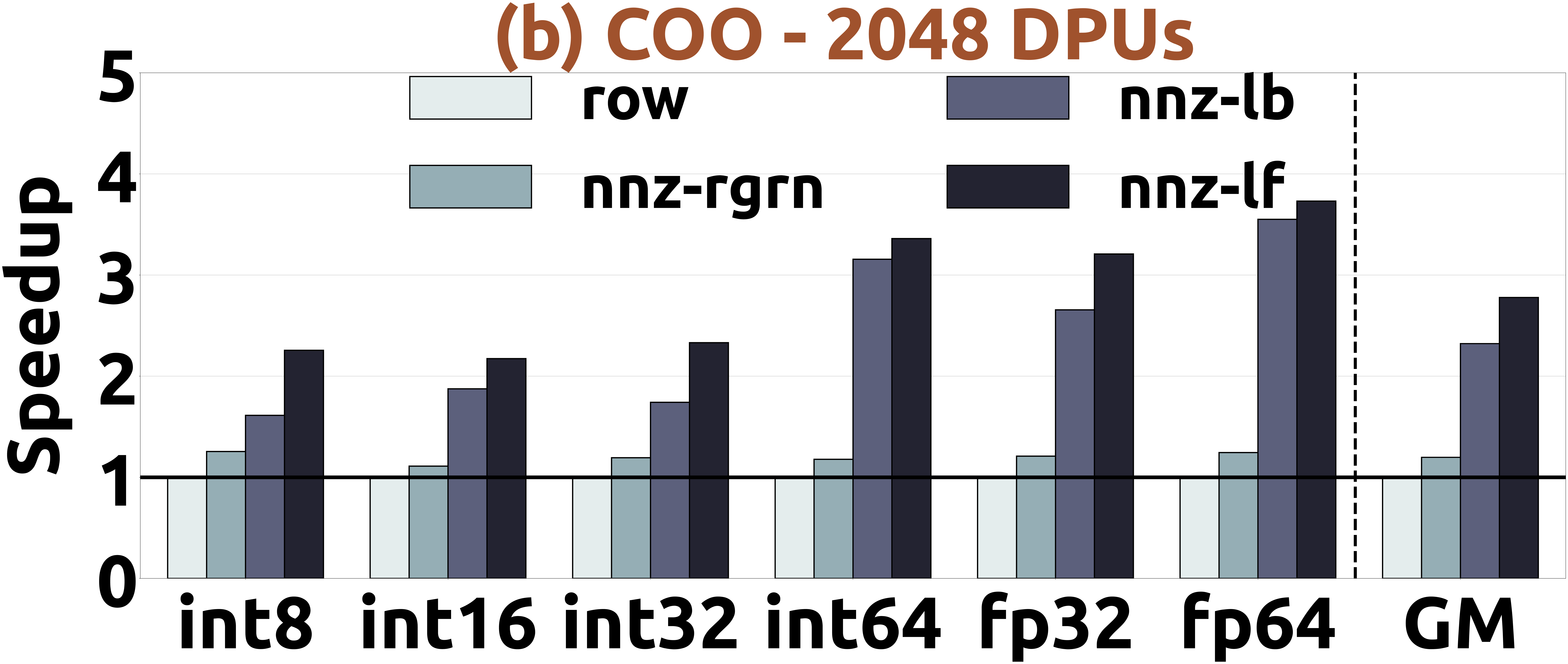} 
\vspace{8pt}
\end{minipage} 
\begin{minipage}{\textwidth}
\centering
\includegraphics[width=.41\textwidth]{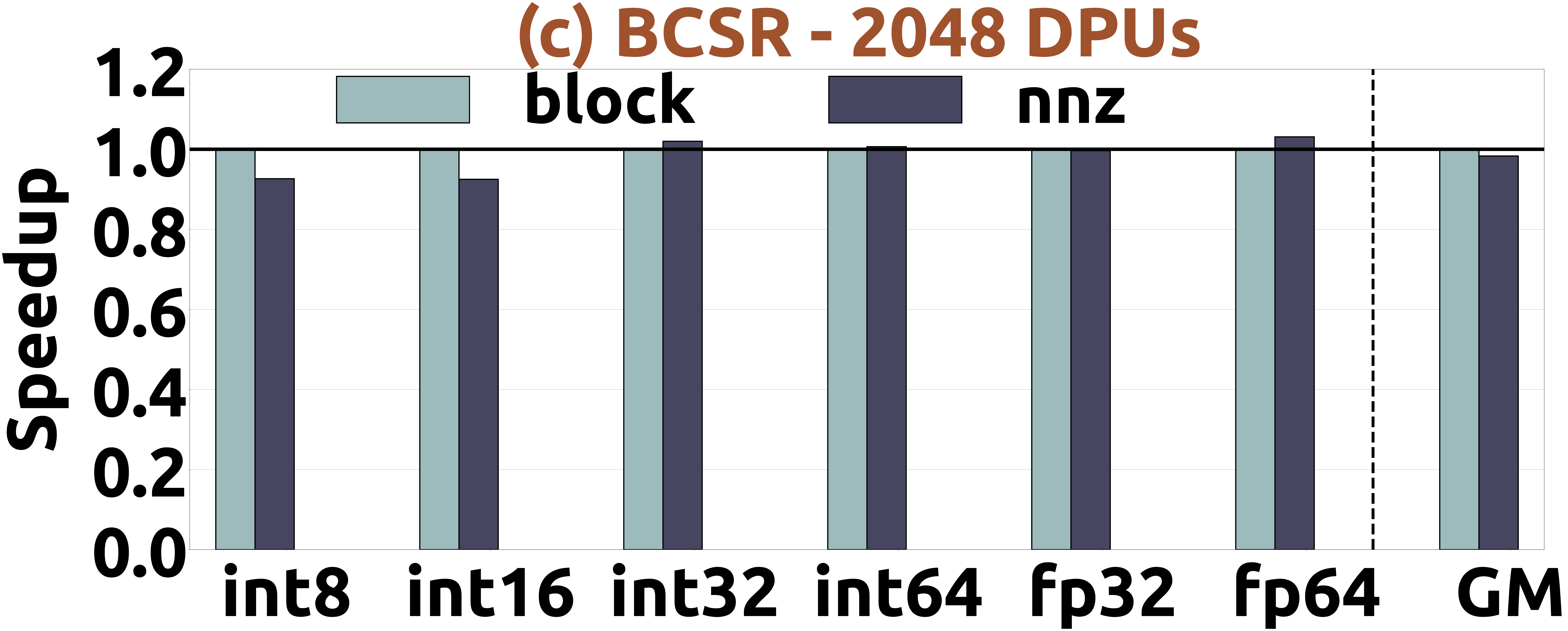} \hspace{10pt}
\includegraphics[width=.41\textwidth]{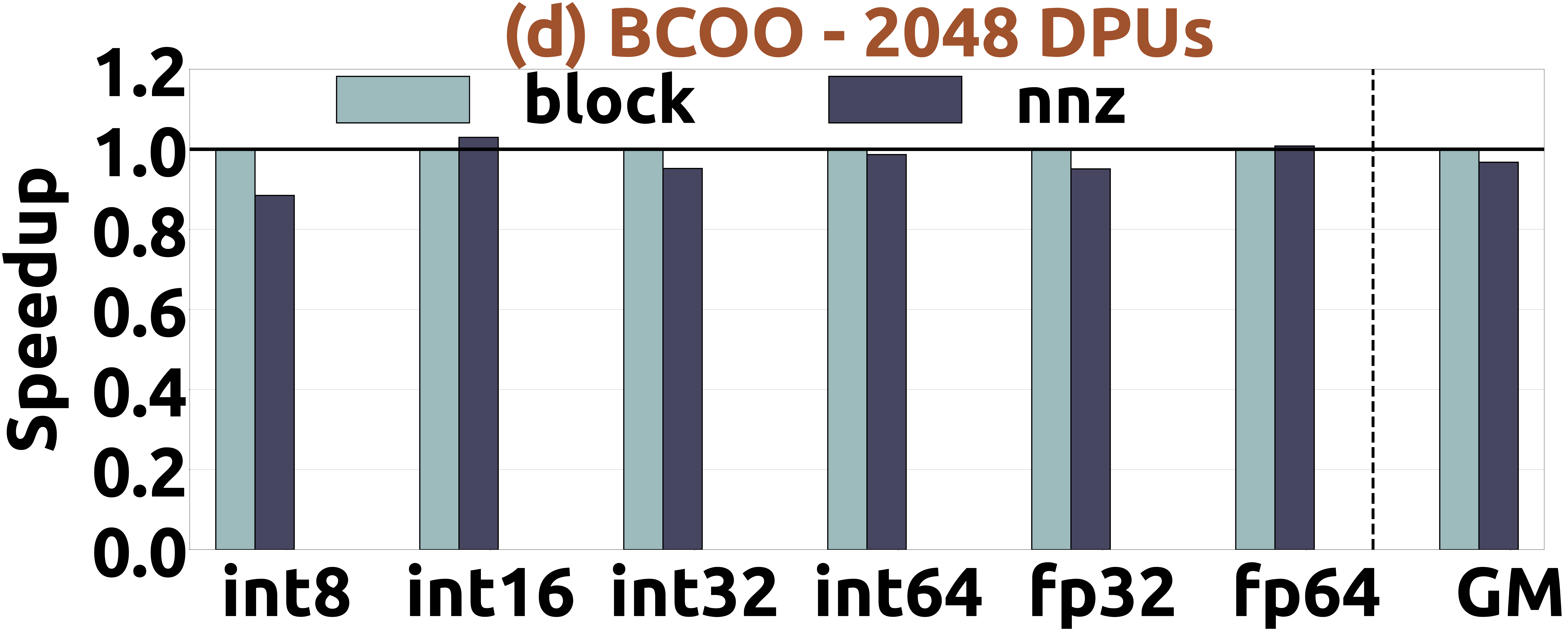}
\end{minipage}  
\vspace{-6pt}
\caption{Performance comparison of load balancing techniques for each data type using 2048 DPUs.}
\label{fig:1D_datatypes}
\vspace{-14pt}
\end{figure}

To further investigate the performance of the various load balancing schemes, Figure~\ref{fig:1D_datatypes} compares them using all the data types. We present the geometric mean of all matrices using 2048 DPUs. In the CSR and COO formats, balancing the non-zero elements \camone{across} DPUs on average outperforms balancing the rows \camone{across} DPUs by 1.18$\times$ and 1.20$\times$, respectively. We observe that in the COO format almost perfectly balancing the non-zero elements \camone{across} DPUs provides significant performance benefits \camone{(2.55$\times$, averaged across all the data types)}, \camone{compared to} balancing the rows, especially when multiplication is not supported by hardware (e.g., for the floating point data types). \camone{In contrast}, in the BCSR and BCOO formats, balancing the blocks \camone{across} DPUs performs only slightly better (on average 2.7\% \camone{across all the data types}) than balancing the non-zero elements.

\begin{figure}[b]
    \centering
    \includegraphics[width=\textwidth]{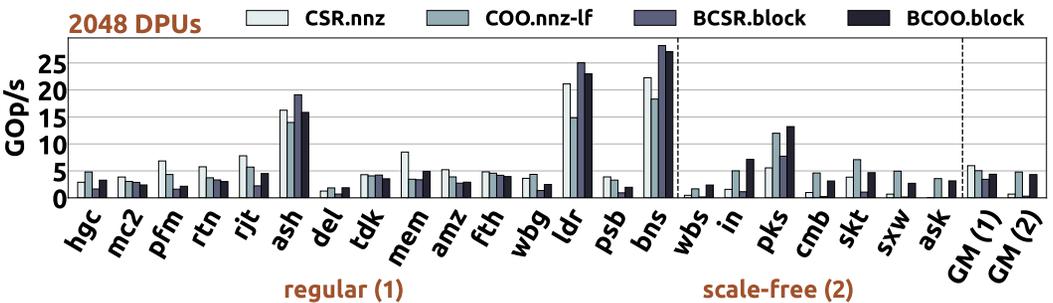}
    \vspace{-20pt}
    \caption{\camone{Throughput of various} compressed formats using 2048 DPUs and the int32 data type.}
    \label{fig:1D_kernel}
\end{figure}

\begin{figure}[t]
    \centering
    \includegraphics[width=\textwidth]{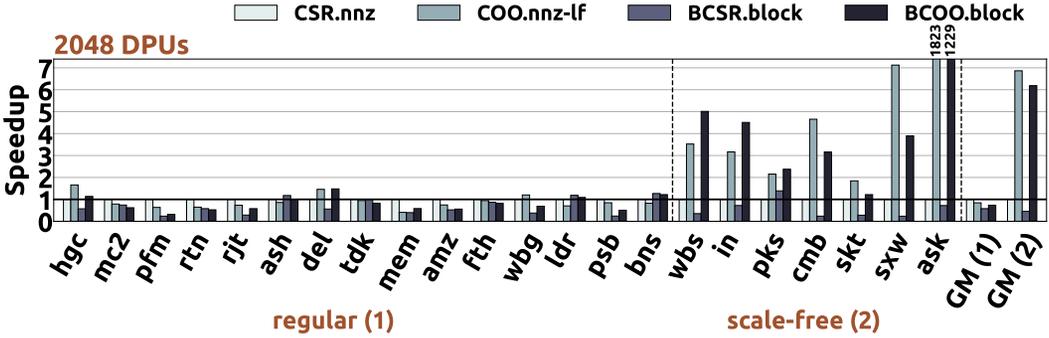}
    \vspace{-20pt}
    \caption{Performance comparison of \camone{various} compressed formats using 2048 DPUs and the int32 data type. \camone{Performance is normalized to that of \texttt{CSR.nnz}.}}
    \label{fig:1D_kernel_perf}
    \vspace{-12pt}
\end{figure}

\noindent{\textbf{Comparison of Compressed Matrix Formats.}}
Figures~\ref{fig:1D_kernel} \camone{and ~\ref{fig:1D_kernel_perf} compare the throughput (in GOperations per second) and the performance, respectively,} achieved by various compressed formats using 2048 DPUs and the int32 data type. For \camone{the} CSR and COO formats, we select balancing the non-zero elements \camone{across DPUs}, and for \camone{the} BCSR and BCOO formats, we select balancing the blocks \camone{across DPUs}, since these are the best-performing schemes for each format averaged across all matrices and data types (Figure~\ref{fig:1D_datatypes}).


We draw four findings. First, matrices that exhibit block pattern \camone{(almost all non-zero elements of the matrix fit in dense sub-blocks)}, i.e., \texttt{ash}, \texttt{ldr}, \texttt{bns}, \texttt{pks} matrices, have the highest throughput, since they leverage higher data locality compared to matrices with non-block pattern. Second, in scale-free matrices, the COO and BCOO formats significantly outperform the CSR and BCSR formats \camone{by 6.94$\times$ and 13.90$\times$}, respectively. This is because they provide better non-zero element balance \camone{across} DPUs. In the CSR and BCSR formats, \camone{the non-zero element balance is \camone{limited to be} performed at row and block-row granularity, respectively, causing performance to be limited by the DPU that processes the largest number of non-zero elements.} Third, we observe that the BCOO format can outperform the CSR format even in \textit{non-blocked} scale-free matrices. Fourth, we find that when the CSR and BCSR formats provide sufficient non-zero element balance \camone{across} DPUs, i.e., in many regular matrices such as \texttt{rtn}, \texttt{tdk}, \texttt{amz}, and \texttt{fth}, they can outperform the COO and BCOO formats, respectively.

\begin{tcolorbox}
\noindent\textbf{OBSERVATION 7:} \\ 
In \textit{scale-free} matrices, the COO and BCOO formats significantly outperform the CSR and BCSR formats, because they provide higher non-zero element balance \camone{across} PIM cores.
\end{tcolorbox}

\subsubsection{\textbf{Analysis of End-To-End \spmv{} Execution}}\label{1D-EndToEnd}  
\begin{figure}[H]
    \centering
    \includegraphics[width=\textwidth]{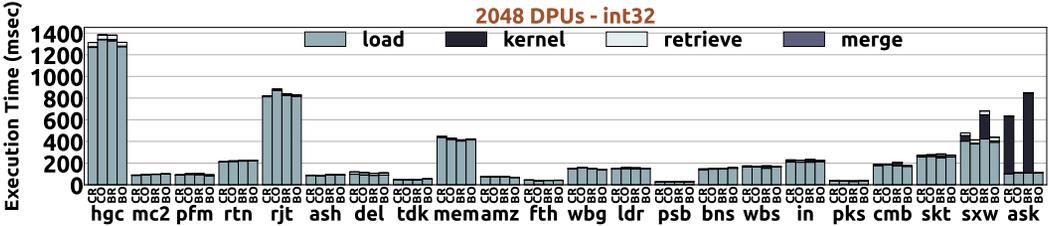}
    \vspace{-12pt}
    \caption{Total execution time when using 2048 DPUs and the int32 data type for CR: \texttt{CSR.nnz}, CO:  \texttt{COO.nnz-lf}, BR: \texttt{BCSR.block} and BO: \texttt{BCOO.block} kernels.}
    \label{fig:1D_transfers}
    \vspace{-6pt}
\end{figure}

\hfill  \\
Figure~\ref{fig:1D_transfers} shows the end-to-end execution time of \camone{1D-partitioned} kernels using 2048 DPUs and the int32 data type. The times are broken down into (i) the time for CPU to DPU transfer to load the input vector \camone{into DRAM} banks (\texttt{load}), (ii) the kernel time on DPUs (\texttt{kernel}), (iii) the time for DPU to CPU transfer to retrieve the results for the output vector (\texttt{retrieve}), and (iv) the time to merge partial results on the host CPU cores (\texttt{merge}).

We draw four findings. First, the \texttt{load} data transfers constitute more than 90\% of the total execution time, because the input vector is replicated and broadcast \camone{into} each DPU, \camone{causing a large number} of bytes \camone{to be} transferred through the narrow off-chip memory bus. An exception is in the CSR and BCSR formats for \texttt{sxw}, \texttt{ask} matrices, which include one very dense row, \camone{and} thus \texttt{kernel} time is highly bottlenecked by one DPU that processes a significantly \camone{larger} number of non-zero elements \camone{than} the rest. Second, the \texttt{kernel} time constitutes on average only 4.3\% of the total execution time, since \spmv{} is effectively parallelized to thousands of DPUs. Third, the \texttt{retrieve} data transfers constitute on average 3.4\% of the total execution time, because the output vector is split across DPUs. Fourth, the \texttt{merge} time on the host CPU is negligible (less than 1\% of the total execution time), since only a few partial results for the elements of the output vector are merged by the host CPU cores in the 1D partitioning techniques.

\begin{tcolorbox}
\noindent\textbf{OBSERVATION 8:} \\
The end-to-end performance of the 1D partitioning techniques is severely bottlenecked by the data transfer costs to replicate and broadcast the whole input vector \camone{into} \textit{each} DRAM bank of PIM cores, which takes place through the narrow off-chip memory bus.
\end{tcolorbox}

\begin{figure}[H]
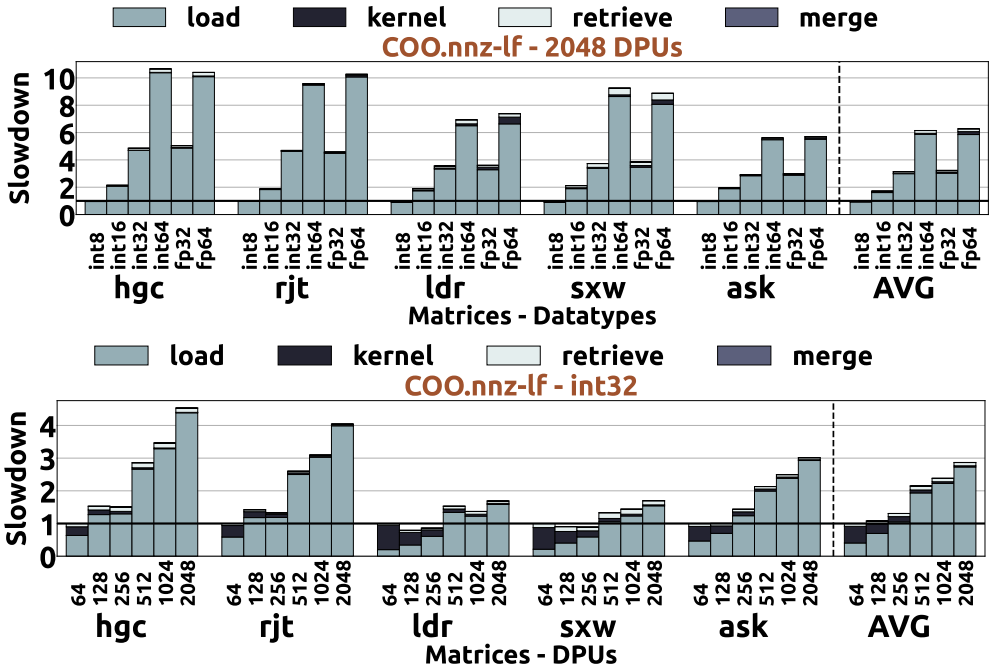

    \includegraphics[width=.94\textwidth]{1D-partitioning/transfers_datatypes_dpus2048.pdf}
    \includegraphics[width=.94\textwidth]{1D-partitioning/transfers_scalability_int32.pdf}
    \vspace{-4pt}
    \caption{End-to-end execution time breakdown achieved by \texttt{COO.nnz-lf} when varying (a) the data type using 2048 DPUs (normalized to \camone{the experiment for the} int8 data type), and (b) the number of DPUs for the int32 data type (normalized to 64 DPUs).}
    \label{fig:1D_transfers_scalability}
\end{figure}

To further investigate on the costs to the load input vector \camone{into} all DRAM banks of PIM-enabled memory, we present in 
Figure~\ref{fig:1D_transfers_scalability} the total execution time achieved by \texttt{COO.nnz-lf} when varying (a) the data type using 2048 DPUs (normalized to \camone{the experiment for the} int8 data type), and (b) the number of DPUs for the int32 data type (normalized to 64 DPUs).

We draw two conclusions. First, the \texttt{load} data transfer costs increase proportionally to the number of bytes of the data type, and still dominate performance even for the data type with the smallest memory footprint (int8). Second, the \texttt{load} data transfer costs and the associated memory footprint for the input vector increase proportionally to the number of DPUs used, and thus the best end-to-end performance is achieved using only a small portion of the available DPUs on the system.

\begin{tcolorbox}
\noindent\textbf{OBSERVATION 9:} \\
\spmv{} execution of the 1D-partitioned schemes cannot scale up to a \camone{large} number of PIM cores due to high data transfer overheads to copy the input vector \camone{into} \textit{each} DRAM bank of PIM-enabled memory.
\end{tcolorbox}

\subsection{Analysis of \spmv{} Execution Using 2D Partitioning Techniques}\label{2D}
We evaluate the 2D-partitioned kernels highlighted in bold in Table~\ref{table:library}. Specifically, for the COO format we use the lock-free approach, and for the BCSR (in the int8 data type) and BCOO formats we use the coarse-grained locking approach. In the \equallyWidth{} and \variableSized{} techniques, for the BCSR and BCOO formats we balance the blocks \camone{across} DPUs of the same vertical partition, since \camone{doing so} performs slightly better than balancing the non-zero elements, as explained in Section~\ref{1D-Kernel}. In all experiments, we balance the non-zero elements \camone{across} 16 tasklets within \camone{a single} DPU.

\subsubsection{\textbf{Sensitivity Studies on 2D Partitioning Techniques}}\label{2D-Studies}  \hfill  \\
We present three characterization studies on the 2D partitioning techniques. First, we evaluate the performance of fine-grained data transfers from/to PIM-enabled memory for the input and output vectors. Second, we evaluate the scalability of the 2D partitioning techniques to thousands of DPUs. Finally, we explore performance implications \camone{on} the number of vertical partitions used in the 2D-partitioned kernels.

\begin{figure}[t]
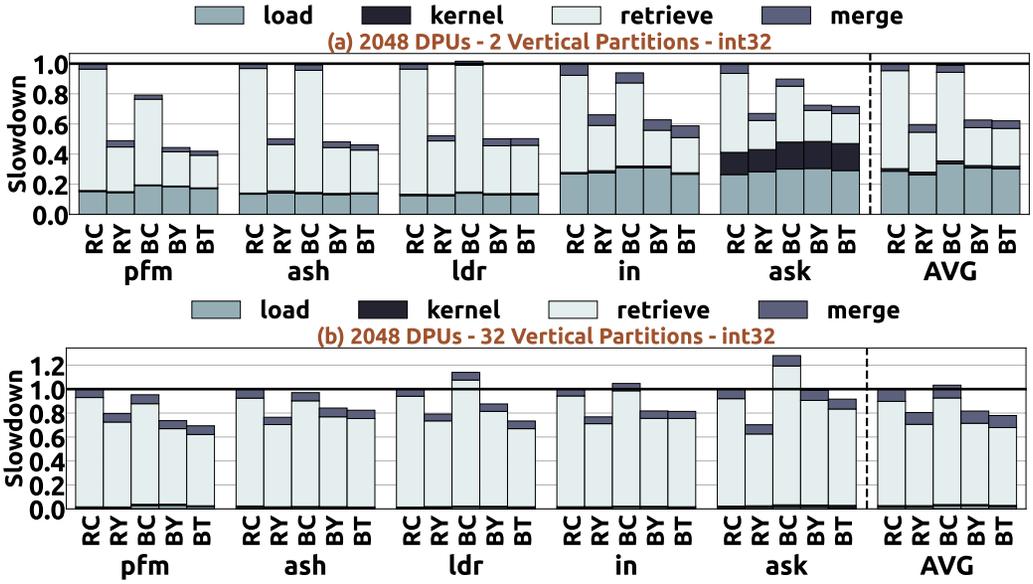

    \centering
    \includegraphics[width=.98\textwidth]{2D-partitioning/trans_reduced_dpus2048_int32_cps2.pdf}
    \includegraphics[width=.98\textwidth]{2D-partitioning/trans_reduced_dpus2048_int32_cps32.pdf}
    \vspace{-8pt}
    \caption{Performance comparison of \texttt{RC}: \texttt{RBDCOO} with coarse-grained transfers, \texttt{RY}: \texttt{RBDCOO} with fine-grained transfers in the output vector, \texttt{BC}: \texttt{BDCOO} with coarse-grained transfers, \texttt{BY}: \texttt{BDCOO} with fine-grained transfers only in the output vector, and \texttt{BT}: \texttt{BDCOO} with fine-grained transfers in both the input and the output vector using the int32 data type, 2048 DPUs and having 2 (left) and 32 (right) vertical partitions. Performance is normalized to that of the \texttt{RC} scheme.}
    \label{fig:2D_fgtransfers}
\end{figure}

\noindent{\textbf{Analysis of Fine-Grained Data Transfers.}}
The UPMEM API~\cite{upmem-guide} has the limitation that \textit{the transfer sizes from/to all DRAM banks involved in the same parallel transfer need to be the same}. The UPMEM API provides \textit{parallel data transfers} either \camone{to} all DPUs of all ranks (henceforth referred to as \textit{coarse-grained} transfers), or at rank granularity, i.e., \camone{to} 64 DPUs of the same rank (henceforth referred to as \textit{fine-grained} transfers). In the first case, parallel data transfers are performed \camone{to} all DPUs used at once, padding with empty bytes \camone{at the granularity of \textit{all} DPUs used, e.g., 2048 DPUs in Figure~\ref{fig:2D_fgtransfers}}. In the latter case, programmers iterate over \camone{the} ranks of PIM-enabled DIMMs, and for \textit{each} rank perform parallel data transfers \camone{to the 64 DPUs of the same rank} padding with empty bytes at the granularity of 64 DPUs.

In \camone{\spmv{} execution, for} the \equallyWidth{} and \variableSized{} techniques the heights and widths of 2D tiles vary, \camone{and thus} padding with empty bytes is necessary for the \texttt{load} and \texttt{retrieve} data transfers of the elements of the input and output vector, respectively. Figure~\ref{fig:2D_fgtransfers} compares coarse-grained data transfers, i.e., performing parallel transfers \camone{to} all 2048 DPUs \camone{at once}, with fine-grained data transfers, i.e., iterating over \camone{the} ranks and \camone{for each rank} performing parallel transfers \camone{to the 64} DPUs of the same rank. We evaluate both the \equallyWidth{} and \variableSized{} techniques using the COO format and with 2 and 32 vertical partitions. Please see Appendix~\ref{sec:appendix-2D-fgtrans} for all matrices.

We draw two findings. First, when the number of vertical partitions is small, e.g., 2 vertical partitions, the \camone{disparity in} widths \camone{across} tiles in the \variableSized{} scheme is low. Thus, \texttt{BT} only slightly outperforms \texttt{BY} by 1\% on average, since in \texttt{BY} \textit{only} a small amount of padding is added \camone{on the \texttt{load} data transfers of} the input vector. \camone{In contrast, the disparity in heights across tiles in the \equallyWidth{} and \variableSized{} schemes is high.} Thus, \texttt{RY} and \texttt{BY} significantly outperform \texttt{RC} and \texttt{BC} by \camone{an average of} 1.68$\times$ and 1.60$\times$, respectively. This is because fine-grained transfers \camone{to retrieve the elements of} the output vector significantly decrease the amount of bytes transferred from PIM-enabled memory to host CPU \camone{over coarse-grained transfers}. Second, when the number of vertical partitions is \camone{large}, e.g., 32 \camone{vertical partitions}, the \camone{disparity in} heights \camone{across} tiles in the \equallyWidth{} and \variableSized{} schemes is lower compared to \camone{when the number of vertical partitions is small}. Thus, \texttt{RY} and \texttt{BY} provide smaller performance benefits over \texttt{RC} and \texttt{BC} (on average 1.24$\times$ and 1.22$\times$, respectively), \camone{respectively,} compared to a small number of vertical partitions. \camone{In contrast, the disparity in heights across tiles in the \equallyWidth{} and \variableSized{} schemes is higher compared to when the number of vertical partitions is small. Thus, \texttt{BT} outperforms \texttt{BY} by 4.7\% on average.} 
Overall, we conclude that fine-grained data transfers (i.e., at rank granularity in the UPMEM PIM system) can significantly improve performance in the \equallyWidth{} and \variableSized{} schemes.

\begin{tcolorbox}
\noindent\textbf{OBSERVATION 10:} \\ 
\textit{Fine-grained} parallel transfers in the \equallyWidth{} and \variableSized{} 2D partitioning techniques, i.e., minimizing the amount of padding with empty bytes in parallel data transfers to/from PIM-enabled memory, can provide large performance gains. 
\end{tcolorbox}

\begin{figure}[H]
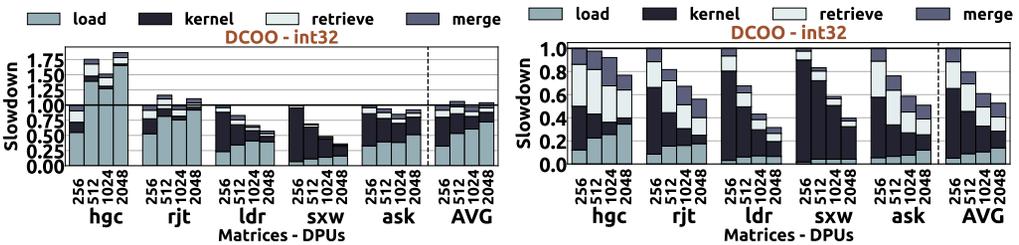

    \centering
    \begin{minipage}{\textwidth}
    \includegraphics[width=.484\textwidth]{2D-partitioning/scalability_fix_int32_col4.pdf}
    \includegraphics[width=.484\textwidth]{2D-partitioning/scalability_fix_int32_col16.pdf}
    \end{minipage}
    \vspace{-8pt}
    \caption{Execution time breakdown of \equallySized{} partitioning technique of the COO format using 4 (left) and 16 (right) vertical partitions when varying the number of DPUs used for the int32 data type. Performance is normalized to that with 256 DPUs.}
    \label{fig:2D_scale_equally_sized}
\end{figure}

\begin{figure}[t]
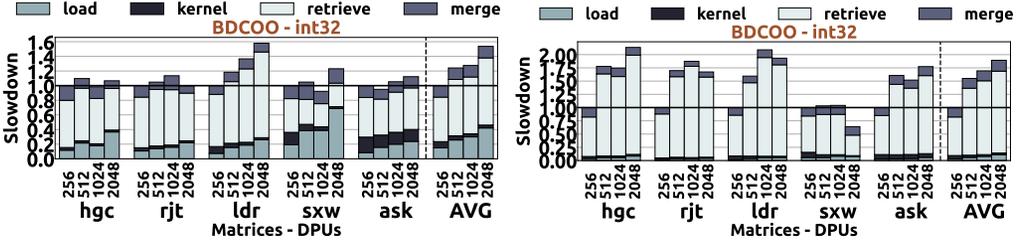

    \centering
    \begin{minipage}{\textwidth}
    \includegraphics[width=.484\textwidth]{2D-partitioning/scalability_bal_int32_col4.pdf}
    \includegraphics[width=.484\textwidth]{2D-partitioning/scalability_bal_int32_col16.pdf}
    \end{minipage}
    \vspace{-8pt}
    \caption{Execution time breakdown of \variableSized{} partitioning technique of the COO format using 4 (left) and 16 (right) vertical partitions when varying the number of DPUs used for the int32 data type. Performance is normalized to that with 256 DPUs.}
    \label{fig:2D_scale_variable_sized}
    \vspace{-14pt}
\end{figure}

\noindent\textbf{Scalability of the 2D Partitioning Techniques.}
We \camone{analyze} scalability \camone{with} the number of DPUs for the 2D partitioning techniques. Figures~\ref{fig:2D_scale_equally_sized}, ~\ref{fig:2D_scale_equally_wide} and ~\ref{fig:2D_scale_variable_sized} compare the performance of the \equallySized{}, \equallyWidth{} and \variableSized{} schemes, respectively, using the COO format and the int32 data type, as the number of DPUs increases.

\begin{figure}[H]
    \centering
    \begin{minipage}{\textwidth}
    \includegraphics[width=.484\textwidth]{2D-partitioning/scalability_rbal_int32_col4.pdf}
    \includegraphics[width=.484\textwidth]{2D-partitioning/scalability_rbal_int32_col16.pdf}
    \end{minipage}
    \vspace{-8pt}
    \caption{Execution time breakdown of \equallyWidth{} partitioning technique of the COO format using 4 (left) and 16 (right) vertical partitions when varying the number of DPUs used for the int32 data type. Performance is normalized to that with 256 DPUs.}
    \label{fig:2D_scale_equally_wide}
    \vspace{-14pt}
\end{figure}

We draw two findings. First, the \equallySized{} scheme (i.e., \texttt{DCOO}) achieves high scalability with a \camone{large} number of vertical partitions. The \texttt{kernel} time of \equallySized{} scheme is mainly limited by the DPU (or a few DPUs) that processes the largest number of non-zero elements. With a \camone{large} number of \textit{static} vertical partitions, the non-zero element \camone{disparity across} DPUs is high, i.e., the \texttt{kernel} time is highly bottlenecked by the DPU that processes the \camone{largest} number of non-zero elements. As a result, increasing the number of DPUs improves performance by decreasing the \texttt{kernel} time via \camone{better} non-zero element balance \camone{across} DPUs. 


\begin{tcolorbox}
\noindent\textbf{OBSERVATION 11:} \\
The \texttt{kernel} time in the \equallySized{} schemes is limited by the PIM core (or a few PIM cores) assigned to the 2D tile with the largest number of non-zero elements. 
\end{tcolorbox}

Second, we observe that the \equallyWidth{} and \variableSized{} schemes (i.e., \texttt{RBDCOO} and \texttt{BDCOO}) are severely bottlenecked by \texttt{retrieve} data transfer costs (a large number of partial results is created on PIM cores), and thus they are \camone{difficult} to scale up to thousands of DPUs. Moreover, when the number of vertical partitions is high, the \camone{disparity in} heights of the tiles is high. Thus, as the number of DPUs increases, the amount of padding needed in \texttt{retrieve} data transfers \camone{becomes very large, causing significant performance degradation.}

\begin{tcolorbox}
\noindent\textbf{OBSERVATION 12:} \\
The scalability of the \equallyWidth{} and \variableSized{} schemes to a \camone{large} number of PIM cores is severely limited by \camone{large} data transfer overheads to retrieve partial results for the elements of the output vector from the DRAM banks \camone{of PIM-enabled memory} to the host CPU via the narrow memory bus.
\end{tcolorbox}

\begin{figure}[H]
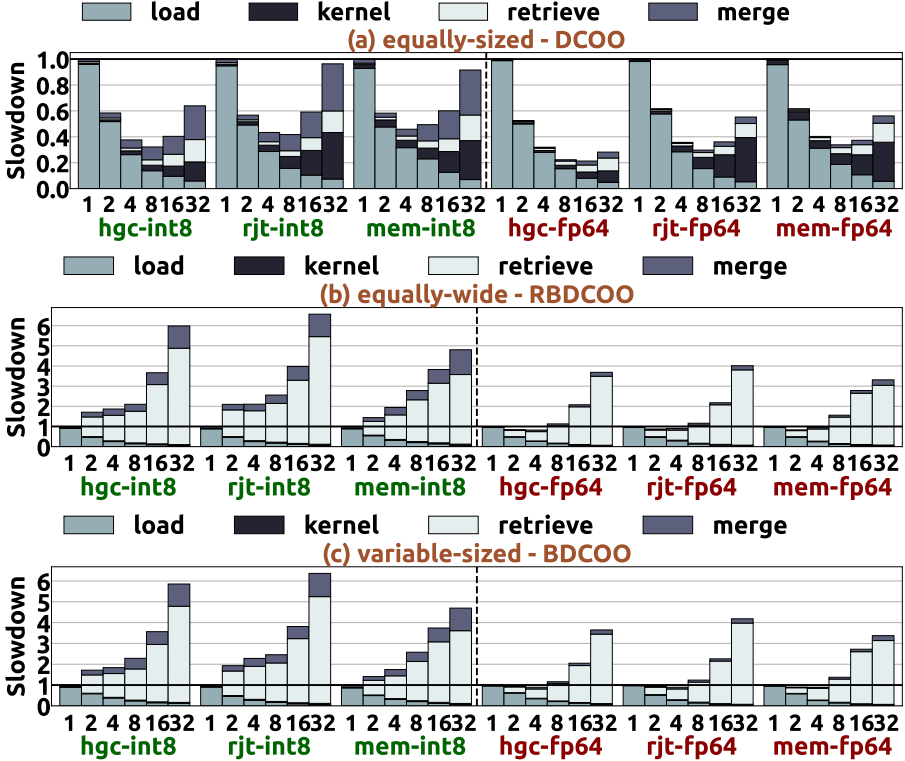

    \centering
    \begin{minipage}{\textwidth}
    \centering
    \includegraphics[width=0.86\textwidth]{2D-partitioning/fixed_vertical_dpus2048.pdf}
    \includegraphics[width=0.86\textwidth]{2D-partitioning/rbal_vertical_dpus2048.pdf}
    \includegraphics[width=0.86\textwidth]{2D-partitioning/bal_vertical_dpus2048.pdf}
    \end{minipage}
    \vspace{-6pt}
    \caption{Execution time breakdown of 2D partitioning schemes using the COO format and 2048 DPUs when varying the number of vertical partitions from 1 to 32 for the int8 and fp64 data types. Performance is normalized to the performance of the experiment with 1 vertical partition.}
    \label{fig:2D_vertpartitions}
    \vspace{-10pt}
\end{figure}

\noindent\textbf{\camone{Effect of} the Number of Vertical Partitions.}
In all experiments presented henceforth, we perform fine-grained data transfers \camone{(at rank granularity, i.e., 64 DPUs in the UPMEM PIM system)} in the 2D partitioning schemes. Figure~\ref{fig:2D_vertpartitions} evaluates performance implications \camone{on} the number of vertical partitions performed in 2D-partitioned kernels. We use the COO format and vary the number of vertical partitions from 1 to 32, \camone{in steps of multiple of 2}. We draw four findings.

First, in the \equallySized{} scheme, as the number of vertical partitions increases, \texttt{kernel} time increases, if there is \camone{\textit{no}} dense row in the matrix. This is because the \camone{disparity in the} non-zero elements \camone{across 2D} tiles increases \camone{as the number of vertical partitions increases.} Thus, performance is limited by one DPU or a few DPUs that process the largest number of non-zero elements.

\begin{tcolorbox}
\noindent\textbf{OBSERVATION 13:} \\
As the number of vertical partitions increases, the \equallySized{} 2D \camone{partitioning} scheme typically increases the non-zero element \camone{disparity across} PIM cores (unless there is one dense row on the matrix), thereby increasing the \texttt{kernel} time. 
\end{tcolorbox}

Second, as the number of vertical partitions increases, \texttt{retrieve} data transfer costs and \texttt{merge} time increase. This is because the partial results created for the output vector increase proportionally with the number of vertical partitions. The performance overheads of \texttt{retrieve} data transfer costs are highly affected by the characteristics of the underlying hardware (e.g., the bandwidth provided on I/O channels of the memory bus between host CPU and PIM-enabled DIMMs). Similarly, the performance cost of the \texttt{merge} step depends on the \camone{hardware} characteristics of the host CPU (e.g., \camone{the} number of \camone{the} CPU cores, the available hardware threads, microarchitecture of CPU cores). \camone{We refer the reader to Appendix~\ref{sec:appendix-2D-vertpartitions} for a comparison of \spmv{} execution using two different UPMEM PIM systems with different hardware characteristics (Table ~\ref{tab:pim-systems}).}

Third, we find that in the \equallyWidth{} and \variableSized{} schemes, there is high \camone{disparity in} heights of 2D tiles, and as a result on the number of partial results created \camone{across} DPUs. Even with fine-grained parallel \texttt{retrieve} data transfers at rank granularity, the amount of padding needed in the \equallyWidth{} and \variableSized{} schemes is at 88.6\% and 88.0\%, respectively, causing high bottlenecks in the narrow memory bus. Therefore, in PIM systems that do not support \camone{very} fine-grained parallel transfers \camone{to gather results from PIM-enabled memory to the host CPU} \textit{at DRAM bank granularity}, execution is highly limited by the amount of padding performed in \texttt{retrieve} data transfers, which can be very \camone{large} in irregular workloads~\cite{Gomez2021Analysis,Gomez2021Benchmarking,Oliveira2021Damov,Lockerman2020Livia,Kanellopoulos2019SMASH,Giannoula2018Combining,besta2017slimsell,dongarra1996sparse,Elafrou2018SparseX,Elafrou2017PerformanceAA,YouTubeGraph,FacebookGraph,Goumas2008Understanding,White97Improving,Helal2021ALTO,Pelt2014Medium,strati2019adaptive} like the \spmv{} kernel.

\begin{tcolorbox}
\noindent\textbf{OBSERVATION 14:} \\
The \equallyWidth{} and \variableSized{} 2D \camone{partitioning} schemes require fine-grained parallel transfers \textit{at DRAM bank granularity} to be supported by the PIM system, \camone{i.e., \textit{zero} padding in \textit{parallel} \texttt{retrieve} data transfers from PIM-enabled memory to the host CPU}, to achieve high performance.
\end{tcolorbox}

Fourth, we find that the number of vertical partitions that provides the best performance depends on the sparsity pattern of the input matrix, the data type, \camone{and the underlying hardware parameters (e.g., number of PIM cores, off-chip memory bus bandwidth, transfer latency costs between main memory and PIM-enabled memory, characteristics and microarchitecture of the host CPU cores that perform the \texttt{merge} step).} For example, with the int8 data type, \texttt{DCOO} performs best for \texttt{hgc} and \texttt{mem} matrices with 8 and 4 vertical partitions, respectively. Instead, with the fp64 data type, \texttt{DCOO} performs best for \texttt{hgc} and \texttt{mem} matrices with 16 and 8 vertical partitions, respectively. \camone{We refer the reader to Appendix~\ref{sec:appendix-2D-vertpartitions} for a characterization study on the number of vertical partitions to perform in the 2D-partitioned kernels using two UPMEM PIM systems with different hardware characteristics. As we demonstrate in Appendix~\ref{sec:appendix-2D-vertpartitions}, the number of vertical partitions that provides best performance on \spmv{} varies across the two different UPMEM PIM platforms. In this work, we leave for future work} the exploration of selection methods for the number of vertical partitions that provide best \spmv{} execution. Overall, based on our analysis we conclude that the parallelization scheme that achieves the best performance in \spmv{} depends on both the input sparse matrix and the hardware characteristics of the PIM system.

\begin{tcolorbox}
\noindent\textbf{OBSERVATION 15:} \\
There is \textit{no one-size-fits-all} parallelization approach for \spmv{} in PIM systems, since the performance of each parallelization scheme depends on the characteristics of the input matrix and the underlying PIM hardware.
\end{tcolorbox}

\subsubsection{\textbf{Analysis of Compressed Formats}}\label{2D-Formats}  
\begin{figure}[H]
    \centering
    \includegraphics[width=0.88\textwidth]{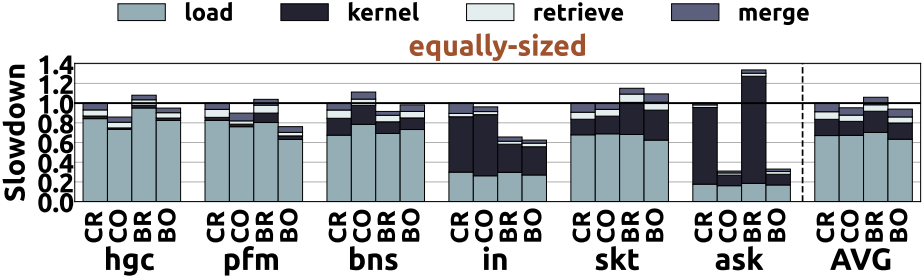}
    \vspace{-10pt}
    \caption{End-to-end execution time breakdown of the \equallySized{} 2D partitioning technique for CR: \texttt{DCSR}, CO: \texttt{DCOO}, BR: \texttt{DBCSR} and BO: \texttt{DBCOO} schemes using 4 vertical partitions and the int32 data type. Performance is normalized to that of \texttt{DCSR}.}
    \label{fig:2D_balance_fixed}
    \vspace{-16pt}
\end{figure}

\begin{figure}[H]
    \centering
    \includegraphics[width=0.88\textwidth]{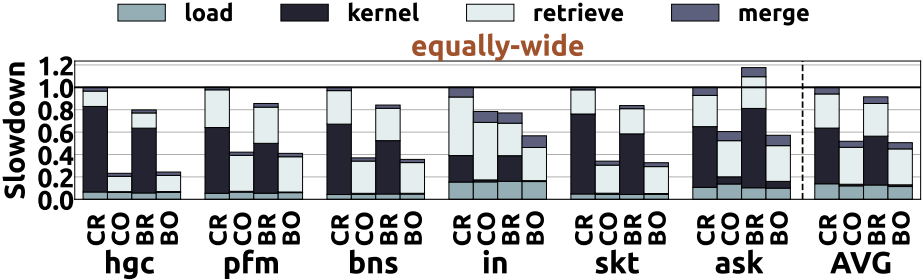}
    \vspace{-10pt}
    \caption{End-to-end execution time breakdown of the \equallyWidth{} 2D partitioning technique for CR: \texttt{RBDCSR}, CO: \texttt{RBDCOO}, BR: \texttt{RBDBCSR} and BO: \texttt{RBDBCOO} schemes using 4 vertical partitions and the int32 data type. Performance is normalized to that of \texttt{RBDCSR}.}
    \label{fig:2D_balance_rbal}
    \vspace{-16pt}
\end{figure}

\begin{figure}[H]
    \centering
    \includegraphics[width=0.88\textwidth]{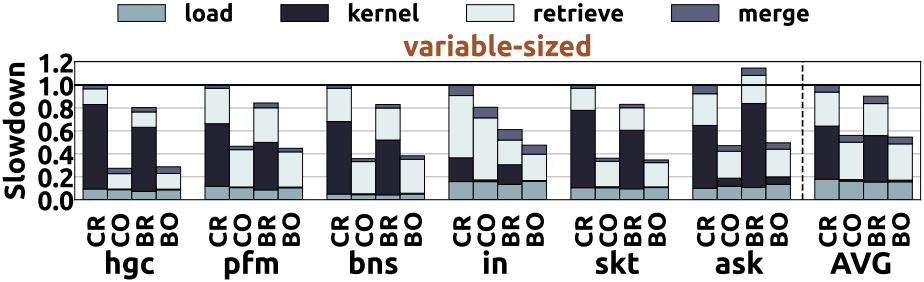}
    \vspace{-10pt}
    \caption{End-to-end execution time breakdown of the \variableSized{} 2D partitioning technique for CR: \texttt{BDCSR}, CO: \texttt{BDCOO}, BR: \texttt{BDBCSR} and BO: \texttt{BDBCOO} schemes using 4 vertical partitions and the int32 data type. Performance is normalized to that of \texttt{BDCSR}.}
    \label{fig:2D_balance_bal}
    \vspace{-10pt}
\end{figure}

\hfill  \\
We compare the performance achieved by various compressed matrix formats for each of the three types of the 2D partitioning technique. The goal of this experiment is to find the best-performing compressed format for each 2D partitioning technique. Figures~\ref{fig:2D_balance_fixed}, ~\ref{fig:2D_balance_rbal}, and ~\ref{fig:2D_balance_bal} compare the performance of compressed matrix formats \camone{for the} \equallySized{}, \equallyWidth{} and \variableSized{} 2D partitioning techniques, respectively. We use 2048 DPUs and the int32 data type having 4 vertical partitions. See Appendix~\ref{sec:appendix-2D-formats} for the complete evaluation on all large \camone{sparse} matrices.

We draw two findings. First, as already explained, \texttt{kernel} time of the \equallySized{} scheme is limited by the DPU (or a few DPUs) assigned to the 2D tile with the largest number of non-zero elements. In scale-free matrices (e.g., \texttt{in} and \texttt{ask}), the \camone{disparity in the} non-zero elements \camone{across} 2D tiles is higher than in regular matrices (e.g., \texttt{pfg} and \texttt{bns}), causing \texttt{kernel} time to be a larger portion of the total execution time. Second, we find that the CSR and BCSR formats perform worse than the COO and BCOO formats, especially in the \equallyWidth{} and \variableSized{} schemes, due to higher \texttt{kernel} times. In the CSR and BCSR formats, data partitioning \camone{across} DPUs and/or \camone{across} tasklets within a DPU is performed at row and block-row granularity, respectively. Thus, the CSR and BCSR formats can cause higher non-zero element imbalance \camone{across} processing units compared to the COO and BCOO formats.
Overall, the COO and BCOO formats outperform the CSR and BCSR formats by 1.59 $\times$ and 1.53 $\times$ (averaged across all three types of 2D partitioning techniques), respectively. 

\begin{tcolorbox}
\noindent\textbf{OBSERVATION 16:} \\
The compressed matrix format used to store the input matrix determines the data partitioning across DRAM banks of PIM-enabled memory. Thus, it affects the load balance \camone{across} PIM cores with corresponding performance implications. Overall, the COO and BCOO formats outperform the CSR and BCSR formats, because they provide higher non-zero element balance across PIM cores.
\end{tcolorbox}

\subsubsection{\textbf{Comparison of 2D Partitioning Techniques}}\label{2D-Comparison}  
\begin{figure}[H]
    \centering
    \includegraphics[width=\textwidth]{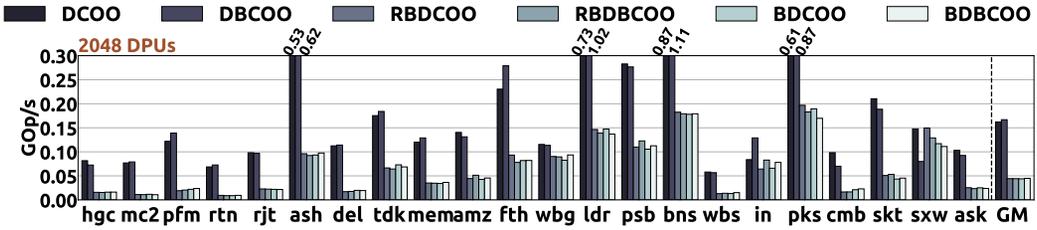}
    \vspace{-17pt}
    \caption{\camone{Throughput of} 2D partitioning techniques using the COO and BCOO formats, 2048 DPUs and the int32 type.}
    \label{fig:2D_best_partition}
    \vspace{-4pt}
\end{figure}

\begin{figure}[H]
    \vspace{-9pt}
    \centering
    \includegraphics[width=\textwidth]{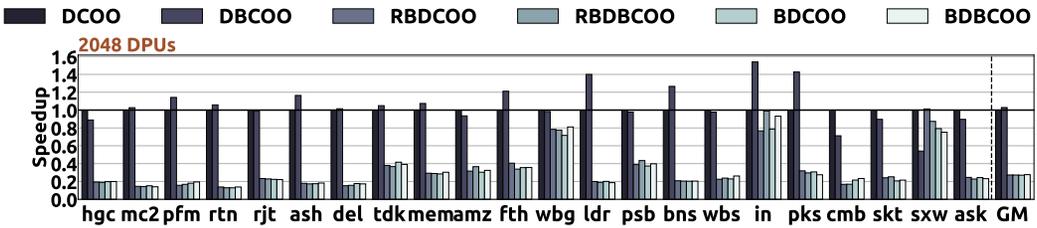}
    \vspace{-17pt}
    \caption{\camone{Performance comparison} of 2D partitioning techniques using the COO and BCOO formats, 2048 DPUs and the int32 type. Performance is normalized to that of \texttt{DCOO}.}
    \label{fig:2D_best_partition_perf}
    \vspace{-4pt}
\end{figure}

\hfill  \\
We compare the best-performing \spmv{} implementations of all 2D partitioning schemes, i.e., using the COO and BCOO formats. \camone{Figures~\ref{fig:2D_best_partition} and ~\ref{fig:2D_best_partition_perf} compare the throughput (in GOperations per second) and the performance, respectively, of \texttt{DCOO}, \texttt{DBCOO}, \texttt{RBDCOO}, \texttt{RBDBCOO}, \texttt{BDCOO}, \texttt{BDBCOO} schemes using 2048 DPUs and the int32 data type.} For each implementation, we vary the number of vertical partitions from 2 to 32, \camone{in steps of multiple of 2}, and select the best-performing \camone{execution throughput}.

We draw two conclusions. First, similarly to 1D-partitioned kernels, matrices that exhibit block pattern (e.g., \texttt{ash}, \texttt{ldr}, \texttt{bns}, \texttt{pks}) have the highest throughput (Figure~\ref{fig:2D_best_partition}). Second, the \equallyWidth{} and \variableSized{} schemes perform similarly, i.e., their performance varies only by $\pm$1.1\% on average. Even though the \variableSized{} technique can improve the non-zero element balance \camone{across} DPUs, and thus \texttt{kernel} time, compared to the \equallyWidth{} technique, the total execution time does not improve. In the UPMEM PIM system, performance of both techniques is severely bottlenecked by data transfer overheads due to a \camone{large} amount of padding needed to retrieve results from PIM-enabled memory to the host CPU. Third, we find that the \equallySized{} technique outperforms the \equallyWidth{} and \variableSized{} techniques by 3.71$\times$ on average, because it achieves lower data transfer overheads. The \equallyWidth{} and \variableSized{} techniques provide near-perfect non-zero element balance \camone{across} DPUs, \camone{but they significantly increase the \texttt{retrieve} data transfer costs due to the large amount  of padding with empty bytes performed}. As a result, we recommend software designers to explore \textit{relaxed} load balancing schemes, i.e., schemes that trade off computation balance \camone{across} PIM cores \camone{for lower amounts of data transfer.}


\subsection{Comparison of 1D and 2D Partitioning Techniques}\label{1D-2D}
\begin{figure}[H]
    \centering
    \includegraphics[width=\textwidth]{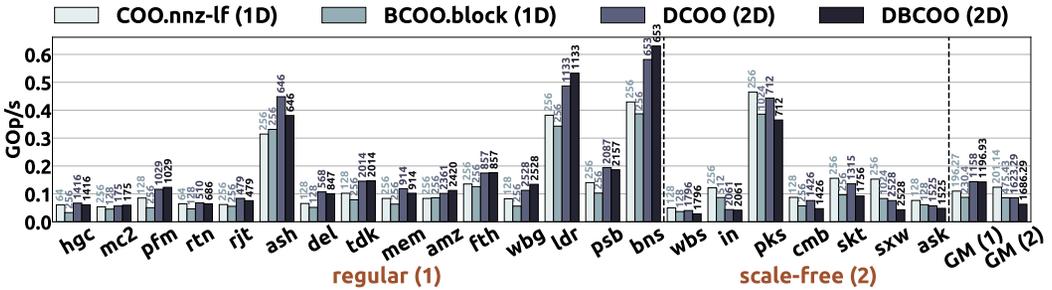}
    \vspace{-20pt}
    \caption{Throughput \camone{of} the best-performing 1D- and 2D-partitioned kernels for the fp32 data type.}
    \label{fig:2D_1D-2D}
    \vspace{-10pt}
\end{figure}

\begin{figure}[H]
    \vspace{-4pt}
    \centering
    \includegraphics[width=\textwidth]{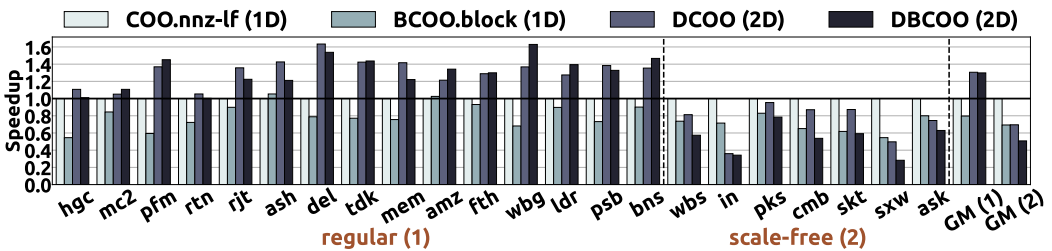}
    \vspace{-18pt}
    \caption{\camone{Performance comparison of} the best-performing 1D- and 2D-partitioned kernels for the fp32 data type. Performance is normalized to that of \texttt{COO.nnz-lf}.}
    \label{fig:2D_1D-2D-perf}
    \vspace{-4pt}
\end{figure}

We compare the \camone{throughput (in GOperations per second) and the performance of the best-performing 1D- and 2D-partitioned kernels in Figures~\ref{fig:2D_1D-2D} and ~\ref{fig:2D_1D-2D-perf}, respectively.} For 1D partitioning, we use the lock-free COO (\texttt{COO.nnz-lf}) and coarse-grained locking BCOO (\texttt{BCOO.block}) kernels. For each matrix, we vary the number of DPUs from 64 to 2528, and select the best-performing end-to-end \camone{execution throughput}. For 2D partitioning, we use the \equallySized{} COO (\texttt{DCOO}) and BCOO  (\texttt{DBCOO}) kernels with \dpuActive{}. For each matrix, we vary the number of vertical partitions from 2 to 32 (\camone{in steps of multiple of 2}), and select the best-performing end-to-end \camone{execution throughput}. \camone{The numbers shown over each bar of Figure~\ref{fig:2D_1D-2D} present the number of DPUs that provide the best-performing end-to-end execution throughput for each input-scheme combination. Please see Appendix~\ref{sec:appendix-1D_2D} for a performance comparison of the best-performing \spmv{} kernels on two UPMEM PIM systems with different hardware characteristics.}

We draw two conclusions. First, we find that best performance is achieved using a much smaller number of DPUs than the available DPUs on the system. In the 1D-partitioned kernels (i.e., \texttt{COO.nnz-lf} and \texttt{BCOO.block}), replicating the input vector \camone{into a large} number of DPUs significantly increases the \texttt{load} data transfer costs. Thus, best performance is achieved using \camone{253 DPUs on average across all matrices}. In the 2D-partitioned kernels (i.e.,  \texttt{DCOO} and  \texttt{DBCOO}), creating \equallySized{} 2D tiles \camone{leads to a large disparity in non-zero element count across tiles,} causing many tiles to be empty, i.e., \camone{without} \textit{any} non-zero element. Thus, best performance is achieved using \camone{1329 DPUs on average across all matrices}, since DPUs associated with empty tiles are idle.

\begin{tcolorbox}
\noindent\textbf{OBSERVATION 17:} \\
Expensive data transfers to PIM-enabled memory performed via the narrow memory bus impose significant performance overhead to end-to-end \spmv{} execution. Thus, it is hard to fully exploit all available PIM cores of the system.
\end{tcolorbox}

Second, we observe that in regular matrices, the 2D-partitioned kernels outperform the 1D-partitioned kernels by 1.45$\times$ on average. This is because the 2D-partitioned kernels use a \camone{larger} number of DPUs, and thus their \texttt{kernel} times are lower. In contrast, in scale-free matrices, the 1D-partitioned kernels outperform the 2D-partitioned kernels by 1.41$\times$ on average. This because the \equallySized{} 2D technique significantly increases the non-zero element \camone{disparity across} DPUs, i.e., \texttt{kernel} time is bottlenecked by only one DPU or a few DPUs that process a much larger number of non-zero elements compared to the rest.

\begin{tcolorbox}
\noindent\textbf{OBSERVATION 18:} \\
In \textit{regular} matrices, 2D-partitioned kernels outperform 1D-partitioned kernels, since \camone{the former} provide a better trade-off between computation and data transfer overheads. In contrast, in \textit{scale-free} matrices, 2D-partitioned kernels perform worse than 1D-partitioned kernels, since \camone{the former's} performance is limited by one DPU or a few DPUs that process the largest number of non-zero elements.
\end{tcolorbox}
\renewcommand\camone[1]{\noindent{\color{black}{#1}}} 
\renewcommand\arxiv[1]{\noindent{\color{black}{#1}}} 

\section{Comparison with CPUs and GPUs}\label{cpu-gpu}

We compare \spmv{} execution on the UPMEM PIM architecture to a state-of-the-art CPU and a state-of-the-art GPU in terms of performance and energy consumption. Our goal is to quantify the potential of the UPMEM PIM architecture on the widely used memory-bound \spmv{} kernel.

We compare the UPMEM PIM system with \dpuActive{} to an Intel Xeon CPU~\cite{intel4110} and an NVIDIA Tesla V100 GPU~\cite{nvidiaTeslaV100}, the characteristics of which are shown in Table~\ref{tab:cpu-gpu}. We use peakperf~\cite{peak-perf} and stream~\cite{stream} for CPU and GPU systems to calculate the peak performance, memory bandwidth, and \camone{Thermal Design Power (TDP)}. For the UPMEM PIM system, we estimate the peak performance as $Total\_DPUs * AT$, where the arithmetic throughput (AT) is \camtwo{calculated for the multiplication operation in Appendix~\ref{sec:appendix-1DPU-AT} (Figure~\ref{fig:1DPU-ai-cloud4})}, the total bandwidth as $Total\_DPUs * Bandwidth\_DPU$, where the $Bandwidth\_DPU$ is 700 MB/s~\cite{Gomez2021Benchmarking,Gomez2021Analysis,devaux2019}, and TDP as $(Total\_DPUs / DPUs\_per\_chip) * 1.2W/chip$ from prior work~\cite{Gomez2021Benchmarking,Gomez2021Analysis,devaux2019}.

\begin{table}[H]
\vspace{2pt}
\begin{center}
\centering
\resizebox{1.0\linewidth}{!}{
\begin{tabular}{|l||c|c|c|c|c|c|c|}
    \hline
    \cellcolor{gray!15} & \cellcolor{gray!15}\raisebox{-0.20\height}{\textbf{Process}} & \cellcolor{gray!15} & \cellcolor{gray!15} & \cellcolor{gray!15}\raisebox{-0.20\height}{\textbf{Peak}} & \cellcolor{gray!15}\raisebox{-0.20\height}{\textbf{Memory}} & \cellcolor{gray!15}\raisebox{-0.20\height}{\textbf{Total}} & \cellcolor{gray!15} \\
     \multirow{-2}{*}{\cellcolor{gray!15}\textbf{System}} & \cellcolor{gray!15}\textbf{Node} & \multirow{-2}{*}{\cellcolor{gray!15}\textbf{Total Cores}} &  \multirow{-2}{*}{\cellcolor{gray!15}\textbf{Frequency}} & \cellcolor{gray!15}\textbf{Performance} & \cellcolor{gray!15}\textbf{Capacity} & \cellcolor{gray!15}\textbf{Bandwidth} & \multirow{-2}{*}{\cellcolor{gray!15}\textbf{TDP}} \\
    \hline \hline
    Intel Xeon 4110 CPU~\cite{intel4110} & 14 nm & 2x8 x86 cores (2x16 threads) & 2.1 GHz & 660 GFLOPS & 128 GB & 23.1 GB/s  & 2x85 W \\ \hline 
    NVIDIA Tesla V100~\cite{nvidiaTeslaV100} & 12 nm & 5120 CUDA cores & 1.25 GHz & 14.13 TFLOPS & 32 GB & 897 GB/s & 300 W \\ \hline 
    PIM System & 2x nm & \dpuActive{} & 350 MHz & 4.66 GFLOPS & 159 GB & 1.77 TB/s & 379 W \\ \hline 

\end{tabular}
}
\end{center}
\vspace{4pt}
\caption{Evaluated CPU, GPU, and UPMEM PIM Systems.}
\label{tab:cpu-gpu}
\vspace{-8pt}
\end{table}

\subsection{Performance Comparison}

For the CPU system, we use the optimized CSR kernel from the TACO library~\cite{Kjolstad2017Taco}. For the GPU system, we use the CSR5 CUDA~\cite{Weifeng2015CSR5,CSR5-cuda} for the int32 data type and cuSparse~\cite{cuSparse} for the other data types. For the UPMEM PIM system, we use the lock-free COO 1D-partitioned kernel (\texttt{\textbf{COO}.nnz-lf}) and the \equallySized{} COO 2D-partitioned kernel (\texttt{\textbf{DCOO}}). In the former, we run experiments from 64 to \dpuActive{}, and in the latter, we use \dpuActive{}, and vary the number of vertical partitions from 2 to 32, \camone{in steps of multiple of 2}. In both schemes, we select the best-performing end-to-end execution \camone{throughput}. We also include the lock-free COO 1D-partitioned kernel using \dpuActive{}, named \texttt{\textbf{COO.kl}}, to evaluate \spmv{} execution using \textit{all} available DPUs of the system.

\begin{figure}[H]
    \begin{minipage}{1.0\textwidth}
    \centering
    \includegraphics[width=.42\textwidth]{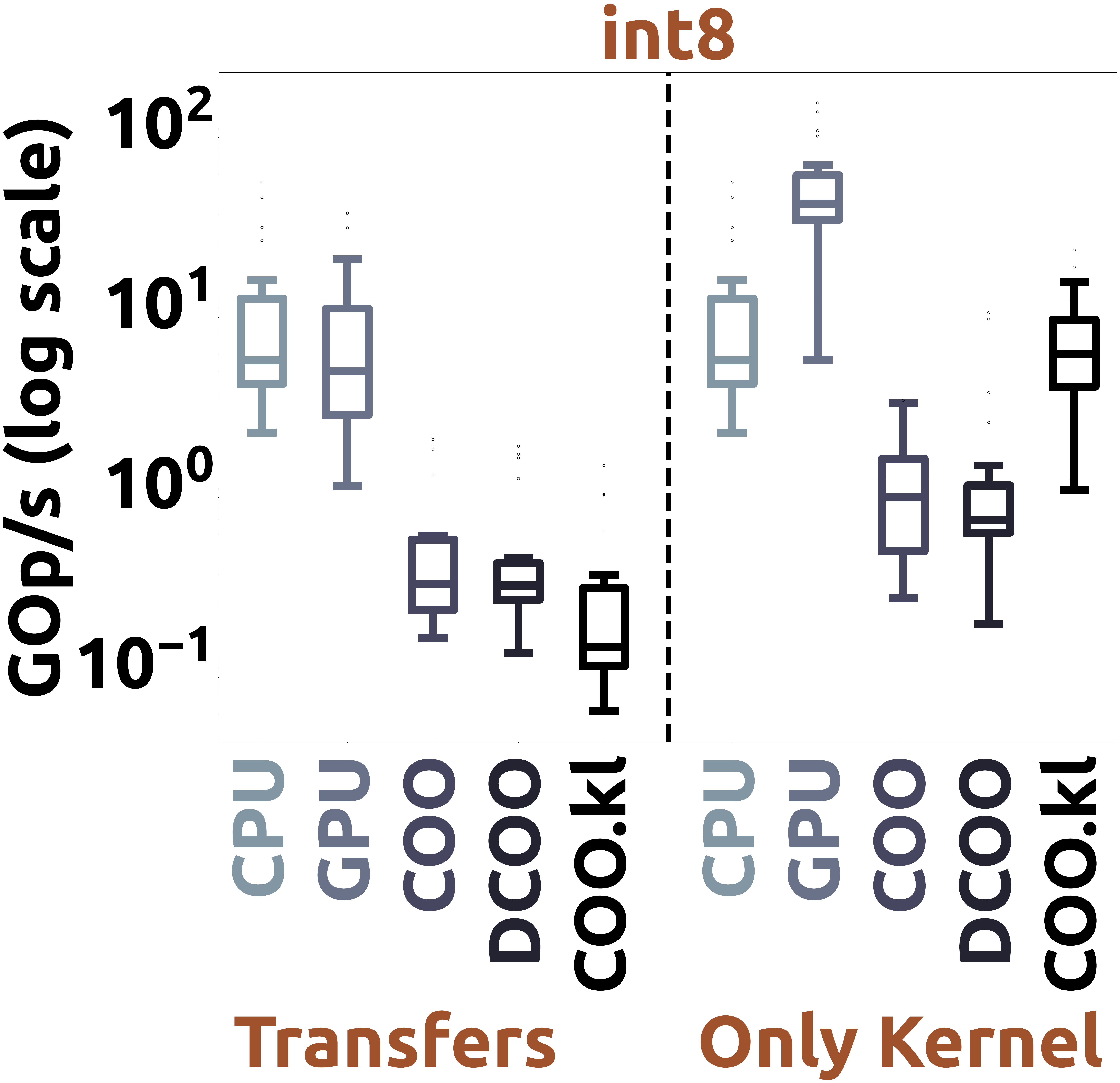}\hspace{16pt}
    \includegraphics[width=.42\textwidth]{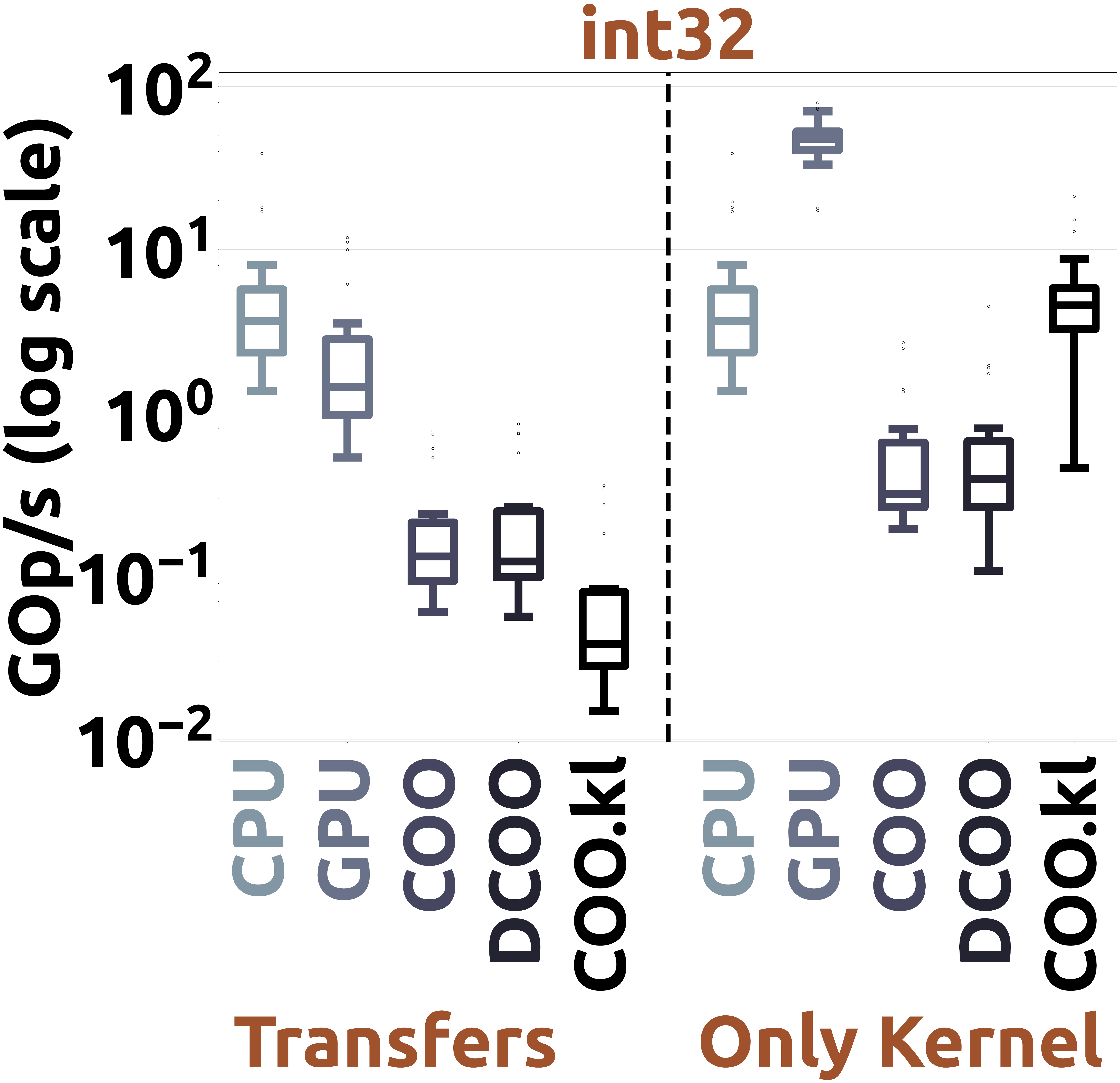}
    \end{minipage}\vspace{18pt}
    \begin{minipage}{1.0\textwidth}
    \centering
    \includegraphics[width=.42\textwidth]{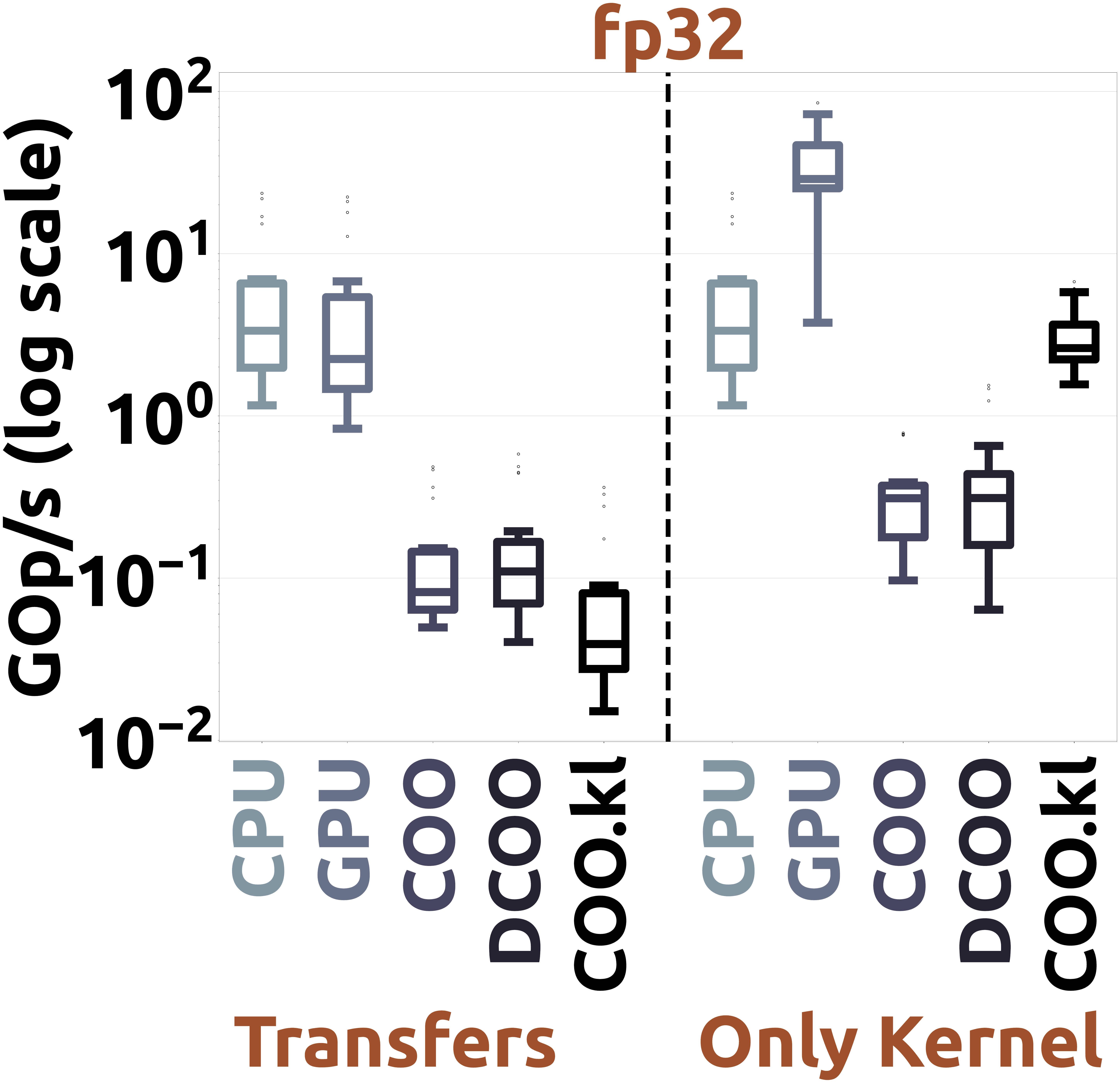}\hspace{16pt}
    \includegraphics[width=.42\textwidth]{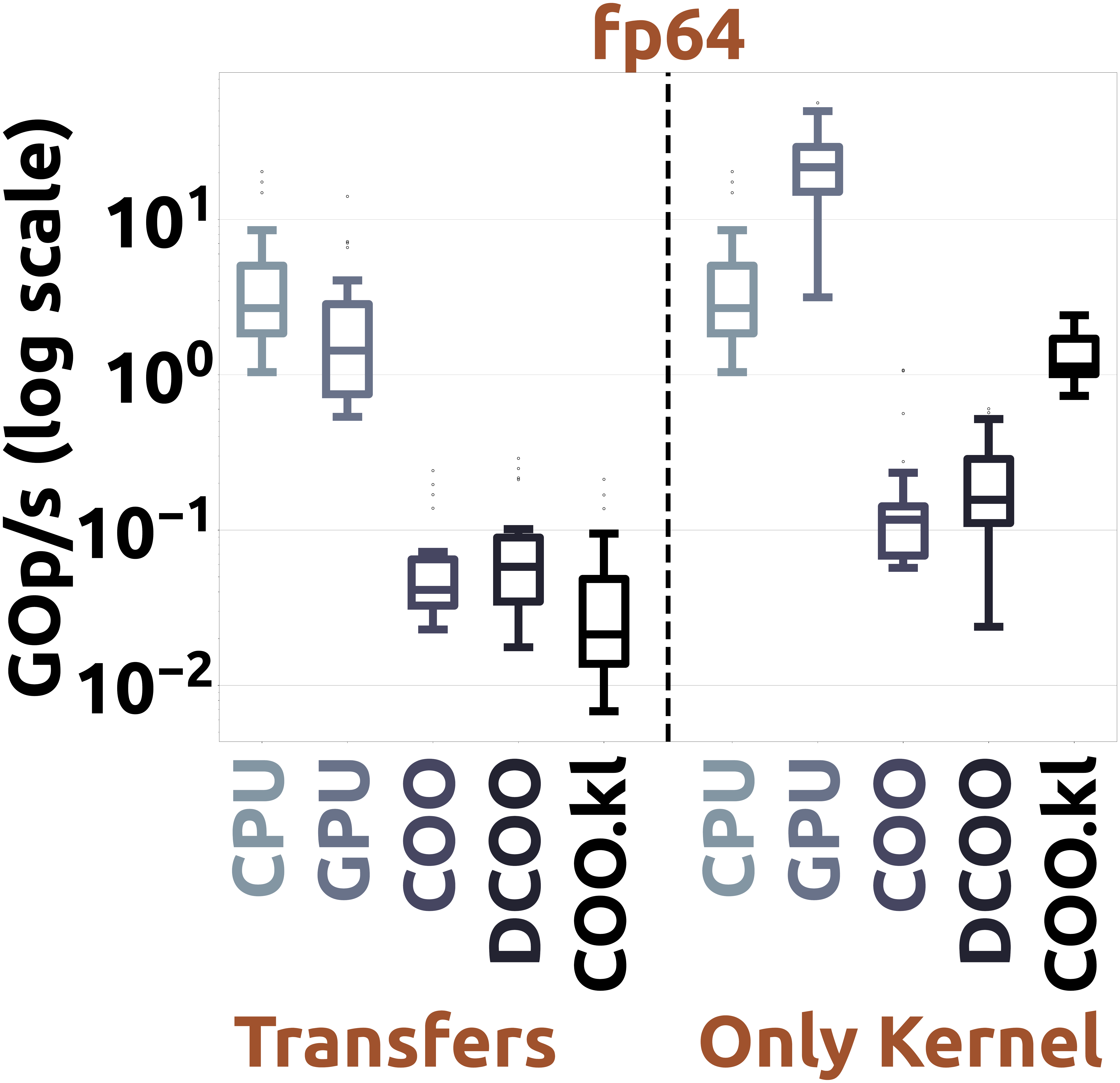}
    \end{minipage}
    \vspace{-6pt}
    \caption{Performance comparison between the UPMEM PIM system, Intel Xeon CPU and Tesla V100 GPU on \spmv{} \camtwo{execution}.}
    \label{fig:cpu-gpu-end_to_end}
\end{figure}

Figure~\ref{fig:cpu-gpu-end_to_end} shows the throughput of \spmv{} (in GOperations per second) in all systems, comparing both the end-to-end execution \camone{throughput} (i.e., including the \texttt{load} and \texttt{retrieve} data transfer costs for the input and output vectors in case of the UPMEM PIM and GPU systems), and only the actual kernel \camone{throughput} (i.e., including the \texttt{kernel} time in DPUs and the \texttt{merge} time in host CPU for the UPMEM PIM system).

We draw three conclusions. First, when data transfer costs to/from host CPU are included, CPU outperforms both \camtwo{the} GPU and UPMEM PIM systems, since data transfers impose high overhead. When only the actual kernel time is considered, GPU performs best, since it is the system that provides the highest computation throughput, e.g., 14.13 TFlops for the fp32 data type. Second, we evaluate the portion of the machine's peak performance achieved on \spmv{} in all systems, and observe that \spmv{} execution on the UPMEM PIM system achieves a much higher fraction of the peak performance compared to CPU and GPU systems. For the fp32 data type, \spmv{} achieves on average 0.51\% and 0.21\% of the peak performance in CPU and GPU, respectively, while it achieves 51.7\% of the peak performance in the UPMEM PIM system \camone{using the \texttt{COO.kl} scheme}. Achieving a high portion of machine's peak performance is highly desirable, since the software highly exploits the computation capabilities of the underlying hardware. This way, it improves the processor/resource utilization, and the cost of ownership of the underlying hardware. Third, we observe that when all DPUs are used, as in \texttt{COO.kl}, \spmv{} execution on the UPMEM PIM outperforms \spmv{} execution on the CPU by 1.09$\times$ and 1.25$\times$ for the int8 and int32 data types, respectively, the multiplication of which is supported by hardware. In contrast, \spmv{} execution on the UPMEM PIM performs 1.27$\times$ and 2.39$\times$ worse than \spmv{} execution on the CPU for the fp32 and fp64 data types, the multiplication of which is software emulated in the DPUs \camtwo{of the UPMEM PIM system.}

\begin{tcolorbox}
\noindent\textbf{OBSERVATION 19:} \\
\spmv{} execution can achieve a \textit{significantly higher} fraction of the peak performance on real memory-centric PIM architectures compared to that on processor-centric CPU and GPU systems, since PIM architectures \camone{greatly} mitigate data movement costs.
\end{tcolorbox}

\subsection{Energy Comparison}

\begin{figure}[t]
    \begin{minipage}{1.0\textwidth}
    \centering
    \includegraphics[width=.444\textwidth]{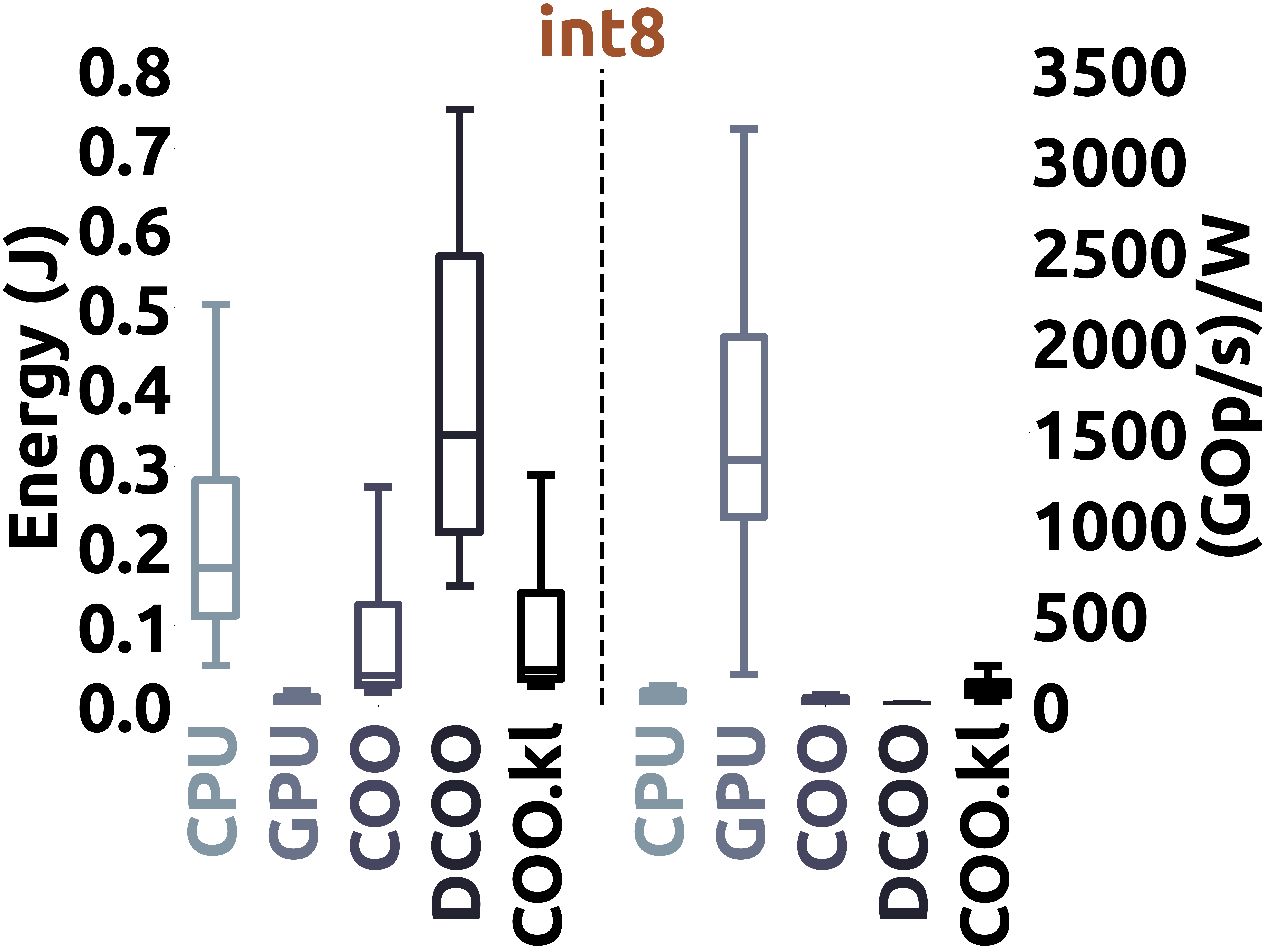}\hspace{16pt}
    \includegraphics[width=.444\textwidth]{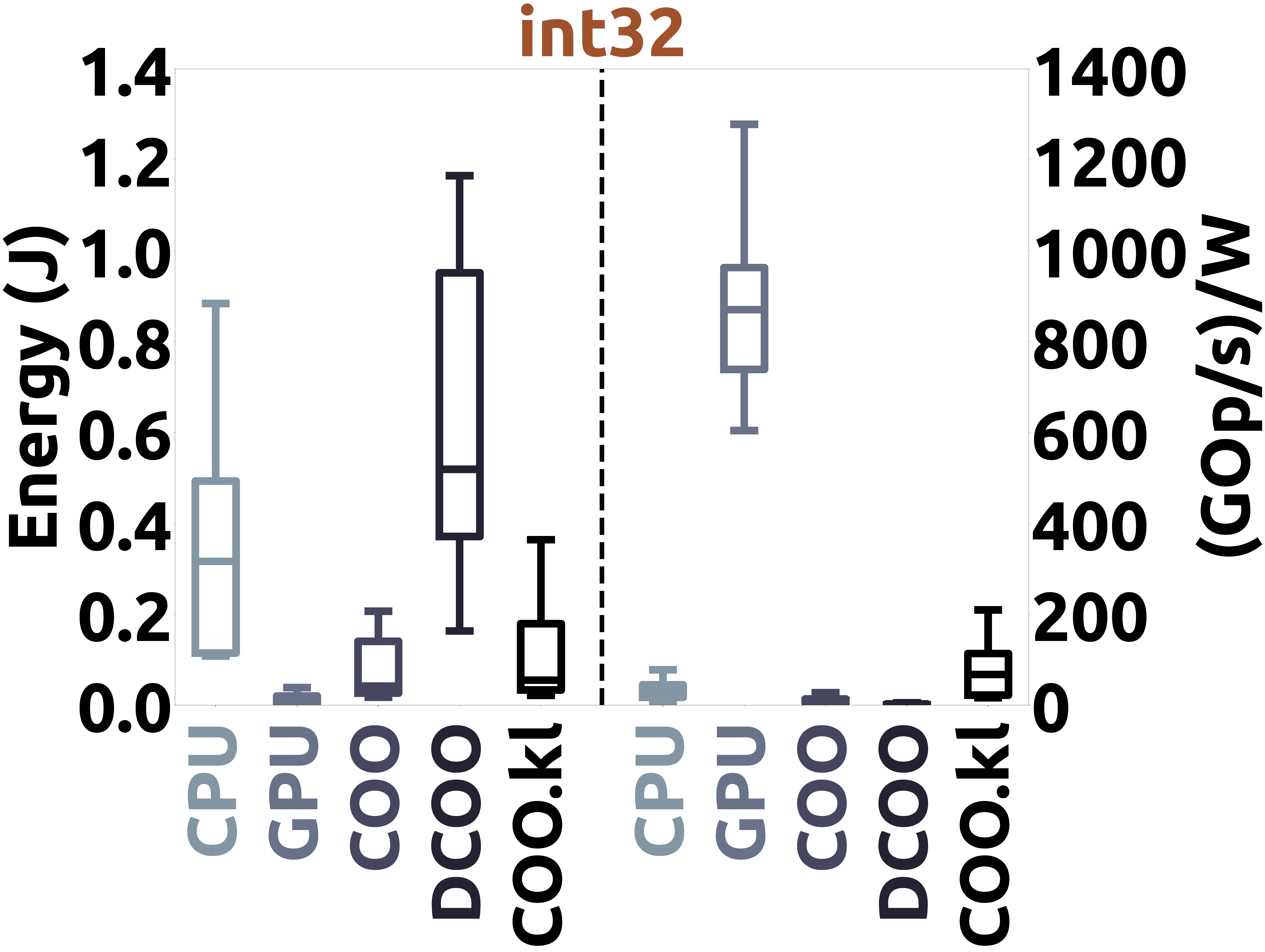}
    \end{minipage} 
    \begin{minipage}{1.0\textwidth}
    \vspace{10pt}
    \centering
    \includegraphics[width=.444\textwidth]{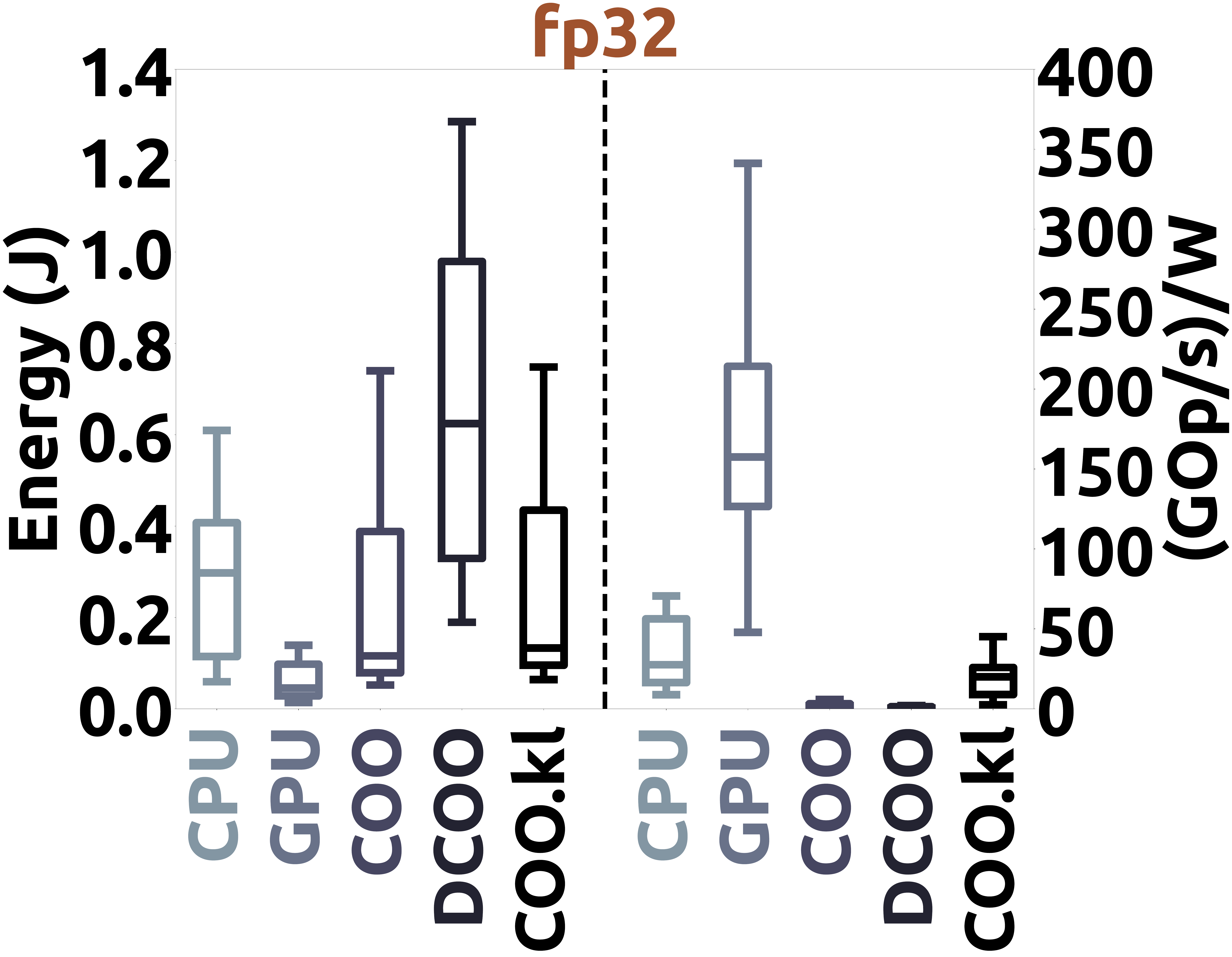}\hspace{16pt}
    \includegraphics[width=.444\textwidth]{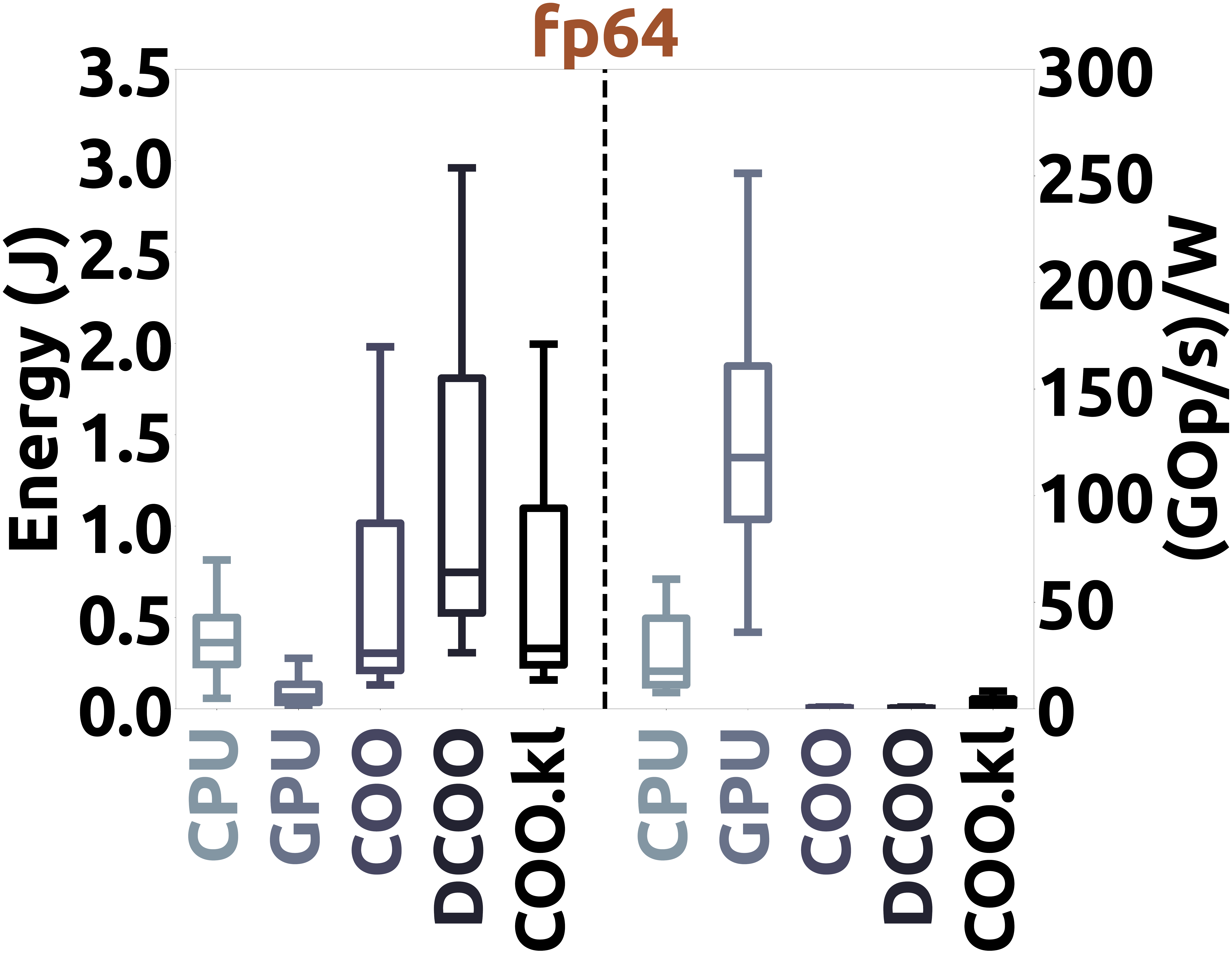}
    \end{minipage}
    \vspace{-4pt}
    \caption{Energy comparison between the UPMEM PIM system, Intel Xeon CPU and Tesla V100 GPU on \spmv{} \camtwo{execution}.}
    \label{fig:cpu-gpu-energy}
    \vspace{-10pt}
\end{figure}

For energy measurements, we consider only the actual kernel time in all systems (in the UPMEM PIM we consider the \texttt{kernel} and \texttt{merge} steps of \spmv{} execution). We use Intel RAPL~\cite{rapl} on the CPU, and NVIDIA SMI~\cite{smi} on the GPU. For the UPMEM PIM system, we measure the number of cycles, instructions, WRAM accesses and MRAM accesses of each DPU, and estimate energy with energy weights provided by the UPMEM company~\cite{upmem}. Figure~\ref{fig:cpu-gpu-energy} shows the energy consumption (in Joules) and performance per energy (in \camtwo{(GOp/s)}/W) for all systems.

We draw three findings. First, GPU provides the \camone{lowest} energy on \spmv{} over the other two systems, since the energy results  \camone{typically} follow the performance results. Second, we find that the 2D-partitioned kernel, i.e., \texttt{DCOO}, consumes more energy than the 1D-partitioned kernels, i.e., \texttt{COO} and \texttt{COO.kl}, \camone{due to} the energy consumed in the host CPU cores. CPU cores merge a \camone{large} number of partial results in the 2D-partitioned kernels to assemble the final output vector, thereby increasing the energy consumption. \camone{Finally, we find that the 1D-partitioned kernels provide better energy efficiency on \spmv{} over the CPU system, when the multiplication operation is supported by hardware. Specifically, 1D-partitioned kernels provide 3.16$\times$ and 4.52$\times$ less energy consumption, and 1.74$\times$ and 1.14$\times$ better performance per energy over the CPU system for the int8 and int32 data types, respectively.}


\begin{tcolorbox}
\noindent\textbf{OBSERVATION 20:} \\
Real PIM architectures can provide high energy efficiency on \spmv{} execution.
\end{tcolorbox}

\subsection{Discussion}
These evaluations are useful for programmers to anticipate how much performance and energy savings memory-centric PIM systems can provide on \spmv{} over commodity processor-centric CPU and GPU systems. However, our evaluated \spmv{} kernels do not constitute the best-performing approaches for \textit{all} matrices. Designing methods to select the best-performing \spmv{} parallelization scheme depending on the particular characteristics of the input matrix would further improve performance and energy savings of \spmv{} execution on memory-centric PIM systems. 
Moreover, the UPMEM PIM hardware is still maturing and is expected to run at a higher frequency in the near future (500 MHz instead of 350 MHz)~\cite{upmem,Gomez2021Benchmarking}. Hence, \spmv{} execution on the UPMEM PIM architecture might achieve even higher performance and energy benefits over the results we report in this comparison. Finally, note that our proposed \SparseP{} kernels can be adapted and evaluated on other current and future real PIM systems with potentially higher computation capabilities and energy efficiency than the UPMEM PIM system.


\section{Key Takeaways and Recommendations}\label{recommendations}

This section summarizes our key takeaways in the form of recommendations to improve multiple aspects of PIM hardware and software.

\noindent\textbf{\textit{Recommendation \#1.}} \textit{Design algorithms that provide high \camone{load} balance \camone{across} threads of a PIM core in terms of computations, loop control iterations, synchronization points and memory accesses.} 
Section~\ref{1DPU} shows that in matrices and formats where the parallelization scheme used causes \textit{high disparity} in the non-zero elements/blocks/rows processed across threads of a PIM core, or the number of lock acquisitions/lock releases/DRAM memory accesses performed across threads,  \spmv{} performance severely degrades in compute-bound DPUs~\cite{Gomez2021Analysis,Gomez2021Benchmarking}. \camone{Therefore, from a programmer’s perspective, providing high operation balance \camone{across} parallel threads is of vital importance in low-area and low-power PIM cores with relatively low \camone{computation} capabilities~\cite{Gomez2021Benchmarking,Gomez2021Analysis}.}


\noindent\textbf{\textit{Recommendation \#2.}} \textit{Design compressed data structures that can be effectively partitioned across DRAM banks, \camone{with the goal of providing} high computation balance \camone{across} PIM cores.} Sections~\ref{1D-Kernel} and ~\ref{2D-Formats} demonstrate that (i) the compressed matrix format used to store the input matrix determines the data partitioning across DRAM banks of PIM-enabled memory, and (ii) \spmv{} execution using the CSR and BCSR formats performs significantly worse than \spmv{} execution using the COO and BCOO formats. 
This is because the matrix is stored in row- or block-row-order for the CSR and BCSR formats, respectively, and thus data partitioning across DRAM banks \camtwo{is limited to be} performed at row or block-row granularity, respectively, leading to high non-zero element imbalance \camone{across} PIM cores. Therefore, we recommend that programmers design compressed data structures that can provide effective data partitioning schemes with high computation balance across thousands of PIM cores.

\noindent\textbf{\textit{Recommendation \#3.}} \textit{Design \textit{adaptive} algorithms that (i) \camone{trade off computation balance across PIM cores for lower data transfer costs to PIM-enabled memory}, and (ii) adapt their configuration to the particular patterns of each input given, as well as the characteristics of the PIM hardware.} Our analysis in Sections~\ref{1D-Kernel}, ~\ref{2D-Studies} and ~\ref{2D-Comparison} demonstrates that the best-performing \spmv{} execution on the UPMEM PIM system can be achieved using algorithms that (i) trade off computation for lower data transfer costs, and (ii) select the load balancing strategy and data partitioning policy based on the particular sparsity pattern of the input matrix. In addition, the performance of each balancing scheme and data partitioning technique for \spmv{} execution highly depends on the characteristics of the underlying PIM hardware, as we explain in Section~\ref{2D-Studies} and Appendix~\ref{sec:appendix-2D-vertpartitions}. To this end, we recommend that software designers implement heuristics and selection methods for their algorithms to adapt their configuration to the underlying hardware characteristics of the PIM system and the input data given.

\noindent\textbf{\textit{Recommendation \#4.}} \textit{Provide low-cost synchronization support and \camone{hardware support to enable concurrent memory accesses by multiple threads to the local DRAM bank} to increase parallelism in a multithreaded PIM core.} Section~\ref{1DPU} shows that (i) lock acquisitions/releases can cause high overheads in the DPU pipeline, and (ii) fine-grained locking approaches to increase parallelism in critical sections do not improve performance over coarse-grained approaches in the UPMEM PIM hardware. This is because the DMA engine of the DPU serializes DRAM memory accesses included in the critical sections. Based on these key takeaways, we recommend that hardware designers provide lightweight synchronization mechanisms for multithreaded PIM cores~\cite{Giannoula2021SynCron}, and enable concurrent access to local DRAM memory arrays to increase execution parallelism. For example, sub-array level parallelism~\cite{Kim2012Case,Chang2014Improving} or multiple DRAM banks per PIM core could be supported in the PIM hardware to improve parallelism.

\noindent\textbf{\textit{Recommendation \#5.}} \textit{Optimize the broadcast collective operation in data transfers from main memory to PIM-enabled memory to minimize overheads of copying the input data \camone{into} all DRAM banks \camtwo{in the PIM} system.} Figures~\ref{fig:1D_transfers} and ~\ref{fig:1D_transfers_scalability} show that \spmv{} execution using the 1D partitioning technique cannot scale up to a \camone{large} number of PIM cores. This is because it is severely limited by data transfer costs to broadcast the input vector into \textit{each} DRAM bank of PIM-enabled DIMMs via the narrow off-chip memory bus. To this end, we suggest that hardware and system designers provide a fast broadcast collective primitive to DRAM banks of PIM-enabled memory modules~\cite{Sun2021ABCDIMMAT}.

\noindent\textbf{\textit{Recommendation \#6.}} \textit{Optimize the gather collective operation \textit{at DRAM bank granularity} for data transfers from PIM-enabled memory to \camtwo{the} host CPU to minimize overheads of retrieving the output results.} Figures~\ref{fig:2D_scale_equally_wide}, ~\ref{fig:2D_scale_variable_sized} and ~\ref{fig:2D_vertpartitions} demonstrate that \spmv{} execution using the \equallyWidth{} and \variableSized{} 2D partitioning schemes is severely limited by data transfers to retrieve results for the output vector from DRAM banks of PIM-enabled DIMMs. This is due to two reasons: (i) 2D-partitioned kernels create a large number of partial results that need to be transferred from PIM-enabled memory to the host CPU via the narrow memory bus in order to assemble the final output vector, and (ii) the UPMEM PIM system has the limitation that the transfer sizes from/to all DRAM banks involved in the same parallel transfer need to be the same, and therefore a large amount of padding with empty bytes is performed in the \equallyWidth{} and \variableSized{} schemes. To this end, we suggest that hardware and system designers provide an optimized \textit{gather} primitive to efficiently collect results from multiple DRAM banks to host CPU~\cite{Sun2021ABCDIMMAT}, and support parallel fine-grained data transfers from PIM-enabled memory to host CPU \textit{at DRAM bank granularity} to avoid padding with empty bytes.

\noindent\textbf{\textit{Recommendation \#7.}} \textit{Design high-speed communication channels and optimized libraries for data transfers to/from thousands of DRAM banks of PIM-enabled memory.} Section~\ref{cpu-gpu} demonstrates that \spmv{} execution on the memory-centric UPMEM PIM system achieves a much higher fraction of the machine's peak performance (on average 51.7\% for the 32-bit float data type), compared to \camtwo{that on} processor-centric CPU and GPU systems. However, the end-to-end performance of both 1D- and 2D-partitioned kernels is significantly limited by data transfer overheads on the narrow memory bus. To this end, we recommend that the hardware architecture and the software stack of real PIM systems be enhanced with low-cost and fast data transfers to/from PIM-enabled memory modules, \camone{and/or with support for efficient direct communication among PIM cores~\cite{Seshadri2013RowClone,Chang2016LISA,Rezaei2020NoM,Wang2020Figaro,Seshadri2017Ambit,Seshadri2017Simple}.}

\section{Related Work}

To our knowledge, this is the first work that (i) extensively characterizes the Sparse Matrix Vector Multiplication (\spmv{}) kernel in a real PIM system, and (ii) presents an open-source \spmv{} library for real-world PIM systems. We briefly discuss closely related prior work.

\noindent\textbf{Processing-In-Memory (\emph{PIM}).}
A large \camone{body of prior work examines Processing-Near-Memory (\emph{PNM})~\cite{Mutlu2019Enabling,Mutlu2019Processing,boroumand2017lazypim,Boroumand2019Conda,Boroumand2018Google,Giannoula2021SynCron,fernandez2020natsa,Alser2020Accelerating,Kim2017GrimFilter,Ahn2015PIMenabled,Cali2020GenASM,Singh2019Napel,Singh2020NEROAN,ahn2015scalable,Hadi2016Chameleon,hashemi2016accelerating,Farmahini2015NDA,Gao2015Practical,Gao2016HRL,hashemi2016continuous,Hsieh2016accelerating,Hsieh2016TOM,choe2019concurrent,liu2017concurrent,Gu2020iPIM,ke2019recnmp,Kwon2019TensorDIMM,huangfu2019medal,kim2016neurocube,Nai2017GraphPIM,Zhang2018GraphP,Youwei2019GraphQ,Drumond2017mondrian,pugsley2014ndc,Zhang2014TOPPIM,Zhu2013Accelerating,upmem,Lee2021HardwareAA,gao2017tetris,Dai2018GraphH,Nair2015Active,Alves2015Opportunities,Nag2021OrderLight,Park2021Trim,Sadredini2021Sunder,Gu2021DLUX,Oliveira2021Damov,Lockerman2020Livia,Sun2021ABCDIMMAT,olgun2021pidram}. PNM integrates processing units near or inside the memory via a 3D PNM configuration (i.e., processing units are located at the logic layer of 3D-stacked memories)~\cite{boroumand2017lazypim,Boroumand2019Conda,Boroumand2018Google,Nai2017GraphPIM,Zhang2018GraphP,ahn2015scalable,Gao2015Practical,liu2017concurrent,choe2019concurrent,Youwei2019GraphQ,Drumond2017mondrian,Zhang2014TOPPIM,Nair2015Active,pugsley2014ndc}, a 2.5D PNM configuration (i.e., processing units are located in the same package as the CPU connected via silicon interposers)~\cite{Singh2020NEROAN,fernandez2020natsa,Giannoula2021SynCron}, a 2D PNM configuration (i.e., processing units are placed inside DDRX DIMMs)~\cite{Hadi2016Chameleon,Gu2020iPIM,Alves2015Opportunities,Nag2021OrderLight,Park2021Trim,Sadredini2021Sunder,Gu2021DLUX,ke2019recnmp,tensordimm,Nider2020Processing,Cho2021Accelerating,Zois2018Massively,Lavenier2016DNA,lavenier2020Variant}, or at the memory controller of CPU \camtwo{systems}~\cite{hashemi2016accelerating,hashemi2016continuous,Lockerman2020Livia}. These works propose hardware designs for irregular applications like graph processing~\cite{Dai2018GraphH,Nai2017GraphPIM,Youwei2019GraphQ,ahn2015scalable,Ahn2015PIMenabled,boroumand2017lazypim,Boroumand2019Conda}, bioinformatics~\cite{Cali2020GenASM,Kim2017GrimFilter,Lavenier2016DNA,lavenier2020Variant,Giannoula2021SynCron}, neural networks~\cite{Choi2010Model,Gu2020iPIM,Boroumand2018Google,gao2017tetris,kim2016neurocube,Boroumand2021Google,fernandez2020natsa,Singh2020NEROAN}, pointer-chasing workloads~\cite{liu2017concurrent,choe2019concurrent,Hsieh2016accelerating,Giannoula2021SynCron}, and databases~\cite{Drumond2017mondrian}. However, \emph{none} of these works examines the \spmv{} kernel in such systems.}

Several prior works enable \camone{Processing-Using-Memory (\emph{PUM})~\cite{Seshadri2017Ambit,Aga2017Compute,Eckert2018Neural,Fujiki2019Duality,Kang2014Energy,Li2016Pinatubo,Seshadri2013RowClone,Angizi2019GraphiDe,Chang2016LISA,Gao2019ComputeDRAM,Xin2020ELP2IM,li2017drisa,Deng2018DrAcc,Hajinazar2021SIMDRAM,Rezaei2020NoM,Wang2020Figaro,Ali2020InMemory,levi2014Loci,Kvatinsky2014Magic,Shafiee2016ISAAC,Kvatinsky2011Memristor,Gaillardon2016Programmable,Bhattacharjee2017ReVAMP,Hamdioui2015Memristor,Xie2015FastBL,Song2018GraphR,Ankit2020Panther,Ankit2019PUMA,Chi2016PRIME,Xi2021Memory,Zheng2016RRAM,Hamdioui2017Memristor,Yu2018Memristive,Kim2018PUF,ferreira2021pluto,Wu2021Sieve,Yuan2021FORMS}. PUM exploits the operational principles of memory cells to perform computation within the memory chip. Prior works propose PUM designs using SRAM~\cite{Aga2017Compute,Eckert2018Neural,Fujiki2019Duality,Kang2014Energy}, DRAM~\cite{Seshadri2017Ambit,Seshadri2013RowClone,Angizi2019GraphiDe,Chang2016LISA,Gao2019ComputeDRAM,Xin2020ELP2IM,li2017drisa,Deng2018DrAcc,Hajinazar2021SIMDRAM,Rezaei2020NoM,Wang2020Figaro,Ali2020InMemory,Kim2018PUF,ferreira2021pluto,Wu2021Sieve}, PCM~\cite{Li2016Pinatubo} or RRAM/memristive memory technologies~\cite{levi2014Loci,Kvatinsky2014Magic,Shafiee2016ISAAC,Kvatinsky2011Memristor,Gaillardon2016Programmable,Bhattacharjee2017ReVAMP,Hamdioui2015Memristor,Xie2015FastBL,Song2018GraphR,Ankit2020Panther,Ankit2019PUMA,Chi2016PRIME,Xi2021Memory,Zheng2016RRAM,Hamdioui2017Memristor,Yu2018Memristive,Yuan2021FORMS}. A few PUM works~\cite{Shafiee2016ISAAC,Chi2016PRIME,Aga2017Compute,Eckert2018Neural,li2017drisa,Deng2018DrAcc,Hajinazar2021SIMDRAM} enable the multiplication operation inside memory cells with the goal of performing efficient matrix vector multiplication at low cost within the memory chip. These works design hardware-based solutions to accelerate the \emph{dense} matrix vector multiplication (GEMV) kernel via PUM. However, there is \emph{no} prior work that leverages PUM to accelerate the \emph{Sparse} Matrix Vector Multiplication (\spmv{}) kernel using state-of-the-art compressed matrix storage formats.}

\noindent\textbf{Sparse Matrix Kernels in PIM Systems.} 
Xie et al.~\cite{Xie2021SpaceA} design heterogenous PIM units to accelerate \spmv{} \camone{via a 3D PNM configuration, i.e., in HMC-based PIM systems.} \camtwo{Sun et al.~\cite{Sun2021ABCDIMMAT} leverage the buffer device space of DIMM modules to add one processing unit per each DIMM module, and design low-cost inter-DIMM broadcast collectives to minimize data transfer overheads on irregular workloads, like \spmv{} and graph processing, executed in 2D PNM configurations.} Zhu et al.~\cite{Zhu2013Accelerating} propose a PIM accelerator for Sparse Matrix Matrix Multiplication via a 3D PNM configuration. Fujiki et al.~\cite{Fujiki2019Near} enhance the memory controllers of GPUs with PIM cores to transform the matrix from the CSR to the DCSR format~\cite{Changwan2018Efficient} on the fly to minimize memory traffic on \spmv{} execution. These works propose hardware designs for sparse matrix kernels. In contrast, our work studies software optimizations and strategies to efficiently map compressed matrix storage formats on real near-bank PIM systems, and accelerate \spmv{} execution on such systems.


\noindent\textbf{\spmv{} in Commodity Systems.} Numerous prior works propose optimized \spmv{} algorithms for CPUs~\cite{Elafrou2018SparseX,Buluc2011Reduced,Elafrou2017PerformanceAA,Kjolstad2017Taco,Merrill2016Merge,Willcock2006Accelerating,Williams2007Optimization,Namashivayam2021Variable,Tang2015Optimizing,Elafrou2019Conflict,Vuduc2005oski,Elafrou2017PerformanceXeon,Rong2016Sparso,Xie2018CVR,Xiao2021CASpMV,Hou2017Auto,Pinar1999Improving,Liu2013Efficient,Mellor2004Optimizing,Oliker2002Effects,Vuduc2005Fast,Toledo1997Improving,Temam1992Characterizing,Aktemur2018ASM,Zhao2020Exploring}, GPUs~\cite{Bolz2003Sparse,Hong2018Efficient,Liu2014AnEfficient,Wu2010Efficient,Guo2014APerformance,su2012ClSpMV,Steinberger2017Globally,Shengen2014YaSpMV,Bell99Implementing,Choi2010Model,Pichel2012Optimization,Sun2011Optimizintg,Vazquez2011New,Yang2011Fast,Elafrou2019BASMAT,Filippone2017Sparse}, heterogeneous CPU-GPU systems~\cite{Yang2017Hybrid,Indarapu2014Architecture,Yang2015Performance,Benatia2020Sparse,Boyer2010Exact,Pichel2013Sparse,Indarapu2013Architecture,Anastasiadis2021CoCoPeLia}, and distributed CPU systems~\cite{Lee2008Adaptive,Bisseling2005Communication,Bylina2014Performance,Page2018Scalability,Kayaaslan2015Semi,Liu2018Towards,Catalyurek1999Hypergraph,Vastenhouw2005Two,Nastea1996Load,Pelt2014Medium,Grandjean2012Optimal,Boman2013Scalable}. Optimized \spmv{} kernels for processor-centric CPU and GPU systems exploit the shared memory model of these systems and data locality in deep cache hierarchies. However, these kernels  cannot be directly mapped to most near-bank PIM systems, which have a distributed memory model and a shallow cache hierarchy. Most well-tuned \spmv{} kernels for distributed CPU and CPU-GPU systems improve performance by overlapping computation with communication among processing units, and exploiting data locality in large cache memories. In contrast, real near-bank PIM architectures are fundamentally different from CPU-GPU systems, since they are \textit{highly distributed}, i.e., there is no direct communication among PIM cores, and include a shallow memory hierarchy. Therefore, \spmv{} kernels designed for common processor-centric systems cannot be directly used in near-bank PIM systems.

\noindent\textbf{Hardware Accelerators for \spmv{}.} Recent works design accelerators for \spmv{}~\cite{Sadi2019Efficient,Fowers2014AHigh,Grigoras2015Accelerating,Lin2010Design,Umuroglu2014Anenergy,Kanellopoulos2019SMASH,Mukkara2018Exploiting,Nurvitadhi2015Sparse} or other sparse kernels~\cite{Asgari2020Alrescha,Hegde2019ExTensor,Zhang2016CambriconX,Pal2018OuterSpace,Nurvitadhi2016Hardware,Eric2020sigma,Zhou2018CambriconS,Mishra2017Fine,zhang2021asplos,zhang2020sparch,parashar2017scnn,Hwang2020Centaur,qin2020sigma}. In contrast, our work proposes software optimizations and provides the first characterization study of \spmv{} on a real PIM system.

\noindent\textbf{Compressed Matrix Storage Formats.} Prior works propose a range of compressed matrix storage formats~\cite{Im1999Optimizing,Langr2016Evaluation,Liu2015CSR5,Pinar1999Improving,Vuduc2005Fast,Yang2014Optimization,Shengen2014YaSpMV,Kourtis2011CSX,Kourtis2008Optimizing,Belgin2009Pattern,LIL,ELL,bjorck1996numerical,Pooch1973Survey,Shubhabrata2007Scan,Changwan2018Efficient,Liu2013Efficient,Monakov2010Automatically,Saad1989Krylov,Buluc2009Parallel,Martone2014251,Martone2010Blas,Kreutzer2012Sparse} and selection methods to find the most efficient compressed format~\cite{su2012ClSpMV,Niu2021TileSpMV,asgari2020copernicus,Zhao2018Overhead,Sedaghati2015Automatic,Benatia2016Sparse,Zhao2018Bridging,Li2013SMAT,Maggioni2013AdELL,Tan2018Design,Li2015Performance,Benatia2018BestSF}. In this work, we extensively explore the four most widely used \emph{general} compressed matrix formats, and observe that the compressed format (i) needs to provide good balance between computation and memory accesses inside the core pipeline, and (ii) affects load balancing across PIM cores, with corresponding performance implications. Therefore, some compressed formats designed for commodity processor-centric systems might not be suitable \camone{or efficient} for real PIM systems. We leave the exploration of other PIM-suitable \camone{compressed matrix storage} formats for future work.

\section{Conclusion}

We  present \SparseP{}, the first open-source \spmv{} library for real Processing-In-Memory (PIM) systems, and conduct the first comprehensive characterization analysis of the widely-used \spmv{} kernel on a real-world PIM architecture. 

First, we design efficient \spmv{} kernels for real PIM systems. Our proposed \SparseP{} software package supports (1) a wide range of data types, (2) two types of well-crafted data partitioning techniques of the sparse matrix to DRAM banks of PIM-enabled memory, (3) the most popular compressed matrix formats, (4) a wide variety of load balancing schemes across PIM cores, (5) several load balancing schemes across threads of a multithreaded PIM core, and (6) three synchronization approaches among threads within PIM core. 

Second, we conduct an extensive characterization study of \SparseP{} kernels on the state-of-the-art UPMEM PIM system. We analyze \spmv{} execution on one single multithreaded PIM core and thousands of PIM cores using 26 sparse matrices with diverse sparsity patterns. We also compare the performance and energy consumption of \spmv{} on the UPMEM PIM system with those of state-of-the-art CPU and GPU systems to quantify the potential of a real memory-centric PIM architecture on the widely used \spmv{} kernel over conventional processor-centric architectures. Our analysis of \SparseP{} kernels provides programming recommendations for software designers, \camone{as well as} suggestions and hints for hardware and system designers of future PIM systems. 

We believe and hope that our work will provide valuable insights to programmers in the development of efficient sparse linear algebra kernels and other irregular kernels from different application domains tailored for real PIM systems, as well as to architects \camone{and system designers} in the development of \camone{future} memory-centric computing systems.

\begin{acks}

We thank the UPMEM company for valuable support. We thank the anonymous reviewers from SIGMETRICS 2022, and our shepherd, Bhuvan Urgaonkar, for their comments and suggestions. \camone{We thank the SAFARI Research Group members for feedback and the stimulating, scholarly and collaborative intellectual environment they provide. We thank the CSLAB Research Group members for continued and undivided support, insightful comments and valuable feedback.} We acknowledge the support of SAFARI Research Group’s industrial partners, especially ASML, Facebook, Google, Huawei, Intel, Microsoft, VMware, the Semiconductor Research Corporation and the ETH Future Computing Laboratory. Christina Giannoula is funded for her postgraduate studies from the Foundation for Education and European Culture. \camtwo{The} \SparseP{} software package is publicly available at \url{https://github.com/CMU-SAFARI/SparseP}.

\end{acks}

\newpage

\bibliographystyle{ACM-Reference-Format}
\bibliography{references,references_arxiv}

\newpage

\section*{{\Large APPENDIX}}
\appendix

\section{Extended Results}

\subsection{Synchronization Approaches in Block-Based \camtwo{Compressed Matrix} Formats}\label{sec:appendix-1DPU-BCOO}

\begin{figure}[H]
    \centering
    \includegraphics[width=\textwidth]{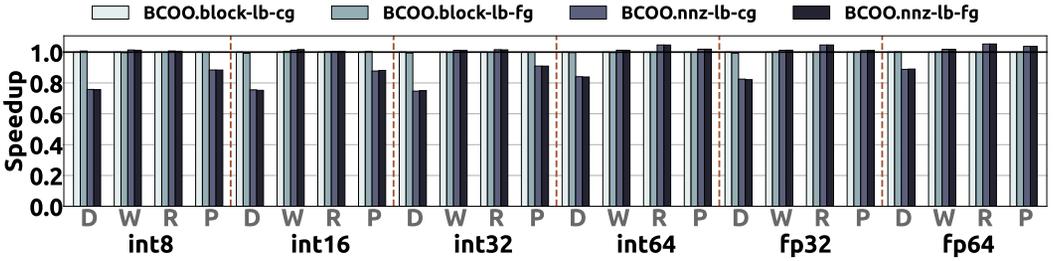}
    \vspace{-15pt}
    \caption{Performance of the BCOO format with various load balancing schemes and synchronization approaches for all the data types and small matrices using 16 tasklets of one DPU.}
    \label{fig:1DPU-bcoo-balance}
    \vspace{-8pt}
\end{figure}

We compare the coarse-grained locking (\textit{lb-cg}) and the fine-grained locking (\textit{lb-fg}) approaches in the BCOO format. Figure~\ref{fig:1DPU-bcoo-balance} shows the performance achieved by the BCOO format for all the data types when balancing the blocks or the non-zero elements \camone{across} 16 tasklets of one DPU. We evaluate all small matrices of Table~\ref{tab:small-matrices}, i.e., delaunay\_n13 (\textbf{D}), wing\_nodal (\textbf{W}), raefsky4 (\textbf{R}) and pkustk08 (\textbf{P}) matrices.

Our key finding is that the fine-grained locking approach performs similarly with the coarse-grained locking approach. The fine-grained locking approach does not increase parallelism in the UPMEM PIM architecture, since memory accesses executed by multiple tasklets to the local DRAM bank are serialized in the DMA engine of the DPU. The same key finding holds independently of the compressed matrix format used.

\subsection{Fine-Grained Data Transfers in 2D Partitioning Techniques}\label{sec:appendix-2D-fgtrans}
\begin{figure}[H]
    \centering
    \includegraphics[width=1\textwidth]{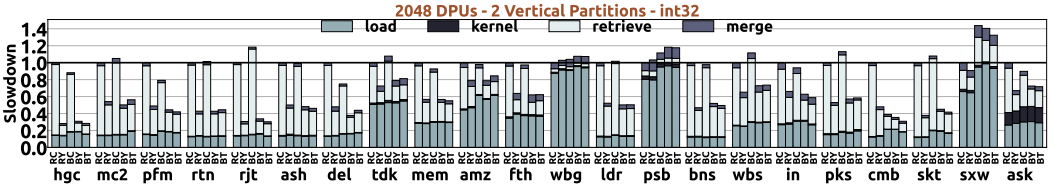}
    \vspace{-15pt}
    \caption{Performance comparison of \texttt{RC}: \texttt{RBDCOO} with coarse-grained transfers, \texttt{RY}: \texttt{RBDCOO} with fine-grained transfers in the output vector, \texttt{BC}: \texttt{BDCOO} with coarse-grained transfers, \texttt{BY}: \texttt{BDCOO} with fine-grained transfers only in the output vector, and \texttt{BT}: \texttt{BDCOO} with fine-grained transfers in both the input and the output vector using the int32 data type, 2048 DPUs and having 2 vertical partitions. Performance is normalized to that of the \texttt{RC} scheme.}
    \label{fig:2D_vp2_fgtransfers}
\end{figure}

\begin{figure}[H]
    \centering
    \includegraphics[width=1\textwidth]{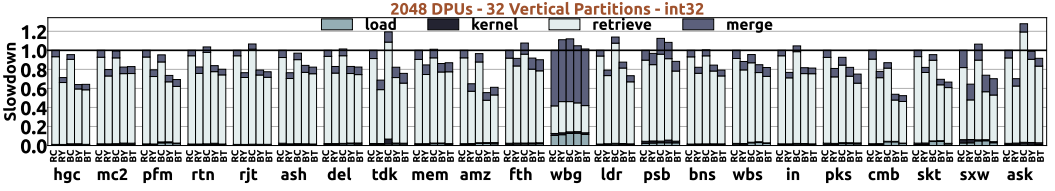}
    \vspace{-15pt}
    \caption{Performance comparison of \texttt{RC}: \texttt{RBDCOO} with coarse-grained transfers, \texttt{RY}: \texttt{RBDCOO} with fine-grained transfers in the output vector, \texttt{BC}: \texttt{BDCOO} with coarse-grained transfers, \texttt{BY}: \texttt{BDCOO} with fine-grained transfers only in the output vector, and \texttt{BT}: \texttt{BDCOO} with fine-grained transfers in both the input and the output vector using the int32 data type, 2048 DPUs and having 32 vertical partitions. Performance is normalized to that of the \texttt{RC} scheme.}
    \label{fig:2D_vp32_fgtransfers}
    \vspace{-4pt}
\end{figure}

Figures~\ref{fig:2D_vp2_fgtransfers} and ~\ref{fig:2D_vp32_fgtransfers} compare coarse-grained data transfers (i.e., performing parallel data transfers to all 2048 DPUs at once, padding with empty bytes at the granularity of 2048 DPUs) with fine-grained data transfers (i.e., iterating over the ranks and for each rank performing parallel data transfers to the 64 DPUs of the same rank, padding with empty bytes at the granularity of 64 DPUs) for all matrices of our large matrix dataset in the \equallyWidth{} and \variableSized{} schemes, respectively. The reported key findings of Figure~\ref{fig:2D_fgtransfers} (Section~\ref{2D-Studies}) apply to all matrices with diverse sparsity patterns.

\subsection{Effect of the Number of Vertical Partitions Using Two Different UPMEM PIM Systems}\label{sec:appendix-2D-vertpartitions}

We compare \spmv{} execution in the two different UPMEM PIM sytems using 2048 DPUs and 16 tasklets for each DPU. Table~\ref{tab:pim-systems} \arxiv{shows the characteristics of two different UPMEM PIM systems. We calculate the available PIM peak performance and PIM bandwidth assuming 2048 DPUs for both PIM systems\footnote{Both UPMEM PIM systems support 20 UPMEM PIM DIMMs with 2560 DPUs in total. However, both UPMEM-based PIM systems include multiple faulty DPUs. Thus, for a fair comparison between two systems we conduct our experiments using 2048 DPUs in both systems.}.} \camtwo{We estimate the PIM peak performance as $Total\_DPUs * AT$, where the arithmetic throughput (AT) is calculated for the multiplication operation by running the arithmetic throughput microbenchmark of the PrIM benchmark suite~\cite{Gomez2021Analysis,Gomez2021Benchmarking} in each of the two UPMEM PIM systems (See Appendix~\ref{sec:appendix-1DPU-AT}). We estimate the PIM bandwidth as $Total\_DPUs * Bandwidth\_DPU$, where the $Bandwidth\_DPU$ is calculated according to prior work~\cite{Gomez2021Analysis, Gomez2021Benchmarking}. Specifically, the theoretical maximum MRAM bandwidth (i.e., $Bandwidth\_DPU$) is 700 MB/s and 850 MB/s at a DPU frequency of 350 MHz (PIM system A) and 425 MHz (PIM system B), respectively.
}

\begin{table}[H]
\begin{center}
\centering
\resizebox{1.0\linewidth}{!}{
\begin{tabular}{|l||c|c|c|c|c|c|c|c|}
    \hline
      \cellcolor{gray!15} & \cellcolor{gray!15}\raisebox{-0.20\height}{\textbf{Avail.}} & \cellcolor{gray!15} & \cellcolor{gray!15}\raisebox{-0.20\height}{\textbf{PIM Peak}} & \cellcolor{gray!15}\raisebox{-0.20\height}{\textbf{PIM}} &  \cellcolor{gray!15}  & \cellcolor{gray!15}\raisebox{-0.20\height}{\textbf{CPU Peak}} & \cellcolor{gray!15}\raisebox{-0.20\height}{\textbf{Bus}}  \\
       \multirow{-2}{*}{\cellcolor{gray!15}\textbf{System}}   & \cellcolor{gray!15} \textbf{DPUs} & \multirow{-2}{*}{\cellcolor{gray!15}\textbf{Frequency}}  & \cellcolor{gray!15}\textbf{Performance} & \cellcolor{gray!15}\textbf{Bandwidth} & \multirow{-2}{*}{\cellcolor{gray!15} \textbf{Host CPU}} & \cellcolor{gray!15}\textbf{Performance} & \cellcolor{gray!15}\textbf{Bandwidth}  \\
    
    \hline \hline
    PIM System A & 2048 DPUs & 350 MHz & 3.78 GFLOPS & 1.43 TB/s &  Intel Xeon Silver 4110 @2.1 GHz & 660 GFLOPS & 23.1 GB/s\\ \hline 
    PIM System B & 2048 DPUs & 425 MHz & 4.63 GFLOPS & 1.74 TB/s &  Intel Xeon Silver 4215 @2.5 GHz & 1016 GFLOPS & 21.8 GB/s \\ \hline 
\end{tabular}
}
\end{center}
\vspace{4pt}
\caption{Evaluated UPMEM PIM Systems.}
\label{tab:pim-systems}
\vspace{-10pt}
\end{table}

\begin{figure}[t]
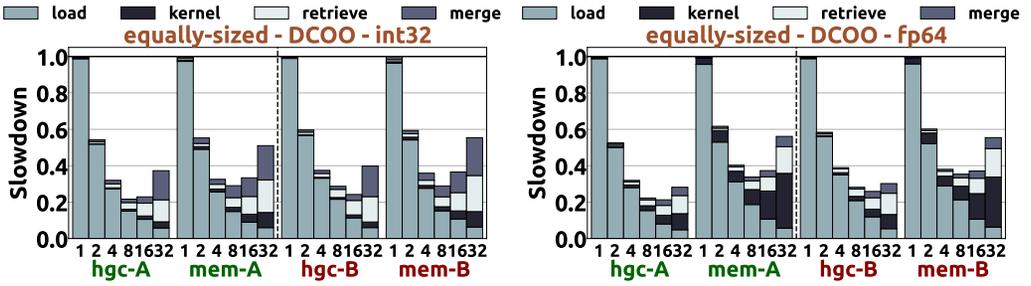

    \centering
    \begin{minipage}{\textwidth}
    \centering
    \includegraphics[width=.49\textwidth]{2D-partitioning/cloud4_cloud7_fixed_vertical_dpus2048_int32.pdf}
    \includegraphics[width=.49\textwidth]{2D-partitioning/cloud4_cloud7_fixed_vertical_dpus2048_dbl64.pdf}
    \end{minipage}
    \vspace{-8pt}
    \caption{Execution time breakdown of \texttt{DCOO} using 2048 DPUs when varying the number of vertical partitions from 1 to 32 for the int32 (left) and fp64 (right) data types on two different UPMEM PIM systems.}
    \label{fig:cloud4-cloud7_fixed_vertpartitions}
\end{figure}

\begin{figure}[t]
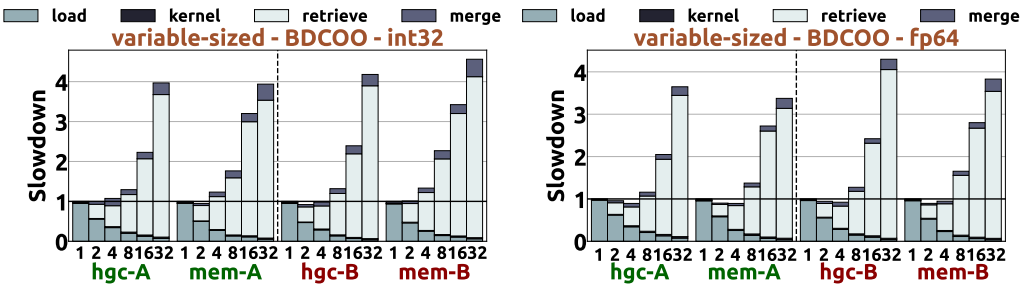

    \centering
    \begin{minipage}{\textwidth}
    \centering
    \includegraphics[width=.49\textwidth]{2D-partitioning/cloud4_cloud7_bal_vertical_dpus2048_int32.pdf}
    \includegraphics[width=.49\textwidth]{2D-partitioning/cloud4_cloud7_bal_vertical_dpus2048_dbl64.pdf}
    \end{minipage}
    \vspace{-8pt}
    \caption{Execution time breakdown of \texttt{BDCOO} using 2048 DPUs when varying the number of vertical partitions from 1 to 32 for the int32 (left) and fp64 (right) data types on two different UPMEM PIM systems.}
    \label{fig:cloud4-cloud7_bal_vertpartitions}
\end{figure}

Figures~\ref{fig:cloud4-cloud7_fixed_vertpartitions}, ~\ref{fig:cloud4-cloud7_rbal_vertpartitions} and ~\ref{fig:cloud4-cloud7_bal_vertpartitions} \arxiv{compare \spmv{} execution in the two different UPMEM PIM systems (2048 DPUs) using 2D-partitioned kernels with the COO format, when varying the number of vertical partitions from 1 to 32 (in steps of multiple of 2) for the int32 (left) and fp64 (right) data types. }

We observe \arxiv{that the number of vertical partitions that provides the best performance on \spmv{} execution varies depending on the input matrix and the PIM system. For example, in PIM system B with the int32 data type, \texttt{DCOO} performs best for the \texttt{hgc} matrix with 16 vertical partitions, while in PIM system A, \texttt{DCOO} performs best for the same matrix with 8 vertical partitions. Similarly, in PIM system A with the fp64 data type, \texttt{BDCOO} performs best for the \texttt{rjt} matrix with 4 vertical partitions. Instead, in PIM system B with the fp64 data type, \texttt{BDCOO}'s performance does not improves for the \texttt{rjt} matrix when having more than 1 vertical partition (i.e., compared to when using the 1D partitioning technique). We conclude that the best-performing parallelization scheme that achieves the best performance in \spmv{} depends on the characteristics of both the input sparse matrix and the underlying PIM system.}

\vspace{6pt}

\begin{figure}[t]
    \centering
    \begin{minipage}{\textwidth}
    \centering
    \includegraphics[width=.49\textwidth]{2D-partitioning/cloud4_cloud7_rbal_vertical_dpus2048_int32.pdf}
    \includegraphics[width=.49\textwidth]{2D-partitioning/cloud4_cloud7_rbal_vertical_dpus2048_dbl64.pdf}
    \end{minipage}
    \vspace{-8pt}
    \caption{Execution time breakdown of \texttt{RBDCOO} using 2048 DPUs when varying the number of vertical partitions from 1 to 32 for the int32 (left) and fp64 (right) data types on two different UPMEM PIM systems.}
    \label{fig:cloud4-cloud7_rbal_vertpartitions}
\end{figure}

\subsection{Performance of Compressed Matrix Formats Using 2D Partitioning Techniques}\label{sec:appendix-2D-formats}
\begin{figure}[H]
    \centering
    \includegraphics[width=1\textwidth]{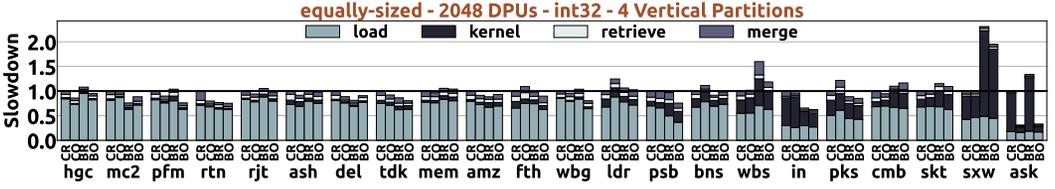}
    \vspace{-16pt}
    \caption{End-to-end execution time breakdown of the \equallySized{} 2D partitioning technique for CR: \texttt{DCSR}, CO: \texttt{DCOO}, BR: \texttt{DBCSR} and BO: \texttt{DBCOO} schemes using 4 vertical partitions and the int32 data type. Performance is normalized to that of \texttt{DCSR}.}
    \label{fig:2D_formats_fixed_all}
\end{figure}

\begin{figure}[H]
    \centering
    \includegraphics[width=1\textwidth]{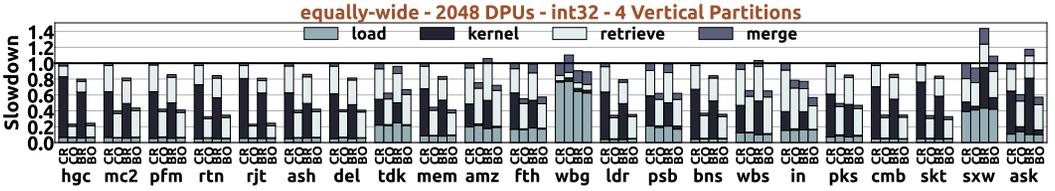}
    \vspace{-16pt}
    \caption{End-to-end execution time breakdown of  the \equallyWidth{} 2D partitioning technique for CR: \texttt{RBDCSR}, CO: \texttt{RBDCOO}, BR: \texttt{RBDBCSR} and BO: \texttt{RBDBCOO} schemes using 4 vertical partitions and the int32 data type. Performance is normalized to that of \texttt{RBDCSR}.}
    \label{fig:2D_formats_rbal_all}
\end{figure}

\begin{figure}[H]
    \centering
    \includegraphics[width=1\textwidth]{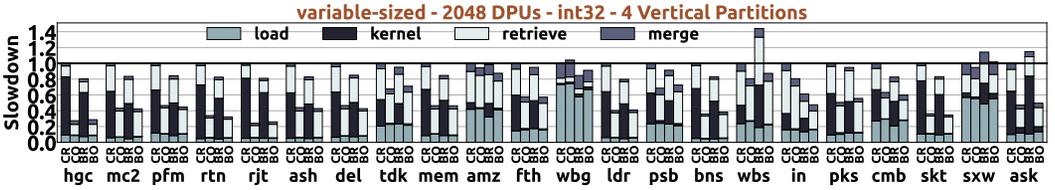}
    \vspace{-16pt}
    \caption{End-to-end execution time breakdown of the \variableSized{} 2D partitioning technique for CR: \texttt{BDCOO}, CO: \texttt{BDCOO}, BR: \texttt{BDBCSR} and BO: \texttt{BDBCOO} schemes using 4 vertical partitions and the int32 data type. Performance is normalized to that of \texttt{BDCSR}.}
    \label{fig:2D_formats_bal_all}
\end{figure}

Figures~\ref{fig:2D_formats_fixed_all},~\ref{fig:2D_formats_rbal_all},~\ref{fig:2D_formats_bal_all} compare the performance achieved by various
compressed matrix formats for each of the three types of the 2D partitioning technique for all matrices of our large matrix dataset. The reported key findings explained in Section~\ref{2D-Formats} apply to all matrices with diverse sparsity patterns.


\subsection{Analysis of 1D- and 2D-Partitioned Kernels in Two  UPMEM PIM Systems}\label{sec:appendix-1D_2D}

\begin{figure}[t]
    \centering
    \includegraphics[width=\textwidth]{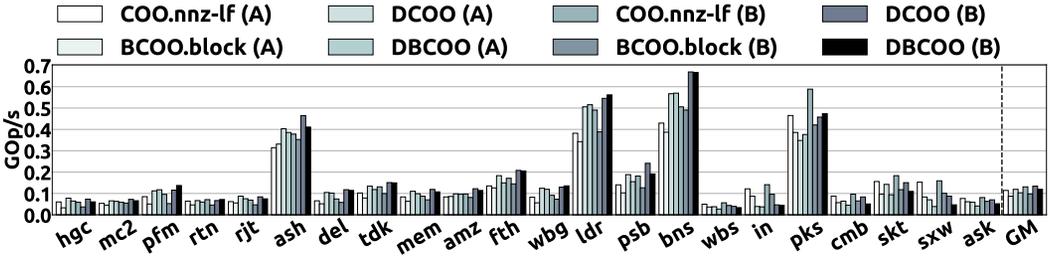}
    \vspace{-18pt}
    \caption{Throughput of 1D- and 2D-partitioned kernels for the fp32 data type using two different UPMEM PIM systems.}
    \label{fig:cloud4-cloud7_ops}
\end{figure}

\begin{figure}[t]
    \centering
    \includegraphics[width=\textwidth]{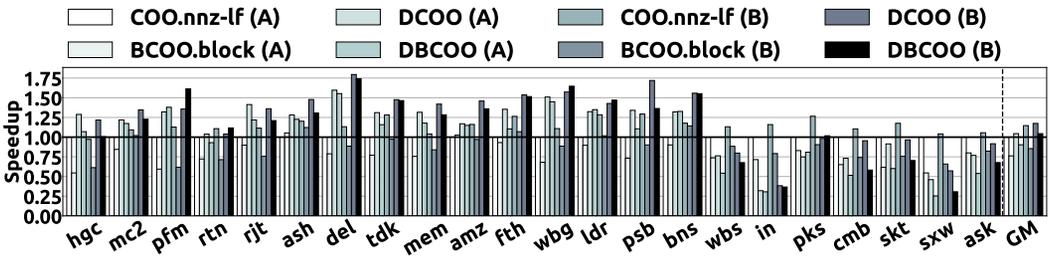}
    \vspace{-16pt}
    \caption{Performance comparison of 1D- and 2D-partitioned kernels for the fp32 data type using two different UPMEM PIM systems. Performance is normalized to that of \texttt{COO.nnz-lf (A)}.}
    \label{fig:cloud4-cloud7_perf}
\end{figure}

Figures~\ref{fig:cloud4-cloud7_ops} and ~\ref{fig:cloud4-cloud7_perf} \arxiv{compare the throughput and the performance, respectively, achieved by the best-performing 1D- and 2D-partitioned kernels in two different UPMEM PIM systems (Table~\ref{tab:pim-systems} presents the characteristics of the two UPMEM PIM systems). For  1D partitioning, we use the lock-free COO (\texttt{COO.nnz-lf}) and coarse-grained locking BCOO (\texttt{BCOO.block}) kernels. For each matrix, we vary the number of DPUs from 64 to 2048 DPUs, and select the best-performing end-to-end execution throughput. For 2D partitioning, we use the \equallySized{} COO (\texttt{DCOO}) and BCOO (\texttt{BCOO}) kernels with 2048 DPUs for both systems. For each matrix, we vary the number of vertical partitions from 2 to 32 (in steps of multiple of 2), and select the best-performing end-to-end execution throughput. 
}

We draw three findings. \arxiv{First, we observe that in both systems the best performance is achieved using a smaller number of DPUs than 2048 DPUs. This is because \spmv{} execution in both UPMEM PIM systems is significantly bottlenecked by expensive data transfers performed via the narrow memory bus. As a result, the best-performing 1D- and 2D-partitioned kernels trade off computation with lower data transfer costs, thus causing many DPUs to be \textit{idle}. Second, we find that in both systems the 2D-partitioned kernels outperform the 1D-partitioned kernels in regular matrices (i.e., from \texttt{hgc} to \texttt{bns} matrices on x axis), while the 1D-partitioned kernels outperform the 2D-partitioned kernels in scale-free matrices, i.e., in matrices that have high non-zero element disparity among rows and columns (i.e., from \texttt{wbs} to \texttt{ask} matrices on x axis). Third, we observe that PIM system B improves performance over PIM system A by 1.14$\times$ (averaged across all matrices). This is because the DPUs of the PIM system B run at a higher frequency than that of PIM system A (425 MHz vs 350 MHz), providing higher peak performance on the system. Specifically, with 2048 DPUs, peak performance of the PIM system A and PIM system B is 3.78 GFlops and 4.63 GFlops, respectively, i.e., PIM system B provides 1.22 $\times$ higher computation throughput than PIM system A.}

\newpage

\section{Arithmetic Throughput of One DPU for the Multiplication Operation}\label{sec:appendix-1DPU-AT}
We evaluate the arithmetic throughput of the DPU for the multiplication (MUL) operation. We use the arithmetic throughput microbenchmark of \camtwo{the PrIM benchmark suite}~\cite{Gomez2021Benchmarking,Gomez2021Analysis} and configure it for the all data types. 

\begin{figure}[H]
\centering
\begin{minipage}{1.0\textwidth}
\centering
\includegraphics[width=.48\textwidth]{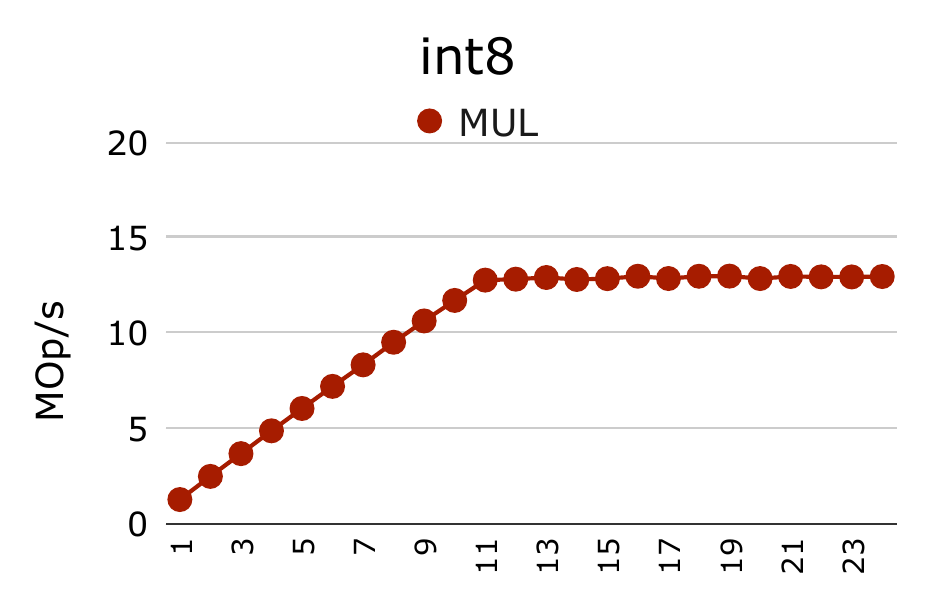}\hspace{12pt}
\includegraphics[width=.48\textwidth]{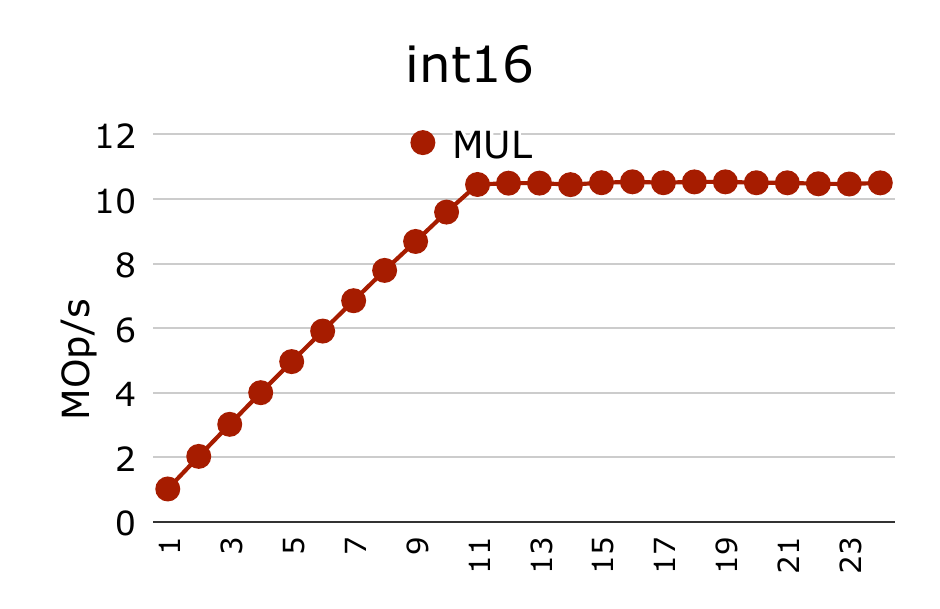}
\end{minipage} %
\begin{minipage}{1.0\textwidth}
\centering
\includegraphics[width=.48\textwidth]{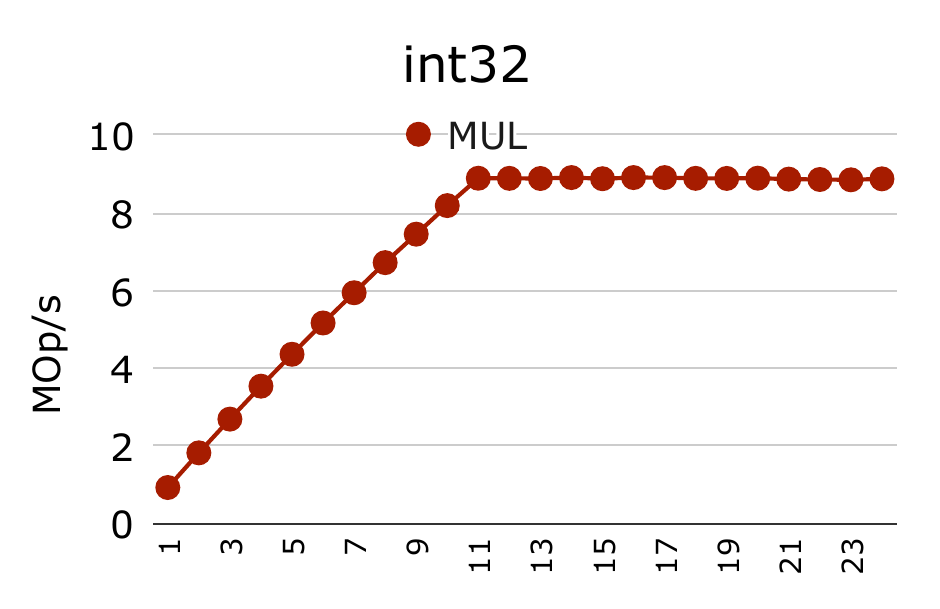}\hspace{12pt}
\includegraphics[width=.48\textwidth]{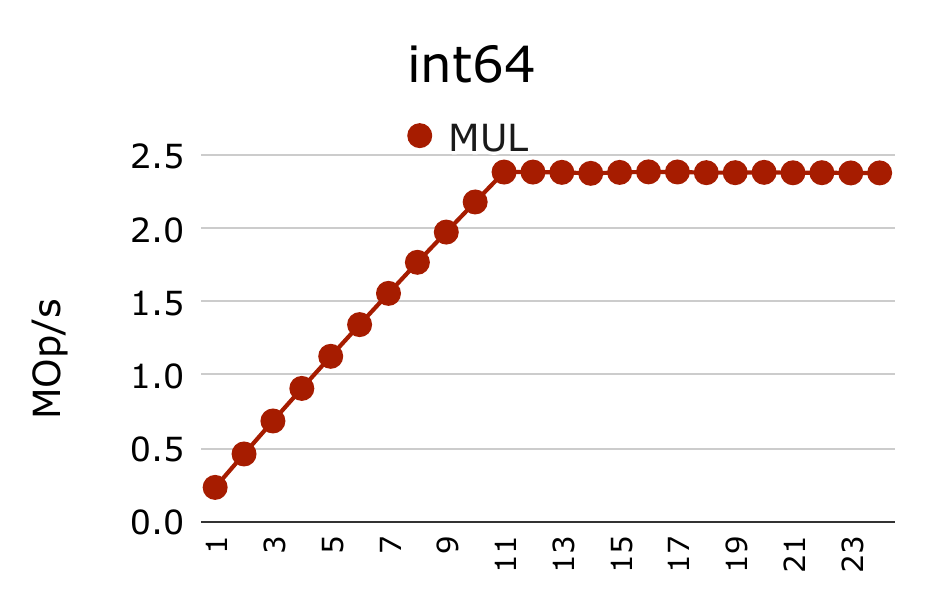}
\end{minipage} %
\begin{minipage}{1.0\textwidth}
\centering
\includegraphics[width=.48\textwidth]{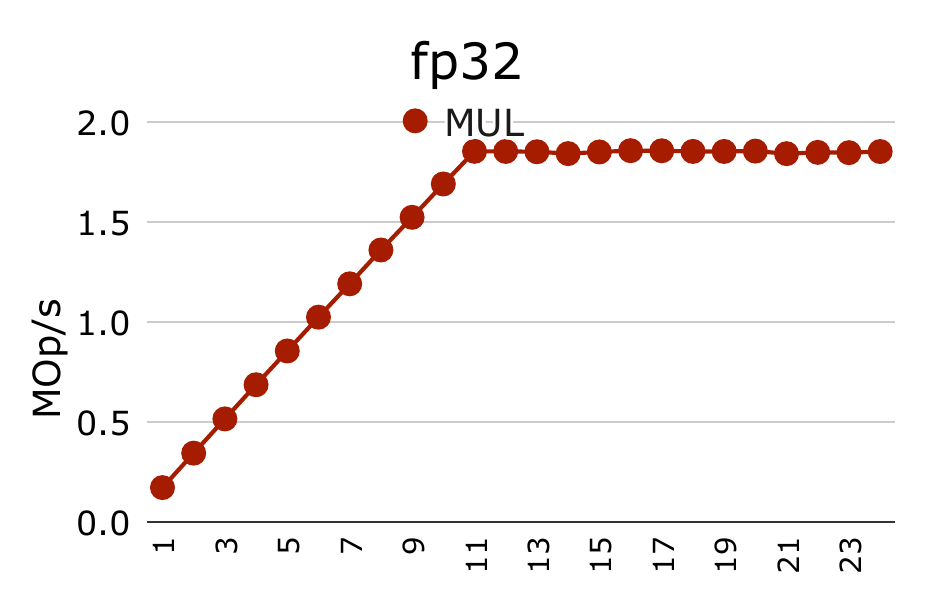}\hspace{12pt}
\includegraphics[width=.48\textwidth]{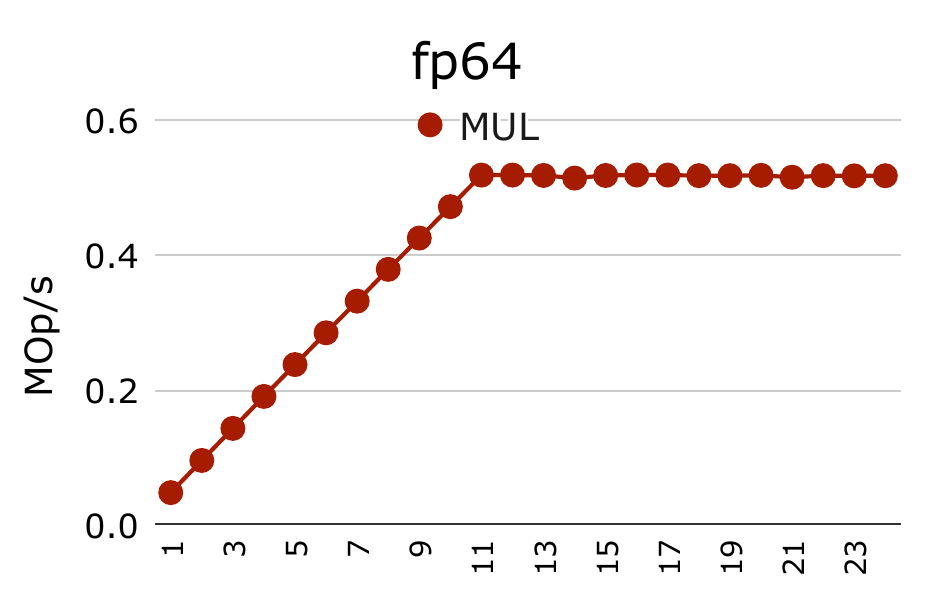}
\end{minipage}
\vspace{-4pt}
\caption{Throughput of the MUL operation on one DPU \camtwo{at 350 MHz} for all the data types.}
\label{fig:1DPU-ai-cloud4}
\end{figure}

Figure~\ref{fig:1DPU-ai-cloud4} shows the measured arithmetic throughput (in MOperations per second) for the MUL operation varying the number of tasklets of one DPU \camtwo{at 350 MHz (PIM system A in Table~\ref{tab:pim-systems})} for all the data types. The arithmetic throughput for the MUL operation is 12.941 MOps, 10.524 MOps, 8.861 MOps, 2.381 MOps, 1.847 MOps, and 0.517 MOps for the int8, int16, int32, int64, fp32 and fp64 data types, respectively.

\newpage

\begin{figure}[H]
\centering
\begin{minipage}{1.0\textwidth}
\centering
\includegraphics[width=.48\textwidth]{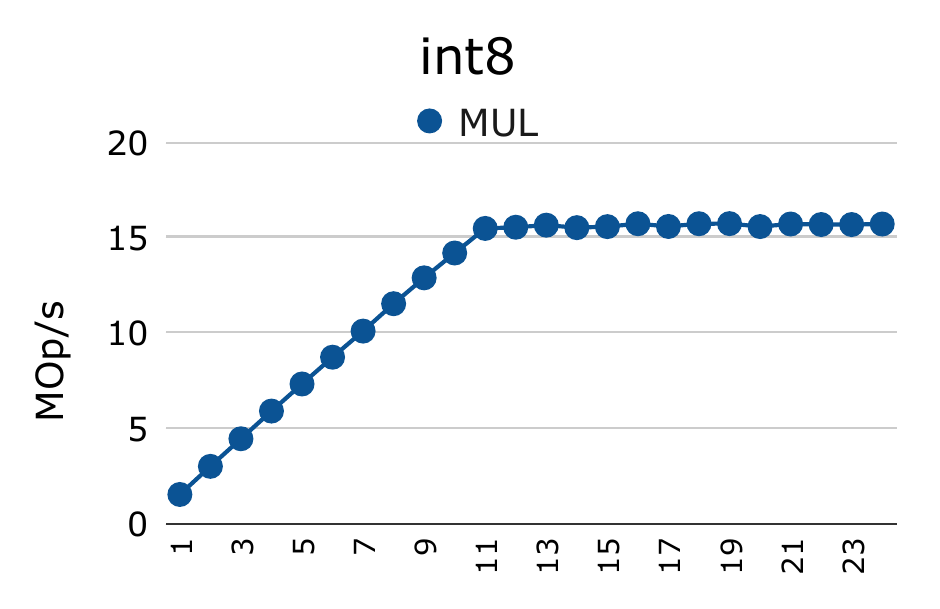}\hspace{12pt}
\includegraphics[width=.48\textwidth]{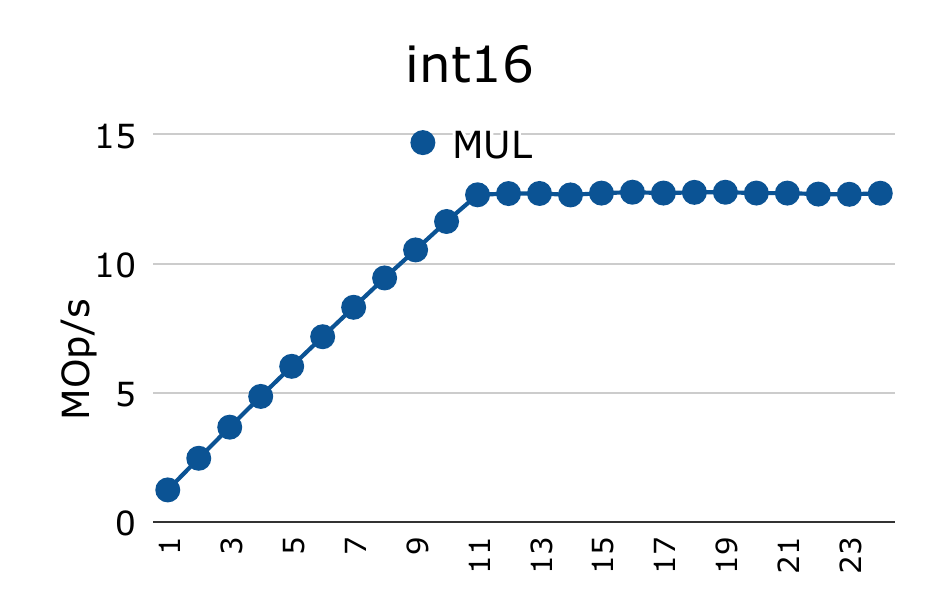}
\end{minipage} %
\begin{minipage}{1.0\textwidth}
\centering
\includegraphics[width=.48\textwidth]{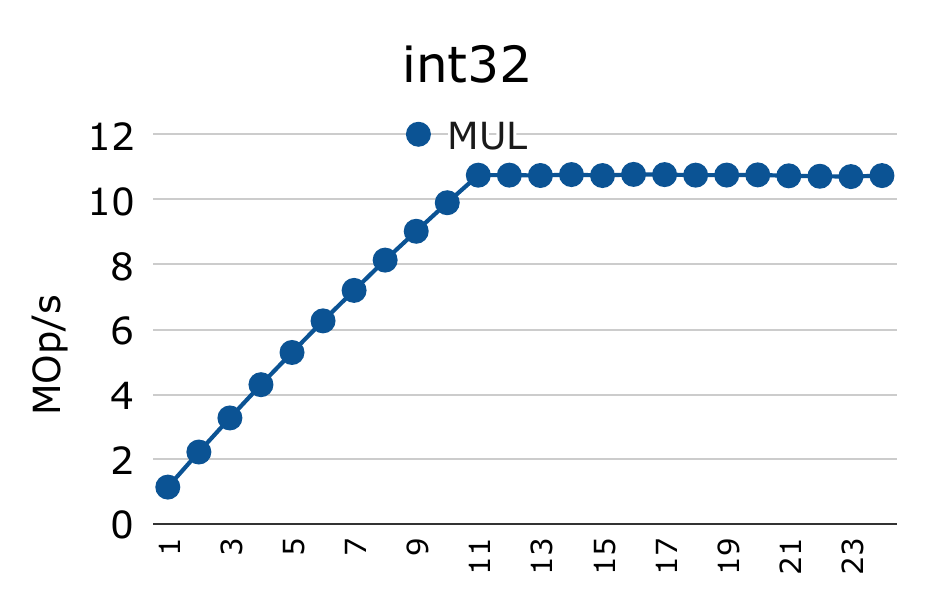}\hspace{12pt}
\includegraphics[width=.48\textwidth]{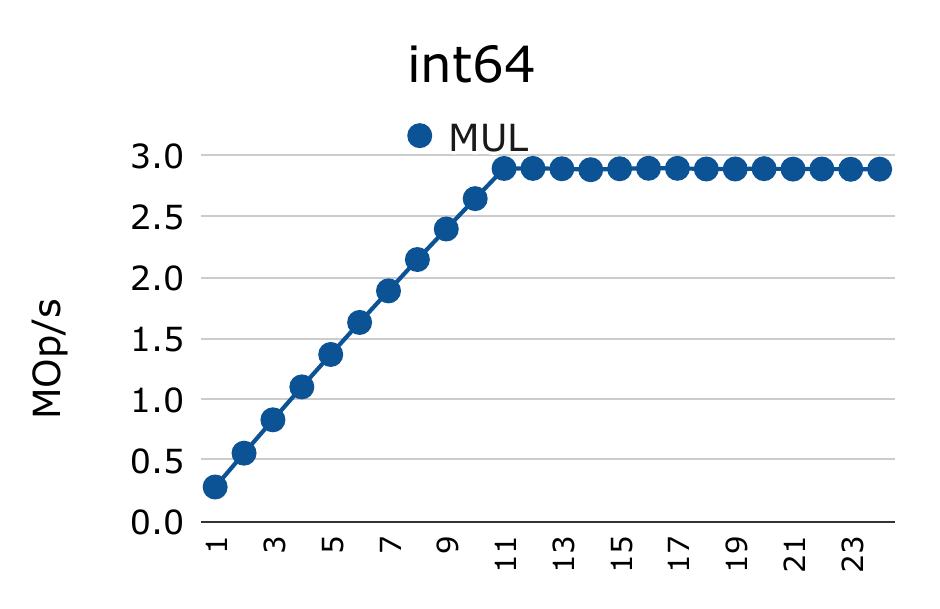}
\end{minipage} %
\begin{minipage}{1.0\textwidth}
\centering
\includegraphics[width=.48\textwidth]{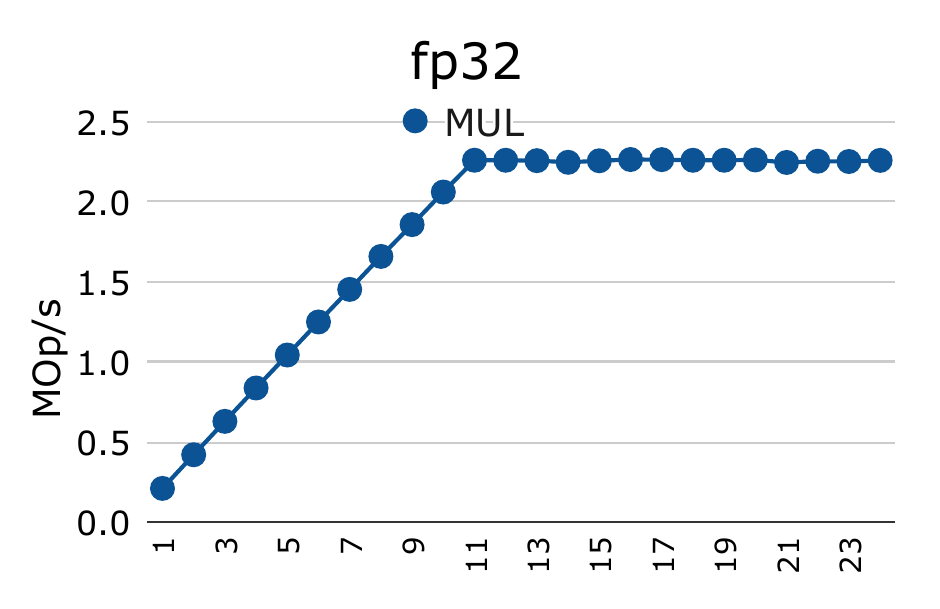}\hspace{12pt}
\includegraphics[width=.48\textwidth]{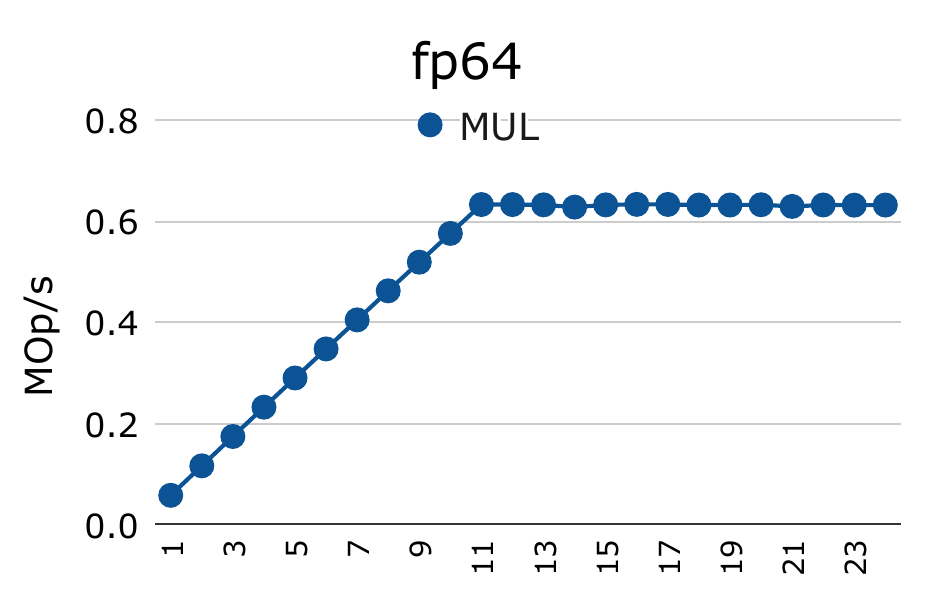}
\end{minipage}
\vspace{-4pt}
\caption{Throughput of the MUL operation on one DPU \camtwo{at 425 MHz} for all the data types.}
\label{fig:1DPU-ai-cloud7}
\end{figure}

Figure~\ref{fig:1DPU-ai-cloud7} \camtwo{shows the measured arithmetic throughput (in MOperations per second) for the MUL operation varying the number of tasklets of one DPU at 425 MHz (PIM system B in Table~\ref{tab:pim-systems}) for all the data types. The arithmetic throughput for the MUL operation is 15.656 MOps, 12.721 MOps, 10.732 MOps, 2.888 MOps, 2.259 MOps, and 0.631 MOps for the int8, int16, int32, int64, fp32 and fp64 data types, respectively.}

\newpage

\section{The \SparseP{} Software Package}\label{sec:appendix-sparseP}
Table~\ref{table:appendix-library} summarizes the \spmv{} PIM kernels provided by the \SparseP{} library. All kernels support a wide range of data types, i.e., 8-bit integer, 16-bit integer, 32-bit integer, 64-bit integer, 32-bit float, and 64-bit float data types.

\begin{table}[H]
\vspace{10pt}
\centering
\begin{minipage}{1\linewidth}
\resizebox{1.0\linewidth}{!}{
\begin{tabular}{| l | l | l | l | l | l |} 
 \hline
\cellcolor{gray!15}\raisebox{-0.20\height}{\textbf{Partitioning}} & \cellcolor{gray!15}\raisebox{-0.20\height}{\textbf{Compressed}} & \cellcolor{gray!15}\raisebox{-0.20\height}{\textbf{Balancing}} & \cellcolor{gray!15}\raisebox{-0.20\height}{\textbf{Balancing}}  & \cellcolor{gray!15}\raisebox{-0.20\height}{\textbf{Synchronization}} \\ 
 \cellcolor{gray!15}\textbf{Technique} & \cellcolor{gray!15}\textbf{Format} & \cellcolor{gray!15}\textbf{Across PIM Cores} & \cellcolor{gray!15}\textbf{Across Threads}  & \cellcolor{gray!15}\textbf{Approach}  \\ [0.5ex] \hline \hline
    \multirow{9}{*}{\hspace*{2.1em}\turnbox{0}{1D}} & \multirow{2}{*}{CSR} &  rows  &  rows, nnz$^{\star}$ & -   \\
    & & nnz$^{\star}$ &  rows, nnz$^{\star}$ & -   \\ \cline{2-5}
    & \multirow{3}{*}{COO} &  rows  &  rows, nnz$^{\star}$ & -   \\
    & & nnz$^{\star}$  &  rows, nnz$^{\star}$ & -   \\
    & & nnz  &  nnz & lb-cg / lb-fg / lf \\  \cline{2-5}
    & \multirow{2}{*}{BCSR} &  blocks$^{\dagger}$  &  blocks$^{\dagger}$, nnz$^{\dagger}$ & lb-cg$^{\ddagger}$ / lb-fg$^{\ddagger}$   \\
    & & nnz$^{\dagger}$  &  blocks$^{\dagger}$, nnz$^{\dagger}$ & lb-cg$^{\ddagger}$ / lb-fg$^{\ddagger}$ \\  \cline{2-5}
    & \multirow{2}{*}{BCOO} &  blocks  &  blocks, nnz & lb-cg / lb-fg / lf \\
    & & nnz &  blocks, nnz & lb-cg / lb-fg / lf  \\ \hline
    \multirow{4}{*}{\shortstack{2D \\ \equallySized}} & CSR & - & rows, nnz$^{\star}$ & -  \\  \cline{2-5}
    &  COO  & - & nnz & lb-cg / lb-fg / lf \\  \cline{2-5}
    &  BCSR  & - & blocks$^{\dagger}$, nnz$^{\dagger}$ & lb-cg$^{\ddagger}$ / lb-fg$^{\ddagger}$  \\  \cline{2-5}
    &  BCOO & - & blocks, nnz & lb-cg / lb-fg  \\ \hline
  \multirow{6}{*}{\shortstack{2D \\ \equallyWidth}} &  CSR & nnz$^{\star}$ & rows, nnz$^{\star}$ & -  \\  \cline{2-5}
    &  COO & nnz & nnz & lb-cg / lb-fg / lf   \\  \cline{2-5}
    & \multirow{2}{*}{BCSR} & blocks$^{\dagger}$ & blocks$^{\dagger}$, nnz$^{\dagger}$ & lb-cg$^{\ddagger}$ / lb-fg$^{\ddagger}$ \\
    &  & nnz$^{\dagger}$ & blocks$^{\dagger}$, nnz$^{\dagger}$ & lb-cg$^{\ddagger}$ / lb-fg$^{\ddagger}$ \\ \cline{2-5}
    & \multirow{2}{*}{BCOO} & blocks  & blocks, nnz & lb-cg / lb-fg   \\ 
    & & nnz  & blocks, nnz & lb-cg / lb-fg   \\ \hline     
  \multirow{6}{*}{\shortstack{2D \\ \variableSized}} &  CSR & nnz$^{\star}$ & rows, nnz$^{\star}$& -  \\  \cline{2-5}
    &  COO & nnz  & nnz & lb-cg / lb-fg / lf \\  \cline{2-5}
    &  \multirow{2}{*}{BCSR} & blocks$^{\dagger}$  & blocks$^{\dagger}$, nnz$^{\dagger}$ & lb-cg$^{\ddagger}$ / lb-fg$^{\ddagger}$   \\
    &  & nnz$^{\dagger}$ & blocks$^{\dagger}$, nnz$^{\dagger}$ & lb-cg$^{\ddagger}$ / lb-fg$^{\ddagger}$   \\ \cline{2-5}
    &  \multirow{2}{*}{BCOO } & blocks  & blocks, nnz & lb-cg / lb-fg  \\ 
    &  &  nnz & blocks, nnz &  lb-cg / lb-fg  \\   \hline   
\end{tabular}
}
\end{minipage} \hspace{2pt}%
\vspace{6pt}
\caption{The \SparseP{} library. $^{\star}$: row-granularity, $^{\dagger}$: block-row-granularity, $^{\ddagger}$: (only for 8-bit integer and small block sizes)}
\label{table:appendix-library}
\end{table}

\newpage

\section{Large Matrix Dataset}\label{sec:appendix-matrix-dataset}
We present the characteristics of the sparse matrices of our large matrix data set. Table~\ref{tab:appendix-large-matrices} 
presents the sparsity of the matrix (i.e., NNZ / (rows x columns)), the standard deviation of non-zero elements among rows (NNZ-r-std) and columns (NNZ-c-std). Table~\ref{tab:appendix-large-matrices-plot} visualizes the sparsity patterns of each sparse matrix of our large matrix data set.

\begin{table}[H]
\begin{center}
\vspace{10pt}
\centering
\begin{minipage}{1\linewidth}
\begin{tabular}{|l||c|r|r|r|r|}
    \hline
    \cellcolor{gray!15}\raisebox{-0.10\height}{\textbf{Matrix Name}} & \cellcolor{gray!15}\raisebox{-0.10\height}{\textbf{Rows x Columns}} & \cellcolor{gray!15}\raisebox{-0.10\height}{\textbf{NNZs}} & \cellcolor{gray!15}\raisebox{-0.10\height}{\textbf{Sparsity}} & \cellcolor{gray!15}\raisebox{-0.10\height}{\textbf{NNZ-r-std}} & \cellcolor{gray!15}\raisebox{-0.10\height}{\textbf{NNZ-c-std}} \\
    \hline \hline
    hugetric-00020 & 7122792 x 7122792 & 21361554 & 4.21e-07 & 0.031 & 0.031 \\ \hline 
    
    mc2depi & 525825 x 525825	 & 2100225 & 7.59e-06 & 0.076 & 0.076 \\ \hline
    
    parabolic\_fem &	525825	 x 525825 & 	3674625 & 	1.33e-05 & 0.153 &  0.153 \\ \hline
    
    roadNet-TX & 	1393383	x 1393383	& 3843320 & 	1.98e-06 &  1.037 & 1.037 \\ \hline
    
    rajat31 & 	4690002	 x 4690002 & 	20316253 &  9.24e-07  & 1.106  & 1.106 \\ \hline
    
    af\_shell1 & 	504855 x 	504855	& 17588875 & 	6.90e-05 & 1.275 & 1.275 \\ \hline  
    
    delaunay\_n19 & 	524288 x 	524288 &	3145646 & 	1.14e-05 &  1.338 & 1.338 \\ \hline
    
    thermomech\_dK  &	204316 x 	204316 & 	2846228 & 6.81e-05 & 1.431 &  1.431 \\ \hline
    
    memchip & 2707524 x 2707524	& 14810202 & 	2.02e-06 &  2.062 & 1.173 \\ \hline
        
    amazon0601 & 403394 x 403394	& 3387388 & 	2.08e-05 &  2.79 & 15.29  \\ \hline
    
    FEM\_3D\_thermal2 & 147900 x 	147900 & 	3489300 & 	1.59e-04 & 4.481  & 4.481 \\ \hline

     
    web-Google	& 916428 x 	916428 & 	5105039 & 	6.08e-06 & 6.557 &  38.366 \\ \hline
    
    ldoor	 & 952203 x 952203	& 46522475 & 	5.13e-05  & 11.951 & 11.951 \\ \hline
    
    poisson3Db & 	85623 x 85623	 & 2374949 & 	3.24e-04  & 14.712  & 14.712 \\ \hline
    
    boneS10 & 	914898  x 	914898 & 	55468422 & 	6.63e-05 &  20.374 &  20.374 \\\hline
    
    webbase-1M	& 1000005 x 1000005	& 3105536 & 	3.106e-06 & 25.345  & 36.890 \\ \hline
    
    in-2004 & 1382908  x 	1382908	 & 16917053 & 	8.846e-06  & 37.230 & 144.062 \\ \hline 
    
    pkustk14 & 	151926  x	151926	 & 14836504 & 	6.428e-04 &  46.508  & 46.508 \\ \hline
    
    com-Youtube & 	1134890	 x  1134890	 & 5975248 & 4.639e-06  & 50.754  & 50.754 \\ \hline

    as-Skitter & 	1696415 x	1696415	& 22190596 & 	7.71e-06  & 136.861 &  136.861 \\ \hline
    
    sx-stackoverflow &	2601977	 x  2601977 & 	36233450 & 	5.352e-06 &  137.849 & 65.367 \\ \hline 

    ASIC\_680 & 	682862 x  682862 & 	3871773 & 	8.303e-06  & 659.807 & 659.807 \\ \hline
 
\end{tabular}
\end{minipage}%
\end{center}
\vspace{6pt}
\caption{Large Matrix Dataset. Matrices are sorted by NNZ-r-std, i.e., based on their irregular pattern. }
\label{tab:appendix-large-matrices}
\end{table}

\newpage

\begin{table}[H]
\begin{center}
\centering
\begin{minipage}{0.5\linewidth}
\centering
 \setlength\lineskip{2pt}
\begin{tabular}{|l|c|}
    \hline
    \cellcolor{gray!15}\raisebox{-0.10\height}{\textbf{Matrix Name}} & \cellcolor{gray!15}\raisebox{-0.10\height}{\textbf{Plot}} \\
    \hline \hline
    
    hugetric-00020 & \raisebox{-0.45\height}{\includegraphics[width=0.25\textwidth]{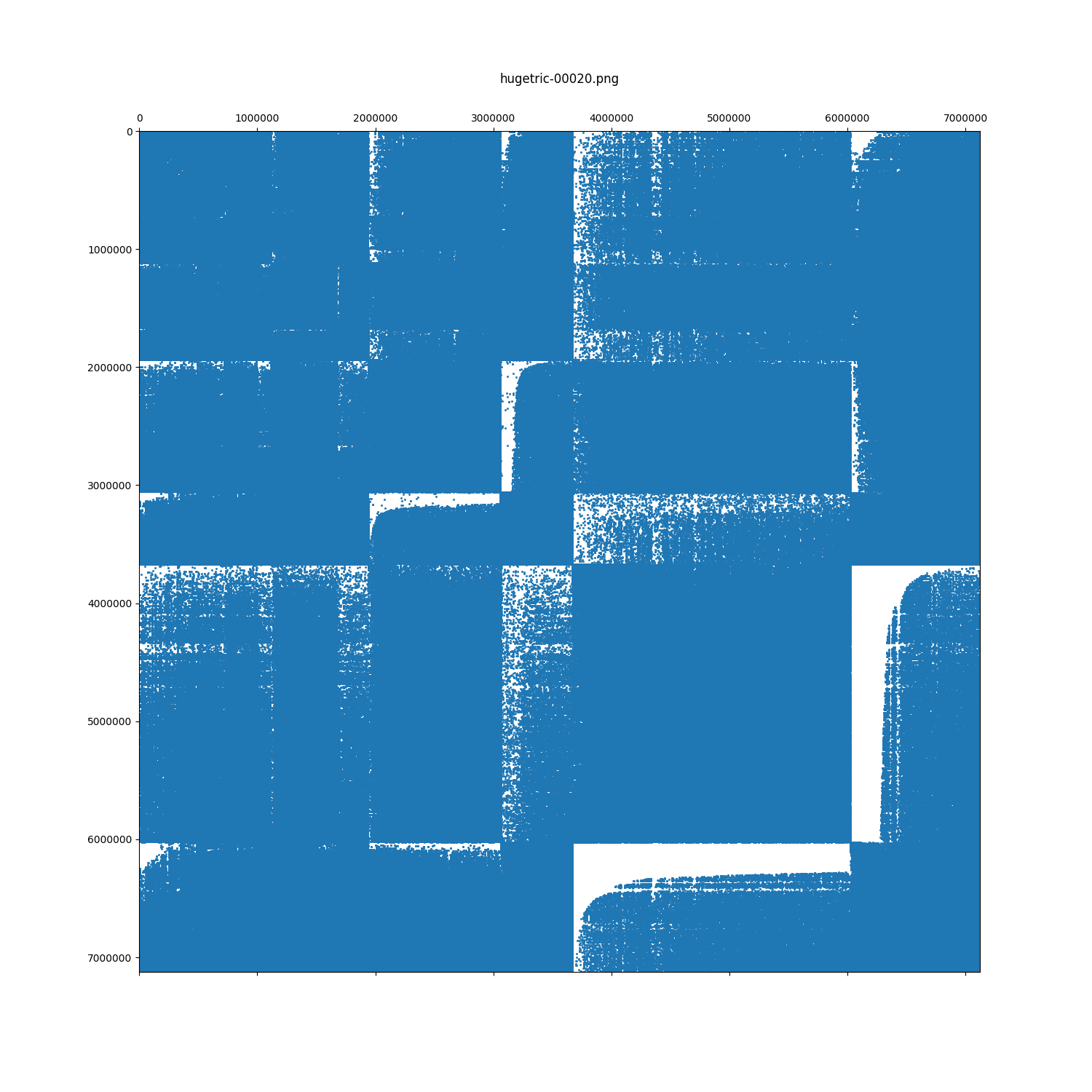}} \\ \hline
    
    mc2depi & \raisebox{-0.45\height}{\includegraphics[width=0.25\textwidth]{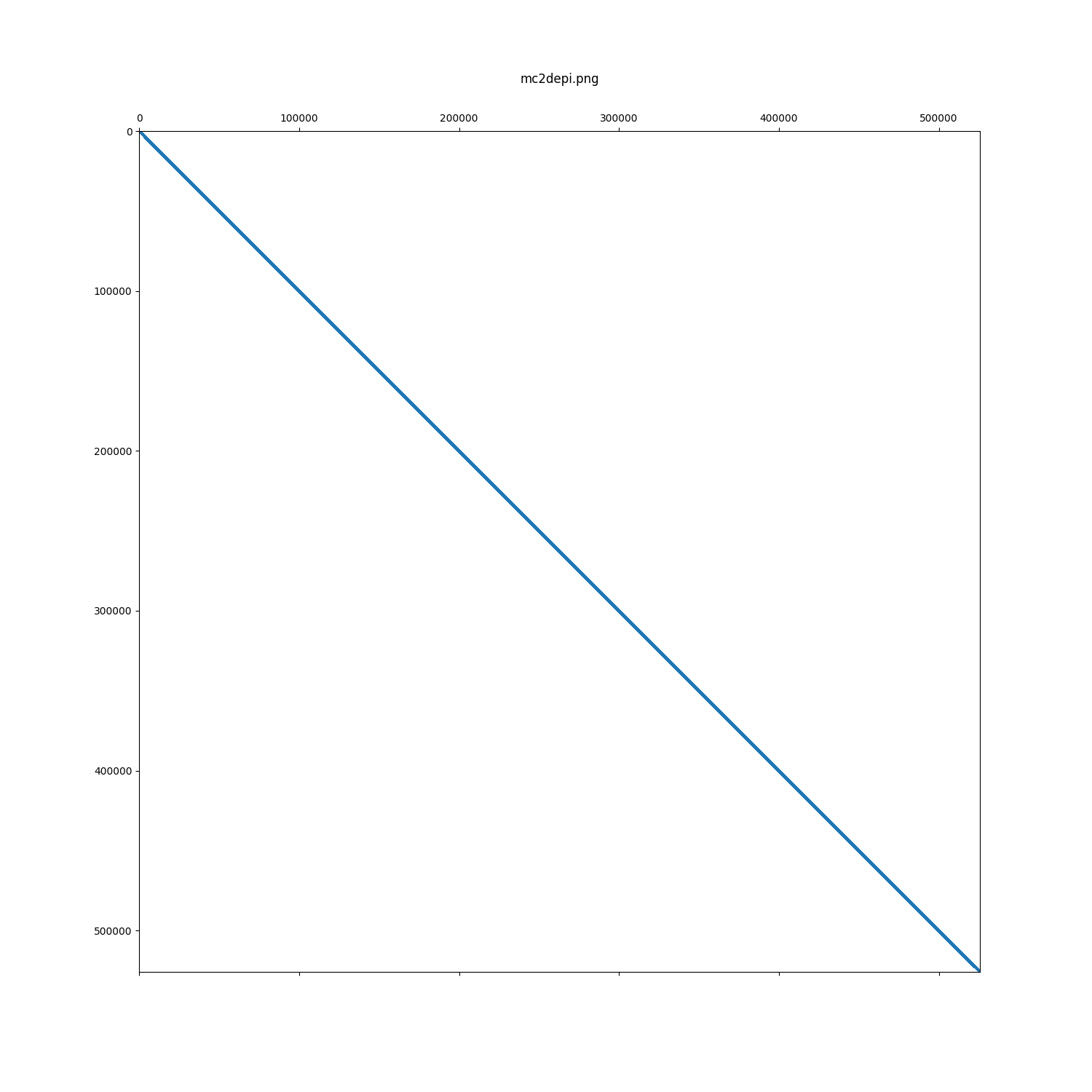}} \\ \hline
    
    parabolic\_fem & \raisebox{-0.45\height}{\includegraphics[width=0.25\textwidth]{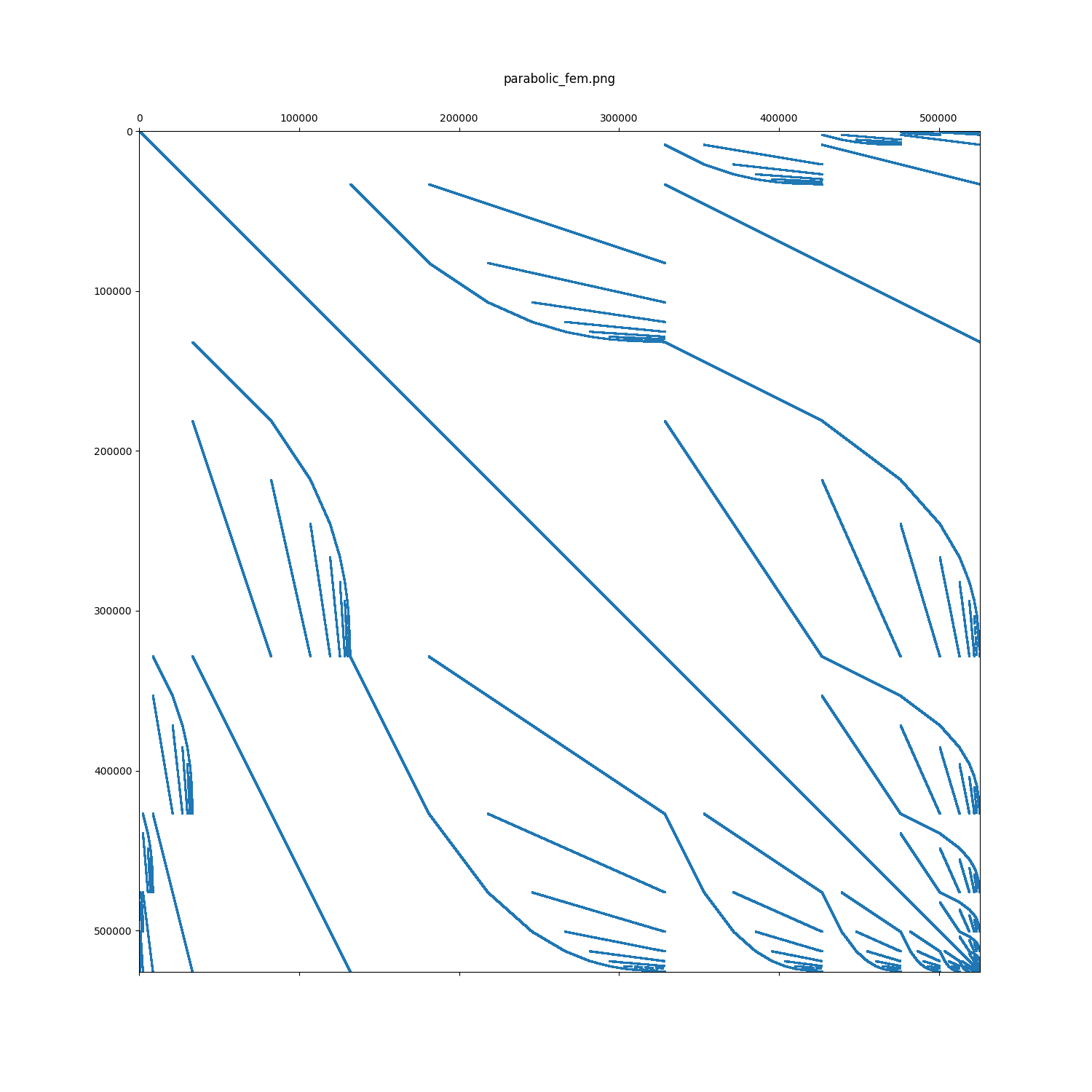}} \\ \hline
    
    roadNet-TX &  \raisebox{-0.45\height}{\includegraphics[width=0.23\textwidth]{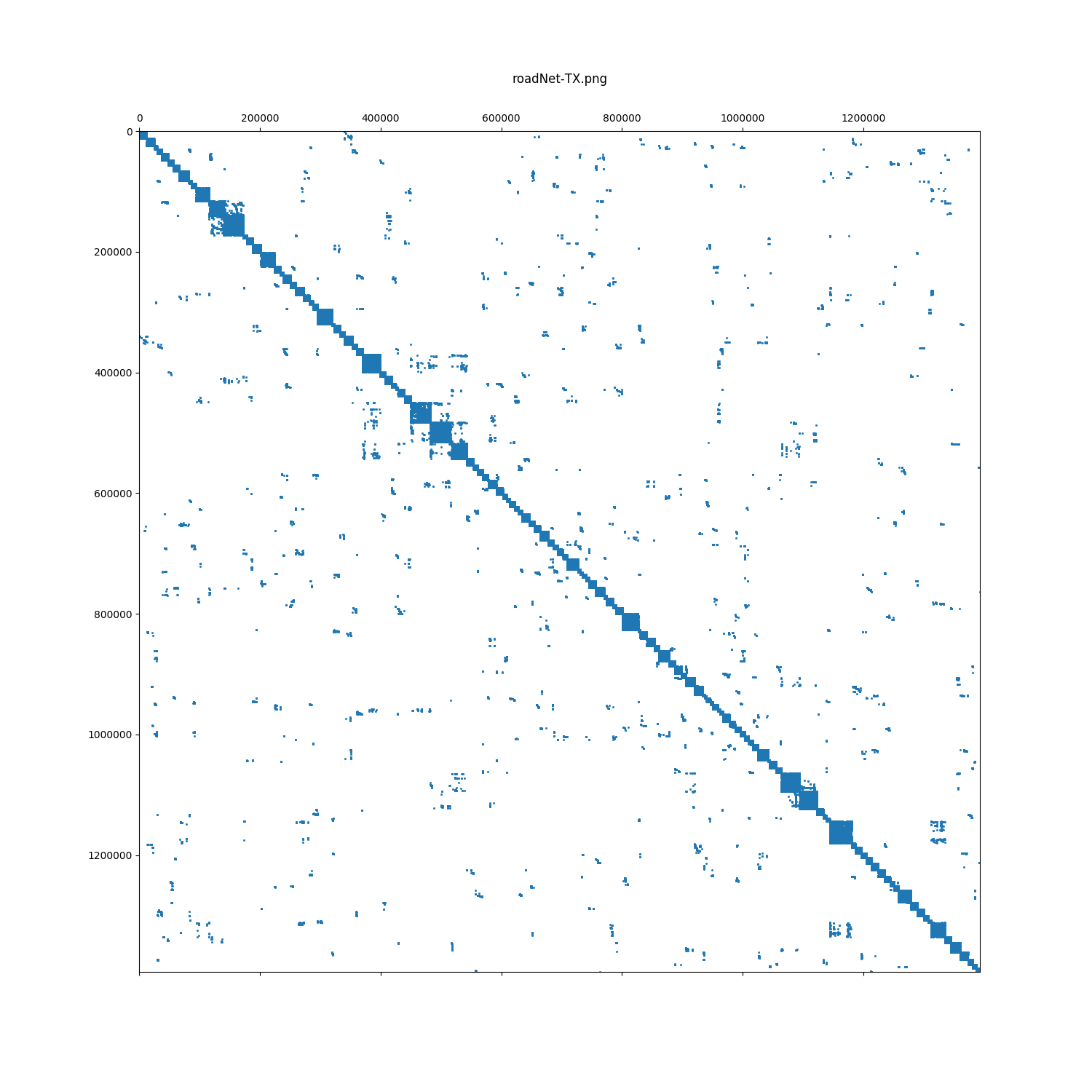}} \\ \hline
    
    rajat31 & 	\raisebox{-0.45\height}{\includegraphics[width=0.25\textwidth]{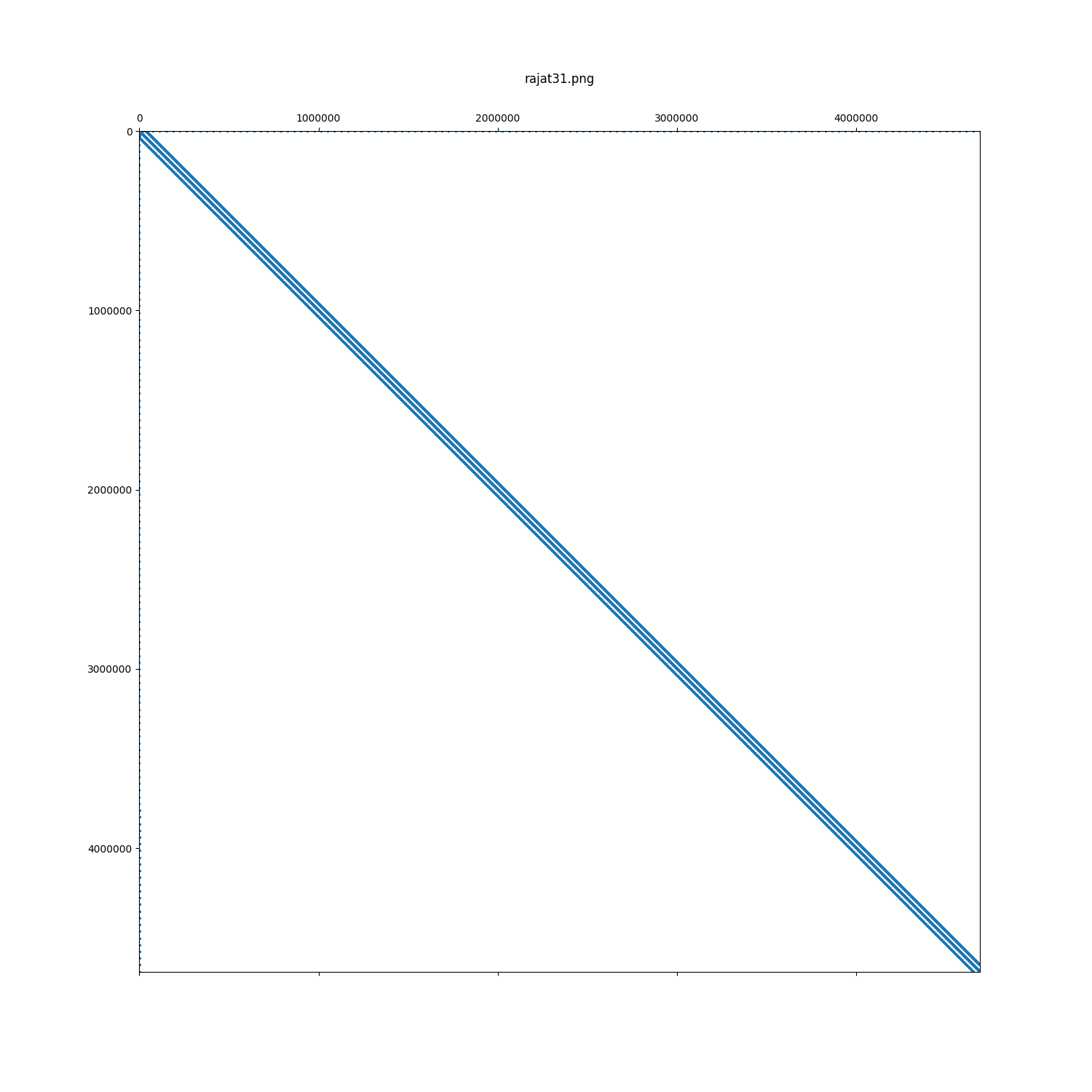}} \\ \hline
    
    af\_shell1 & 	\raisebox{-0.45\height}{\includegraphics[width=0.25\textwidth]{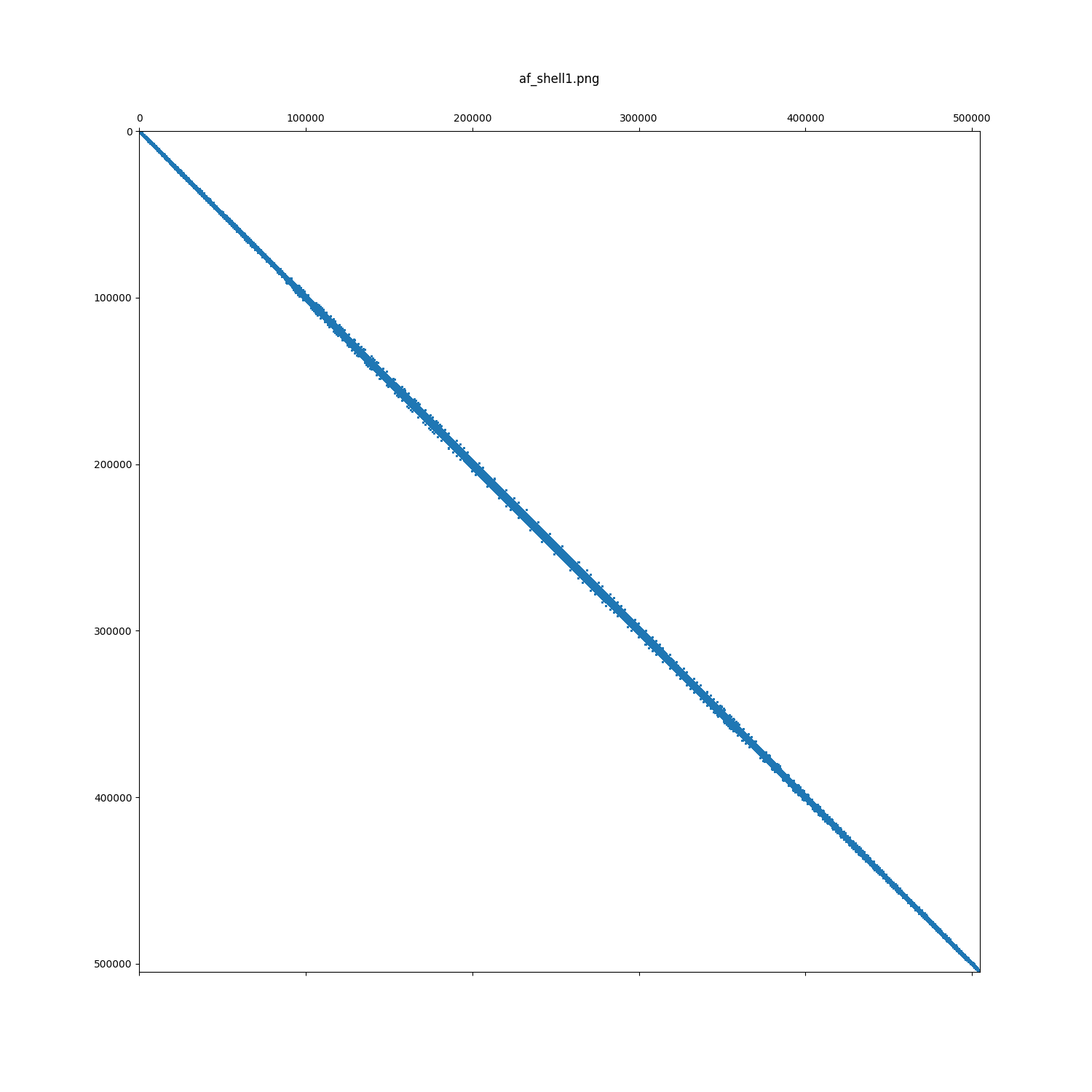}} \\ \hline 
    
    delaunay\_n19 & \raisebox{-0.45\height}{\includegraphics[width=0.25\textwidth]{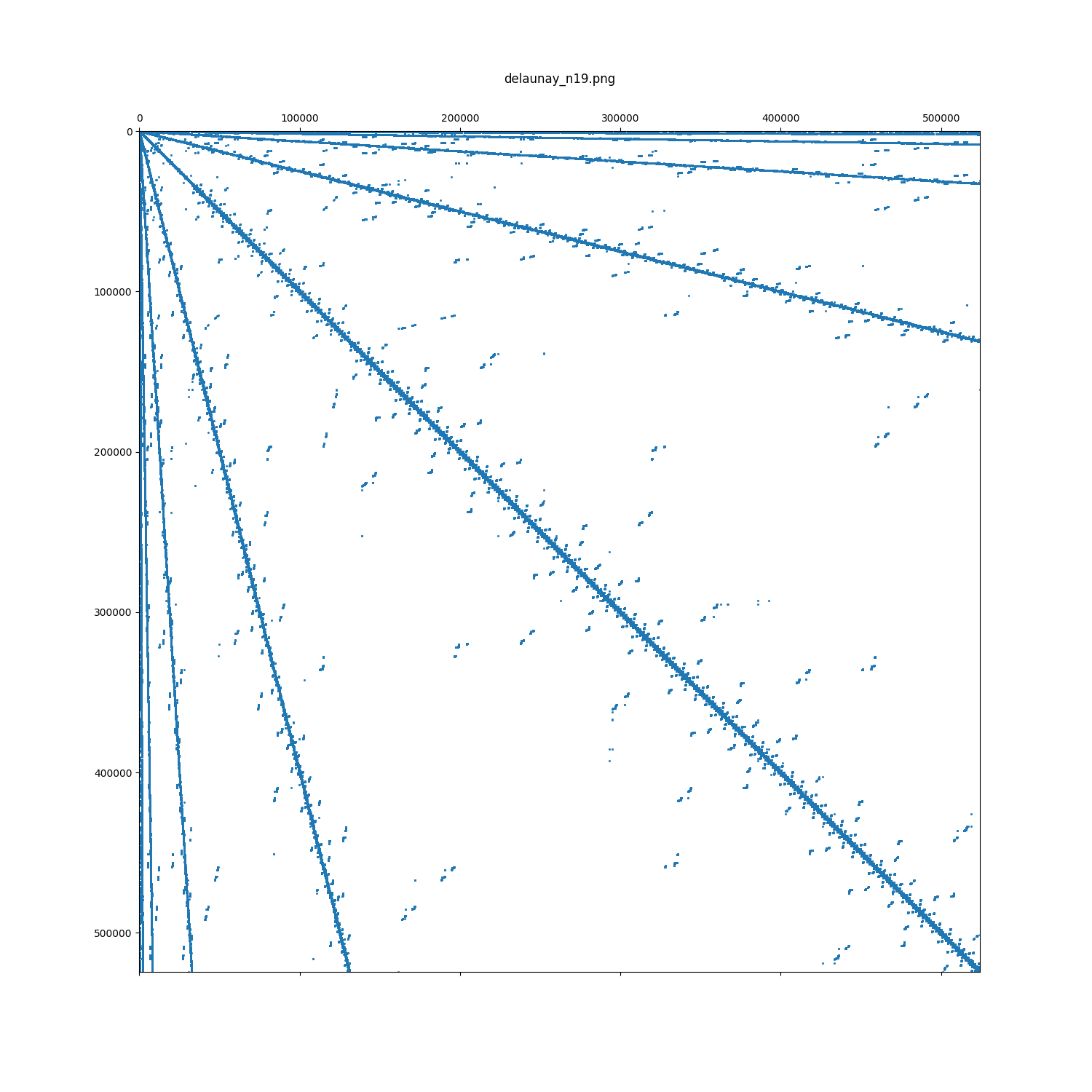}}  \\ \hline 
    
    thermomech\_dK  & \raisebox{-0.45\height}{\includegraphics[width=0.25\textwidth]{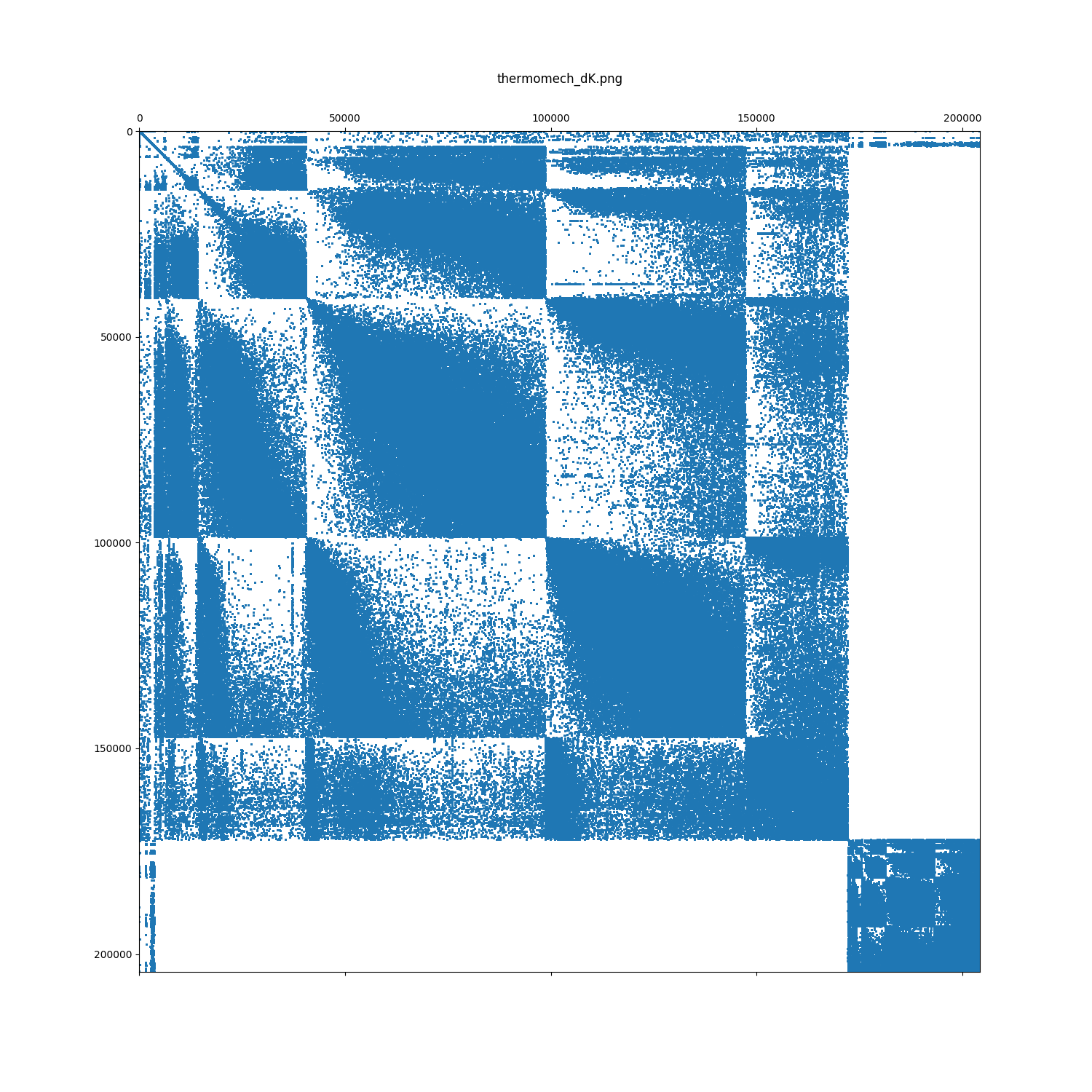}} \\ \hline
    
    memchip &  \raisebox{-0.45\height}{\includegraphics[width=0.25\textwidth]{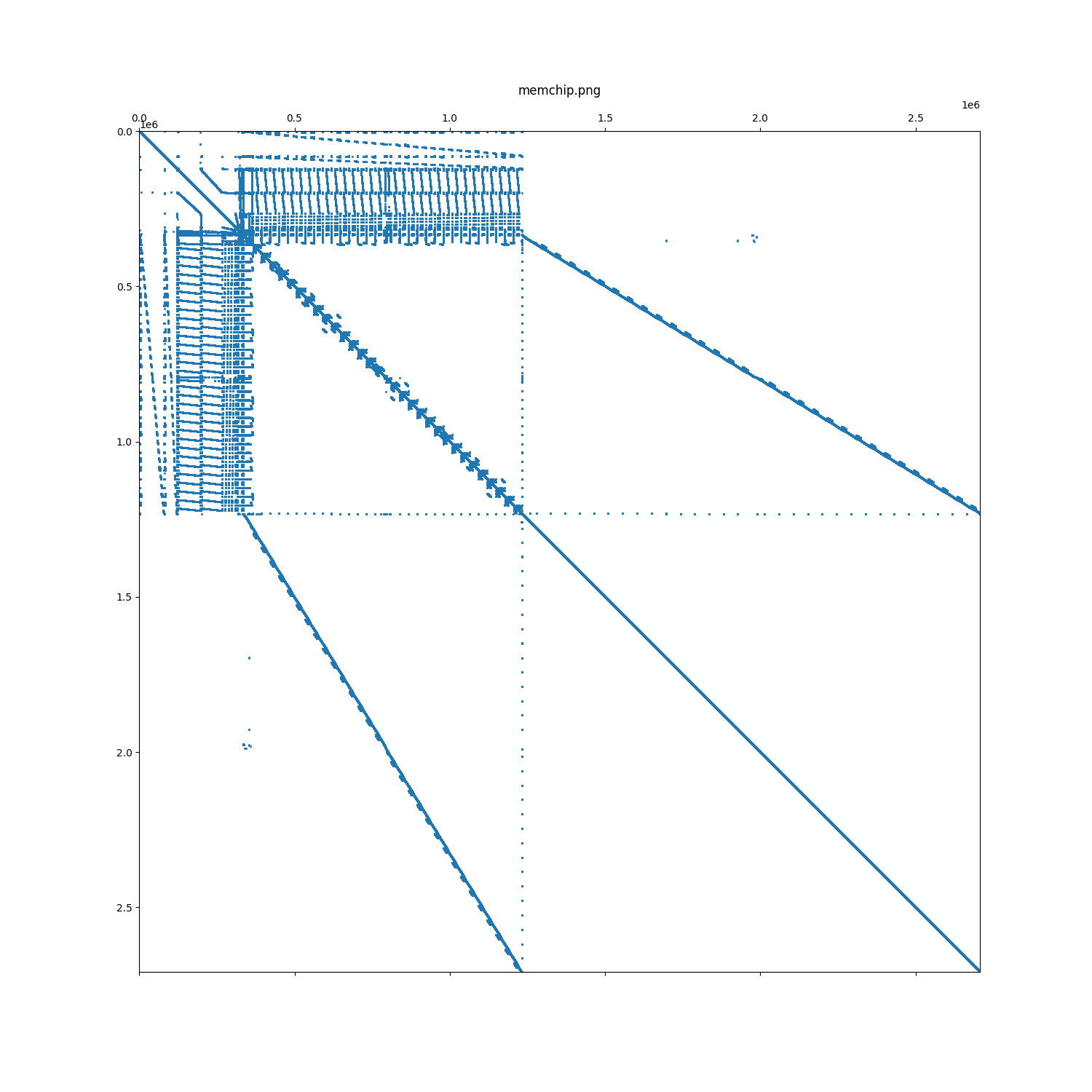}} \\ \hline
        
    amazon0601 & \raisebox{-0.45\height}{\includegraphics[width=0.25\textwidth]{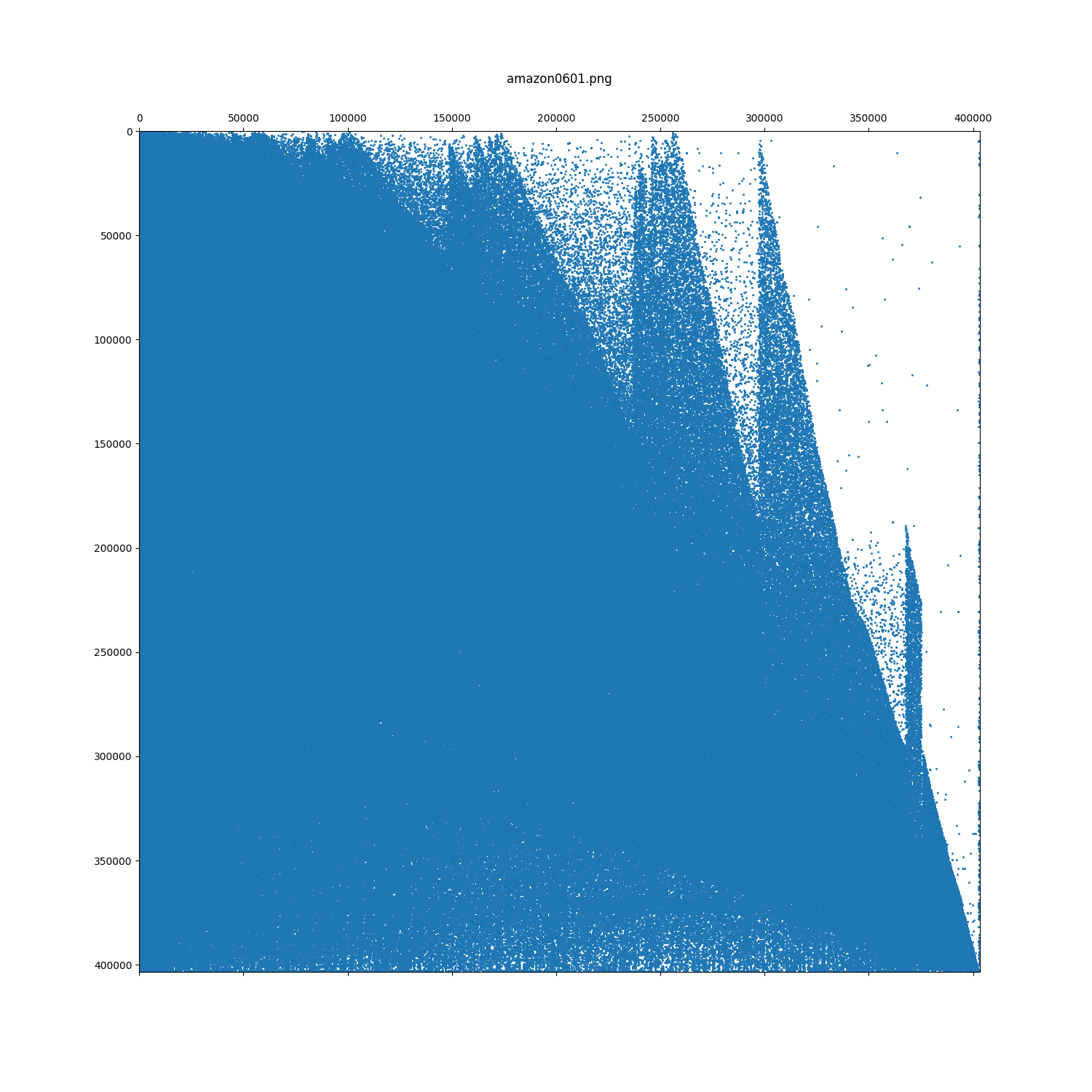}} \\ \hline
    
    FEM\_3D\_thermal2 & \raisebox{-0.45\height}{\includegraphics[width=0.25\textwidth]{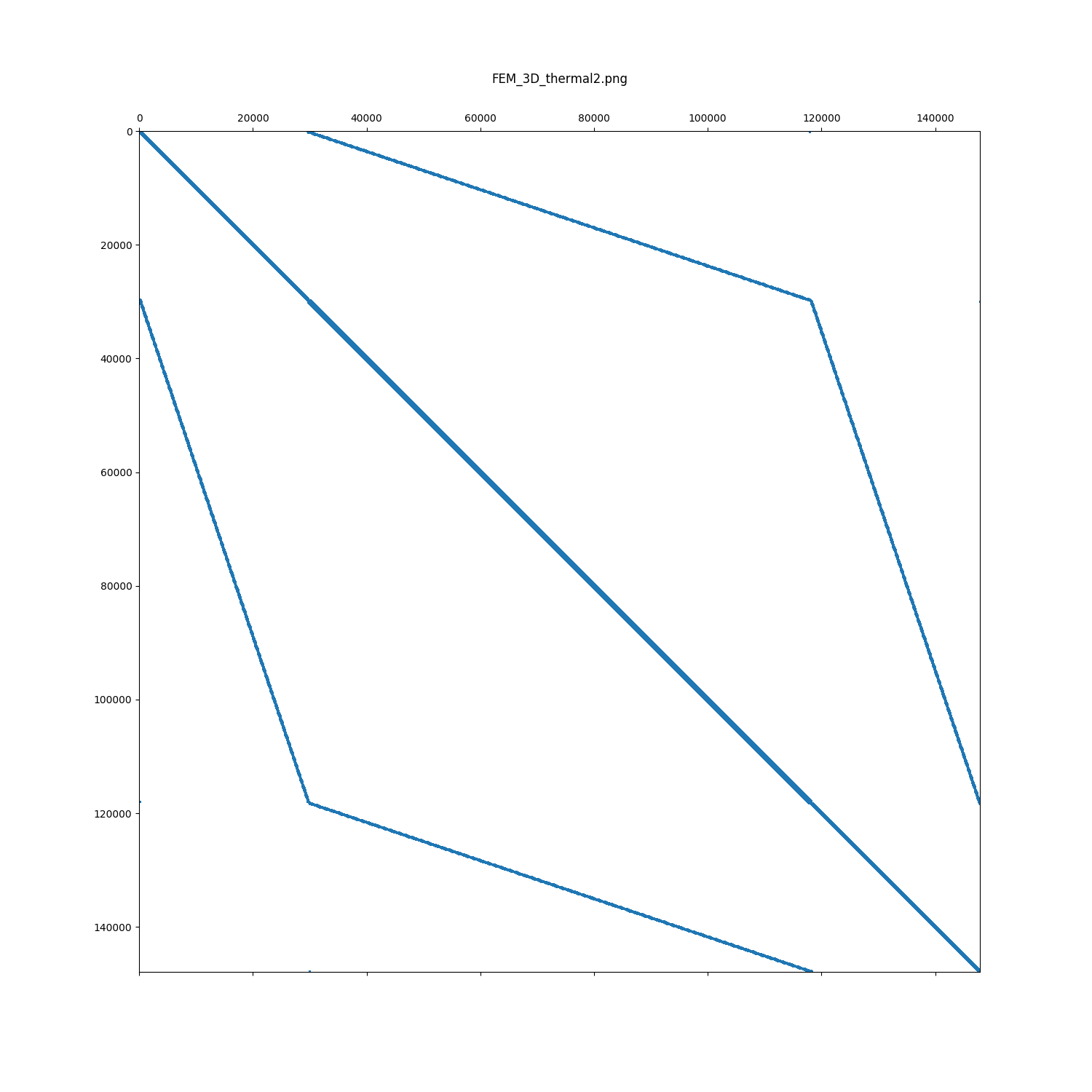}}\\ \hline
\end{tabular}
\end{minipage}%
\begin{minipage}{0.5\linewidth}
\centering
\begin{tabular}{|l|c|}
    \hline 
    \cellcolor{gray!15}\raisebox{-0.10\height}{\textbf{Matrix Name}} & \cellcolor{gray!15}\raisebox{-0.10\height}{\textbf{Plot}} \\
    \hline \hline
    web-Google	& \raisebox{-0.45\height}{\includegraphics[width=0.25\textwidth]{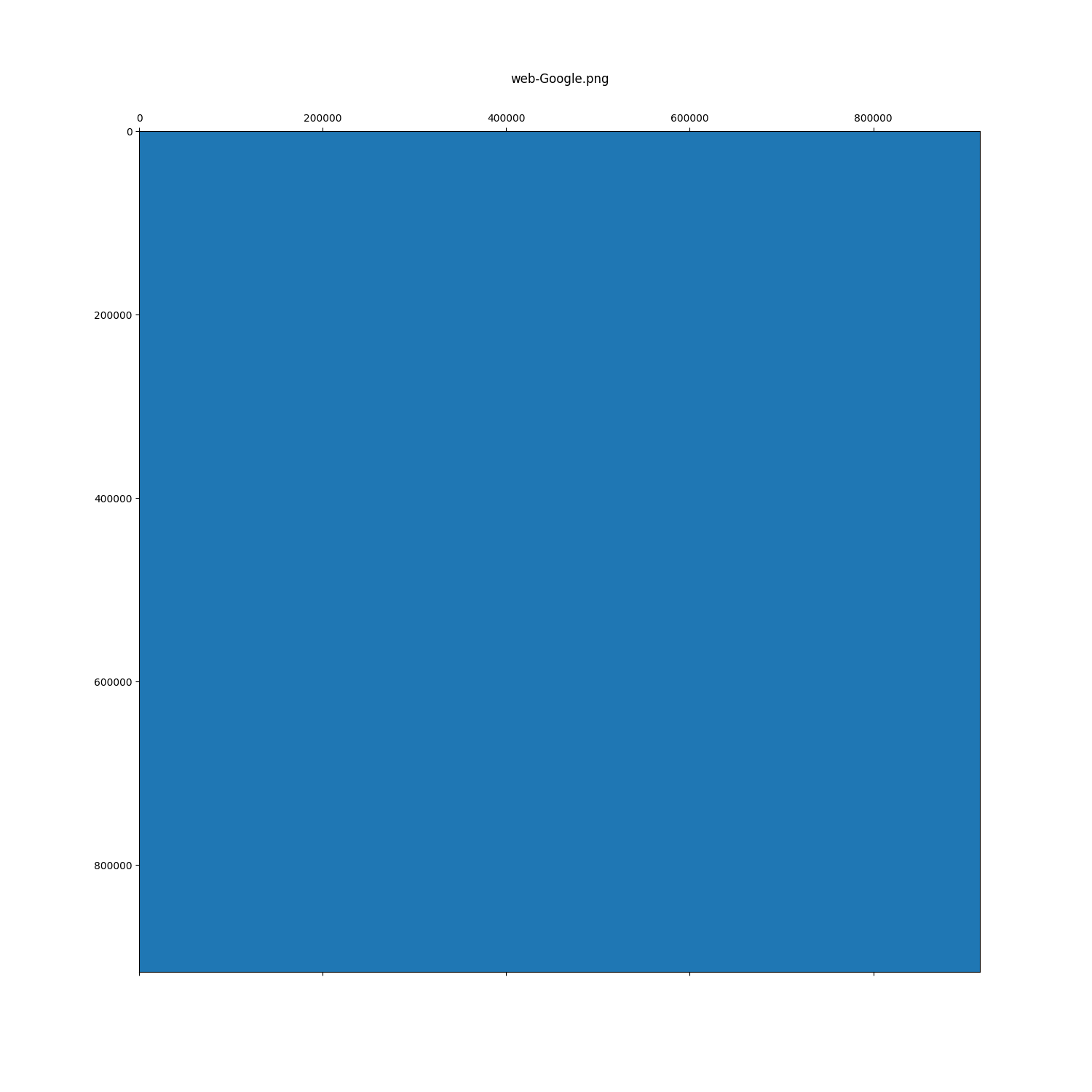}}  \\ \hline 
    
    ldoor	 & \raisebox{-0.45\height}{\includegraphics[width=0.25\textwidth]{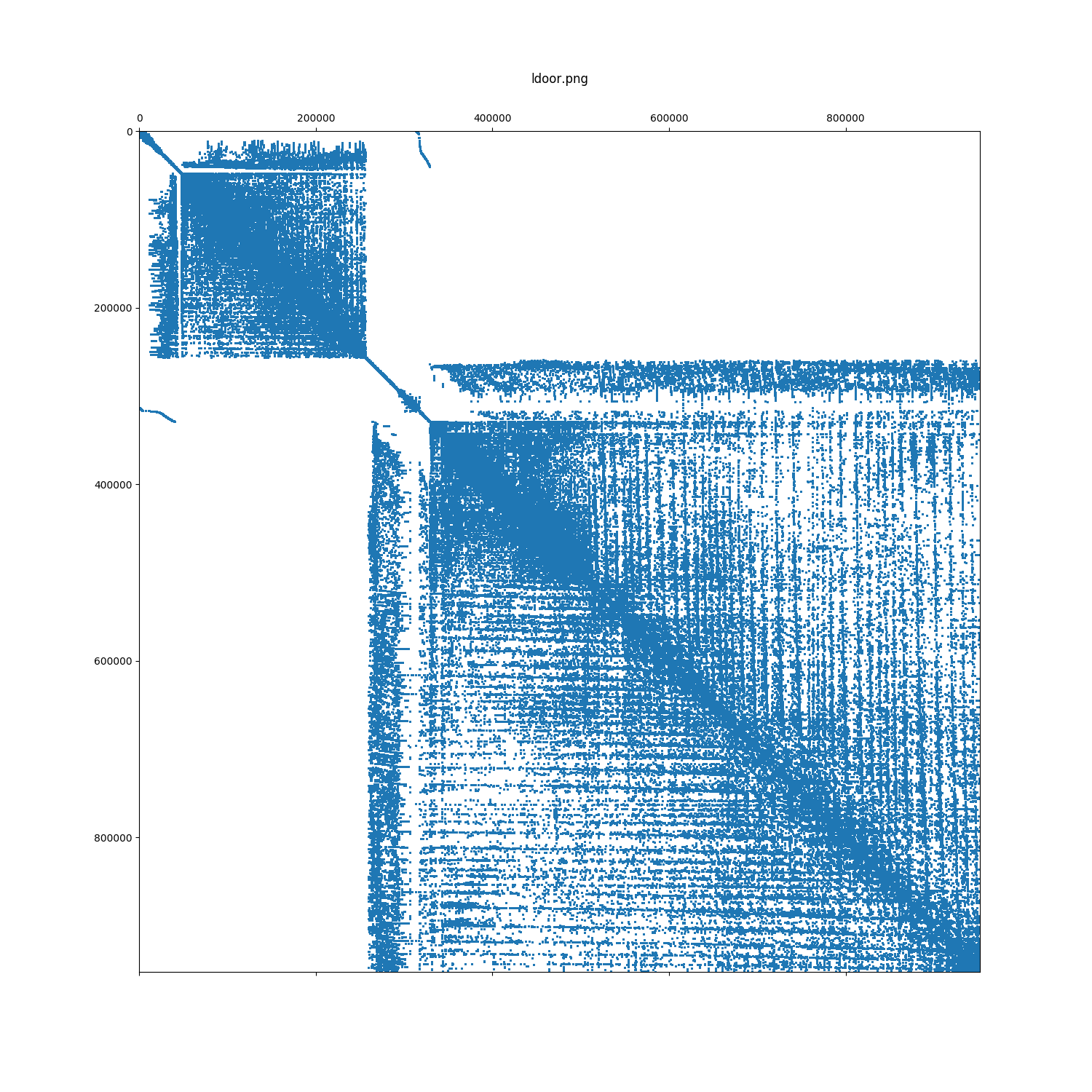}}  \\ \hline 
    
    poisson3Db & \raisebox{-0.45\height}{\includegraphics[width=0.25\textwidth]{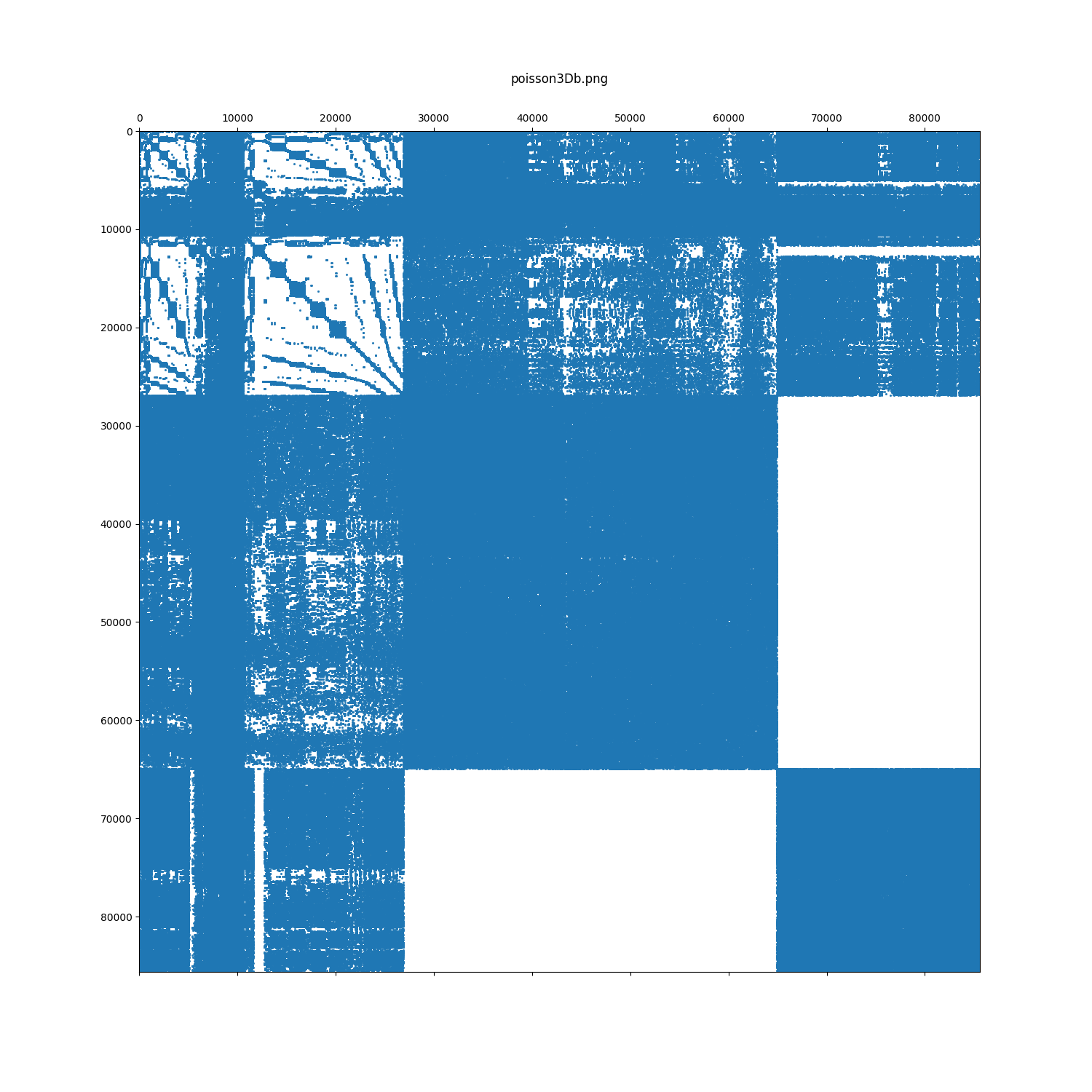}}	 \\ \hline 
    
    boneS10 & 	\raisebox{-0.45\height}{\includegraphics[width=0.25\textwidth]{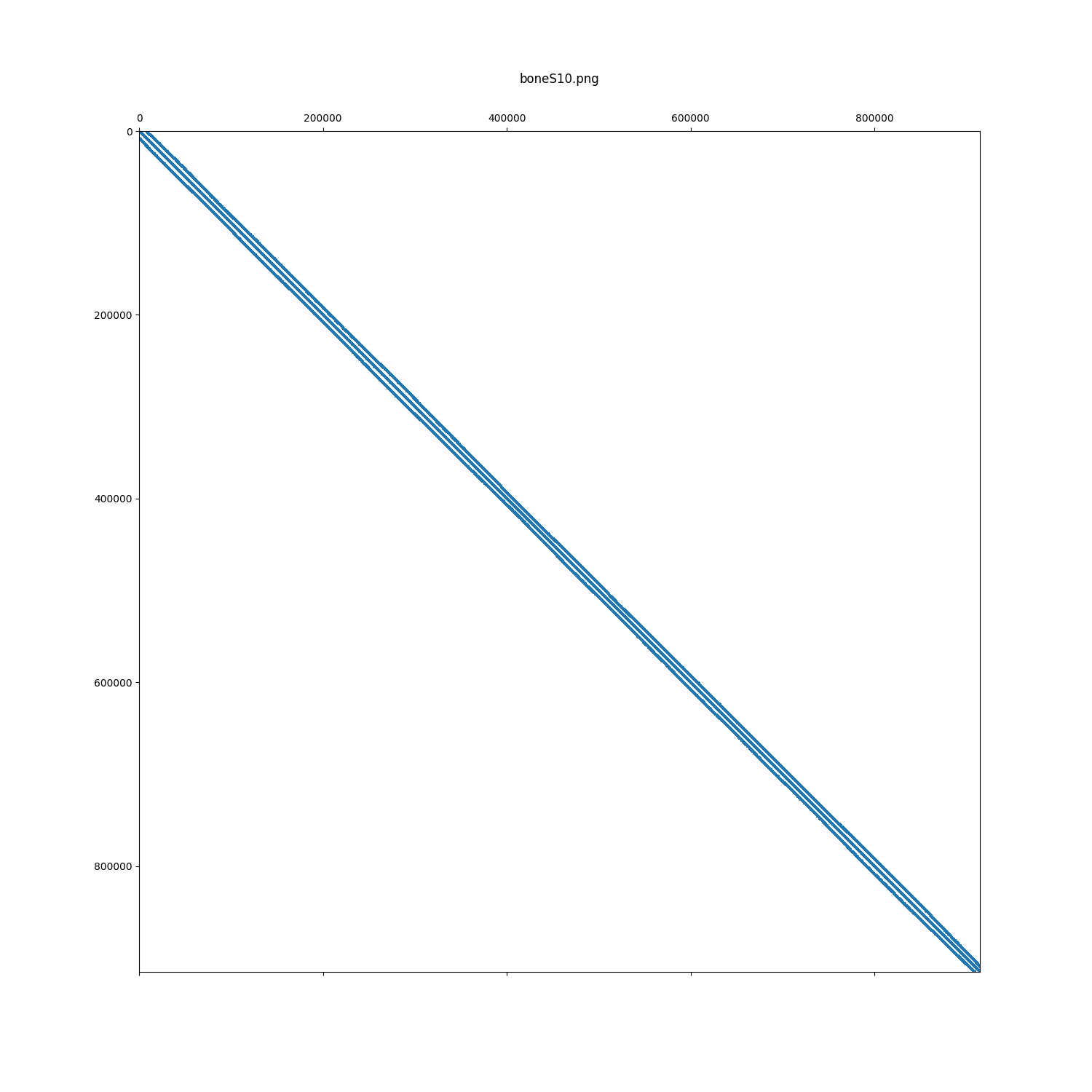}}  \\ \hline 
    
    webbase-1M	& \raisebox{-0.45\height}{\includegraphics[width=0.25\textwidth]{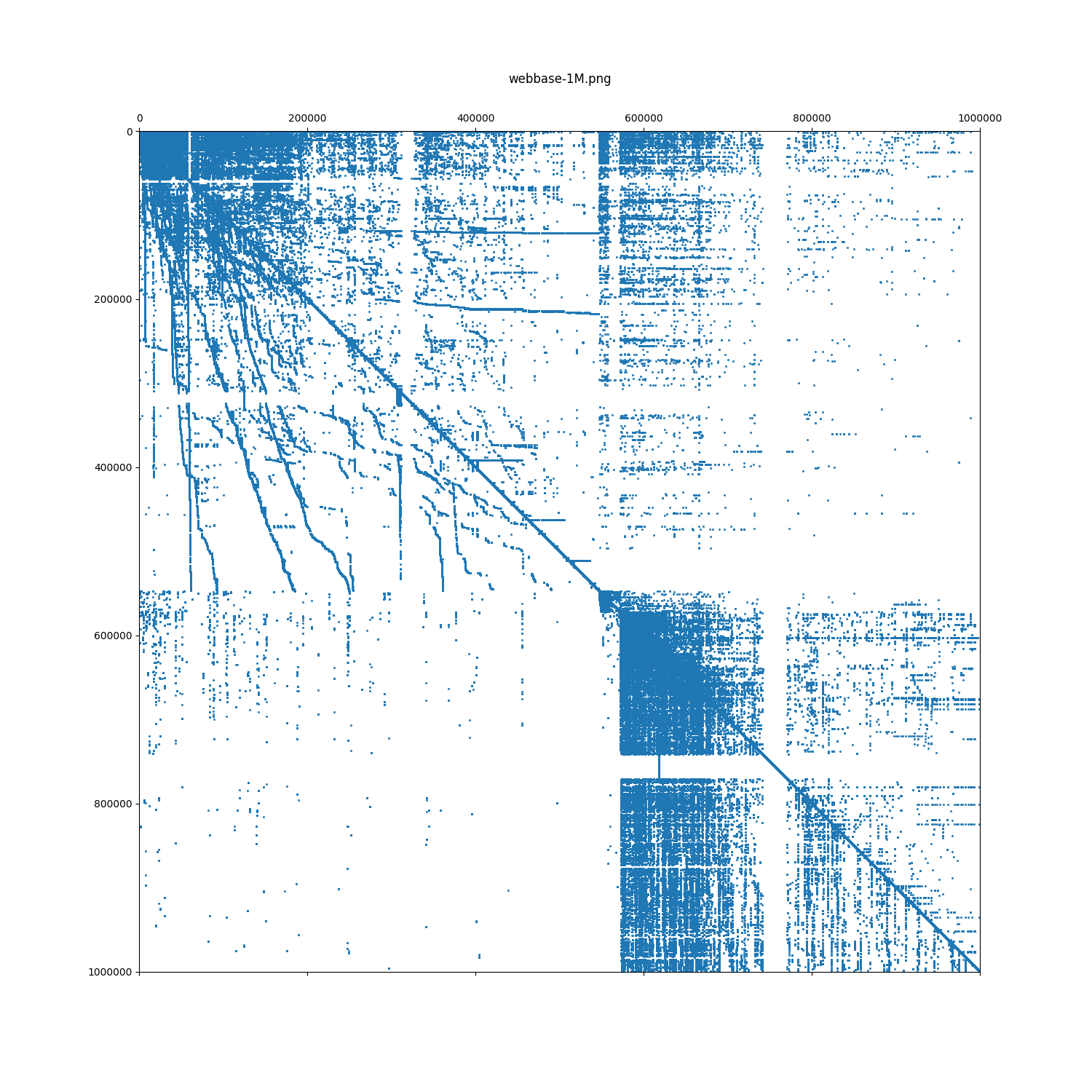}} \\ \hline  
    
    in-2004 & \raisebox{-0.45\height}{\includegraphics[width=0.25\textwidth]{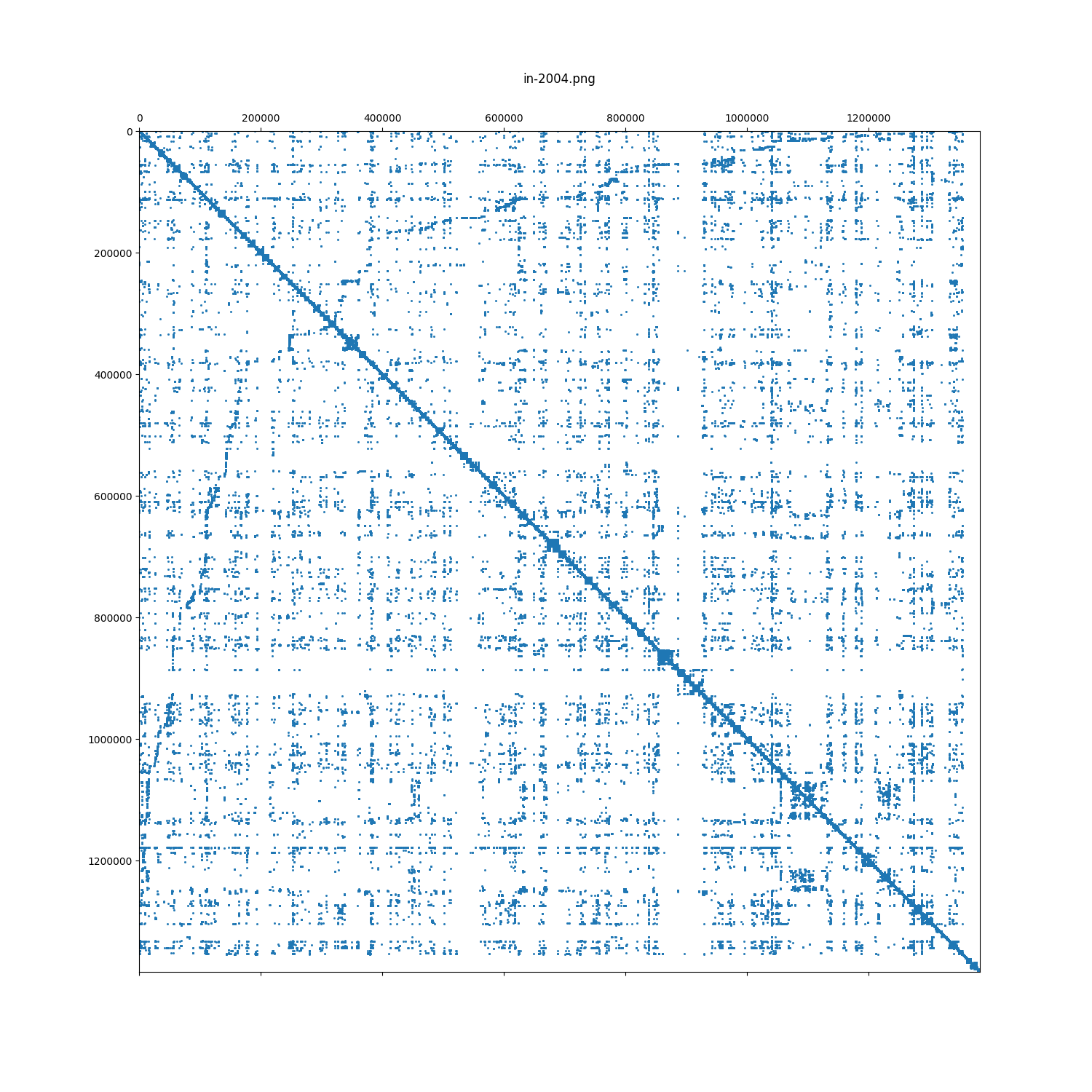}} \\ \hline  
    
    pkustk14 & \raisebox{-0.45\height}{\includegraphics[width=0.25\textwidth]{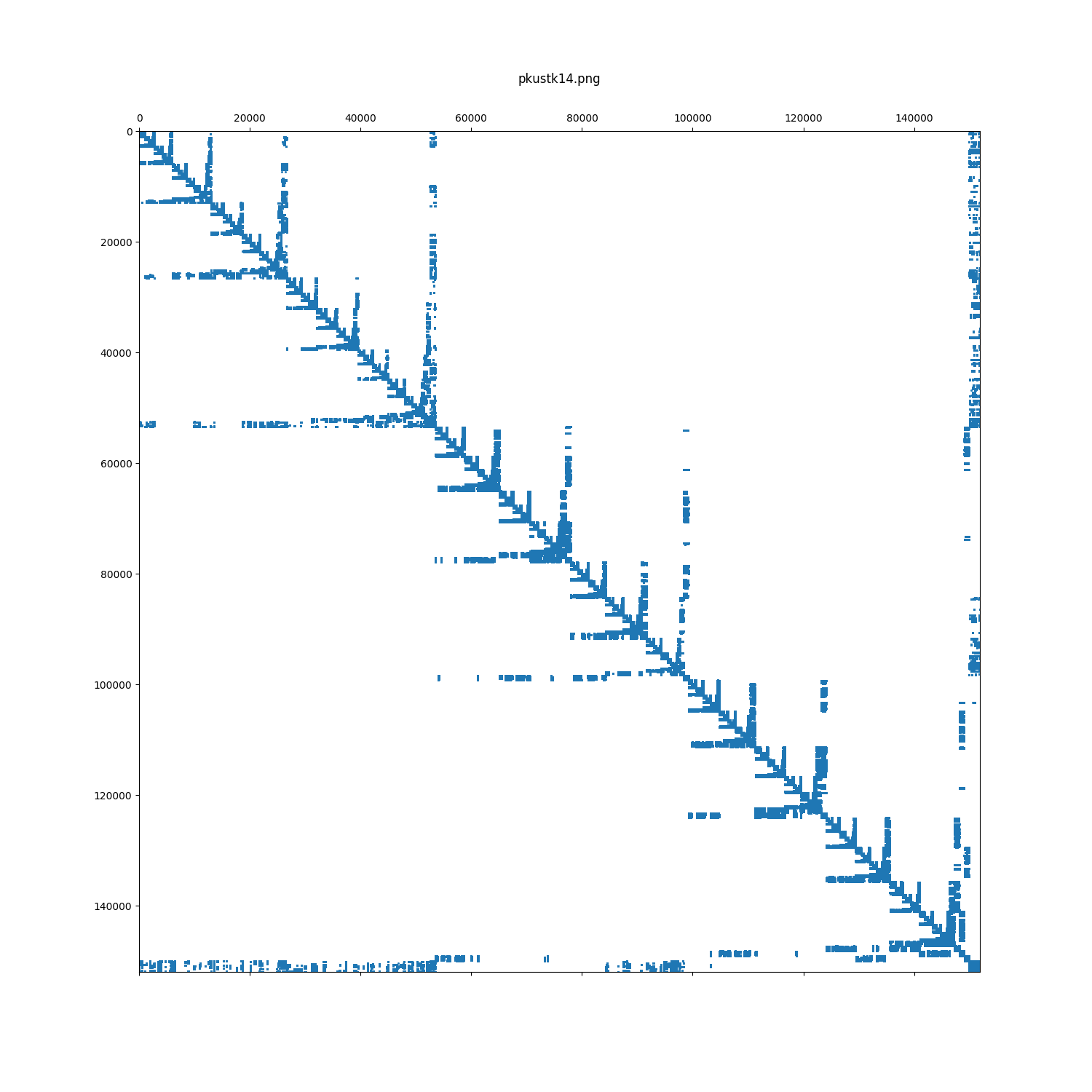}}	\\ \hline   
    
    com-Youtube & \raisebox{-0.45\height}{\includegraphics[width=0.25\textwidth]{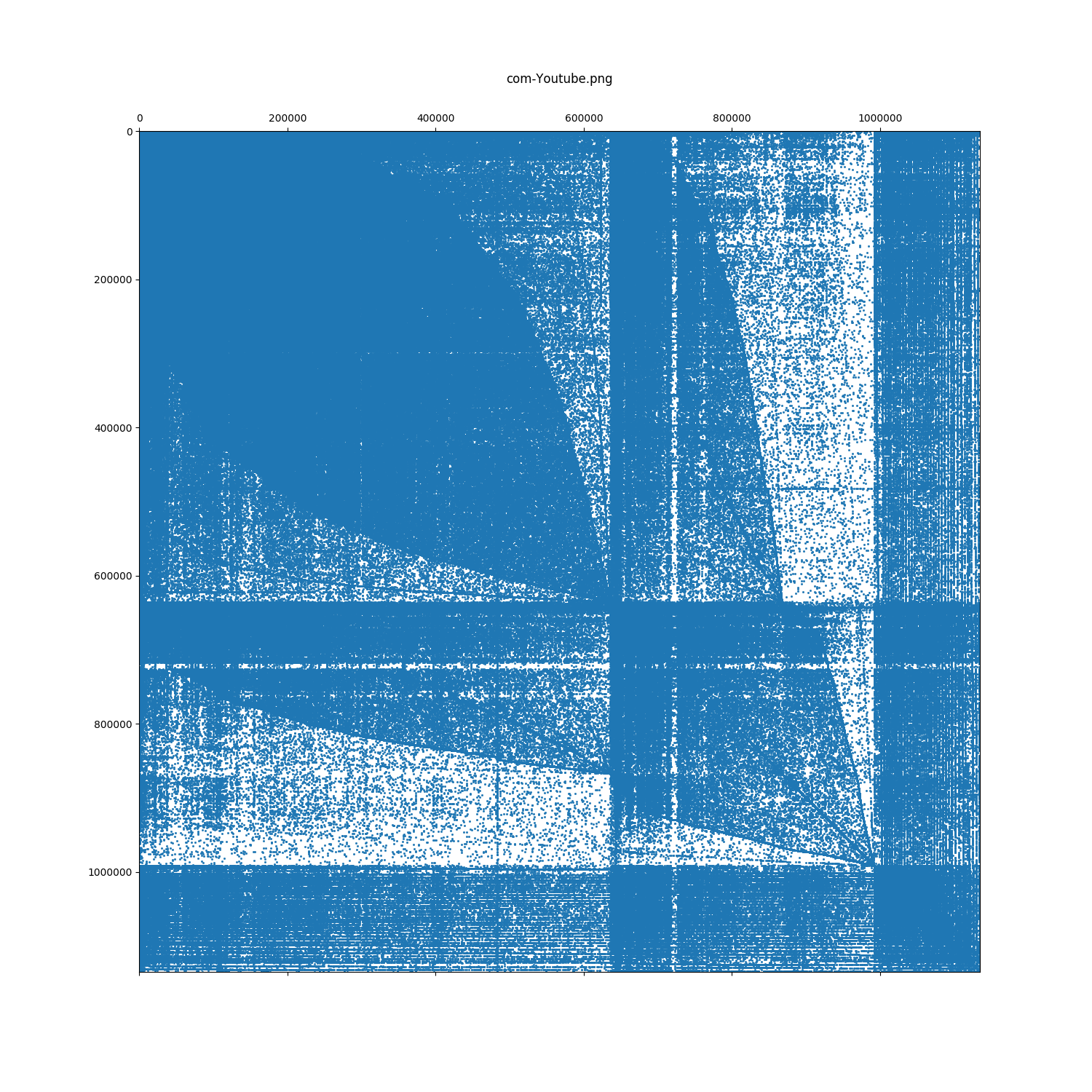}} \\ \hline  

    as-Skitter & \raisebox{-0.45\height}{\includegraphics[width=0.25\textwidth]{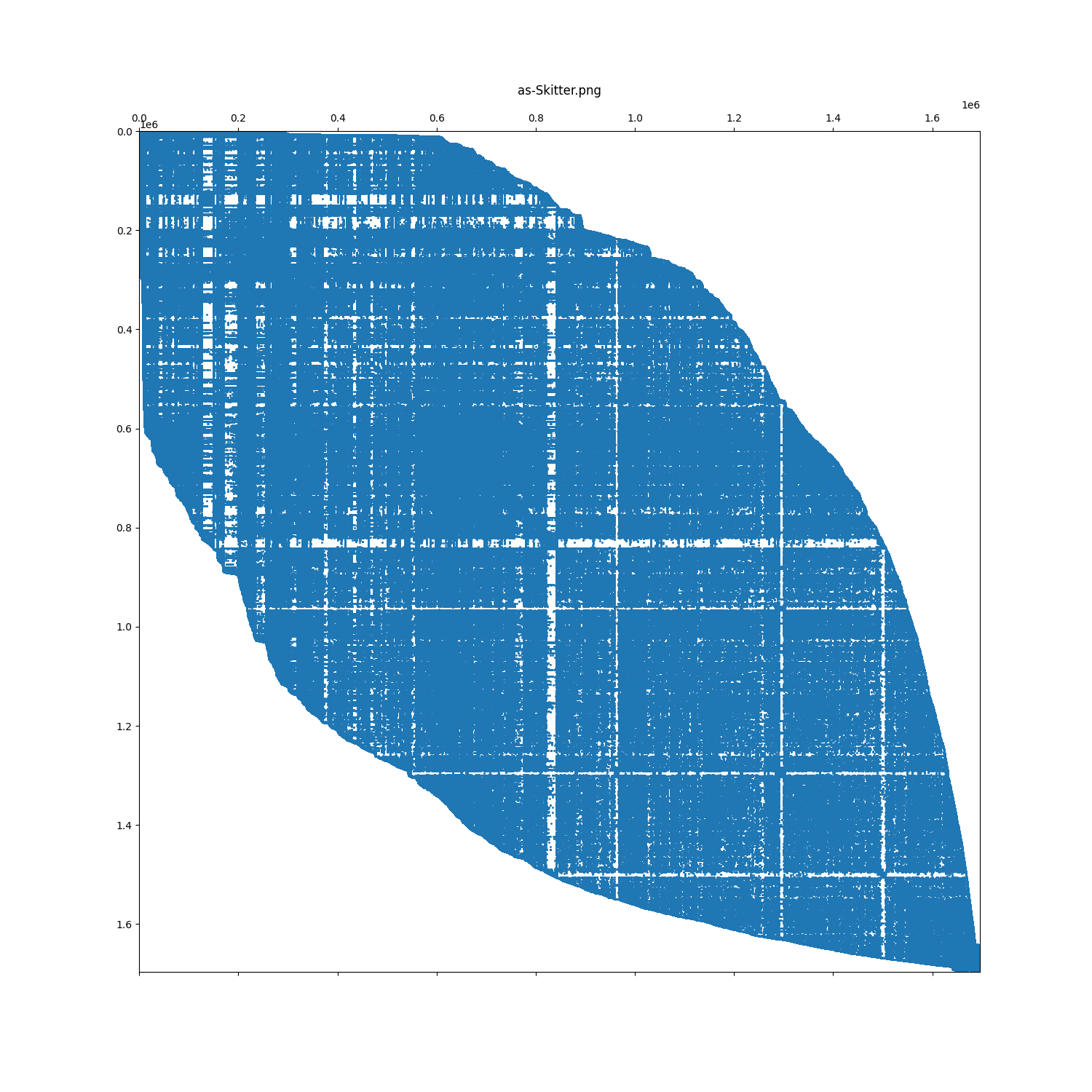}} \\ \hline  
    
    sx-stackoverflow & \raisebox{-0.45\height}{\includegraphics[width=0.25\textwidth]{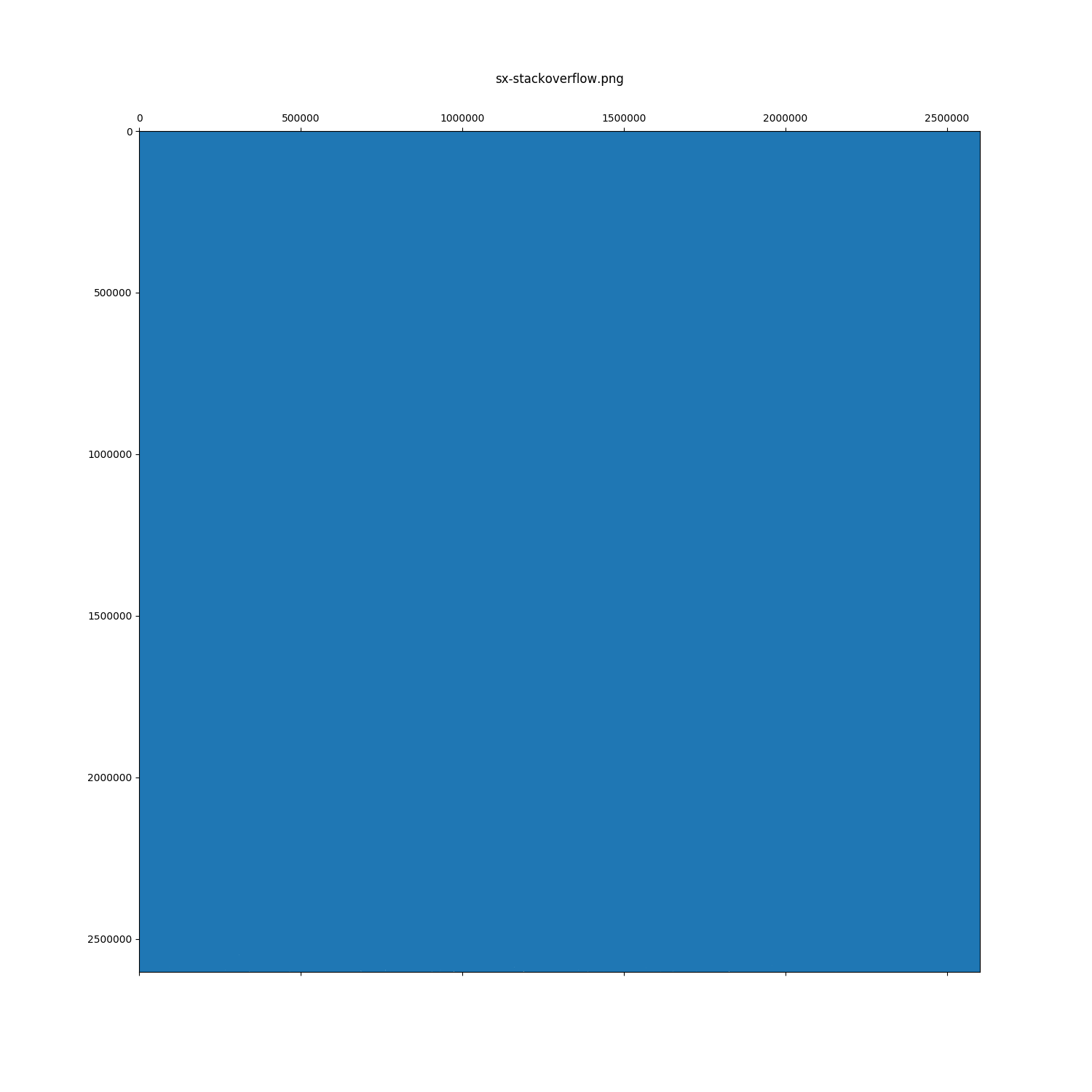}} \\ \hline

    ASIC\_680 & \raisebox{-0.45\height}{\includegraphics[width=0.25\textwidth]{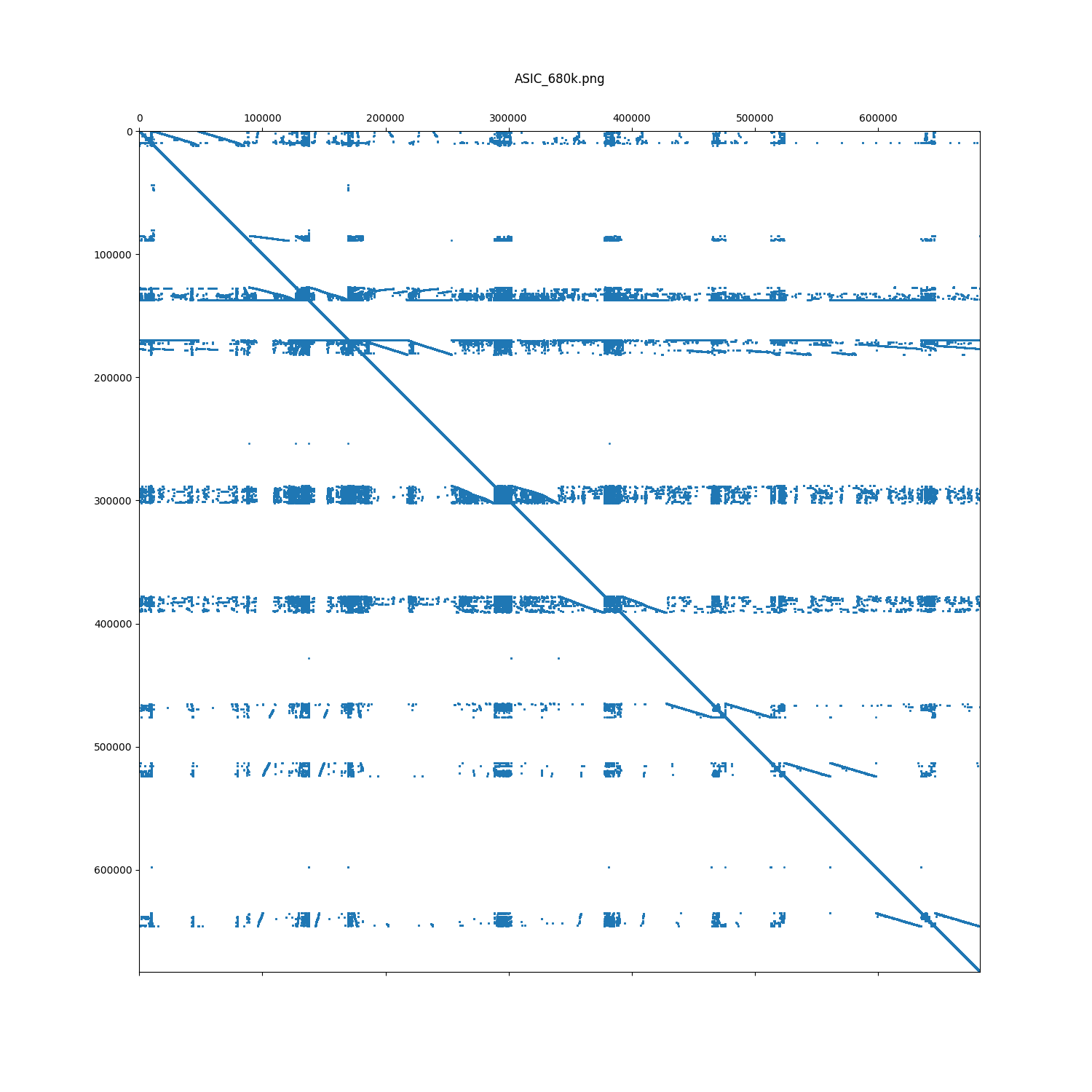}} \\ \hline 
 
\end{tabular}
\end{minipage}%
\end{center}
\vspace{4pt}
\caption{Sparsity patterns of the sparse matrices of our large matrix data set.}
\label{tab:appendix-large-matrices-plot}
\end{table}

\end{document}